%% file: jetgradients.tex
\documentclass{aa}  

\usepackage{graphicx}
\usepackage{xcolor}
\usepackage{natbib}
\usepackage{ulem}
\usepackage{txfonts}
\usepackage{booktabs}

\begin{document} 


\title{Dual-high-frequency VLBI study of blazar-jet brightness-temperature gradients and collimation profiles}

\titlerunning{Dual-high-frequency VLBI study of blazar-jet brightness-temperature gradients and collimation profiles}

\author{P.\,R.\,Burd\inst{1}, M.\,Kadler\inst{1}, K.\,Mannheim\inst{1}, A.-K.\,Baczko\inst{2}, J.\,Ringholz\inst{1}, E.\,Ros\inst{2}
          }

   \institute{Institut für  Theoretische Physik and Astrophysik, Julius-Maximilians-Universität Würzburg,
              Emil-Fischer-Straße 31, D-97074 Würzburg, Germany  \email{paul.r.burd@astro.uni-wuerzburg.de}  
 \and             
Max-Planck-Institut für Radioastronomie,
              Auf dem Hügel 69, D-53121 Bonn, Germany
              }

             
\authorrunning{Burd et al.}

   \date{Version: \today 
    Received 04/10/2021; accepted 06/12/2021}

 
  \abstract
   {
  On the kiloparsec scale, extragalactic radio jets show two distinct morphologies related to their power:  collimated high-power jets ending in a bright termination shock and low-power jets opening up close to the core and showing a more diffuse surface brightness distribution.  The emergence of this morphological dichotomy on the parsec scale at the innermost jet regions can be studied with very-long-baseline interferometry (VLBI) radio observations of blazars in which the jet emission is strongly Doppler boosted due to relativistic bulk motion at small angles between the jet direction and the line of sight.
   }
   {
   We seek to characterize the geometry and emission profiles of the parsec-scale radio jets of
   flat-spectrum radio quasars (FSRQs) and BL\,Lacertae objects (BL\,Lacs) on parsec scales to derive properties of the magnetic field, environment, and energetics for different classes of extragalactic jets.
   }
   {We analyze the VLBI radio data of 15 FSRQs, 11 BL\,Lacs, and two radio galaxies contained in both the Monitoring Of Jets in Active galactic nuclei with VLBA Experiments (MOJAVE) data archive and the Boston University (BU) blazar group sample archive at 15\,GHz and 43\,GHz, repectively. 
   We derived the brightness-temperature and jet-width gradients along the jet axis  from parameterizations of the jets using 2D Gaussian brightness distributions.}
{
In most BL\,Lac objects, the diameter and brightness-temperature gradients along the jet axis can generally be described well by single power laws, while the jets of FSRQs show more complex behavior and remain more strongly collimated on larger physical scales. We find evidence for a transition of the global jet geometry from a parabolic to a conical shape in the BL\,Lac objects 3C\,66A, Mrk\,421 and BL\,Lacertae, the radio galaxy 3C\,111 and the FSRQs CTA\,26,  PKS\,0528+134, 4C\,+71.07, 4C\,+29.45, and 3C\,279 outside the Bondi sphere.}
{
Our results combined with findings from kinematic VLBI studies that the jets of FSRQs exhibit larger bulk Lorentz factors than BL\,Lacs are in agreement with  relativistic magnetohydrodynamical jet-disk simulations in which the flattening of the jet magnetization profile due to magnetic fields from the accretion disk leads to a more persistent collimation in high-accretion-rate blazars.
}


   \keywords{Galaxies:active --
                Galaxies:jets--
                Jets:parsec-scale
               }

   \maketitle
%
\input{intro.tex}
\input{sample.tex}
\input{method.tex}

\input{results_gradients.tex}
\input{discussion.tex}

\input{conclusion_2.tex}

\begin{acknowledgements}
We are grateful to Till Steinbring, who did important preliminary work on which this study partly has been based.
We appreciate valuable scientific discussions with the MOJAVE team, especially Matthew L.\ Lister, Yuri Y.\ Kovalev, and Evgeniya V.\ Kravchenko, which helped us to improve this paper. 

This research has made use of data from the MOJAVE database that is maintained by the MOJAVE team \citep{lister2009,lister2018_pol}.
This study makes use of VLBA data from the VLBA-BU Blazar Monitoring Program (BEAM-ME and VLBA-BU-BLAZAR; \url{http://www.bu.edu/blazars/BEAM-ME.html}), funded by NASA through the \textit{Fermi} Guest Investigator Program. The VLBA is an instrument of the National Radio Astronomy Observatory. The National Radio Astronomy Observatory is a facility of the National Science Foundation operated by Associated Universities, Inc.\newline
\end{acknowledgements}


\bibliographystyle{jwaabib}
\bibliography{aa-abbrv,mnemonic,bib}

\input{Appendix.tex}

\end{document}

%% file: intro.tex
\section{Introduction}
Parsec-scale properties of jets in radio-loud active galactic nuclei (AGN) have been studied in much detail using the radio-astronomical technique of very long baseline interferometry (VLBI).
In particular, the monitoring program MOJAVE \citep{lister2009} has been highly successful in exploring the jet kinematics of sources at a frequency of $15~\mathrm{GHz}$ \citep{lister2016_kinematic}. These data have been used to  study the acceleration and collimation of parsec-scale jets \citep{Homan2015_acc_col}, their opening angles \citep{pushkarev2017_opang}, polarization properties \citep{lister2018_pol,pushkarev2017_pol}, spectral distributions \citep{Hovatta2014_spec_distr}, and other applications. \newline
By employing the model of a freely expanding relativistic jet  \citep{BK79,K81}, the underlying physical properties such as the electron density, magnetic field strength, or power of the jets can be inferred from such observations. The strong observational biases due to relativistic bulk motion, such as aberation and Doppler boosting, can be corrected for by using statistical methods \citep{cohen2007}. Beyond scaling models, VLBI observations can also
be compared with detailed relativistic magnetohydrodynamical simulations \citep{2011ApJ...737...42P,2006ApJ...651..272F}, which describe the launching of  jets from the black-hole accretion disk \citep{blandford_payne1982} and ergosphere \citep{blandford_znajek}, as well as their acceleration and collimation. The simulations describe the fluid-dynamical evolution of the Poynting-flux dominated jets on subparsec scales 
all the way to the scales where they become visible on (multi) parsec scales as typical kinetic-energy-dominated jets to VLBI and connected radio interferometers. On VLBI scales, shock waves are commonly believed to accelerate particles to ultrarelativistic energies by the Fermi mechanism, causing moving knots in the jet flow and multiwavelength flares \citep[e.g.,][]{2008Natur.452..966M}.

In addition to  brightness temperature ($T_{\rm B}$) observations of the VLBI cores of AGN jets \citep{Homan2021}, the measurement of
brightness-temperature gradients along their jets is a powerful independent observational tool. This approach can probe the internal and external physical conditions in AGN jets and has been employed in several single-source jet studies. For 3C\,111, \citet{Kadler2008_3C111} measured brightness-temperature gradients and their evolution to study the
particle and magnetic field density associated with individual jet components during flares.
In the case of NGC\,1052 (a twosided source), \citet{Kadler2004} and \citet{Baczko2019_tb} studied the $T_{\rm B}$ gradients along both jets. \citet{Kadler2004} find a break of the $T_{\rm B}$-gradients near the cores which they attribute to possible free-free absorption associated with a dusty torus. \citet{Baczko2019_tb} further find that single power-law fits do not sufficiently describe the measured $T_{\rm B}$ behavior  and adopt a two-zone model with a break in the brightness-temperature and diameter gradient of the jet. 
Studies of the brightness temperatures of VLBI jets and their gradients with distance from the core are still rare for larger samples.

In order to characterize jets close to their collimation zone, \citet{Pushkarev_Kovalev2012} carried out dual-frequency observations at $2\, \mathrm{GHz}$ and $8\,\mathrm{GHz}$, respectively, using a sample of 30 AGN jets. They found that at $8\,\mathrm{GHz}$ the jet diameter gradient shows flatter slopes than expected for freely expanding jets, while at $2\, \mathrm{GHz}$ the jet geometry is consistent with a conical geometry. They found brightness temperature gradients of $T_{\rm B}\propto r^{-2.2}$. Furthermore,  \citet{Pushkarev_Kovalev2012} found indications for the presence of a transition zone in blazar jets where the jet flow switches from a confined, accelerating jet (parabolic geometry) to a freely expanding jet (conical geometry). In the case of M87, \citet{Asada_2012} showed that such a geometry transition can be observed in the vicinity of the Bondi-sphere \citep{Russel2015} where the external medium is expected to change its pressure gradient. More recent works also found these transitions in several different sources, NGC\,6251 \citep{Tseng2016}, NGC\,4261 \citep{Nakahara2018}, NGC\,1052 \citep{Nakahara2020, Baczko2021_arx}, NGC 315\,\citep{Boccardi2020,Park2020} and 1H0323+342 \citep{Hada2018}. In the cases of M87 \citep{Mertens2016,Hada2018, Park2019}, NGC\,315 \citep{Park2020} and 1H0323+342 \citep{Hada2018} the jets also show accelerating jet components near the geometry-transition zones.
\citet{Kovalev2020} systematically searched for these geometry transitions in 331 MOJAVE sources at $15\, \mathrm{GHz}$ and $1.4\, \mathrm{GHz}$. They demonstrated the presence of geometry transitions from a parabolic to a conical geometry in 10 nearby sources on scales of several thousand Schwarzschild radii $R_\textrm{S}$. \citet{Potter_Cotter_2015} applied parameters, obtained from 42 simultaneous blazar spectral energy distributions (SED) to a 1d-relativistic fluid-dynamical jet model to predict a region where the jets experience a transition between a magnetically-dominated region (closer to the central engine), where the jet geometry is parabolic and a particle-dominated region of the jet (farther downstream) where the jet geometry is conical. \citet{Potter_Cotter_2015} predict a  correlation between the radii of the jets in the transition zone and the jet power as well as the maximum bulk Lorentz factors. Consequently, they also predict a dichotomy of the blazar classes, BL Lacertae objects (BL Lacs) and flat-spectrum radio quasars (FSRQs), regarding the jet radii in the transition zone, where BL Lacs are expected to show smaller jet radii than FSRQs. \citet{Kovalev2020} argue that external pressure gradients most likely are not solely responsible for the geometry transitions within the VLBI jets and that the mechanism might
involve a transition of the jets from a Poynting-flux to a particle-dominated configuration. \\
In order to investigate the $T_{\rm B}$-gradients and morphological changes of parsec-scale jets at higher frequencies, we conducted a systematic dual-high-frequency ($15\, \mathrm{GHz}$ and $43\,\mathrm{GHz}$)  analysis of a sample of blazars using data obtained with the Very Long Baseline Array (VLBA) data as part of the MOJAVE\footnote{http://www.physics.purdue.edu/astro/MOJAVE/} (15\, GHz) and Boston University\footnote{https://www.bu.edu/blazars/VLBAproject.html} (BU, 43\, GHz)  jet monitoring programs.   The following cosmological constants were adopted throughout the paper: $h = 0.71, \Omega_\Lambda = 0.73,\, \Omega_M = 0.27$.

%% file: sample.tex
\section{Sample}
\label{sec_sample}
The objects and their observation epochs (i.e., characterizing the time interval during which the observation was carried out) for this study were selected by the following criteria:
\begin{itemize}
\item[a)] The sources are in the MOJAVE data archive  \citep{lister2018_pol} and the BU sample \citep{Jorstad2016}.
\item[b)] Data exists for epochs between 2003 and 2013
\item[c)] The imaged jet brightness distribution can be sufficiently represented and parametrized with at least three 2D-Gaussian jet components in at least 5 epochs\footnote{We exclude the peculiar, complex-structure radio galaxy 0316+413 (3C84)}.

\end{itemize}
We chose the earliest epoch considered in our study to be at the onset of the new MOJAVE program \citep{lister2005} at which the overall image fidelity and typical image noise levels improved systematically over earlier data sets from the progenitor 2cm-Survey program \citep{kellermann1998}. The resulting sample consists of 15 flat-spectrum radio quasars (FSRQs), 11 BL Lacertae objecs (BL\,Lacs) and two radio galaxies (RGs), in total 28 objects. The object names, their classification,  redshifts. and black hole masses are listed in Table~\ref{table1}. 

\begin{table}
\caption{Source sample and source properties}
\label{table1}
\begin{small}
\begin{tabular}{@{}l@{~~}c@{}c@{}c@{}c@{~}c@{~}c@{}}
\toprule
Name & Alt. name & Class. & $z$ & $\mathrm{log}_{10}\left(M_\textrm{BH} /M_{\odot} \right) $ & $\frac{\mathrm{pc}}{\mathrm{mas}}$ & Epochs \\
\midrule
0219+428 &  3C\,66A & B & $0.34^{\textrm{a}}$ & $(8.47-8.65)^{\textrm{I}}$ & $4.8$ & $19/38$ \\
0336$-$019 & CTA\,26 & Q & $0.852^{\textrm{b}}$ & $(7.55-7.73)^{\textrm{I}}$ & $7.7$&  $18/15$\\
0415+379 & 3C\,111 & G & $0.0491^{\textrm{c}}$ & $(8.18-8.38)^{\textrm{II}}$ & $0.95$ & $59/20$ \\
0430+052 & 3C\,120 & G & $0.033^{\textrm{d}}$ & $(7.08-7.58)^{\textrm{III}}$ & $0.65$ & $59/24$\\
0528+134 & PKS\,0528+134 & Q & $2.07^{\textrm{e}}$ & $(8.35-8.52)^{\textrm{I}}$ &$8.5$ & $18/79$ \\
0716+714 & S5\,0716+71 & B & $0.803^{\textrm{f}}$ & $(7.51-7.68)^{\textrm{I}}$ & $7.5$ & $59/71$\\
0735+178 & OI\,158 & B & $0.45^{\textrm{g}}$ & $(8.35-8.51)^{\textrm{I}}$ & $5.8$ & $14/60$ \\
0827+243 & OJ\,248 & Q & $0.94^{\textrm{h}}$ & $(6.94-7.12)^{\textrm{I}}$ & $7.9$& $14/14$\\
0829+046 & OJ\,049 & B & $0.174^{\textrm{i}}$ & $(8.55-8.71)^{\textrm{I}}$ & $2.9$ & $14/64$\\
0836+710 & 4C\,+71.07 & Q & $2.218^{\textrm{j}}$ & $(7.8-7.97)^{\textrm{I}}$ & $8.4$& $14/14$ \\
0851+202 & OJ\,287 & B & $0.306^{\textrm{k}}$ & $(7.12-7.31)^{\textrm{I}}$ & $4.5$ & $26/28$ \\
0954+658 & S4\,0954+65 & B & $0.45^{\textrm{l}}$ & $(8.6-8.77)^{\textrm{I}}$ & $5.7$ & $ 21 / 60 $ \\
1101+384 & Mrk\,421 & B & $0.0308^{\textrm{m}}$ & $(7.28-7.49)^{\textrm{I}}$ & $ 0.61$ & $ 27 / 40 $\\
1127$-$145 & PKS\,1127$-$14 & Q & $1.184^{\textrm{n}}$ & $(7.77-7.94)^{\textrm{I}}$ & $8.3$& $ 15 / 44 $ \\
1156+295 & 4C\,+29.45 & Q & $0.725^{\textrm{i}}$ & $(8.2-8.38)^{\textrm{I}}$ & $7.3$& $ 20 / 14 $ \\
1219+285 & W\,Comae & B & $0.102^{\textrm{o}}$ & $(7.17-7.37)^{\textrm{I}}$ & $1.9$& $ 11 / 47 $\\
1222+216 & 4C\,+21.35 & Q & $0.433^{\textrm{i}}$ & $(8.38-8.55)^{\textrm{I}}$ & $5.6$& $ 23 / 17 $ \\
1226+023 & 3C\,273 & Q & $0.1583^{\textrm{p}}$ & $(8.13-8.3)^{\textrm{I}}$ & $2.7$& $ 32 / 13 $\\
1253$-$055 & 3C\,279 & Q & $0.6^{\textrm{q}}$ & $(8.65-8.83)^{\textrm{I}}$ & $6.7$& $ 24 / 69 $\\
1308+326 & OP\,313 & Q & $0.997^{\textrm{i}}$ & $(7.64-8.56)^{\textrm{IV}}$ & $8.0$& $ 19 / 28 $ \\
1633+382 & 4C\,+38.41 & Q & $1.814^{\textrm{h}}$ & $(8.36-8.54)^{\textrm{I}}$ & $8.5$& $ 27 / 79 $ \\
1652+398 & Mrk\,501 & B & $0.0337^{\textrm{r}}$ & $(8.02-8.22)^{\textrm{I}}$  & $0.66$& $ 12 / 6 $\\
1730$-$130 & NRAO\,530 & Q & $0.902^{\textrm{s}}$ & $(8.35-8.52)^{\textrm{I}}$ & $7.8$& $ 24 / 71 $\\
1749+096 & OT\,081 & B & $0.322^{\textrm{t}}$ & $(8.66-8.66)^{\textrm{V}}$ & $4.6$ & $ 26 / 49 $\\
2200+420 & BL\,Lac & B & $0.0686^{\textrm{u}}$ & $(8.02-8.17)^{\textrm{I}}$ & $1.3$& $ 64 / 80 $ \\
2223$-$052 & 3C\,446 & Q & $1.404^{\textrm{v}}$ & $-$ & $8.5$& $ 15 / 14 $\\
2230+114 & CTA\,102 & Q & $1.037^{\textrm{w}}$ & $(8.4-8.57)^{\textrm{I}}$ & $8.1$& $ 18 / 13 $ \\
2251+158 & 3C\,454.3 & Q & $0.859^{\textrm{x}}$ & $(7.21-7.4)^{\textrm{I}}$ & $7.7$& $ 22 / 14 $ \\
\hline 
\end{tabular}
\end{small}
\tablefoot{Columns: (1) IAU B\,1950 name, (2) Alternative (common) name, (3) Classification, (4) Redshift, (5) Black-hole mass, (6) Angular to linear scale conversion factor, (7) Number of epochs ($15~\mathrm{GHz}/43~\mathrm{GHz}$)}
\tablebib{
a) \citet{0219+428},
b) \citet{0336-019},
c) \citet{0415+379},
d) \citet{0430+052}, 
e) \citet{0528+134},
f) \citet{0716+714},
g) \citet{0735+178},
h) \citet{Paris2017},
i) \citet{Schneider2010}, 
j) \citet{0836+710},
k) \citet{0851+202},
l) \citet{0954+658},
m) \citet{1101+384},
n) \citet{1127-145}, 
o) \citet{1219+285},
p) \citet{1226+023},
q) \citet{1253-055},
r) \citet{1652+398},
s) \citet{1730-130},
t) \citet{1749+096},
u) \citet{2200+420},
v) \citet{2223-052},
w) \citet{2230+114},
x) \citet{2251+158},
I) \citet{Fan2008},
II) \citet{3C111_mass_halpha}
III) \citet{3C120_reverb}
IV) \citet{GUPTA20128_1308+236_mass}
V) \citet{Falomo2003_1749+096_mass}
}
\end{table}

%% file: method.tex
\section{Method}
\label{sec_method}
\subsection{Parameter fitting}
\begin{figure}
\centering
	 \includegraphics[width=\hsize]{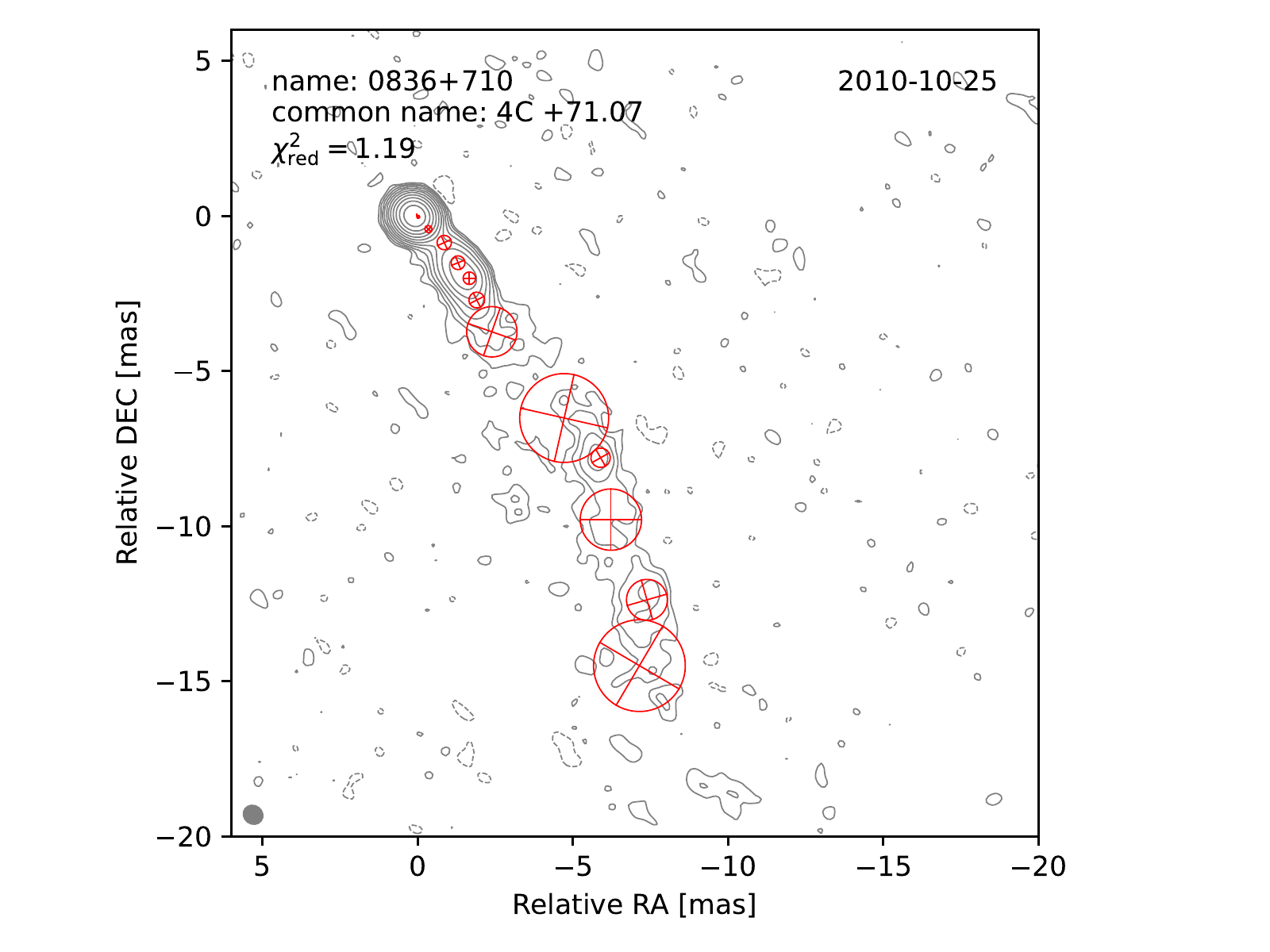}
	\includegraphics[width=\hsize]{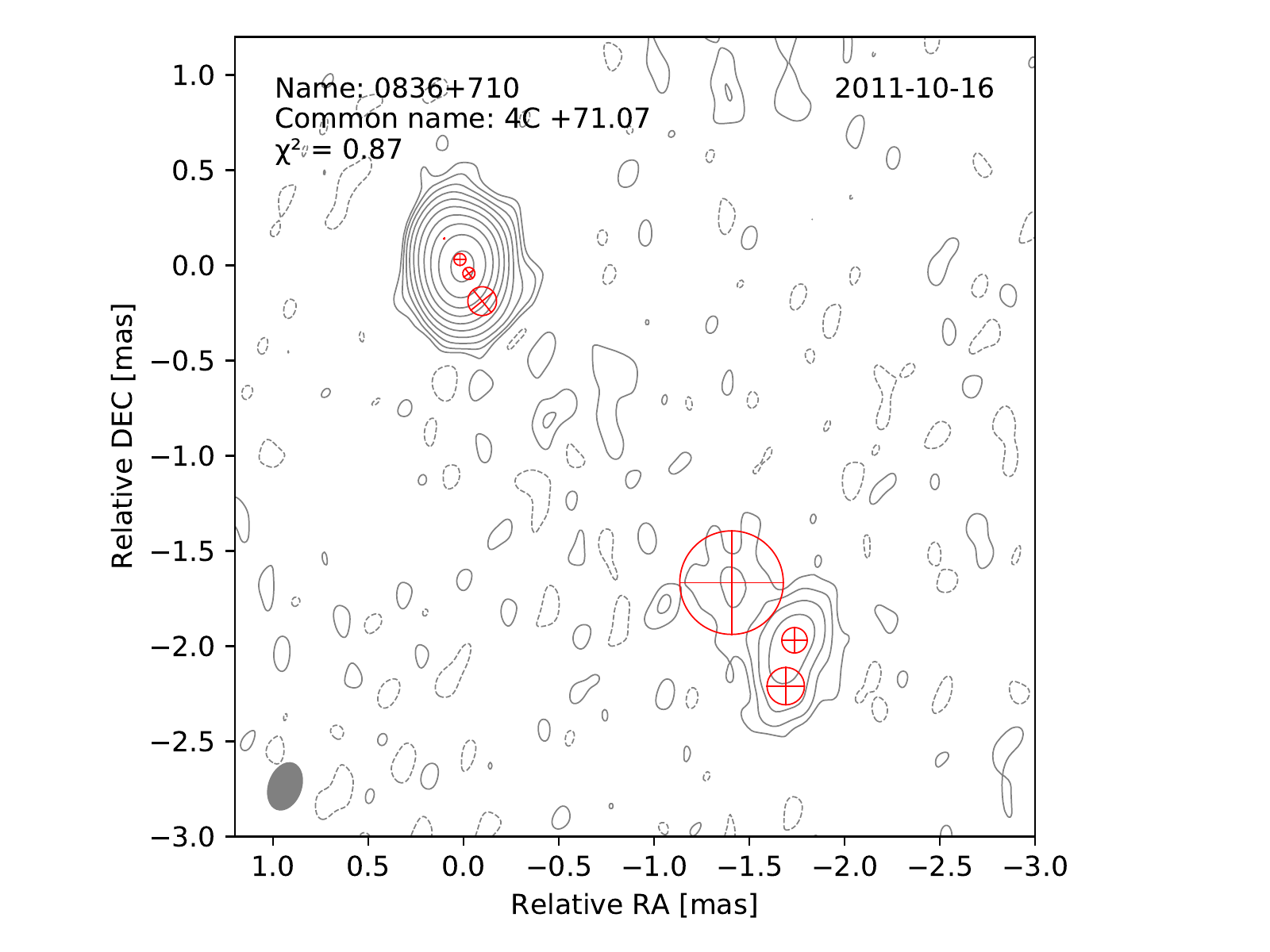}
      \caption{Fitted Gaussian-component representation of 4C\,+71.07 (0836+710) at $15\,\mathrm{GHz}$ (top) and at $43\,\mathrm{GHz}$ (bottom). The contours for both frequencies are shown at a $5~\sigma$ level and are separated by a factor of 2 . In the $15\,\mathrm{GHz}$ image the map noise level is $\sim 0.2 ~\mathrm{mJy}$. The contour levels are $[-0.85,0.85,1.7,3.4,6.7,13.7,27.5,54.9,109.8,219.6,430.2,878.5]~\mathrm{mJy}$).
      At $43\,\mathrm{GHz}$ the map noise level is $\sim 0.8 ~\mathrm{mJy}$. The contour levels are $[-3.8,3.8,7.5,15.1,30.2,60.3,120.6,241.2,284.5,482.5,964.9,\\1929.8,3859.6]~\mathrm{mJy}$.}
       \label{modelfitting}
\end{figure}

The optically thin synchrotron emission regions in the jet were fitted with circular Gaussians using \textsc{difmap}\footnote{\url{https://science.nrao.edu/facilities/vlba/docs/manuals/oss2013a/post-processing-software/difmap}} \citep{DIFMAP}. Initially, the visibility map is opened and a starting model consisting of  a circular Gaussian component in the center is fitted to best match the data. If the residual map still shows nonzero intensity after subtracting the established model, further Gaussians are added and fitted until the residual map shows only (white) noise. As an example, the corresponding jet model resulting from a single epoch of 4C\,+71.07 (0836+710) observations is shown in Fig.\ref{modelfitting} for both frequencies. Our model fits are optimized to represent accurately the local surface brightness and jet width along the jet rather than individual moving knots within the jet. For this reason, our model fits differ from the ones used in standard MOJAVE kinematics studies.
The diameter $d$, flux $F = I \times \Omega _G$ ($I$ is the intensity and $\Omega _G$ the solid angle of the fit Gaussian component) , and the (angular) distance from the core $r$ are measured for each component. Unresolved components are filtered according to
\begin{equation}
 r_{\mathrm{lim}} = b_{\Psi} \sqrt{\frac{4\,\mathrm{ln}(2)}{\pi} \mathrm{ln}\left(\frac{\mathrm{S/N}}{\mathrm{S/N}-1} \right)} 
\end{equation}
which describes the minimal radius a resolved fitted component can achieve for a given signal-to-noise ratio ($\mathrm{S/N}$) and beam size ($b_{\Psi} $), \citep[cf.][]{Kovalev2005}. The uncertainties of the parameters of each component are approximated with the post-fit RMS $\sigma_{rms}$ information \citep{Fomalont1999}. The uncertainty of the flux density is further altered by also geometrically adding a systematic value of $5\%$ of the flux density to account for absolute flux-scale uncertainties:   
\begin{gather}
\sigma_{F} = \sqrt{\sigma_{\mathrm{rms}}^2 + \sigma_{\mathrm{rms}}^2 \mathrm{S/N} + \left(0.05  F\right)^2} \\
\sigma_{d} = d \frac{\sigma_{F}}{F} \quad .
\end{gather}
 The diameters of the remaining components from each epoch are plotted together into a single graph against the distance from the radio core, at a 1-$\sigma$ level (Figs.~\ref{0219+428} ff.). The radio core itself is the unresolved component closest to the (0/0)-position in the map. All distances are measured against the position of the respective core component. This ensures that the set of model components characterizing the jet is representative of the long-term time-averaged intensity of the evolving jet. By contrast, individual model components at a given epoch are representative of the local jet brightness and width at a given time. The components in these sources typically have considerable proper motions of up to  several milliarcseconds along the jet per year \citep[e.g.,][]{lister2016_kinematic}, evolving in brightness and in width during this time. Our long-term multi-epoch data base allows us to parameterize the jet geometry and brightness temperature for each source in our sample at two separate high radio frequencies and at submilliarcsecond resolution. To this end, we fit the jet diameter with $d\propto r^l$, that is as a power-law of distance $r$ with index $l$. The brightness temperature is also fitted as a power law $T_{\rm B} \propto r^s$, characterized by the index $s$. This procedure is performed for all sources at both frequencies, $15\,\mathrm{GHz}$ and $43\,\mathrm{GHz}$, along the respective entire jet length and in the overlap regions of both frequencies. To determine a physical overlap region, the core shift \citep{Pushkarev_coreshifts_2012} has to be taken into account since the turnover in the synchrotron spectrum from optically thick to optically thin emission depends on the frequency, that is, the radio core appears closer to the jet apex at higher frequencies. The core shift $\Delta r_{cs}$ usually has to be calculated on a source-by-source basis between two frequencies. At lower frequencies ($\sim 8-15 \,\mathrm{GHz}$), \citet{Pushkarev_coreshifts_2012} studied the nuclear opacity of radio jets on a larger sample of 191 sources. For higher frequencies, \citet{Osullivan2009} analyzed the core shift between $\sim 15$ and $43\,\mathrm{GHz}$ for three sources, $1418+546$, $2007+777$ and $2200+420$. The core shifts are calculated assuming conical jet geometries and are measured to be between $0.04 \lesssim \Delta r_{cs}/\mathrm{mas}\lesssim 0.1$, which corresponds to $5-10\%$ of the respective beam size of the used data.
 Typical beam sizes for the data used in this work are $\sim 0.5\, \mathrm{mas}$ for $15 \, \mathrm{GHz}$ and  $\sim 0.3\, \mathrm{mas}$ for $43 \, \mathrm{GHz}$, and thus by at least a factor of 3 larger than reported core shifts between these frequencies. The effect of such small core shifts in this study are negligible, however, in the case of 2200+420 we tested the impact of this core shift on the results which will be discussed in the corresponding paragraph in Sect.~\ref{par.:2200}.


\subsection{Scaling model for the jets}
\label{Sect.:Jetmodel}
For the parameterization of the jet diameter and brightness temperature, we employ the relativististic jet model introduced by \citet{BK79} and \citet{K81}. The basic assumptions of the model are:
\begin{itemize}
	\item[a)] The jet has a semi-opening angle $\Phi$ and an inclination angle $\vartheta$ to the line of sight. The diameter of the jet scales along the jet-axis $\boldsymbol{r}$, $d_{\mathrm{jet}} \propto r^l$. The corresponding jet morphologies relate to a free expansion with constant advance speed $(l=1)$, accelerated motion $(l<1)$ or jet deceleration $(l>1)$. 
	\item[b)] Electrons are injected continuously into the jet and the electron energy density ($n_e$) decreases as a power law along the jet axis according to $n_e \propto r^n$. 
	\item[c)] The magnetic field scales with a power law along the jet axis according to $B \propto r^b$. 
	A purely toroidal (azimuthal) magnetic field scales with the jet transverse radius $R$ as $B\propto R^{-1}$, resulting in a scaling along the jet axis of $B\propto r^{-l}$. A purely poloidal magnetic field (parallel to the jet axis) would scale with the jet radius according to $B\propto R^{-2}$ which translates to scaling along the jet-axis according to $B\propto r^{-2\,l}$.
\end{itemize}
The observed flux $F_\nu$ of a circular component with a diameter  $d_{\mathrm{jet}}$ and radiating area $A\propto d_{\mathrm{jet}}^2$ is proportional to
\begin{equation}
	\label{flux}
	F_{\nu} \propto A \cdot d_{\mathrm{jet}} \cdot j_{\nu}
\end{equation}
where $j_{\nu} \propto n_e \cdot B \cdot \left(\frac{\nu}{\nu_B}\right)^{\alpha}\, , (\nu_B \propto B),$ is the emissivity of the synchrotron emission \citep[cf.][Sect. 9.2.1]{Krolik1999}. The properties of the jet geometry, magnetic field, and the electron energy density a)-c), Eq.~(\ref{flux}) implies that the flux $F_\nu$ of a model component has a power law dependency, scaling along the jet axis
\begin{equation}
	\label{flux_2}
	F_{\nu} \propto r^{3l + n + b(1-\alpha)}
\end{equation}
The brightness temperature $T_{\rm B}$ of a Gaussian-shaped component is given by
\begin{equation}
	\label{tb_kov}
	T_{\rm B} = \frac{2\,\mathrm{ln}(2)}{\pi k} \frac{F_\nu\lambda ^2 (1+z)}{R^2}
\end{equation}
where $k$ denotes the Boltzmann constant, $\lambda$ the observed wavelength, and $r=d_{\mathrm{jet}}/2$ the radius of the component\footnote{Note: If an emission region is fitted with an elliptical Gaussian component, $r^2$ must be replaced by $\theta _{\mathrm{maj}} \theta _{\mathrm{min}}$, where $\theta _{\mathrm{maj}}$ and $\theta _{\mathrm{min}}$ are the major and minor axis of the elliptical Gaussian, respectively.}. 
Eq.\ref{tb_kov} implies that $T_{\rm B} \propto F_{\nu}/R^2$ ($R^2\propto r^{2l}$) and therefore 
\begin{equation}
	\label{def_s}
	T_{\rm B} \propto r^{l+n+b(1-\alpha)} \quad .
\end{equation}
For convenience, we define the index $s$ that describes the scaling of $T_{\rm B}\propto r^s$ by
\begin{equation}
    s=l+n+b(1-\alpha) \quad .
\end{equation}
Now, the complete set of scaling indices is given by the shape index $l$, the density index $n$, the magnetic field index $b$, and the brightness-temperature index $s$. Their interdependence depends on the degree of fulfillment of adiabatic expansion, particle number conservation, and magnetic-field equipartition.

\subsubsection{Comparing the modelfit method with the stacked-image analysis}
\label{method_comp}
We compared the method of estimating the jet widths at a given position, described above with the stacked-image analysis  as employed by \citet{pushkarev2017_opang} and \citet{Kovalev2020}. 
It is not a-priori clear that both methods should yield the same results, since it is known that stacked images map out a broad profile  while at any given time traveling jet features do not fill out this entire stacked-profile range \citep{pushkarev2017_opang,Lister2021}. On small scales, a varying jet nozzle can lead to strong variations of component position angles in the inner jet, especially in high-resolution data at high frequencies, which biases the apparent jet geometry in stacked images. 
In individual sources, further differences can arise when a source exhibits a strongly resolved wide jet emission region far downstream from the core. The modelfit method will typically yield one single large component to represent such an extended region in each individual epoch while image stacking and image-plane fitting of stacked images will benefit from enhanced signal-to-noise ratio and cover a large area. In result, a diameter gradient fit with the modelfit method will put much stronger weight on the (typically many) components from the compact jet and little weight on the single extended Gaussian component representing the diffuse emission. The extended jet regions in the image-stacking method are also weighted less because the method is sensitive to the signal to noise ratio \citep[e.g., ][]{Kovalev2020,Casadio2021}. However, since the jet is split into equidistant sections along the jet axis, the larger area of extended emission will yield more data points to be fit in these regions as the modelfit method. For such cases the modelfit method can derive flatter diameter gradients than the stacked image-plane method. We emphasize that a modelfit component in a single epoch at a given position in some cases will not fill the physical jet width. Therefore it is important to not derive jet parameters from single epoch fits. Therefore when studýing the gradients all epochs are considered at once.
In consideration of the arguments made above, we consider the modelfit method in general as the most robust way to characterize the jet geometry.

A further difference between our analysis and the approach of \citet{Kovalev2020} is that we are not attempting to fit an additional parameter $r_0$ to consider the separation of the VLBI jet core from the true jet base. Median value of this parameter determined on large samples are of the order of 0.1\,mas (e.g., Kravchenko et al., in prep.) but the generalized fitting of both $r_0$ and $l$ does not converge in case of all of our data sets. We therefore choose to neglect this offset, understanding that this tends to lead to somewhat flattened apparent jet gradients. We tested the strength of this effect in case of 18 sources for which a generalized fitting was possible and found differences in $|l|$ in the range of $1.4\times 10^{-3}$ to $0.27$. In case of sources with well-determined and realistically small $r_0$ values,
 we found that both methods, model fitting and image stacking, yield comparable results and are overall in good agreement.

\subsubsection{Changes in the geometrical shape}\label{sect:method_breaks}
To measure possible changes in the geometrical shape of the jets, we subdivided each jet along the distance from the core into $10^4$ slices. The center of each slice represents a possible break point $x_b$ in the jet. One of these break points is picked randomly, then two power-law fits are performed left and right of $x_b$, respectively. If the fit parameters are plausible, namely $\chi _{\mathrm{red}} ^2 \sim 1$, the errors of the inner and outer jet geometry fits ($l_{in}, l_{out}$) are of the same order of magnitude and also if $l_{in} < l_{out}$, the solution is added to the sum of solutions. This is done 5000 times. To then find the best ($l_{in}, l_{out},x_b$), the mean value of the respective parameter is calculated over all runs and weighted with the respective $\chi _{\mathrm{red}} ^2$ value. This set of parameters ($l_{in}, l_{out},x_b$) is used as initial values for a smooth broken-power-law fit of the form
\begin{equation}
    f(x) = A~2^{(l_{in}-l_{out})/\Delta} ~ \left(\frac{x}{x_b}\right)^{l_{in}} ~ \left[1+\left(\frac{x}{x_b}\right)^{\Delta}\right]^{(l_{out}-l_{in})\Delta}
\end{equation}
\citep{Nakahara2020,astropy:2013,astropy:2018}, where $A=d_j(x_b)$ is the jet diameter at the break point $x_b$, $l_1$ is the diameter gradient index for the inner jet (closer to the radio core) and $l_2$ is the diameter gradient of the outer jet.
This procedure turned out to be necessary because the fit of the smoothly broken power law is sensitive to properly guessed initial values. In order to not having to find them by visual inspection, we performed the search algorithm described above. There are different ways this function can be fit to data. \citet{Baczko2021_arx} deploy the orthogonal distance regression while \citet{Casadio2021} use a gnuplot implementation of the Levenberg-Marquardt algorithm. In this work we use scipy \citep{scipy19} in order to fit the smoothly broken power law with a classical least square method. We used the trust-region reflective algorithm rather than the Levenberg-Marquardt algorithm as in \citet{Boccardi2020}. The problem at hand is a bound problem in the sense that we let the fit algorithm vary the initial conditions. These initial conditions are obtained from the numerical splitting of the jet. A comparison between the least square fit results and the orthogonal distance regression can be found in Appendix~\ref{appendix}. We tested the parameter range of $\Delta$ between 10 \citep{Baczko2021_arx} and 100 \citep{Boccardi2020}. Within the errors the results are not distinguishable, therefore we set the parameter constant to 100. 



%% file: results_gradients.tex
\section{Results}
\label{results}
The results of the diameter and brightness-temperature gradient fits are shown in the appendix \ref{appendix_gradfit} in Figs.~\ref{0219+428}-\ref{2251+158} and summarized in Table~\ref{tab-fitresults}.  Results were obtained for the full jet length at each frequency and for the overlap regions probed at both frequencies.

\begin{table*}

\caption{Fitting results of jet geometry and brightness-temperature gradients}
\label{tab-fitresults}
\centering
\resizebox{1.5\columnwidth}{!}{.  
\begin{tabular}{lcccccc}
\toprule
Name & $l_{\mathrm{15}}$ & $l_{\mathrm{43}}$ & $s_{\mathrm{15}}$ & $s_{\mathrm{43}}$  \\
\midrule
0219+428 & $1.02 \pm 0.17$ $(1.0, 151)$ & $0.57 \pm 0.16$ $(0.9, 179)$ & $-2.05 \pm 0.43$ $(2.1, 151) $ & $-1.6 \pm 0.35$ $(1.1, 179)$\\ & $0.63 \pm 0.12$ $(0.4, 97)$ & $0.56 \pm 0.17$ $(1.1, 149)$ & $-2.08 \pm 0.38$ $(1.1, 97) $ & $-1.45 \pm 0.36$ $(1.2, 149)$\\ \midrule
0336-019 & $0.7 \pm 0.17$ $(3.5, 68)$ & $0.81 \pm 0.25$ $(2.7, 55)$ & $-1.12 \pm 0.38$ $(3.8, 68) $ & $-1.87 \pm 0.48$ $(3.1, 55)$\\ & $0.68 \pm 0.18$ $(3.9, 58)$ & $0.84 \pm 0.26$ $(3.0, 49)$ & $-1.18 \pm 0.36$ $(3.1, 58) $ & $-1.7 \pm 0.51$ $(3.6, 49)$\\ \midrule
0415+379 & $0.62 \pm 0.19$ $(2.5, 515)$ & $0.53 \pm 0.2$ $(1.2, 151)$ & $-1.83 \pm 0.59$ $(5.3, 515) $ & $-1.51 \pm 0.55$ $(2.5, 151)$\\ & $0.58 \pm 0.2$ $(2.8, 420)$ & $0.6 \pm 0.18$ $(1.3, 131)$ & $-1.57 \pm 0.63$ $(6.0, 420) $ & $-1.56 \pm 0.53$ $(2.7, 131)$\\ \midrule
0430+052 & $0.56 \pm 0.23$ $(5.6, 491)$ & $0.46 \pm 0.17$ $(1.0, 112)$ & $-1.54 \pm 0.54$ $(6.5, 491) $ & $-1.67 \pm 0.48$ $(2.3, 112)$\\ & $0.56 \pm 0.19$ $(3.0, 223)$ & $0.43 \pm 0.18$ $(1.2, 89)$ & $-1.42 \pm 0.54$ $(5.0, 223) $ & $-1.44 \pm 0.51$ $(2.5, 89)$\\ \midrule
0528+134 & $0.79 \pm 0.21$ $(1.8, 98)$ & $0.45 \pm 0.19$ $(1.5, 242)$ & $-2.65 \pm 0.46$ $(2.5, 98) $ & $-1.23 \pm 0.43$ $(2.4, 242)$\\ & $0.6 \pm 0.18$ $(1.8, 51)$ & $0.51 \pm 0.19$ $(1.5, 194)$ & $-2.34 \pm 0.59$ $(4.5, 51) $ & $-1.94 \pm 0.43$ $(2.4, 194)$\\ \midrule
0716+714 & $0.84 \pm 0.16$ $(0.6, 248)$ & $0.78 \pm 0.21$ $(1.8, 194)$ & $-3.0 \pm 0.35$ $(0.3, 248) $ & $-2.6 \pm 0.52$ $(1.0, 194)$\\ & $0.79 \pm 0.14$ $(0.6, 185)$ & $0.78 \pm 0.19$ $(1.5, 178)$ & $-2.84 \pm 0.35$ $(0.3, 185) $ & $-2.56 \pm 0.46$ $(0.7, 178)$\\ \midrule
0735+178 & $0.85 \pm 0.08$ $(0.6, 55)$ & $0.77 \pm 0.21$ $(1.2, 132)$ & $-2.46 \pm 0.37$ $(2.3, 55) $ & $-1.75 \pm 0.36$ $(1.0, 132)$\\ & $0.83 \pm 0.08$ $(0.7, 46)$ & $0.75 \pm 0.21$ $(1.2, 110)$ & $-1.85 \pm 0.33$ $(1.6, 46) $ & $-1.77 \pm 0.29$ $(0.8, 110)$\\ \midrule
0827+243 & $0.82 \pm 0.19$ $(1.2, 57)$ & $0.56 \pm 0.17$ $(0.7, 37)$ & $-2.61 \pm 0.47$ $(1.6, 57) $ & $-2.55 \pm 0.48$ $(1.6, 37)$\\ & $0.83 \pm 0.17$ $(1.0, 36)$ & $0.54 \pm 0.14$ $(0.6, 30)$ & $-3.11 \pm 0.5$ $(2.2, 36) $ & $-2.74 \pm 0.44$ $(1.4, 30)$\\ \midrule
0829+046 & $0.98 \pm 0.13$ $(1.0, 76)$ & $0.7 \pm 0.23$ $(1.7, 176)$ & $-2.56 \pm 0.3$ $(0.9, 76) $ & $-1.87 \pm 0.54$ $(1.8, 176)$\\ & $0.7 \pm 0.12$ $(0.6, 51)$ & $0.79 \pm 0.24$ $(1.8, 144)$ & $-2.0 \pm 0.33$ $(0.9, 51) $ & $-1.89 \pm 0.56$ $(2.0, 144)$\\ \midrule
0836+710 & $0.77 \pm 0.18$ $(2.7, 99)$ & $0.37 \pm 0.22$ $(3.9, 44)$ & $-1.96 \pm 0.49$ $(14.2, 99) $ & $-1.97 \pm 0.7$ $(15.8, 44)$\\ & $0.46 \pm 0.18$ $(1.2, 58)$ & $0.3 \pm 0.22$ $(4.7, 40)$ & $-1.98 \pm 0.52$ $(14.1, 58) $ & $-1.63 \pm 0.66$ $(15.9, 40)$\\ \midrule
0851+202 & $0.88 \pm 0.19$ $(1.5, 110)$ & $0.82 \pm 0.34$ $(5.1, 263)$ & $-2.87 \pm 0.46$ $(2.9, 110) $ & $-2.79 \pm 0.81$ $(8.3, 263)$\\ & $0.69 \pm 0.17$ $(1.3, 75)$ & $0.94 \pm 0.33$ $(5.7, 234)$ & $-2.1 \pm 0.49$ $(3.0, 75) $ & $-3.05 \pm 0.77$ $(8.7, 234)$\\ \midrule
0954+658 & $1.2 \pm 0.21$ $(1.2, 102)$ & $1.09 \pm 0.25$ $(2.8, 184)$ & $-3.39 \pm 0.47$ $(1.3, 102) $ & $-2.97 \pm 0.52$ $(2.4, 184)$\\ & $1.37 \pm 0.21$ $(1.2, 77)$ & $1.17 \pm 0.26$ $(3.1, 161)$ & $-3.98 \pm 0.5$ $(1.6, 77) $ & $-3.32 \pm 0.52$ $(2.7, 161)$\\ \midrule
1101+384 & $0.89 \pm 0.19$ $(1.8, 151)$ & $0.61 \pm 0.26$ $(1.9, 106)$ & $-2.19 \pm 0.26$ $(0.9, 151) $ & $-1.57 \pm 0.46$ $(1.5, 106)$\\ & $0.75 \pm 0.15$ $(1.9, 97)$ & $0.65 \pm 0.27$ $(2.0, 92)$ & $-2.2 \pm 0.2$ $(1.1, 97) $ & $-1.57 \pm 0.45$ $(1.5, 92)$\\ \midrule
1127-145 & $0.48 \pm 0.24$ $(6.3, 95)$ & $0.35 \pm 0.26$ $(7.1, 96)$ & $-1.14 \pm 0.66$ $(10.7, 95) $ & $-1.52 \pm 0.62$ $(6.1, 96)$\\ & $0.41 \pm 0.27$ $(7.5, 38)$ & $0.71 \pm 0.25$ $(5.8, 74)$ & $-1.03 \pm 0.61$ $(9.2, 38) $ & $-2.18 \pm 0.6$ $(6.1, 74)$\\ \midrule
1156+295 & $0.76 \pm 0.16$ $(3.6, 61)$ & $0.73 \pm 0.18$ $(1.6, 27)$ & $-2.47 \pm 0.39$ $(1.1, 61) $ & $-2.24 \pm 0.49$ $(2.5, 27)$\\ & $0.55 \pm 0.14$ $(0.5, 31)$ & $0.73 \pm 0.18$ $(1.6, 26)$ & $-2.21 \pm 0.47$ $(1.4, 31) $ & $-2.44 \pm 0.49$ $(2.7, 26)$\\ \midrule
1219+285 & $0.77 \pm 0.2$ $(2.1, 84)$ & $0.73 \pm 0.23$ $(1.3, 104)$ & $-1.78 \pm 0.38$ $(1.3, 84) $ & $-1.76 \pm 0.5$ $(1.9, 104)$\\ & $0.98 \pm 0.18$ $(0.8, 36)$ & $0.74 \pm 0.26$ $(1.7, 71)$ & $-2.58 \pm 0.32$ $(0.5, 36) $ & $-1.56 \pm 0.52$ $(2.2, 71)$\\ \midrule
1222+216 & $0.85 \pm 0.19$ $(2.2, 173)$ & $0.55 \pm 0.17$ $(0.7, 66)$ & $-2.17 \pm 0.58$ $(3.5, 173) $ & $-2.5 \pm 0.58$ $(2.3, 66)$\\ & $0.93 \pm 0.14$ $(0.4, 50)$ & $0.53 \pm 0.18$ $(0.8, 51)$ & $-3.47 \pm 0.52$ $(2.1, 50) $ & $-2.89 \pm 0.61$ $(2.7, 51)$\\ \midrule
1226+023 & $0.68 \pm 0.21$ $(12.9, 267)$ & $0.5 \pm 0.23$ $(4.7, 134)$ & $-1.86 \pm 0.47$ $(9.1, 267) $ & $-1.65 \pm 0.47$ $(3.6, 134)$\\ & $0.81 \pm 0.24$ $(9.1, 172)$ & $0.48 \pm 0.23$ $(5.2, 119)$ & $-1.93 \pm 0.5$ $(6.8, 172) $ & $-1.63 \pm 0.46$ $(3.8, 119)$\\ \midrule
1253-055 & $0.85 \pm 0.3$ $(10.1, 123)$ & $0.44 \pm 0.24$ $(3.7, 271)$ & $-2.33 \pm 0.63$ $(9.2, 123) $ & $-1.95 \pm 0.69$ $(4.6, 271)$\\ & $0.92 \pm 0.22$ $(4.0, 49)$ & $0.6 \pm 0.24$ $(3.6, 210)$ & $-2.96 \pm 0.5$ $(4.2, 49) $ & $-2.53 \pm 0.69$ $(4.3, 210)$\\ \midrule
1308+326 & $0.9 \pm 0.25$ $(2.8, 118)$ & $0.87 \pm 0.25$ $(2.6, 81)$ & $-2.58 \pm 0.61$ $(3.6, 118) $ & $-2.81 \pm 0.66$ $(3.7, 81)$\\ & $0.69 \pm 0.21$ $(1.9, 46)$ & $0.87 \pm 0.25$ $(2.6, 81)$ & $-2.88 \pm 0.56$ $(3.1, 46) $ & $-2.81 \pm 0.66$ $(3.7, 81)$\\ \midrule
1633+382 & $0.74 \pm 0.22$ $(4.8, 149)$ & $0.71 \pm 0.22$ $(2.9, 351)$ & $-2.16 \pm 0.47$ $(5.3, 149) $ & $-1.9 \pm 0.45$ $(4.2, 351)$\\ & $0.75 \pm 0.23$ $(5.2, 139)$ & $0.72 \pm 0.22$ $(3.0, 300)$ & $-1.84 \pm 0.49$ $(7.1, 139) $ & $-1.92 \pm 0.44$ $(4.6, 300)$\\ \midrule
1652+398 & $0.64 \pm 0.14$ $(6.5, 60)$ & $0.51 \pm 0.22$ $(1.4, 17)$ & $-1.72 \pm 0.18$ $(1.2, 60) $ & $-1.33 \pm 0.39$ $(2.0, 17)$\\ & $0.53 \pm 0.15$ $(0.6, 22)$ & $0.81 \pm 0.16$ $(0.8, 13)$ & $-1.37 \pm 0.17$ $(0.3, 22) $ & $-1.51 \pm 0.16$ $(0.4, 13)$\\ \midrule
1730-130 & $0.68 \pm 0.2$ $(9.4, 140)$ & $1.04 \pm 0.26$ $(2.8, 165)$ & $-1.63 \pm 0.46$ $(6.5, 140) $ & $-2.78 \pm 0.48$ $(2.9, 165)$\\ & $0.85 \pm 0.21$ $(5.8, 66)$ & $1.05 \pm 0.26$ $(2.9, 149)$ & $-2.64 \pm 0.39$ $(3.7, 66) $ & $-2.79 \pm 0.44$ $(2.8, 149)$\\ \midrule
1749+096 & $0.98 \pm 0.16$ $(0.9, 82)$ & $0.79 \pm 0.26$ $(2.5, 145)$ & $-3.1 \pm 0.34$ $(1.0, 82) $ & $-3.19 \pm 0.64$ $(3.7, 145)$\\ & $0.99 \pm 0.17$ $(1.0, 65)$ & $0.95 \pm 0.27$ $(2.5, 107)$ & $-3.34 \pm 0.38$ $(1.2, 65) $ & $-3.47 \pm 0.54$ $(2.6, 107)$\\ \midrule
2200+420 & $1.59 \pm 0.3$ $(5.1, 446)$ & $0.84 \pm 0.18$ $(3.4, 616)$ & $-2.98 \pm 0.46$ $(6.0, 446) $ & $-2.4 \pm 0.49$ $(3.6, 616)$\\ & $0.91 \pm 0.17$ $(3.6, 354)$ & $0.96 \pm 0.17$ $(3.2, 530)$ & $-2.45 \pm 0.4$ $(3.5, 354) $ & $-2.58 \pm 0.46$ $(3.5, 530)$\\ \midrule
2223-052 & $0.61 \pm 0.24$ $(2.8, 76)$ & $0.89 \pm 0.24$ $(2.5, 54)$ & $-2.49 \pm 0.42$ $(2.5, 76) $ & $-2.45 \pm 0.45$ $(3.7, 54)$\\ & $0.63 \pm 0.25$ $(3.7, 41)$ & $0.96 \pm 0.23$ $(2.3, 46)$ & $-2.75 \pm 0.48$ $(3.8, 41) $ & $-2.44 \pm 0.41$ $(3.2, 46)$\\ \midrule
2230+114 & $0.76 \pm 0.23$ $(6.8, 161)$ & $0.58 \pm 0.26$ $(2.3, 49)$ & $-1.91 \pm 0.41$ $(4.3, 161) $ & $-1.7 \pm 0.53$ $(2.5, 49)$\\ & $0.71 \pm 0.13$ $(1.2, 41)$ & $0.69 \pm 0.27$ $(2.8, 42)$ & $-2.48 \pm 0.35$ $(2.1, 41) $ & $-1.83 \pm 0.54$ $(2.8, 42)$\\ \midrule
2251+158 & $0.77 \pm 0.2$ $(13.6, 132)$ & $0.48 \pm 0.26$ $(3.6, 52)$ & $-1.94 \pm 0.39$ $(8.3, 132) $ & $-1.96 \pm 0.63$ $(5.5, 52)$\\ & $0.88 \pm 0.2$ $(9.7, 72)$ & $0.48 \pm 0.28$ $(4.0, 46)$ & $-2.46 \pm 0.4$ $(6.4, 72) $ & $-1.75 \pm 0.67$ $(5.9, 46)$\\ 
\bottomrule
\end{tabular}
}
\tablefoot{For each source, the top row refers to the full jet length probed and the bottom row refers to the overlap region probed at both frequencies; Columns: (1) IAU B\,1950 name, (2) Fit parameter of jet-geometry at 15\,GHz $l$ (reduced $\chi ^2$, degrees of freedom), (3) Fit parameter of jet-geometry at 43\,GHz $l$ (reduced $\chi ^2$, degrees of freedom), (4) Fit parameter of brightness-temperature gradient at 15\,GHz $l$ (reduced $\chi ^2$, degrees if freedom), (3) Fit parameter of brightness-temperature gradient at 43\,GHz $l$ (reduced $\chi ^2$, degrees of freedom) }

\end{table*}

\begin{figure*}[htpb]
\centering
\includegraphics[width=0.85\textwidth]{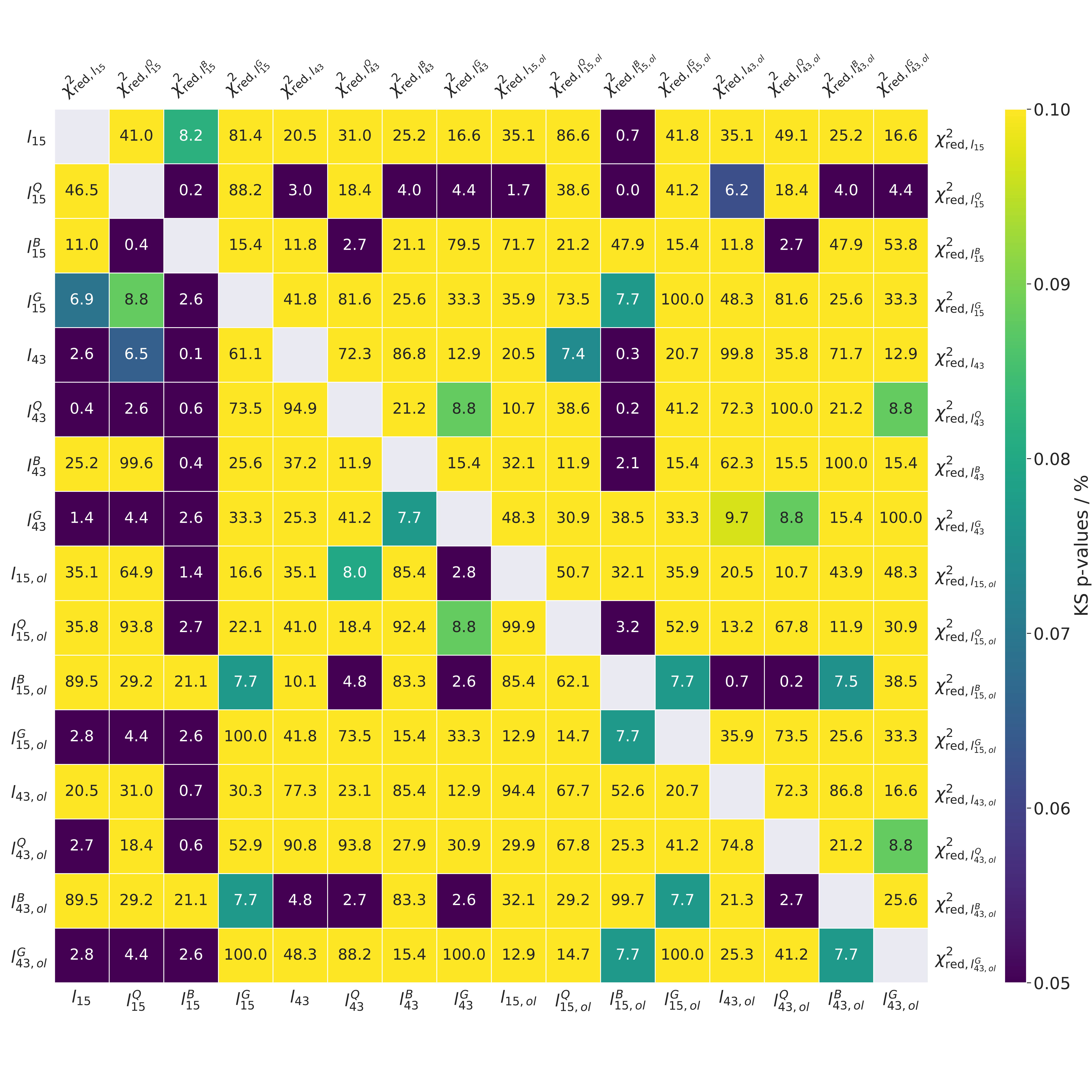}\hfill
\caption{The p-values for each pair of parameters involving the jet geometry compared with the two-sided KS test. 
 The fields are color coded ranging from $p<5\,\%$ (dark blue) to $p\geq 10\,\%$ (yellow). The lower triangle, where all $l$ values are compared and the upper triangle where all $\chi^2 _{\mathrm{red}}$-values are compared, in the plot have to be read separately. The indices (Q,B,G) denote FSRQs, BL Lacs and radio galaxies, respectively. The index "ol" denotes values that are calculated in the overlap region in contrast to the entire jet length. These were introduced in order to make the plots more readable
 }
\label{plot:KS_matrices1}
\end{figure*}

\begin{figure*}[htpb]
\centering
\includegraphics[width=0.85\textwidth]{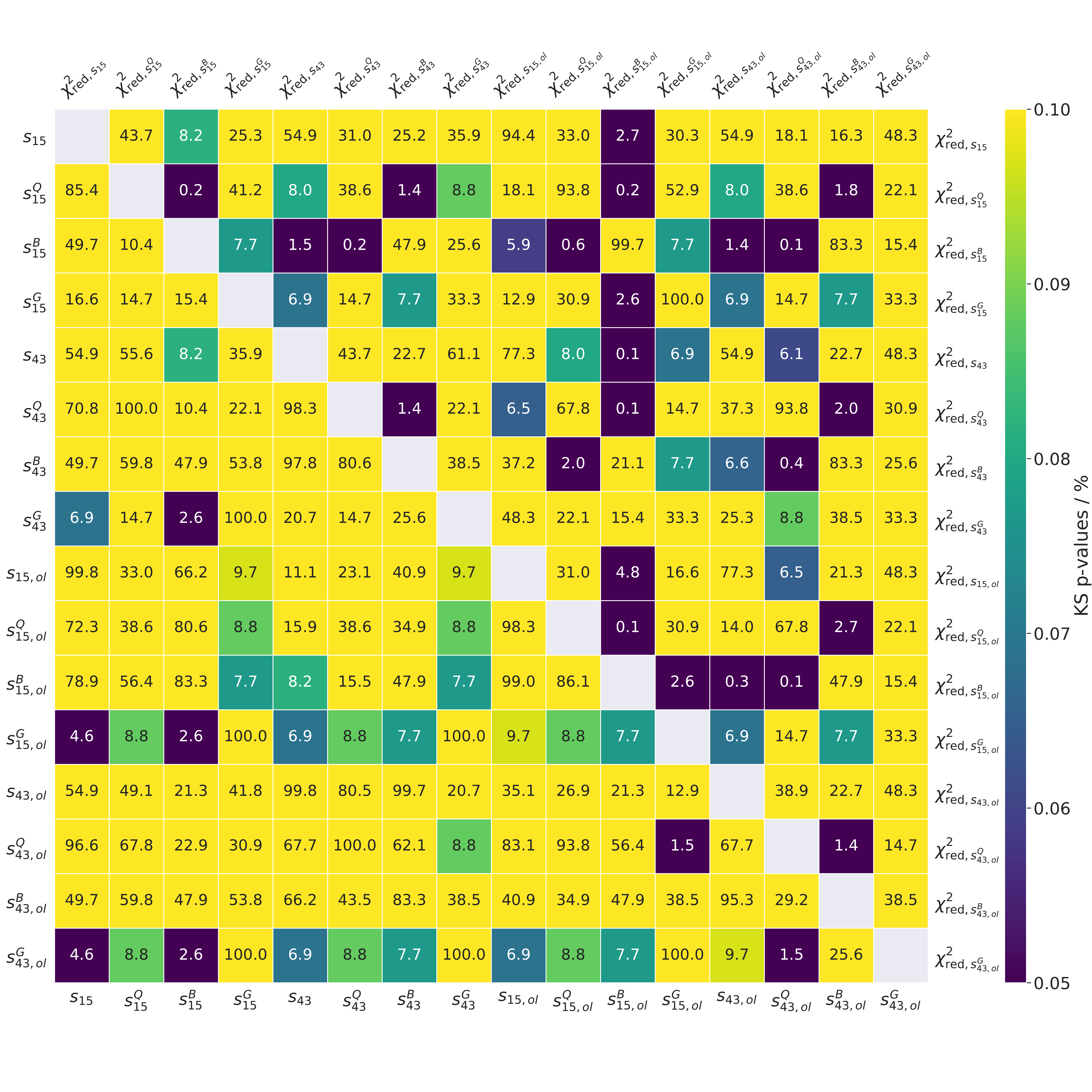}\hfill
\caption{The p-values for each pair of parameters involving the $T_\mathrm{B}$ gradients compared with the two-sided KS test. 
 The fields are color coded ranging from $p<5\,\%$ (dark blue) to $p\geq 10\,\%$ (yellow). The lower triangle, where all $s$ values are compared and the upper triangle where all $\chi^2 _{\mathrm{red}}$-values are compared, in the plot have to be read separately. 
 }
\label{plot:KS_matrices2}
\end{figure*}

A statistical comparison was carried out to find whether the scaling parameters show systematic differences between the FSRQ and BL Lac classes.  
The two RGs in the sample, 3C 111 and 3C 120, show a behavior similar to the FSRQs with a pronounced complex substructure consisting of bright radio knots. They will be discussed in more detail in Sect.~\ref{sect:individual_sources}.
The distributions of the scaling parameters are shown in Figs.~\ref{plot:l-values}-\ref{plot:s-values_sc}. All histograms are shown with a standard binning and a binning method based on Bayesian block analysis \citep{Scargle_bayblock}, implemented in astropy (\citeauthor{astropy:2013} \citeyear{astropy:2013}, \citeauthor{astropy:2018} \citeyear{astropy:2018}). The Bayesian binning method allows us to pinpoint significant differences between histograms at a glance. For all plots the indices (Q,B,G) denote FSRQs, BL Lacs and radio galaxies, respectively. The index "ol" denotes values that are calculated in the overlap region in contrast to the entire jet length. These were introduced in order to make the often densely packed plots more readable.
In cases of uneven sample sizes, only the Kolmogorov-Smirnov (KS) test was performed. For samples of equal size, {\it viz.} the comparison of both frequencies, also the Spearman, Kendall and Pearson correlation coefficients were calculated. In cases where a difference between two classes are indicated by the KS-test, we checked the robustness of the difference by repeatedly performing the KS test while randomly excluding sources from each class (15 FSRQs, 11 BL\,Lacs) to find out  which and how many sources are driving the possible differences. This was done by randomly excluding up to \sout{ten} \textbf{two} sources from each sample.
The results of the KS-test analysis is shown in Fig.~\ref{plot:KS_matrices1} and Fig.~\ref{plot:KS_matrices2}. 

\subsection{Diameter gradients}
\begin{figure}
\centering
	 \includegraphics[width=0.5\textwidth]{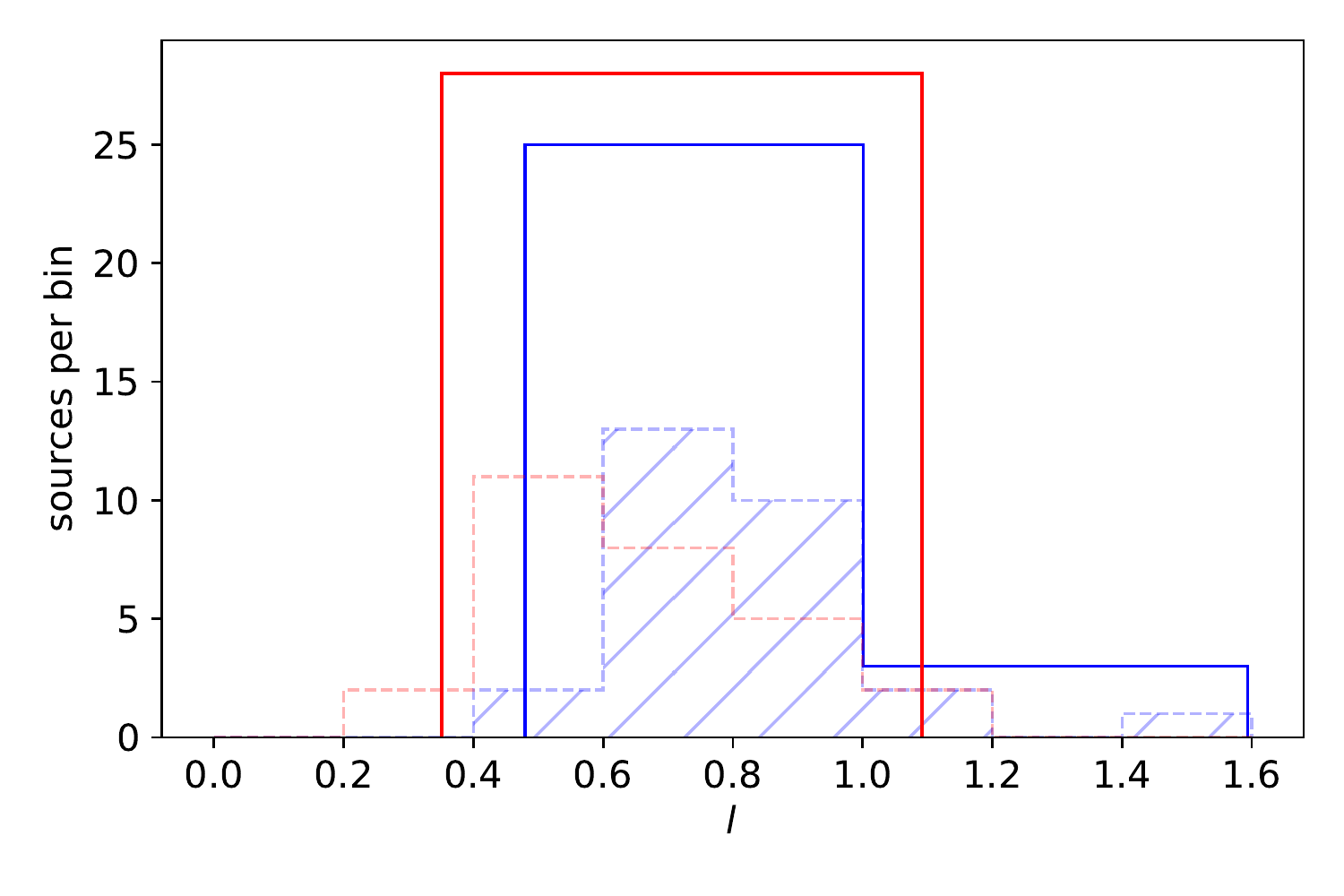}
		\includegraphics[width=0.5\textwidth]{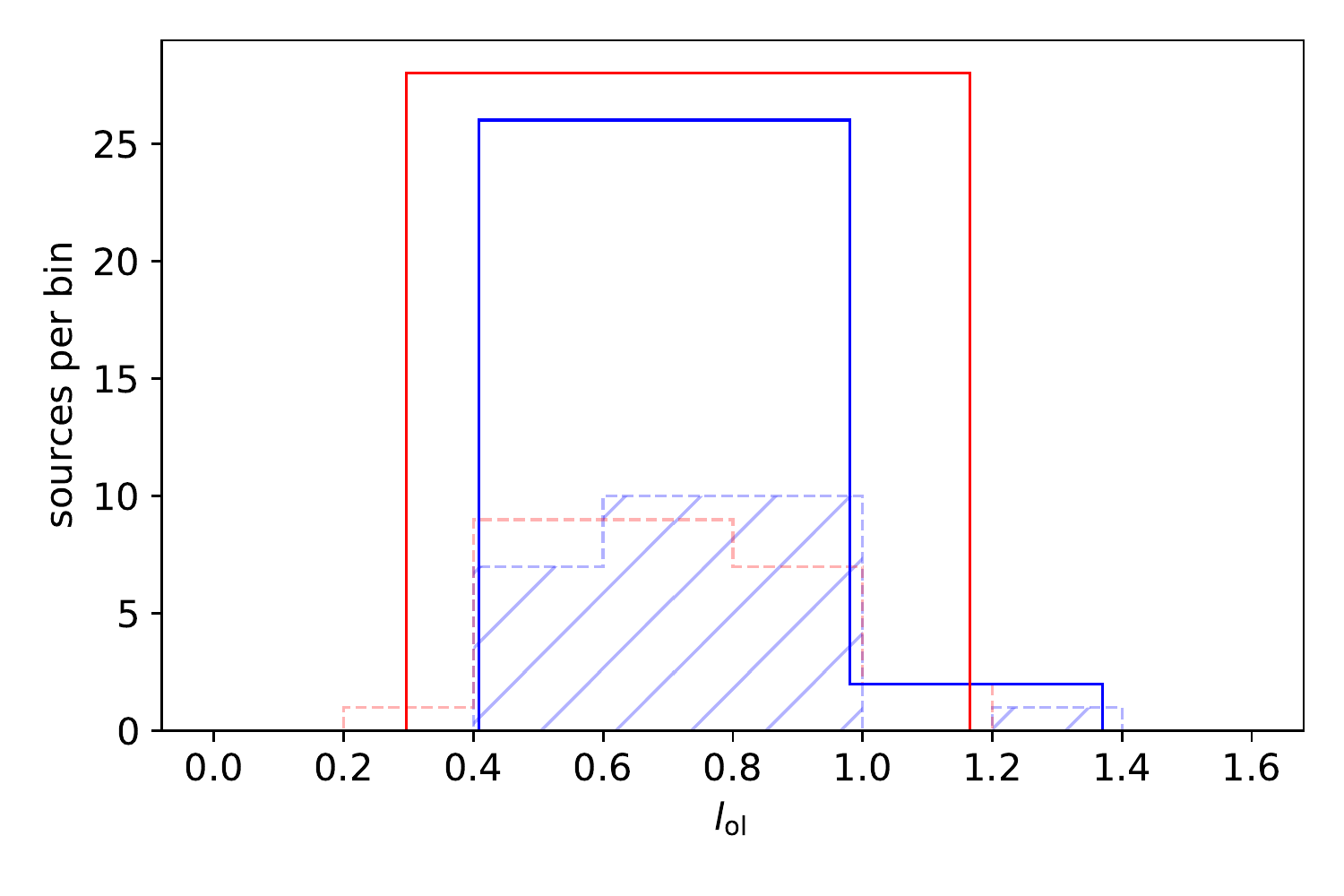}
      \caption{Distribution of $l$-values along the entire jet (top panel) and along the overlap-region (bottom). Values at 15\,GHz are shown in blue and 43\,GHz in red. The histograms are shown for equal bin width of $\Delta l = 0.2$ (dashed lines) and for Bayesian binning (solid lines; see text)}
       \label{plot:l-values}
\end{figure}

For the diameter gradients the results of the statistical comparisons between the frequencies are listed in Table~\ref{tab3} and the results of the comparison of the source classes are shown in Fig~\ref{plot:KS_matrices1}. We find typical values of $l$ in the range between $0.5$ to $1$, i.e., collimated jet geometries. Uncertainties range between $\sim 0.08$ in case of the apparently most uniform jet geometries to $\sim 0.3$ for some complex cases where deviations from a simple $r^l$ power-law behavior are most pronounced. The most common $l$ value at 43\,GHz is around $0.5$ with typically somewhat larger values at 15\,GHz.
\paragraph{Comparison of both frequencies:}
Figure~\ref{plot:l-values} shows the $l$-value distribution at both frequencies along the entire jet detected at the respective frequency (top) and the overlap region (bottom). Along the entire jets  there is a hint that the jets at $43\,\mathrm{GHz}$ are more collimated than the jets at $15\,\mathrm{GHz}$ (KS: $p\sim 2.6\%$ ). This tentative difference vanishes when comparing the jets in the overlap region (KS: $p\sim 94.4\%$).

\begin{figure*}[htpb]
\centering

	 \includegraphics[width=0.47\textwidth]{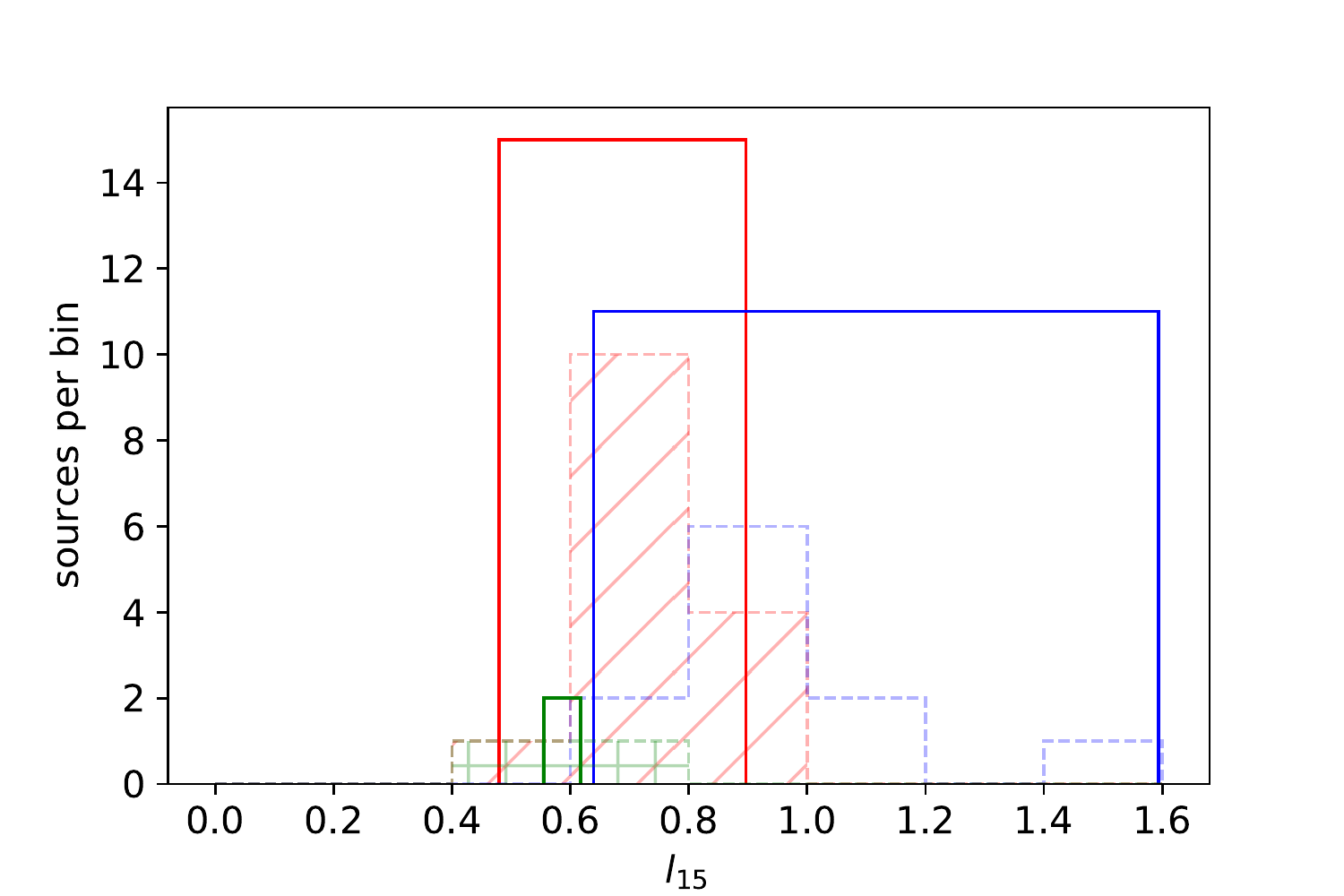}\hfill
	 \includegraphics[width=0.47\textwidth]{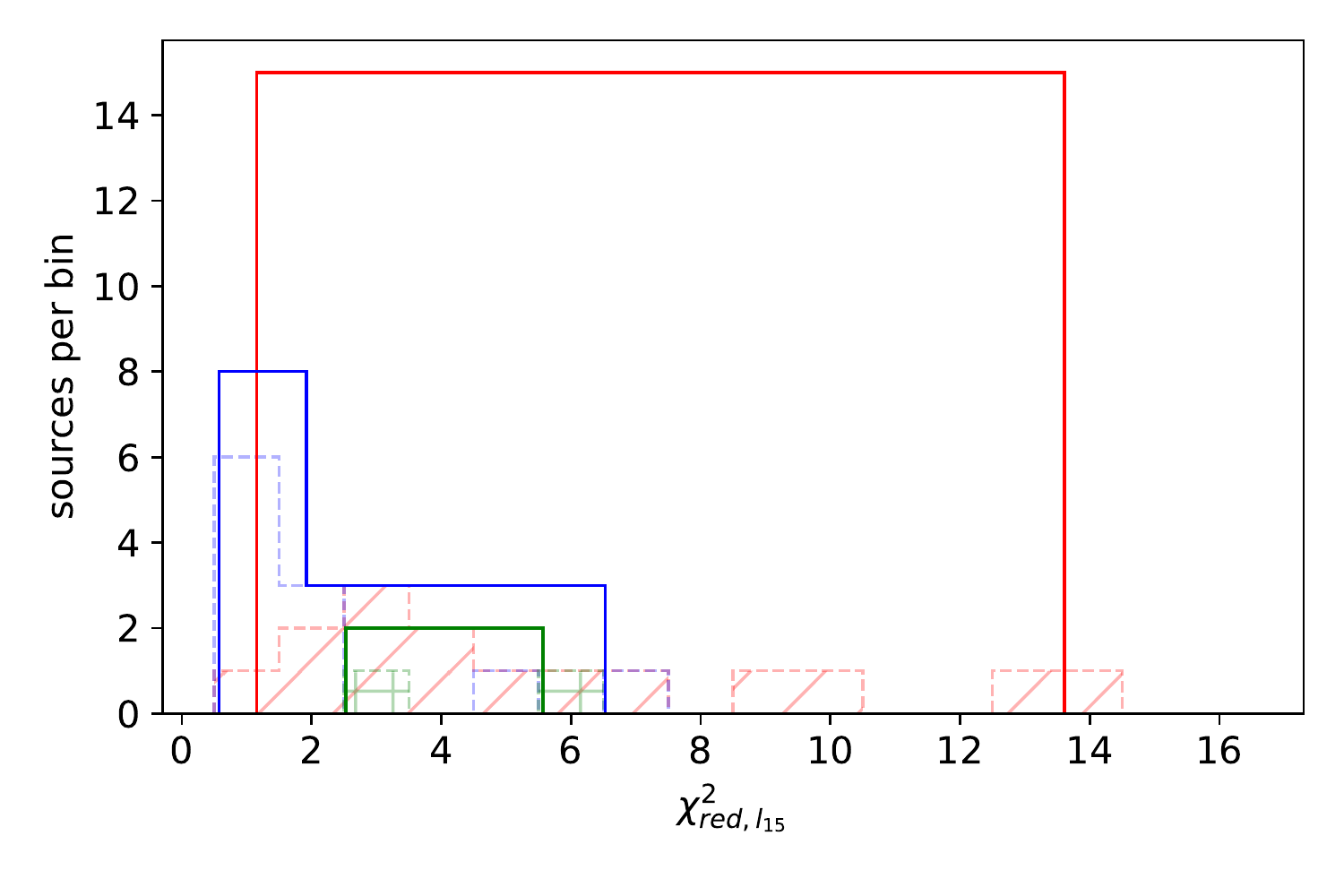}
	 \includegraphics[width=0.47\textwidth]{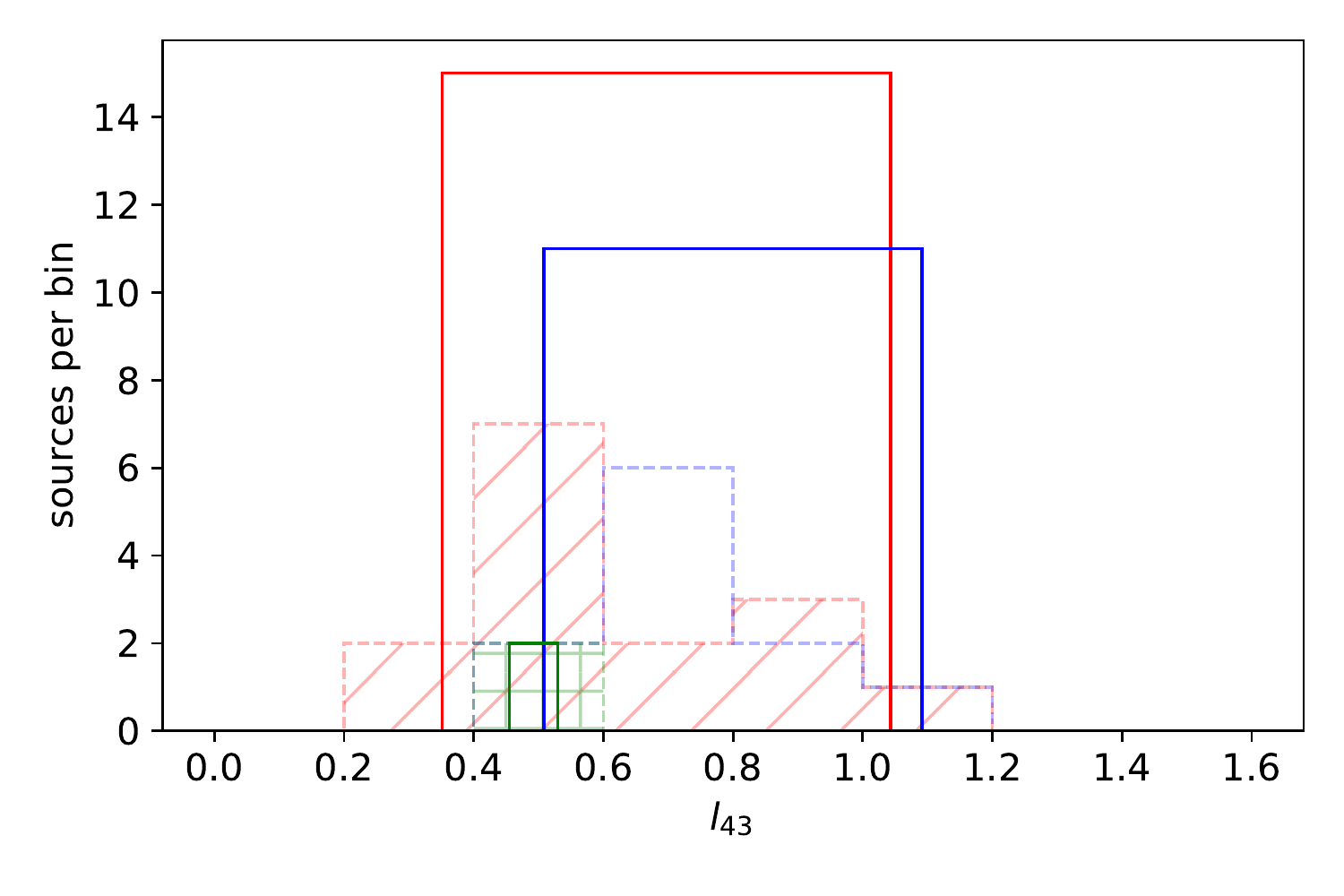}\hfill
	 \includegraphics[width=0.47\textwidth]{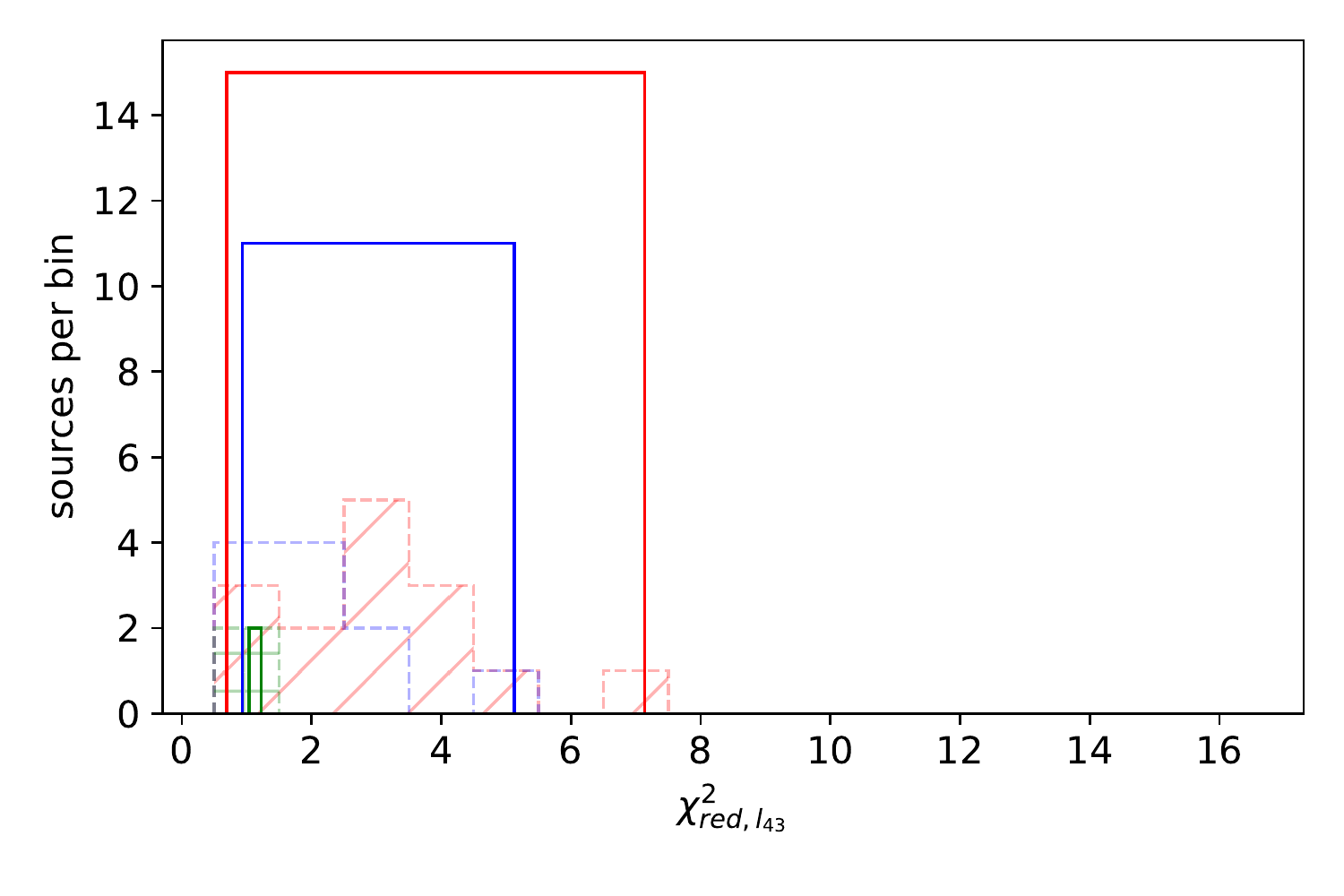}
	
	\includegraphics[width=0.47\textwidth]{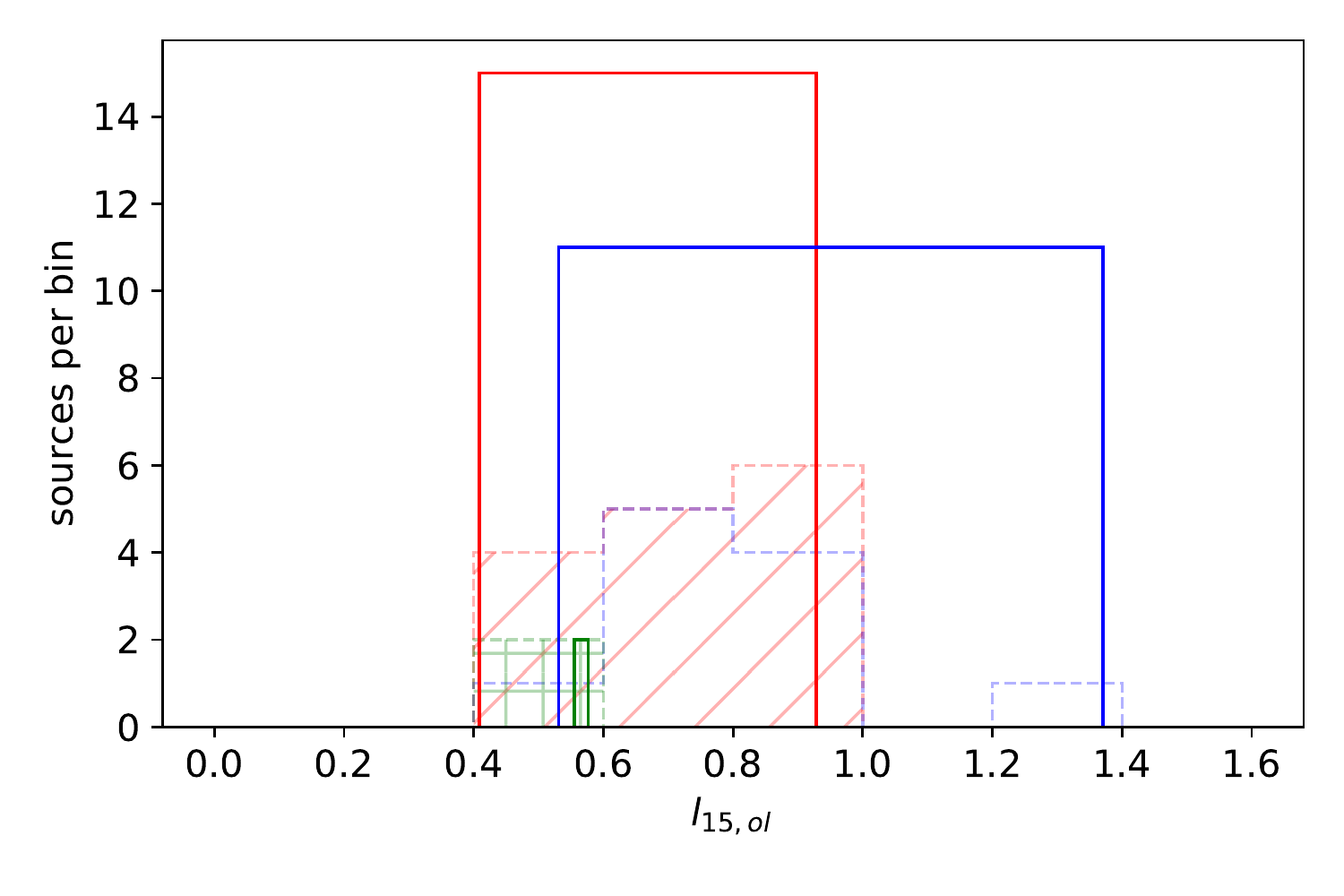}\hfill
	\includegraphics[width=0.47\textwidth]{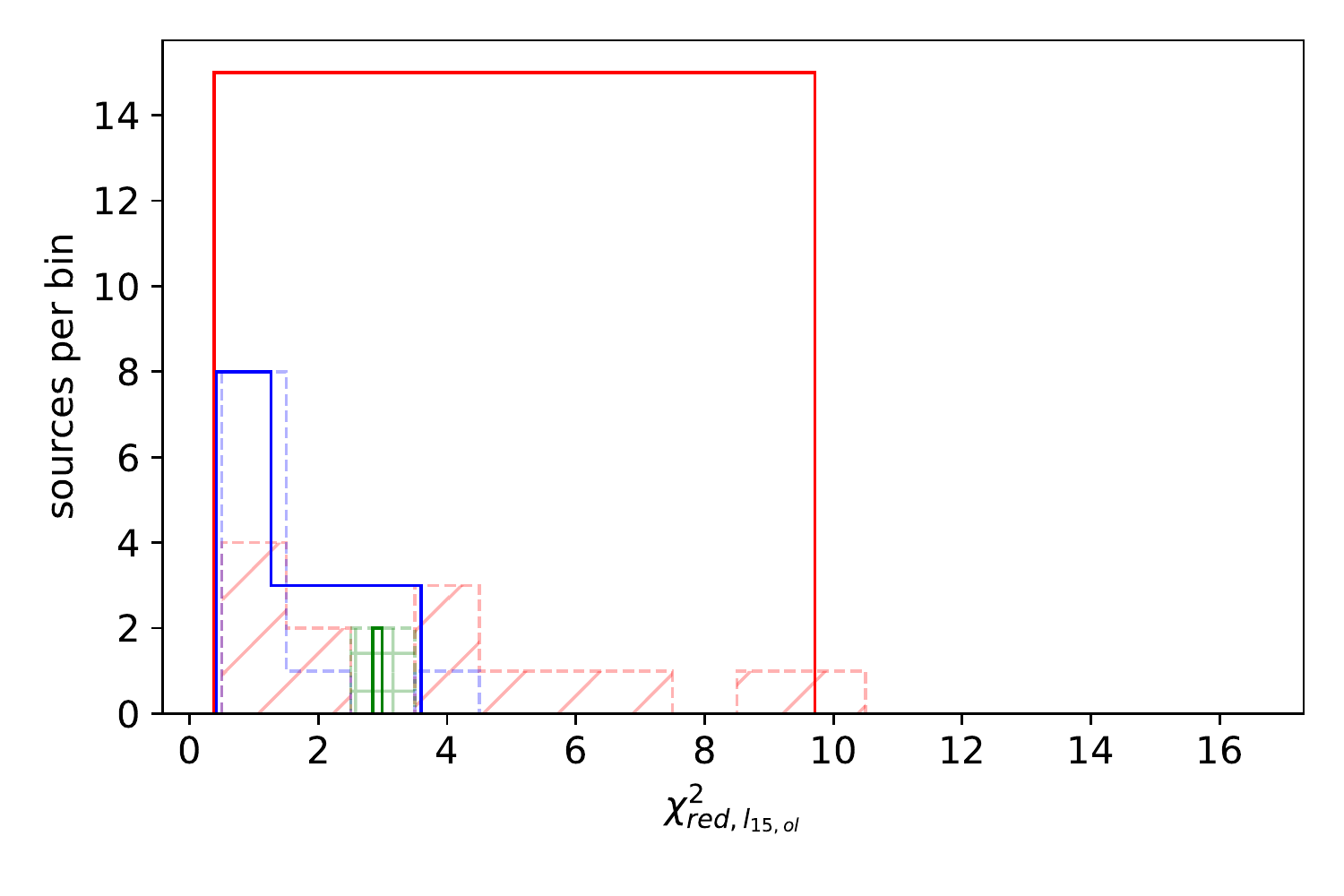}

	\includegraphics[width=0.47\textwidth]{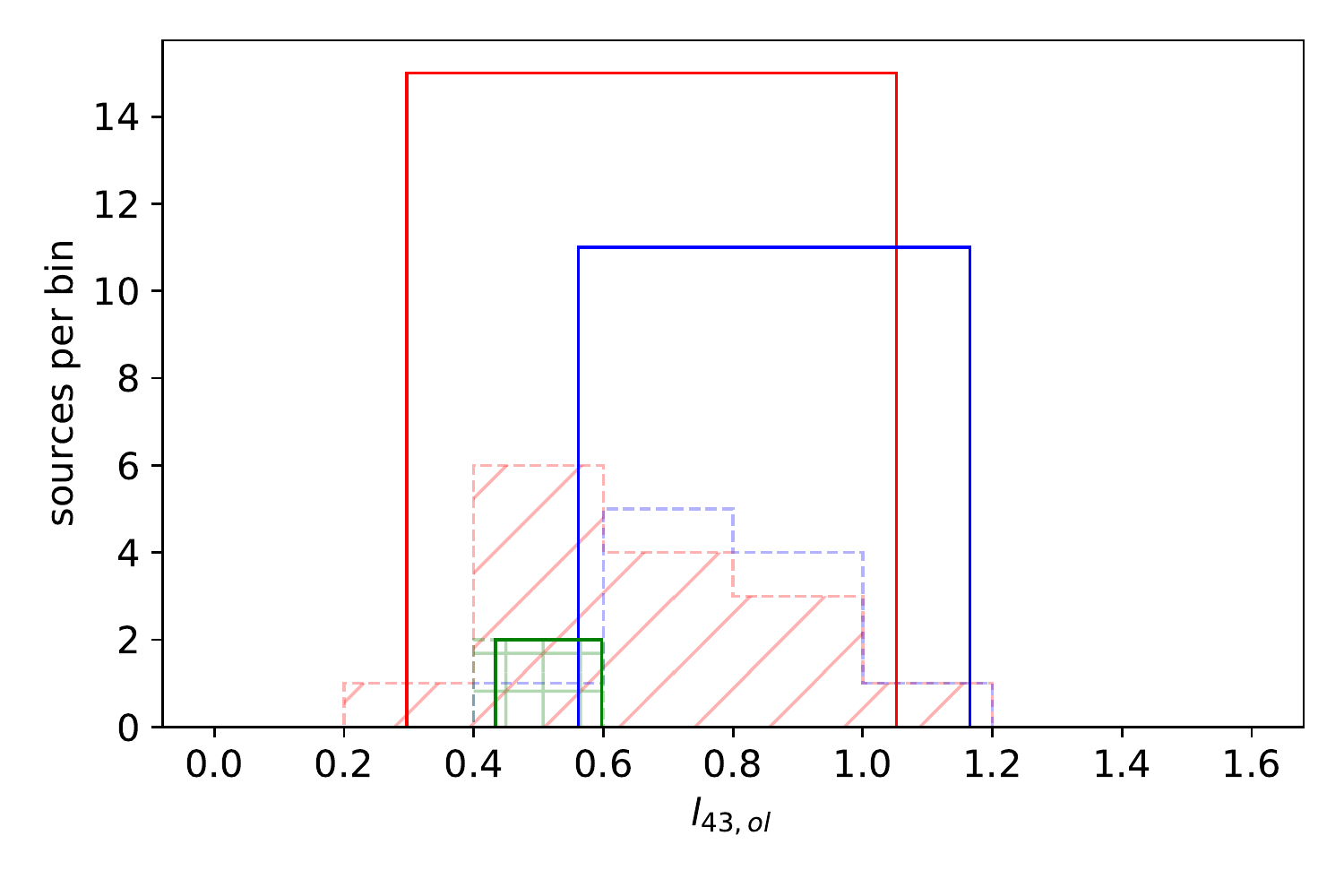}\hfill
	\includegraphics[width=0.47\textwidth]{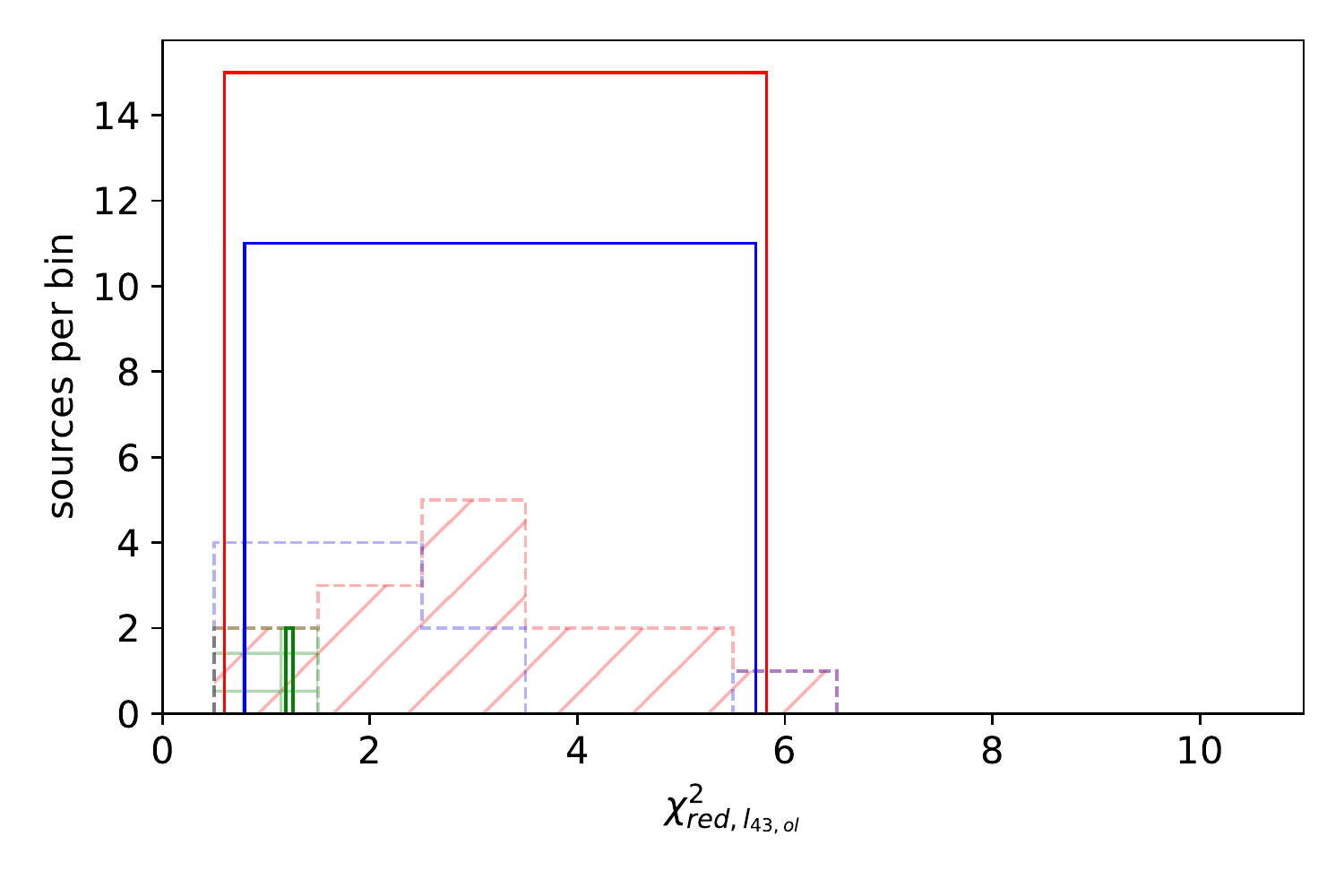}

      \caption{Distributions of the geometry scaling parameter $l$ (\textit{left}) and its variance $\chi^2$ (\textit{right}) shown with Bayesian  (solid lines) and uniform binning (dashed lines) for FSRQs (red) BL\,Lacs (bue), and RGs (green). \textit{Upper four panels}: Distributions for the entire jet at both frequencies; \textit{Lower four panels}: Distributions for the overlap-region at both frequencies. Bin sizes were chosen as 0.2 for the $l$ distributions and 1.0 for the variance distributions.
     }

       \label{plot:l-values_sc}
\end{figure*}

\paragraph{Comparison of the source classes at $15\,\mathrm{GHz}$:}
Figure~\ref{plot:l-values_sc} shows the $l-$value distribution for the 15 GHz jets. The measurements along the entire jet (first row) indicate a difference in jet geometry between the FSRQs and the BL\,Lac objects (KS: $p\sim 0.4\%$).
The FSRQs show smaller $l$-values and thus stronger collimated jets than the BL\,Lac objects. The two radio galaxies 3C\,111 and 3C\,120 show collimated jets in the same regime as the FSRQs. As before in the case of the 15\,GHz to 43\,GHz comparison, however, when considering the overlap regions (third row), the differences between FSRQs and BL\,Lacs in terms of $l$  become  insignificant (KS: $p\sim  62.1\%$). When studying the distribution of the $\chi ^2 _{\mathrm{red}}$ values for the jet-geometry power-law fits (Fig.~\ref{plot:l-values_sc}, top row, right column) we find a pronounced difference between the FSRQs and the BL\,Lacs (KS: $p\sim 0.2\%$). A simple power law fit is a systematically worse description for the FSRQs than for the BL\,Lacs, regarding the $\chi ^2 _{\mathrm{red}}$ values. This difference does not disappear when comparing the overlap regions in the jet (third row right column, KS: $p\sim 3.2\%$).

\paragraph{Comparison of the source classes at $43\,\mathrm{GHz}$:}
At $43\, \mathrm{GHz}$ the $l$-values measured along the entire jet (Fig.~\ref{plot:l-values_sc}, second row, left column) yield no significant distinction between FSRQs and BL\,Lacs (KS: $p~\sim 11.9\%$) nor between FSRQs and RGs (KS: $p~\sim 41.2\%$). Also the comparison of BL\,Lacs and RGs yield no hint for a difference between the source classes (KS: $p~\sim 7.7\%$). 
In the overlap region, however, BL\,Lacs behave differently than FSRQs (KS: $p\sim 2.7\%$), showing a tendency for larger $l$-values. 
Regarding the $\chi^2_{\mathrm{red}}$-distributions, no difference between FSRQs and BL Lacs can be constituted along the entire jet length and the overlap region, see the right column in Fig~\ref{plot:l-values}, second and fourth row.

\begin{table}
\caption{Correlation coefficients for different statistical tests}
\label{tab3}
\begin{tabular}{cccc}
\toprule
Pair of  & Statistical & Correlation  & p-value   \\
 parameters  & test &  Coefficient &   \\
\midrule
\textbf{diameter gradients}& & & \\
$l_{15}-l_{43}$ & KS& & $0.026$ \\ 
 				& Pearson & 0.42& 0.027\\
  				& Spearman & 0.37& 0.056\\
  				& $\tau _{\mathrm{kendall}}$ & 0.27& 0.044\\
&&&\\
$l_{15,\mathrm{overlap}}-l_{43,\mathrm{overlap}}$ & KS& & $0.94$ \\ 
 				& Pearson & 0.43& $0.021$\\
  				& Spearman & 0.25& $0.20$\\
  				& $\tau _{\mathrm{kendall}}$ & 0.15& $0.27$\\
  				
\textbf{$T_\mathrm{B}$ gradients}& & & \\
$s_{15}-s_{43}$ & KS& & $0.55$ \\ 
 				& Pearson & 0.61& $5.3\times 10^{-4}$\\
  				& Spearman & 0.59& 0.0011\\
  				& $\tau _{\mathrm{kendall}}$ & 0.45& $7.8 \times10^{-4}$\\
&&&\\
$s_{15,\mathrm{overlap}}-s_{43,\mathrm{overlap}}$ & KS& & $0.35$ \\ 
 				& pearson & 0.73& $1.0\times 10^{-5}$\\
  				& spearman & 0.70& $4.0\times 10^{-5}$\\
  				& $\tau _{\mathrm{kendall}}$ & 0.51& $1.5\times 10^{-4}$\\

\bottomrule
\end{tabular}
\end{table}

\subsection{Brightness-temperature gradients}
\label{sec:overall_s_results}
For the power-law brightness-temperature gradient index $s$ the results of the statistical comparisons between the frequencies are listed in Table~\ref{tab3} and the results of the comparison of the source classes are shown in Fig~\ref{plot:KS_matrices2}. Most sources show $T_b$ power law indices in the range of $-3\lesssim s \lesssim -1.5$ with the width of the distribution somewhat increasing when only the overlap region at both frequencies is considered. Uncertainties of individual $s$-values range from 0.2 to 0.7, reflected in the large Bayesian-block widths. Most commonly, the slopes along the full jet widths lie in the range between $-2$ and $-1.5$.

\begin{figure}
\centering
	 \includegraphics[width=0.49\textwidth]{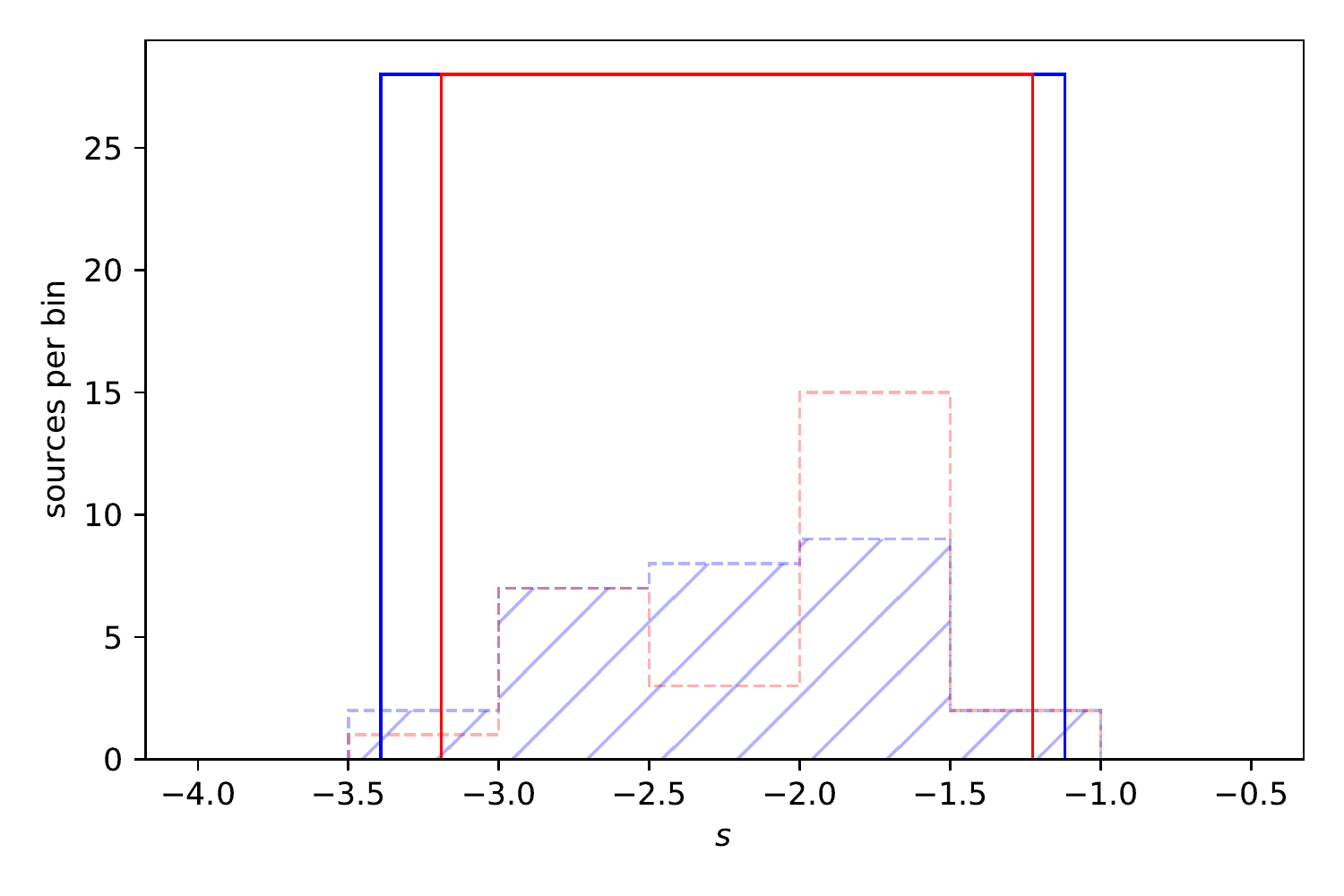}
		\includegraphics[width=0.49\textwidth]{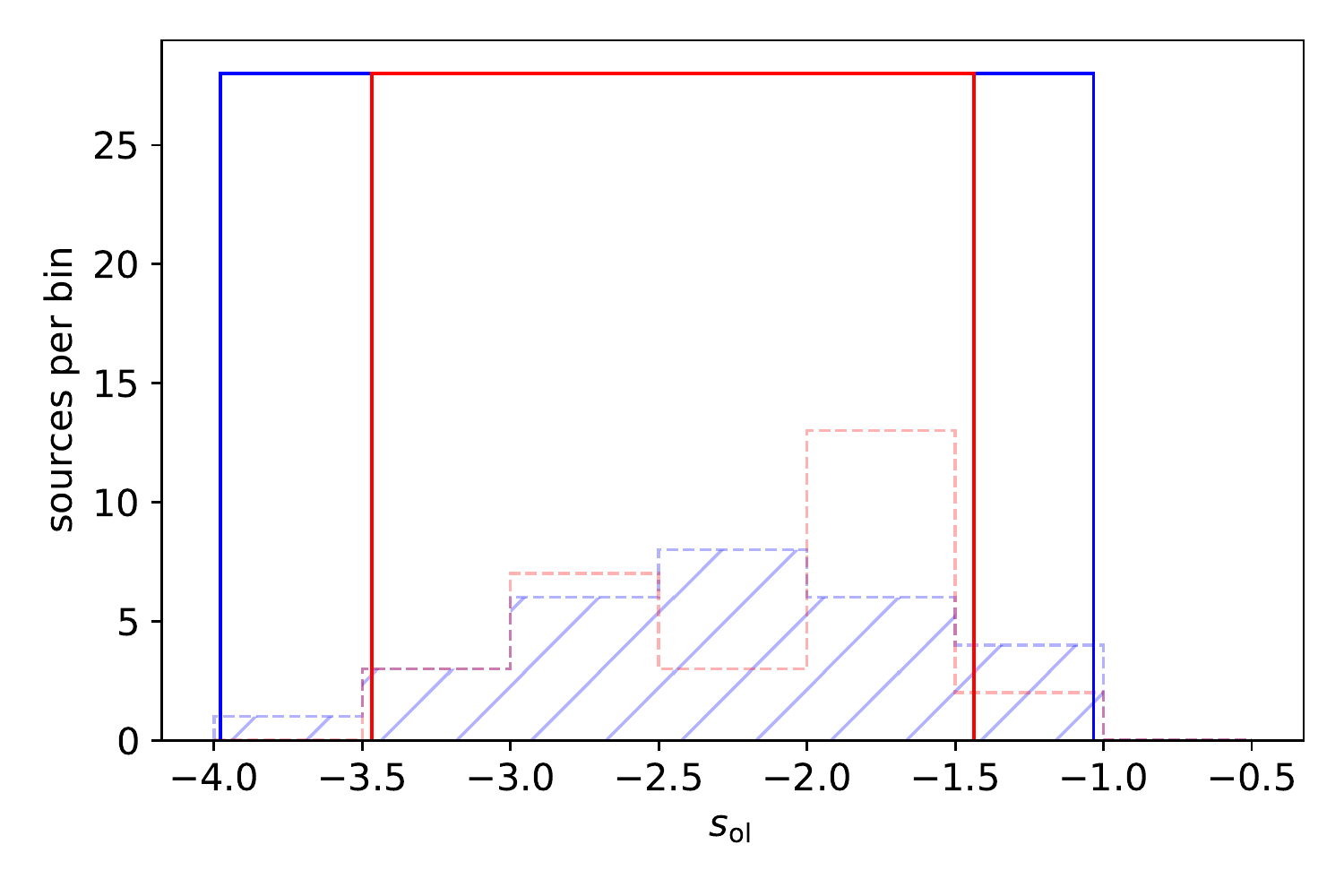}
      \caption{Distribution of power-law slopes $s$ of the brightness-temperature gradient along the entire jet (top panel) and along the overlap-region (bottom) at 15\,GHz (blue) and 43\,GHz (red). The histograms are shown for uniform bin width $\Delta s = 0.5$ (dashed lines) and for Bayesian binning (solid lines; see text)}
       \label{plot:s-values}
\end{figure}

\paragraph{Comparison of both frequencies:}
Figure~\ref{plot:s-values} shows the $s$-value distribution of both frequencies evaluated along the entire detected jet at the respective frequency (top) and along the overlap region (bottom). The $s$-values are indistinguishable for both frequencies in the entire jet and overlap regions, respectively.

\begin{figure*}[htpb]
\centering

	 \includegraphics[width=0.47\textwidth]{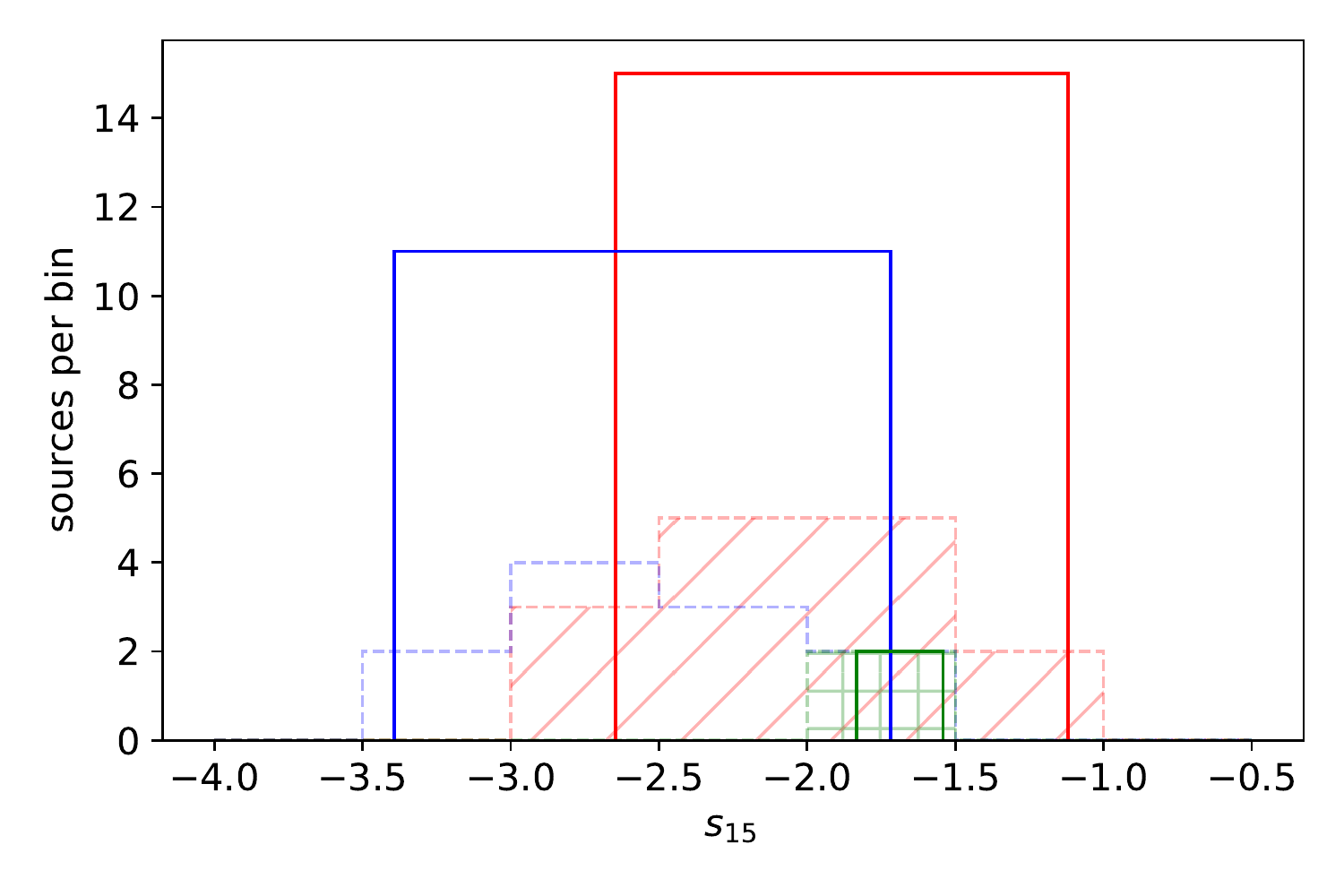}\hfill
	 \includegraphics[width=0.47\textwidth]{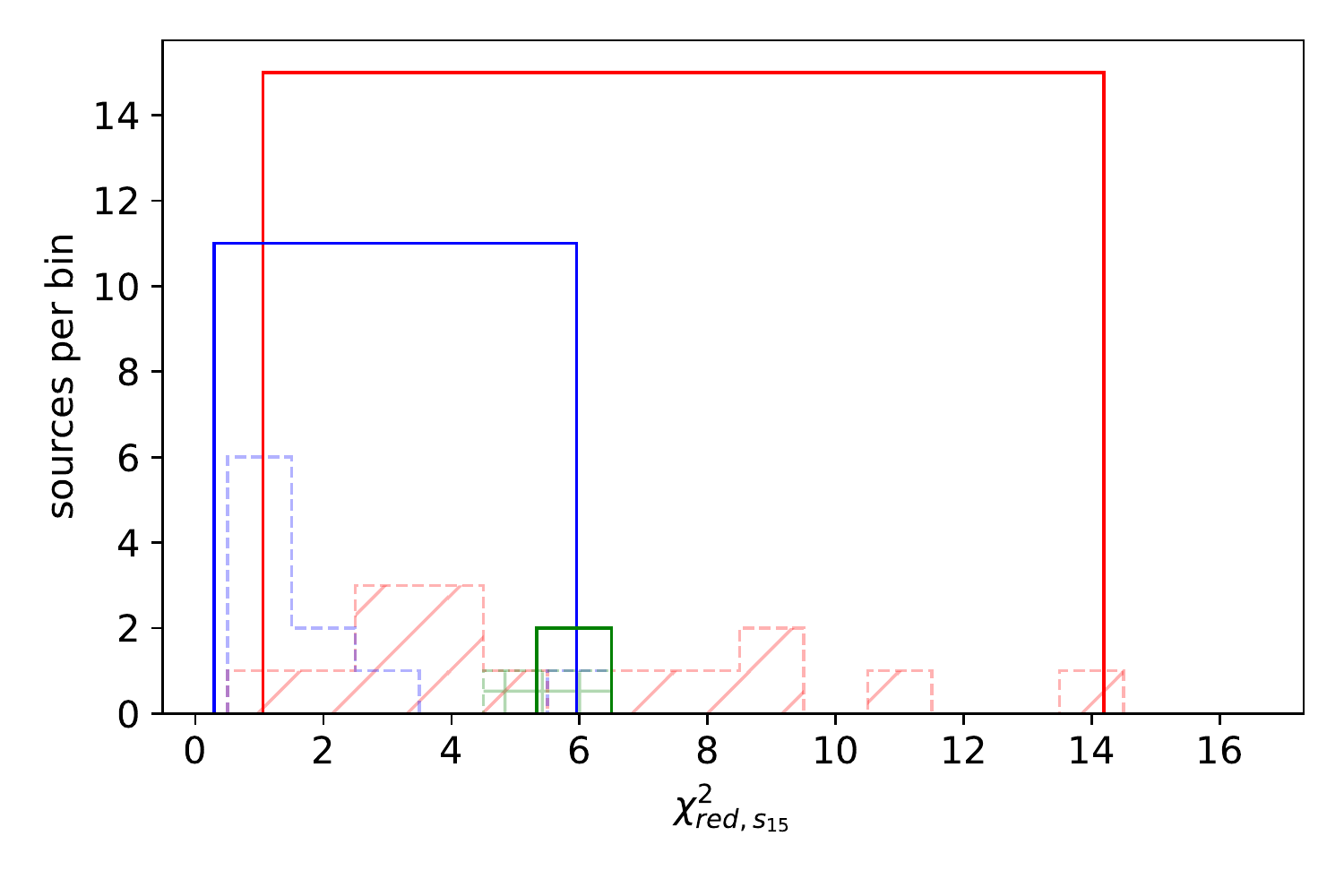}
	 \includegraphics[width=0.47\textwidth]{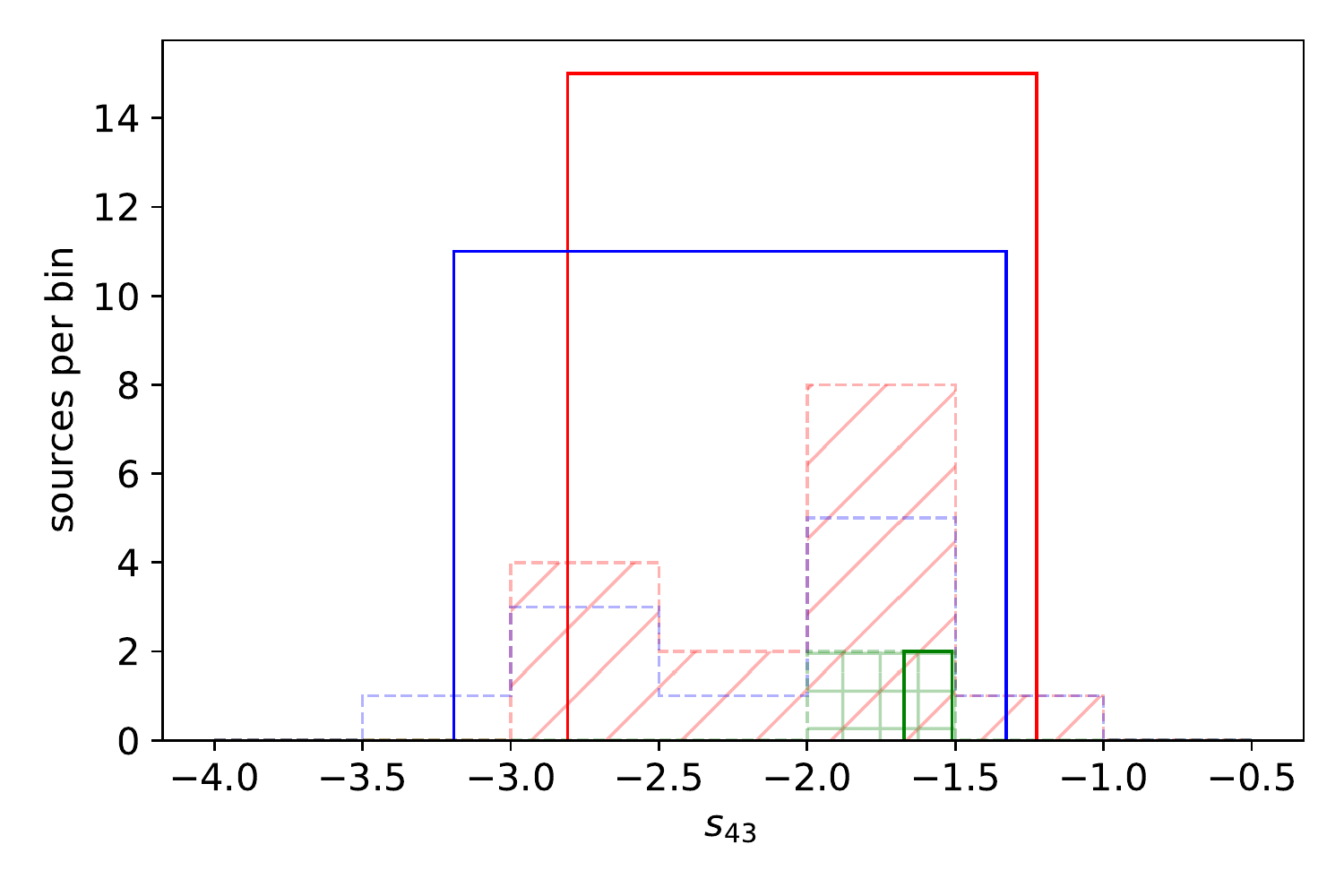}\hfill
	 \includegraphics[width=0.47\textwidth]{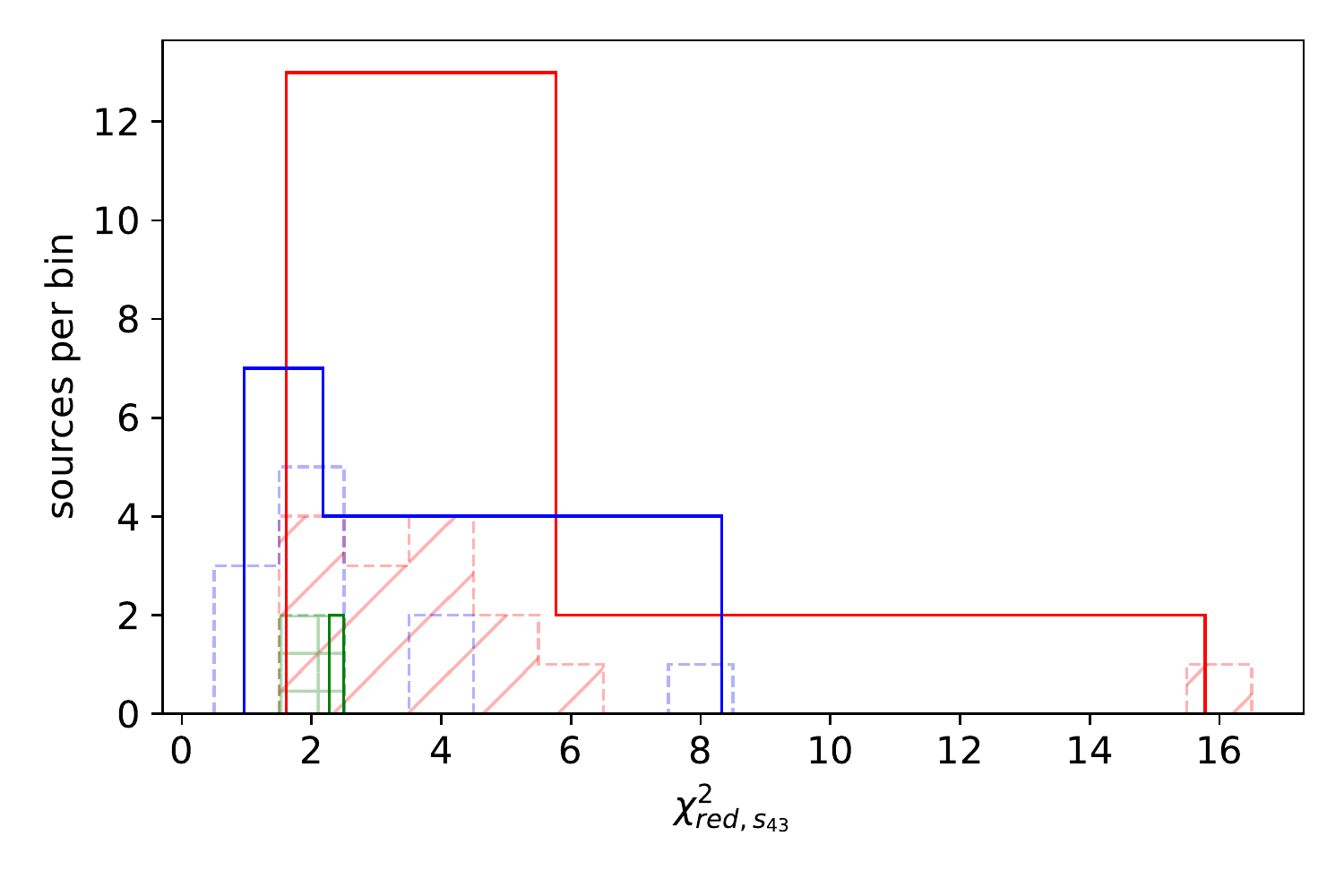}
	
		\includegraphics[width=0.47\textwidth]{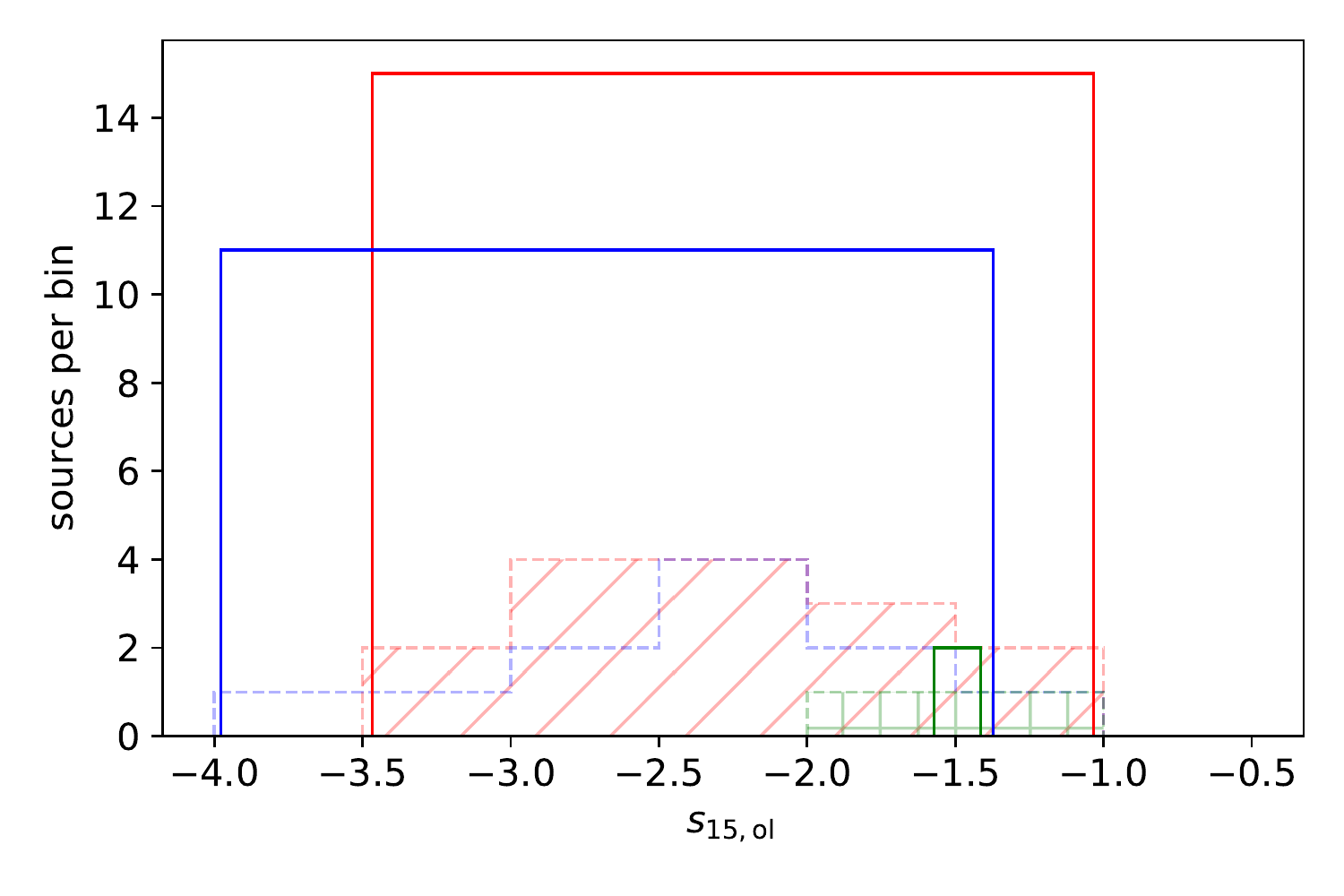}\hfill
		\includegraphics[width=0.47\textwidth]{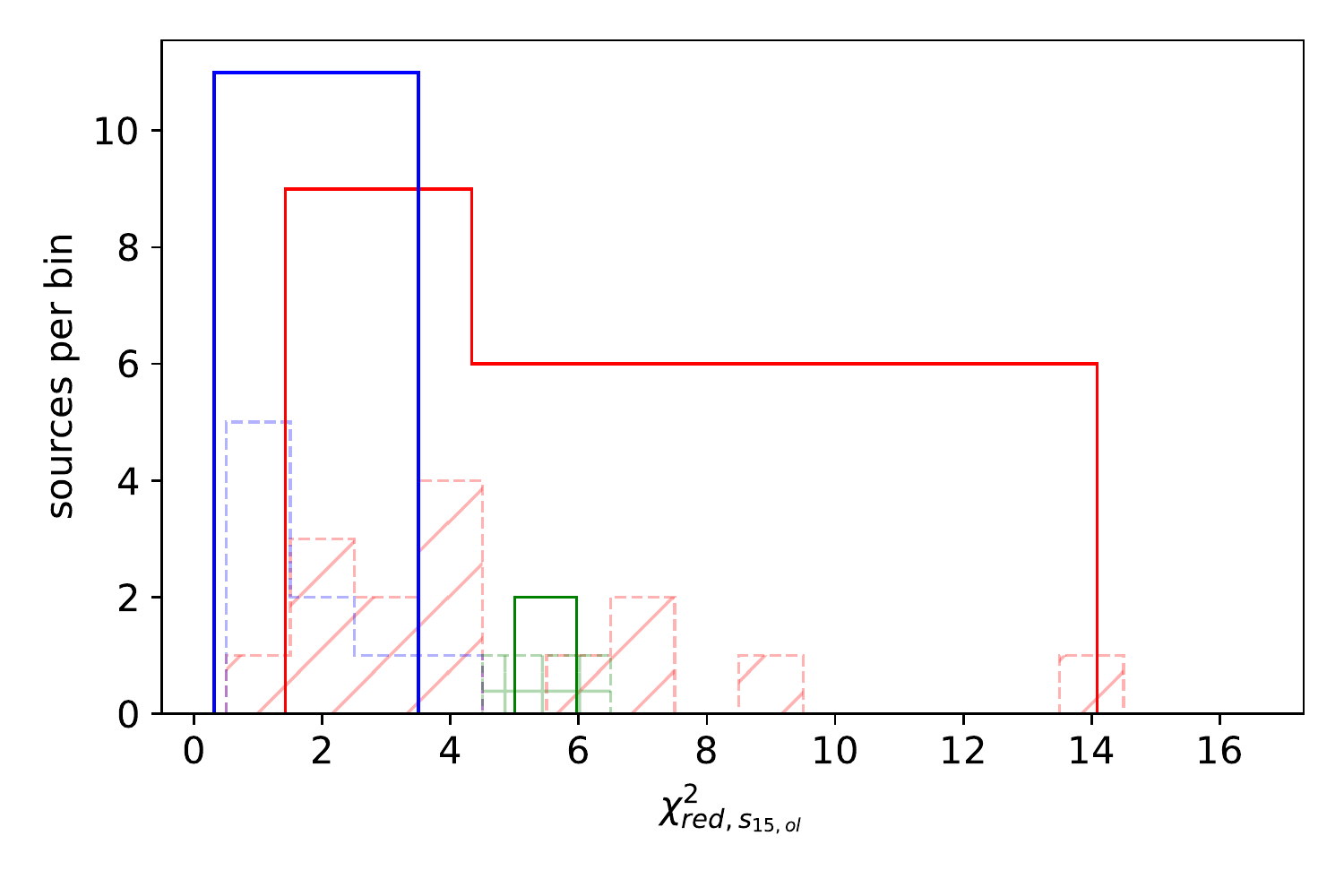}

		\includegraphics[width=0.47\textwidth]{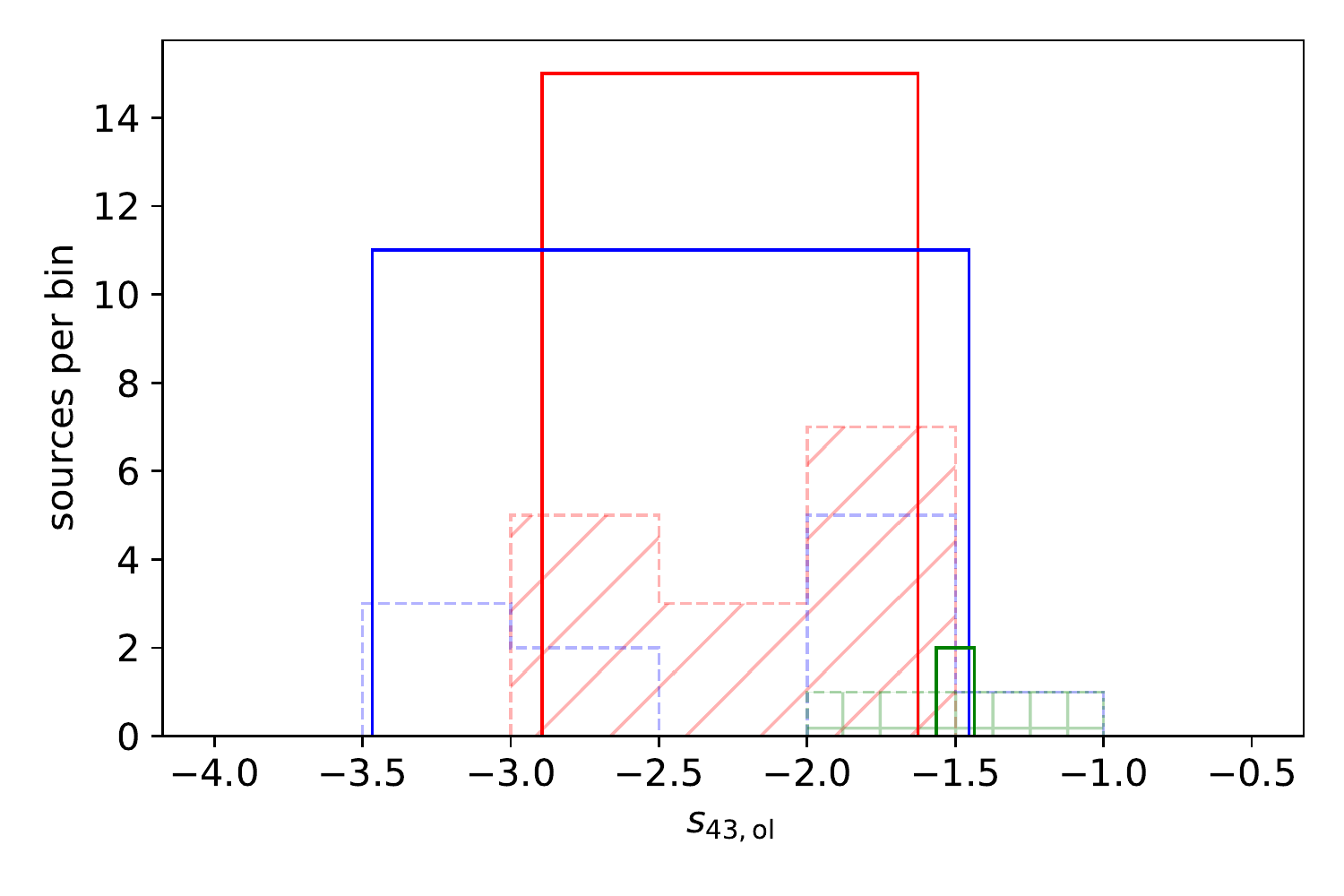}\hfill
		\includegraphics[width=0.47\textwidth]{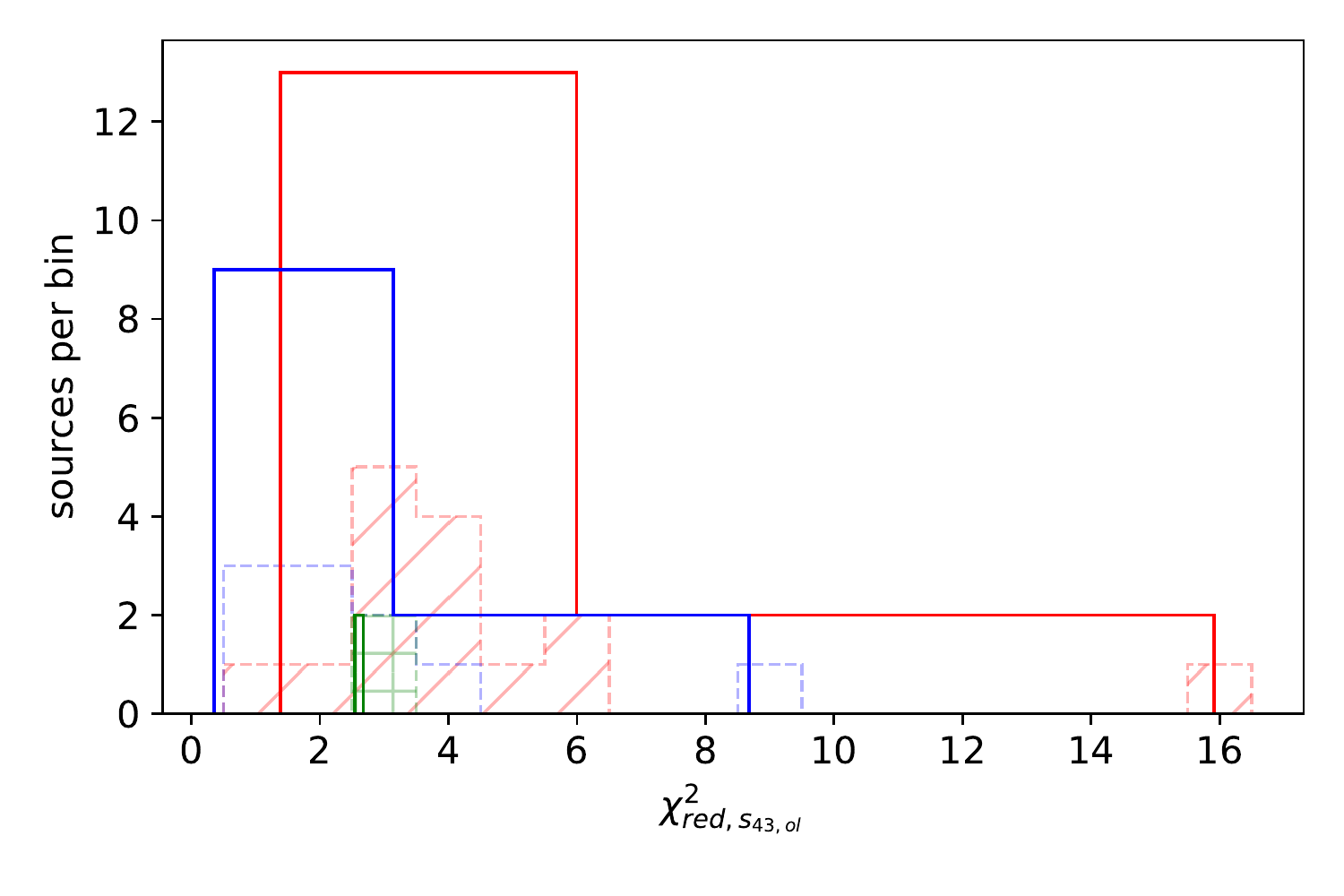}

      \caption{Bayesian binning (solid lines) and uniform binning (dashed lines) of the $T_{\rm B}$ power-law slope distributions. The colors indicate the source classes (red: FSRQs, blue: BL\,Lacs, green: RGs). \textit{Top two rows}: Distributions along the entire jet for both frequencies; \textit{Bottom two lines}: Distributions along the overlap-region of both frequencies. \textit{Left column:} Distribution of $s$-values. For both frequencies, the bin size is 0.5. \textit{Right column} $\chi ^2 _\mathrm{red}$ value distribution describing the brightness temperature. For both frequencies, a unit bin size was chosen.}

       \label{plot:s-values_sc}
\end{figure*}

\paragraph{Comparison of the source classes at $15\,\mathrm{GHz}$:}
Figure~\ref{plot:s-values_sc} (top row) shows the $s$-value distribution at $15\,\mathrm{GHz}$ along the full jet length and the corresponding overlap region, where the colors indicate the source classes. Regarding the distribution of the brightness temperature power law indices, the different source classes are indistinguishable, both along the entire jet length and the overlap region.
The $\chi ^2 _{\mathrm{red,s_{15}}}$ distribution (Fig.~\ref{plot:s-values_sc}, top row), however, indicates a difference between FSRQs and BL\,Lacs along the entire jet length. The BL\,Lac objects show a tendency for smaller $\chi ^2 _{\mathrm{red,s_{15}}}$ values than FSRQs, indicating that a single power law is a better description for the brightness-temperature gradient for BL\,Lacs than for FSRQs (KS: $p\sim 0.2\%$). With reference to the overlap region, this difference also becomes apparent (KS: $p\sim 0.1\%$). 
\paragraph{Comparison of the source classes at $43\,\mathrm{GHz}$:}
Figure~\ref{plot:s-values_sc} shows the $s$-value distribution at $43\,\mathrm{GHz}$ along the full jet length (second row) and the corresponding overlap region (fourth row). The $s$-value distributions are statistically indistinguishable both along the entire jet and the overlap region. As in the case of the jet-geometry power-law fits, the $\chi ^2 _{\mathrm{red,s_{43}}}$ value distributions (Fig.~\ref{plot:s-values_sc}) indicate smaller values for BL\,Lac object than FSRQs along the entire jet lengths and the overlap regions (KS: $p\sim 1.4\%$ in both cases). 

%% file: discussion.tex
\section{Discussion}
In this section, we discuss the general implications of the gradient fits at $15$~GHz and $43$~GHz that we have obtained in the previous section for our sample of 28 blazars and radio galaxies.
A discussion of our findings regarding the individual sources will follow in the next section.

\subsection{Brightness-temperature gradients}
\citet{Pushkarev_Kovalev2012} analyzed the $T_{\rm B}$-gradients for 30 sources at $2\, \mathrm{GHz}$ and $8\,\mathrm{GHz}$ and found a mean value of $\langle s \rangle \sim -2.2$. Furthermore Kravchenko et al (in prep., 2021 priv. com.) find a similar value analyzing the entire MOJAVE data archive at $15~\mathrm{GHz}$.  This is consistent with our smaller dual-frequency sample results (cf. Sect.~\ref{sec:overall_s_results}). The $T_{\rm B}$-gradients are determined by the jet geometry (see below), particle and magnetic-field gradients. We conclude that studies at different frequencies in general lead to consistent results for these physical quantities.
Specifically, for both our frequencies, the $s$ distributions are not clearly distinguishable  along the entire jets and the overlap regions. Moreover, the different source classes of BL\,Lacs and FSRQs do not show any strong differences in terms of the average $T_{\rm B}$-gradient trends.
However, the $\chi ^2 _{\mathrm{red}}$-distribution at both frequencies show that a single power law fit is more appropriate for BL Lacs than for FSRQs, indicating that the latter show a more complex $T_{\rm B}$-distribution along the jet axis due to local surface brightness enhancements, leading to more scatter in our global fits. 
This is consistent with the findings of \citet{Jorstad2017} who find high $T_{\rm B}$-values to be more common in knots of FSRQs as compared to BL\,Lacs. This gives support to the idea of systematically higher Doppler factors and stronger shocks in FSRQ jets and/or possibly more rapid expansion in the jets of BL\,Lacs. 

\subsection{Jet geometry}
Measured along the entire jet lengths there is a systematic difference in collimation between the jets measured at $15\,\mathrm{GHz}$ and $43\, \mathrm{GHz}$. The jets at $43\, \mathrm{GHz}$ appear more collimated (tend to show smaller $l$-values) than at $15\,\mathrm{GHz}$. 
In the overlap region, this difference is not significant.
It is therefore plausible, that the differences arise from the geometrical properties of the jet at larger scales, i.e. that the $l$-values change with $r$. Indeed, \citet{Kovalev2020} showed that in many cases, it is possible to measure a transition from parabolic to conical jet geometries.

At $15\,\mathrm{GHz}$, the source classes show a notable difference when comparing them along the entire jet. FSRQs systematically show lower $l$-values than BL Lacs indicating a stronger jet collimation. This is also found by Kravchenko et al. (in prep., 2021 priv. com.) on a larger 15\,GHz AGN-jet sample. In our study, this difference is again not significantly found in the (smaller) overlap region of the images at both frequencies indicating again that the differences arise further down the jet and reinforcing the idea of a geometry transition on parsec scales.
Analyzing the $\chi ^2 _{\mathrm{red}}$ of the single power-law fits to the jet geometry also yields interesting results. The FSRQs tend to show larger $\chi ^2 _{\mathrm{red}}$ values, both along the respective entire jet and in the overlap regions compared to BL Lacs. This indicates that the geometric structure in FSRQs is more complicated than the smooth power law observed
in BL Lacs. The complexity could arise from regions where the full jets recollimate and widen again locally. Alternatively, some of the model components in FSRQ jets might indeed be representative of local brightness enhancements that are smaller than the full jet width at this distance. An example where this seems to be the case can be seen in Fig.~\ref{modelfitting} (top panel) at the relative coordinates ($-6, -7$).

At $43\, \mathrm{GHz}$ the difference between the source classes is not significantly detected along the entire jet nor in the overlap region. This is true for both, the $l$-values distribution and the $\chi ^2 _{\mathrm{red}}$ - distribution.
This could be simply a result of reduced image quality at the higher frequency. Alternatively, the high-frequency emission could be associated with plasma closer to the inner jet axis (the spine of the jet) while the lower-frequency emission might be dominated by outer layers (sheath). In such a stratified-jet scenario, the difference between FSRQ and BL\,Lac jet geometry that we find at 15\,GHz can be interpreted as a process that is largely associated with the sheath while
the spine remains unaffected. 

\subsection{Measurement constraints on jet-internal gradients}
\begin{figure}
\centering
	 \includegraphics[width=\hsize]{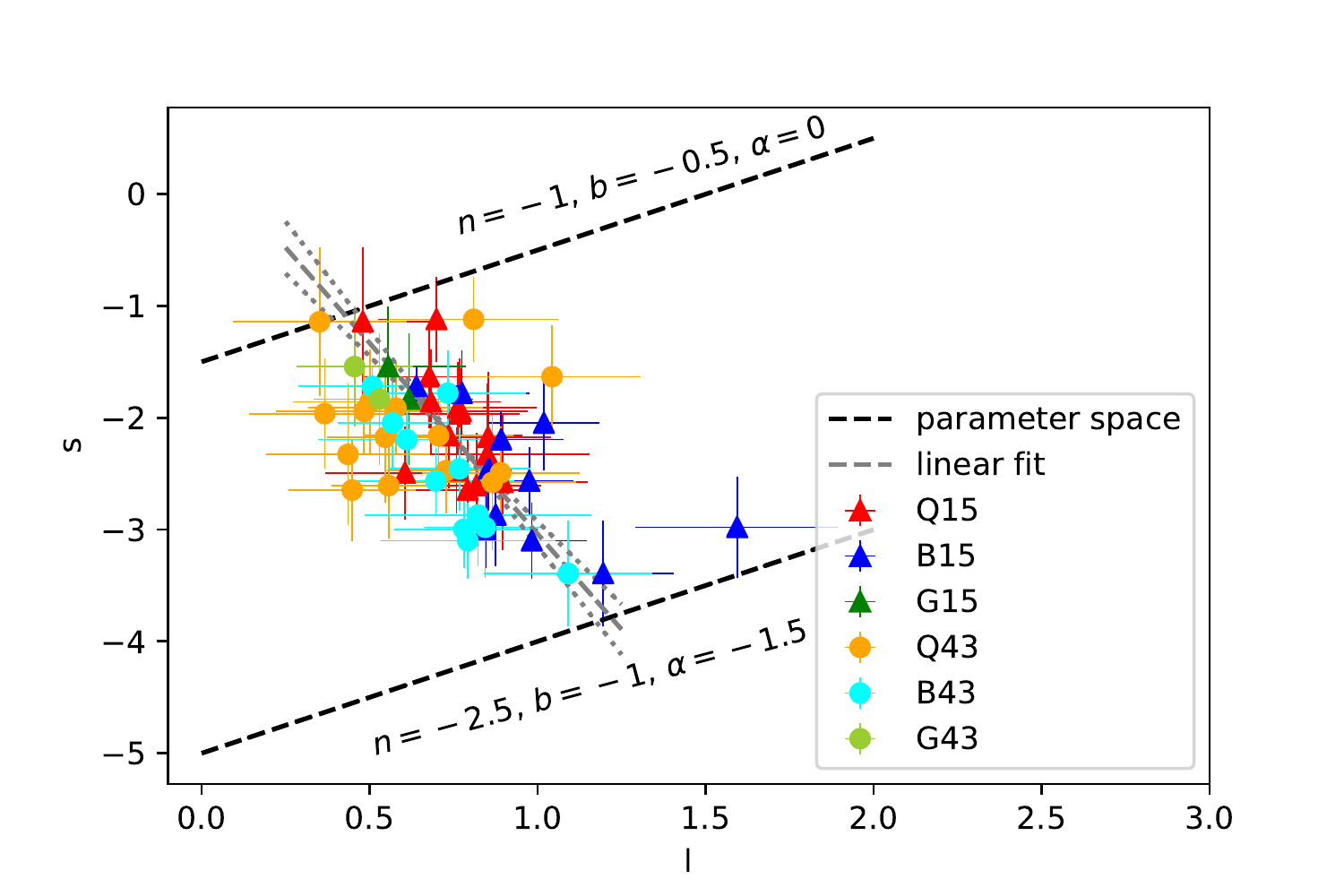}
      \caption{The parameter space of $n,b$ and $\alpha$, spanned by the measured $s$ and $l$ values at both frequencies.}
       \label{parameter_space}
\end{figure}

Figure~\ref{parameter_space} illustrates the allowed parameter space for the jet-internal gradients of the electron density $n$, the magnetic field strength $b$, and the spectral index $\alpha$ constrained by the measured values of the power-law slopes of the brightness temperature $s$ and jet diameter $l$. 

To investigate these dependencies in detail, we first reconsider some theoretical standard scenarios \citep[cf.][]{Kadler2004} in order to estimate plausible $s$-values, depending on $l,n,b$ and $\alpha$. We assume equipartition between the electron energy density and magnetic field strength density, thus $n_e~\propto B^2$ and therefore the scaling parameter of the electron energy density $n$ depends on the scaling of the magnetic field along the jet axis ($n = 2b$).
\begin{itemize}
\item Conical jet ($l = 1$):
    \begin{itemize}
        \item torodial magnetic field ($b=-1$),equipartition ($n=-2$), $\alpha=-0.5$ $\rightarrow$ $s=-2.5$
        \item polodial magnetic field ($b=-2$), equipartition ($n=-4$), $\alpha=-0.5$ $\rightarrow$ $s=-6$
    \end{itemize}
    \end{itemize}
In the case of a conical jet, equipartition, and a reasonable spectral index of $\alpha=-0.5$ the $s$-values thus have a domain of reasonable values of $-6\leq s\leq -2.5$

\begin{itemize}
\item Parabolic jet ($l = 0.5$):
    \begin{itemize}
        \item torodial magnetic field ($b=-0.5$),equipartition ($n=-1$), $\alpha=-0.5$ $\rightarrow$ $s=-1.25$
        \item polodial magnetic field ($b=-1$), equipartition ($n=-2$), $\alpha=-0.5$ $\rightarrow$ $s=-3$
    \end{itemize}
\end{itemize}
In the case of a parabolic jet, equipartition, and a reasonable spectral index of $\alpha=-0.5$ the $s$-values have a domain of reasonable values of $-3\leq s\leq -1.25$ \\

The measured $s$ and $l$ values for both frequencies seen in Fig.~\ref{parameter_space} show an anticorrelation ($r_\mathrm{spearman} \sim -0.73$, $r_\mathrm{pearson} \sim -0.69$) in the sense that larger $l$ values lead to more negative $s$ values. The $r_\mathrm{pearson}$ even indicates a simple linear connection of both parameters. This can also be constituted for the frequencies separately $(r_\mathrm{spearman,15},r_\mathrm{pearson,15},r_\mathrm{spearman,43},r_\mathrm{pearson,43}) \sim  (-0.73,-0.69,-0.64,-0.69)$. Further splitting the values into source classes, ignoring the radio galaxies by number within the sample also keeps the anti-correlation measurable 
$(r_\mathrm{spearman,15,Q},r_\mathrm{pearson,15,Q},r_\mathrm{spearman,43,Q},r_\mathrm{pearson,43,Q}) \sim  (-0.6,-0.62,-0.56,-0.66)$ and $(r_\mathrm{spearman,15,B},r_\mathrm{pearson,15,B},r_\mathrm{spearman,43,B},r_\mathrm{pearson,43,B}) \sim  (-0.56,-0.58,-0.86,-0.79)$.
The anti-correlation of the $s$ and $l$ values suggests that more collimated jets imply flatter brightness-temperature gradients. It shows that the main impact of a resulting $s$-value comes from the jet geometry because other wise an uncorrelated data cloud would be expected. However the anti-correlation is not as clear, as the $s$ values seem not to depend solely on $l$.
The rank correlation indices show ranges of $-0.8\lesssim r_{s/p} \lesssim -0.5$, which hints to a possible impact of the parameters $n$ and $b$ to the resulting $s$-values. This will be discussed source by source in the next section.
The data are fit with a linear model (gray dashed lines) to measure the slope of the anti correlation which is $\sim -3.4$. Equation~\ref{def_s} however suggests that this slope should be $~-1$ if $n,b$ and $\alpha$ where completely independent of the jet geometry. However as seen in the standard scenarios above and Sect.\ref{Sect.:Jetmodel}, at least the scaling of the magnetic field strength depends on the jet geometry. Depending on how $n$ and $b$ are de-coupled, the scaling of the electron energy density also depends on the jet geometry. The black dashed lines indicate the parameter space of the parameters $(n,b,\alpha)_{min} =(-2.5,-1.25,-1) $ and $(n,b,\alpha)_{max} =(-1,-0.5,0)$. This can be concluded to be the parameter space of jet-internal $l$, $b$, and $n$ gradients that is in agreement with the observational constraints for our sample analysis of VLBI jets.

\subsection{Tests for redshift biases}
The redshifts of the sources in our sample range from 0.0308 (Mrk\,421) to 2.218 (4C\,+71.07). Moreover, the angular-to-linear scale conversion also strongly depends on the orientation of the jets to the line of sight. Thus, the angular scales probed with the VLBA in different sources can correspond to very different linear scales.
To test whether our conclusions are affected by these effects and whether the overlap regions are comparable for the different sources, 
we first compute the apparent length of the jet from its maximum observed angular
distance from the radio core. Second, we deproject the jet length from the
plane of the sky using the inclination angle $\theta$ between the jet direction and the line of sight given in \citet{Pushkarev2009} to obtain its radial length in parsecs. For sources for which \citet{Pushkarev2009} does not list an inclination angle, we searched the literature in order to obtain an inclination angle, which is discussed on a source by source basis in Sect.\,\ref{sect:individual_sources}. Assuming that the  distance between the radio core and the jet base is small compared to the jet length, we find the overlap region typically at jet lengths $5\lesssim d_c/\mathrm{pc}\lesssim 100$. Along the entire jet length and the overlap region, respectively, the brightness temperature and diameter gradients for each jet at $15\,\mathrm{GHz}$ were fitted. We tested the interesting pairs from table\,\ref{tab3} and and Fig.~\ref{plot:KS_matrices1},\ref{plot:KS_matrices2}, regarding the jet collimation and complexity. The results from these tests are consistent with the findings from the projected $\mathrm{mas}$-scale jets. Along the entire jet length, FSRQs are more collimated than BL Lac objects (KS: $p\sim 1.7\%$). Within the overlap region this difference is not significant (KS: $p\sim 28.7\%$). The interpretation is the same as before. The jets close to the base are similar while a possible difference in collimation can be measured further down stream. Regarding the $\chi ^2 _{\mathrm{red}}$ distribution for the $l_{15}$ fits, FSRQs show a more complex structure along the entire jet length than BL Lac objects (KS: $p\sim 0.44\%$) which is again not found significant in the overlap region (KS: $p\sim 25.9\%$). Regarding the $\chi ^2 _{\mathrm{red}}$ distribution from the $s_{15}$-fits, FSRQs consistently show larger $\chi ^2 _{\mathrm{red}}$-values than BL Lacs, both along the entire jet length (KS: $p\sim 2.4\%$) and the overlap region (KS: $p\sim 0.04\%$). 
We conclude that using the lengths of the deprojected jets to describe the scaling behavior rather than their angular diameters does not alter our findings from the previous sections.
\subsection{Possible biases form geometry transition sources}
With the mean value and the variance of the $l$ and $s$-value and the $\chi ^2 _{\mathrm{red}}$ distributions, we simulated normal distributions with different class sizes. We tested, how much each class can be reduced in size in order for the KS-test to yield reliable results and therefore find the differences between the known distributions. We find, that each class should contain at least 8-10 values in order for the KS test to reliably find differences in known distributions. Knowing this, we excluded all sources that show a geometry transition (3 BL Lacs, 5 FSRQs and one radio galaxy). This reduces the sample sizes to 8 BL Lacs, 10 FSRQs  and one radio galaxy. With the reduced sample, the difference in jet collimation at $15\,\mathrm{GHz}$ between FSRQs and BL Lacs become less significant along the entire jet length (KS: $p\sim 7\%$), however the complexity of the jet is still larger for FSRQs than for BL Lacs, for both the diameter and brightness temperature gradient fit, when comparing the $\chi ^2 _{\mathrm{red}}$-values of the source classes for the different fits (KS: p-values in the range of $(0.1-0.5)\%$). We want to emphasize, that this has to be taken with caution, since the sample sizes are reduced to an extent to where it is not entirely clear whether the KS-test is even able to yield interpretable results. Overall these findings however do not contradict the found differences between the source classes. 

\subsection{Geometry transition within the VLBI jets}
\label{gt_sec}
A number of VLBI jets are known to exhibit transitions from a parabolic ($l\sim 0.5$) to a conical ($l\sim$ 1) geometry, for example for M87  \citep{Asada_2012}, NGC\,6251 \citep{Tseng2016}, NGC\,4261 \citep{Nakahara2018}, NGC\,1052 \citep{Nakahara2020}, NGC 315\,\citep{Boccardi2020,Park2020} and 1H0323+342 \citep{Hada2018}.
M87 \citep{Mertens2016,Hada2018, Park2019}, NGC\,315 \citep{Park2020} and 1H0323+342 \citep{Hada2018} also feature accelerating jet components near the geometry transition zones.
By comparing $2~\mathrm{GHz}$ and $8~\mathrm{GHz}$ data, \citet{Pushkarev_Kovalev2012} found hints for a transition in the diameter scaling in jets.
\citet{Kovalev2020} showed that geometry transitions from a parabolic to a conical geometry can be observed on scales of several thousand Schwarzschild radii $R_\textrm{S}$ in individual AGN jets.

 According to \citet{Potter_Cotter_2015},  the four-momentum of a relativistic jet is composed of \textsc{i)} the internal, kinetic and rest mass energy, \textsc{ii)} the magnetic energy and \textsc{iii)} the synchrotron, inverse Compton, and adiabatic losses. Depending on which term is larger than the magnetic term, the jet is said to be particle dominated or magnetically dominated, respectively.
 Transitions from parabolic to conical geometry as observed in the cases listed above may be explained by the jet switching from a magnetically dominated to a kinetically dominated flow. In the following, we use our systematic modeling of the 28 sources in our sample to put constraints on possible geometry breaks seen with this dual-frequency approach in AGN jets.

\subsubsection{Geometry breaks}
\begin{figure}
\centering
    \includegraphics[width=\hsize]{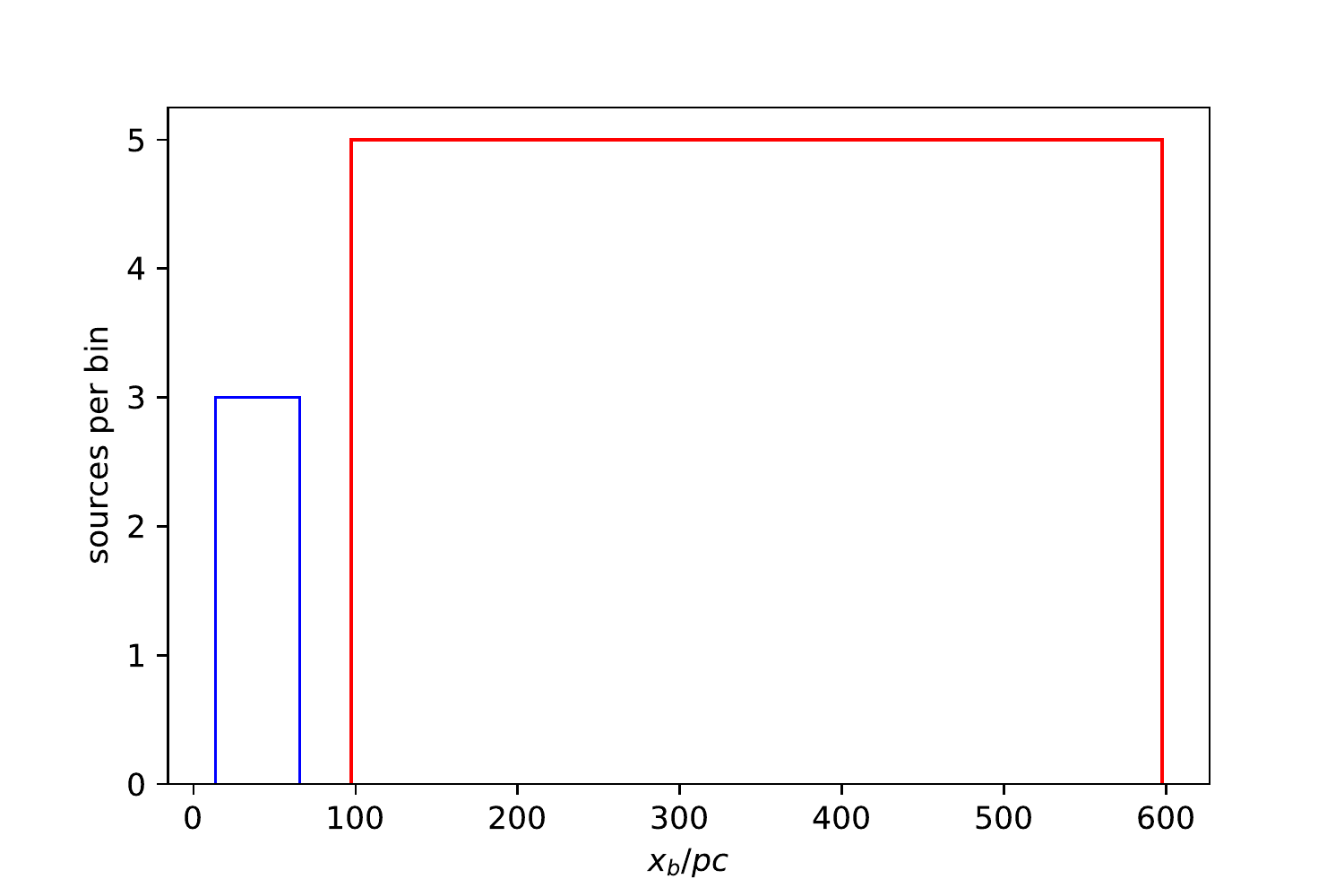}
	 \includegraphics[width=\hsize]{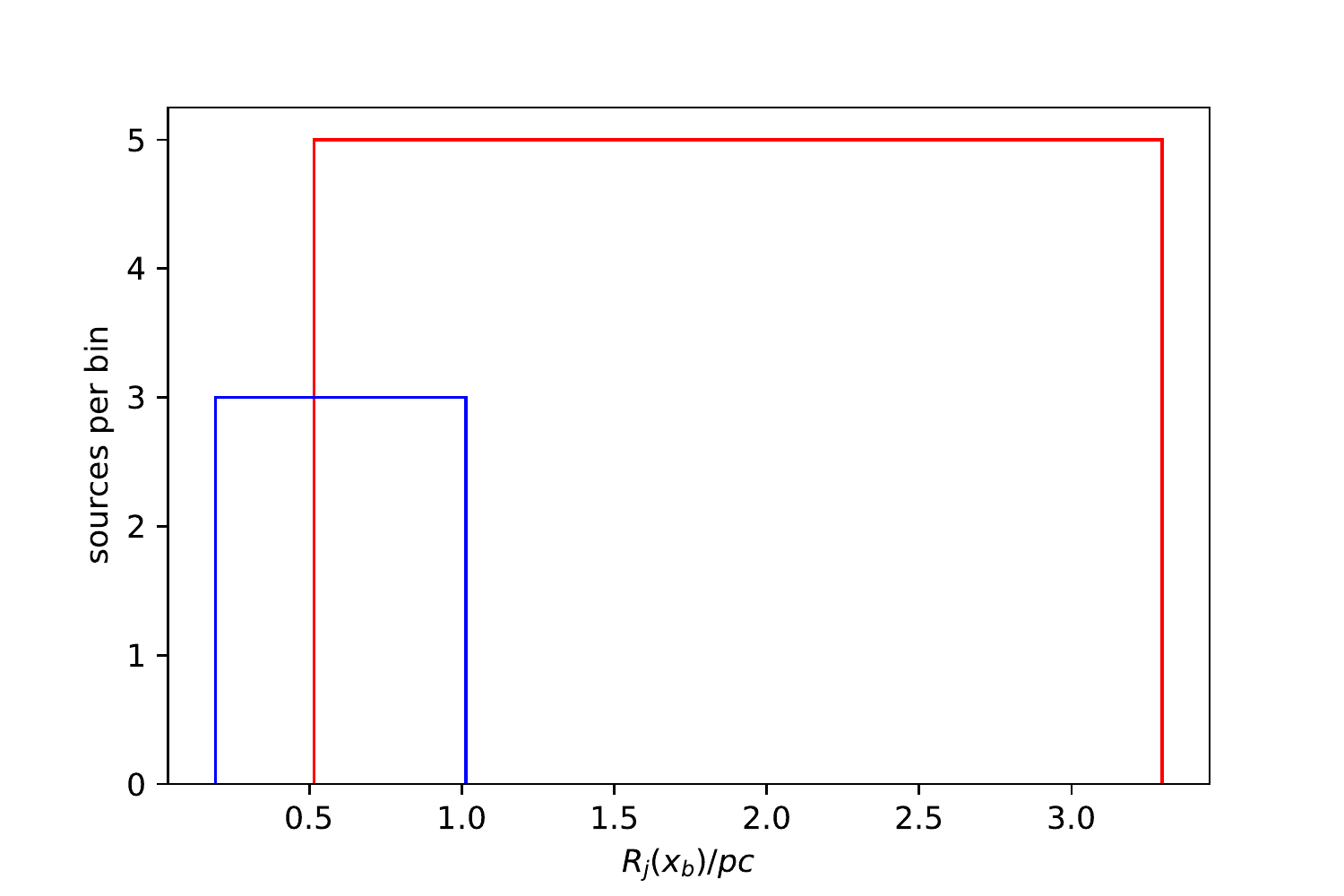}
      \caption{FSRQs are shown in red, BL Lacs are shown in blue. \textit{Top:} Distribution of the break points in linear scale and deprojected.\textit{Bottom:} The distribution of the jet radii at the respective geometry break point. }
       \label{R_xb}
\end{figure}

 Figure.~\ref{smooth0219}-\ref{smooth2200}, show the geometry breaks inferred from the fit procedure described above along with estimated positions of the Bondi radius $r_b$ for each source. In order to obtain $r_b$, the mass of the SMBH and gas temperature of the matter accreting onto the SMBH has to be known: \citep{Bondi1952,Russel2015}.
\begin{equation}
\frac{r_b}{\mathrm{pc}} = 31 \times \left(\frac{k_bT}{\mathrm{keV}}\right)^{-1} \times \left(\frac{M}{10^9 M_\odot}\right) \quad .
\end{equation} 
In the case of M87, the black-hole mass was determined by \citet{Gebhardt2011} ($M = (6.6\pm 4.4) \times 10^{9} M_\odot$) and \citet{walsh2013} ($M = 3.5^{+0.9} _{-0.7}\, 10^{9} M_\odot$). Later the \citet{EHT2019_mass} found the black-hole mass to be $M =  (6.5\pm 0.9)\, 10^{9} M_\odot$ in accordance with \citet{Gebhardt2011}. \citet{Russel2015} determined a gas temperature of $T \sim 0.9 \, \mathrm{keV}$ in the center of M87 resulting in a Bondi radius of $r_b \sim 224\,\mathrm{pc}$ from X-ray observations. For blazars, the temperature of the coronal plasma near the SMBH must be estimated taking into account that the observed X-ray emission is dominated by the nonthermal emission from the jet. For a sample low-power radio galaxies, \citet{Balmaverde2008} measured gas temperatures between $0.4\lesssim E/\mathrm{keV}\lesssim 1.7$. For high-power sources, typical gas temperatures near the SMBH are up to $\sim 7\,\mathrm{keV}$, \citep[e.g.]{Yaji2010,Gaspari2013}. Since we did not measure the Bondi radius for the sample, discussed in this work, we allow the gas temperature to vary within the bounds $0.4\lesssim E/\mathrm{keV}\lesssim 7$ for each source and compare the region of plausible Bondi radii with the geometry transition region.
The respective jet-inclination (viewing) angle is taken into account by using the measured angles from \citet{Pushkarev2009} for the cases of CTA\,26 (0336$-$019, $\theta \sim 3.2\, \mathrm{deg}$), PKS\,0528+134 (0528+134, $\theta \sim 1.7\,\mathrm{deg}$), 4C\,+71.07 (0836+710, $\theta \sim 3.2 \, \mathrm{deg}$), 4C\,+71.07 (1156+295, $\theta \sim 2 \, \mathrm{deg}$) and BL\,Lacertae (2200+420, $\theta \sim 7.5 \, \mathrm{deg}$). In the other cases, we took values of the inclination angles from
the references cited in Sect.\,\ref{sect:individual_sources}. The results of the broken geometry fits are listed in Table~\ref{table:geo breaks}. In each case we compare the $\chi^2 _{\mathrm{red}}$ of smoothly broken power law fit with a single power law fit to both the $15\,\mathrm{GHz}$ and $43\,\mathrm{GHz}$ data. In all cases an improvement is achieved by using the smoothly broken power law fit to both frequencies.  For each source we bracket the Bondi radius by the minimum/maximum black hole mass from Table~\ref{table1}, the minimum/maximum gas plausible temperature and the maximum/minimum inclination angle. 
For 3C\,66A (0219+428), CTA\,26 (0336-019), PKS\,0528+134 (0528+134), 4C\,+71.07 (0836+710),  Mrk\,421 (1101+384), 4C\,+29.45 (1156+295), 3C\,279 (1255-055) and BL Lac (2200+420) the transition zone lies beyond plausible Bondi radius ranges. In such cases \citet{Kovalev2020} argue that the geometry transition could mark the transition of a magnetically dominated to a particle-dominated jet. The transition zone of 3C\,111 (0415+379) lies within the plausible Bondi radius ranges. In this case, the lowest gas temperatures has to be assumed in order for the Bond sphere to extent beyond the transition region.
For CTA\,26 (0336$-$019) the jet features a bright, stationary component in the brightness temperature gradient between $d_c\sim 1-2 ~\mathrm{mas}$ during the ten year observation period, see Fig.~\ref{0336-019}. 
This possibly indicates stationary shocks or jet curvature, where the jet in parts crosses the line of sight, creating the seemingly stationary features. 
The distribution of the break points shown in Fig.\,\ref{R_xb} (top) 
tentatively suggests relatively low values for BL Lacs as compared to a wider distribution ranging to about a factor of ten larger maximum values in the case of FSRQs. This is found to be significant by a two-sample KS test ($p\sim 3.7\%$). Furthermore, a KS-test with the $H_0$-hypothesis, that $x_{b,Q} < x_{b,B}$ yields $p\sim1.7\%$ which suggests, that the break point of the geometry transition is further down the jet stream for FSRQs than for BL Lacs. This is also found on the entire MOJAVE data sample at $15~\mathrm{GHz}$ by Kravchenko et al. (in prep., 2021 priv. com.)  A difference might indeed be expected according to \citet{Potter_Cotter_2015} based on the assumption that the geometry breaks result from the magnetic-to-kinetic transition. Figure\,\ref{R_xb} (bottom) shows the distribution of the jet radii at the transition zone (break point $x_b$). Again, we do not find a significant dichotomy between the two source classes 
(KS: $p\sim 14.3\%$). Again an alternative KS-test with $H_0: R_j(x_{b,Q}) < R_j(x_{b,B})$ yields $p\sim 7\%$. This suggests that the jet radii at the break points of FSRQs and BL Lacs are likely drawn from the same distribution however FSRQs show a tendency for larger radii at the break point.  It must be emphasized, however, that the sample contains only ten sources that do show a geometry break, of which only three are of the BL\,Lac type.  
Furthermore, we cautiously remark that different methods can lead to somewhat different results.
For example, \citet{Boccardi2020} analyzed the jet of NGC315  in the visibility domain at different epochs and frequencies finding a transition from parabolic to conical jet geometry already at a distance of about half a parsec. \citet{Park2020} performing a multifrequency single-epoch study in the image plane, found the transition to occur at a distance about 10 times larger.
In order to further shed light onto the issue of systematic jet geometry transitions, it would be interesting to study more sources from even higher ($>43~\mathrm{GHz}$) frequencies to lower frequencies ($\sim 1~\mathrm{GHz}$) in order to study the collimation of a jet across a large scale and determine whether geometry breaks might happen on all scales or whether the transitions found are interesting physical points along the physical flow of the jet.

\begin{table*}
\caption{Geometry transition properties of individual sources}
\label{table:geo breaks}
\begin{small}
\centering
\resizebox{2\columnwidth}{!}{. 
\begin{tabular}{lccccccccc}
\toprule
Name & Class & $x_b/\mathrm{mas}$ ($x_b/\mathrm{pc}$)&$R_j(x_b)/\mathrm{mas}$ &  $l_{\mathrm{in}}$ & $l_{\mathrm{out}}$ & $\theta / \mathrm{deg} $ & $\chi^2_{\mathrm{red, broken}}$ & $\chi^2_{\mathrm{red, single}}$ & $r_B / \mathrm(mas)\, (r_B / \mathrm(pc))$ \\
\midrule
3C\,66A & B &$2.31 \pm 0.32$ $(65)$&$0.42 \pm 0.14$ & $0.56 \pm 0.13$ & $1.33 \pm 0.11$ & $9.8^{\textrm{a}}$ & $4.0$ $(326)$ & $8.1 (328)$ & $0.046-1.255$ $(1.31-35.51)$ \\
CTA\,26 & Q & $2.12 \pm 0.26$ $(293)$ & $0.68 \pm 0.15$ &$0.62 \pm 0.12$ & $1.78 \pm 0.27$ & $3.2^{\textrm{b}}$ &$16.8$ $(119)$ & $21.3$ $(121)$ & $0.0011 - 0.0310$ $(0.16-4.27)$  \\
3C\, 111 & G & $5.42 \pm 0.29$ $(16)$ & $0.56 \pm 0.09$ & $0.60 \pm 0.08$ & $1.69 \pm 0.24$ & $19.0^{\textrm{c}}$ &$12.1$ $(662)$ & $12.8$ $(664)$ & $0.23-6.55$ $(0.67-19.07)$ \\
PKS\,0528+134 & Q & $2.10 \pm 0.38$ $(596)$ & $0.56 \pm 0.16$ & $0.54 \pm 0.11$ & $1.24 \pm 0.25$ & $1.7^{\textrm{b}}$ & $10.0$ $(336)$ & $10.9$ $(338)$ & $0.0035-0.0925$ $(0.99-26.32)$ \\
4C\,+71.04 & Q & $2.75 \pm 0.41$ $(413)$ & $0.35 \pm 0.13$ & $0.48 \pm 0.14$ & $1.15 \pm 0.15$ & $3.2^{\textrm{b}}$ &$15.7$ $(139)$ & $19.0$ $(141)$ & $0.0018 -  0.0495$ $(0.28-7.42)$ \\
Mrk\,421 & B & $1.93 \pm 0.38$ $(13-34)$ & $0.63 \pm 0.19$ & $0.54 \pm 0.13$& $1.02 \pm 0.13$ & $(2 - 5)^{\textrm{d}}$ &$10.1$ $(253)$ & $11.5$ $(255)$& $0.0048-0.3522$ $(0.084-2.456)$ \\
4C\,+29.45 & Q & $2.52 \pm 0.60$ $(522)$ & $0.91 \pm 0.37$ & $0.73 \pm 0.17$ & $1.23 \pm 0.21$ & $2.0^{\textrm{b}}$ & $8.5$ $(84)$ &$9.3$ $(86)$& $0.0034-0.0918$ $(0.084-2.456)$ \\
3C\,279 & Q & $0.61 \pm 0.22$ $(97)$ & $0.15 \pm 0.09$ & $0.49 \pm 0.15$ & $0.78 \pm 0.09$ & $2.4^{\textrm{b}}$ & $33.1$ $(390)$ & $33.8$ $(392)$ &  $0.012-0.337$ $(1.98-53.74)$ \\ 
BL\,Lac & B & $1.97 \pm 0.13$ $(20)$& $0.37 \pm 0.07$ & $0.78 \pm 0.09$  & $1.97 \pm 0.08$ & $7.5^{\textrm{b}}$ &$9.6$ $(1058)$ & $17.0$ $1060$ & $0.046-1.184$ $(0.46-11.76)$  \\
\hline 
\end{tabular}
}
\end{small}
\tablefoot{Columns: (1) Common name, (2) Classification, (3) geometry break point in angular scale (geometry break point in linear scale deprojected), (4) Jet radius at the break point, (5) $l_{\mathrm{in}}$, (5) $l_{\mathrm{out}}$, (6) inclination angle, (7) $\chi^2 _{\mathrm{red}}$ for the broken power law fit, (8) $\chi^2 _{\mathrm{red}}$ for the single power law fit.(9) range of plausible Bondi radii in angular scale (range of plausible Bondi radii in linear scale deprojected)}   
\tablebib{
a) \citet{Hovatta2009},
b) \citet{Pushkarev2009},
c) \citet{Kadler2008_3C111},
d) \citet{Lico2012},
}
\end{table*}

\section{Comments on individual sources}
\label{sect:individual_sources}
\paragraph{$0219+428$ (3C\,66A):} The brightness-temperature along the jet of this source shows a local excess over the general power-law decrease between $2\lesssim d_{\mathrm{core}}/\mathrm{mas} \lesssim 3$, which is visible at both frequencies (Fig.~\ref{0219+428}). This excess indicates a local deviation from the simple jet model (see Eq.~\ref{def_s}). The jet-diameter does decrease as a power law as described by the model but shows a dip at the same distance from the radio core where the $T_\mathrm{B}$ excess is observed. The dip is persistent over the epochs, indicated by the different colors, represented within the dip. This might hint at a stationary region where the jet recollimates and the local surface brightness temperature rises. 

\begin{figure}
\centering
	 \includegraphics[width=\hsize]{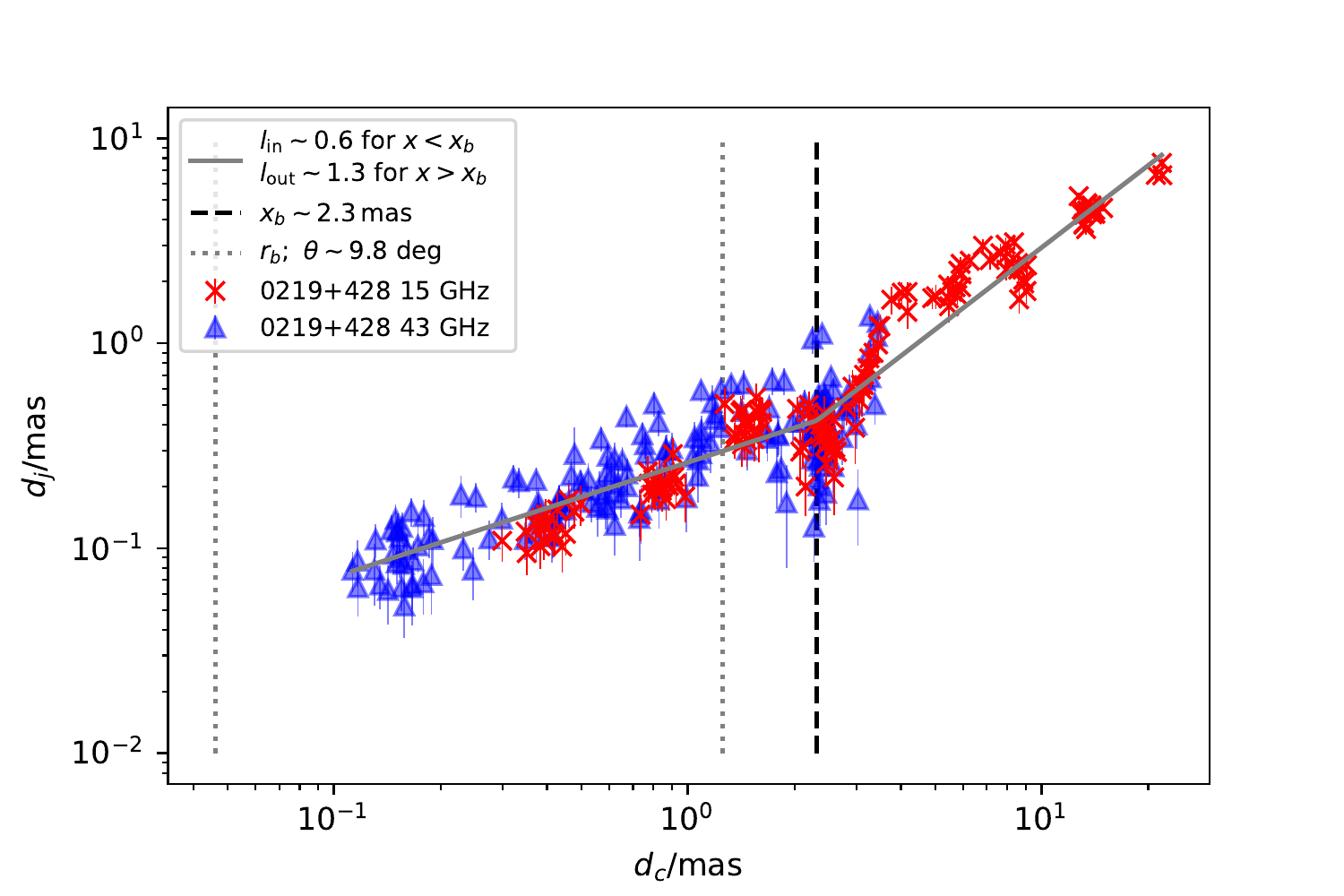}
      \caption{Jet diameter vs. core distance for 3C\,66A (0219+428). The jet shows a transition from a parabolic shape to a conical shape at $x_b \sim  2.3 \, \mathrm{mas}$ (dashed line). The dotted lines indicate the estimated location of the Bondi radius.}
       \label{smooth0219}
\end{figure}

The $s$-values at both frequencies are indistinguishable within the errors, however the $l$-values differ along the respective entire jet length, while remaining indistinguishable in the overlap region. We find a transition of the geometry in this jet, which is shown in Fig.~\ref{smooth0219}, featuring a parabolic jet shape up to $d_c \sim 2.3\, \mathrm{mas}$ that transitions to a conical shape jet downstream. 

To estimate the angle to the line of sight, the variability Doppler factor can be used to estimate $\theta \sim 9.8 \, \mathrm{deg}$ as discussed in \citet{Hovatta2009}. 
The scaled Bondi radius is shown in Fig.~\ref{smooth0219} projected for the derived angle, respective SMBH mass and the gas plausible gas temperature range (dotted lines). The break point $x_b$ lies outside the Bondi-sphere, which is located close to the possible recollimation region.

\begin{figure}
\centering
	\includegraphics[width=\hsize]{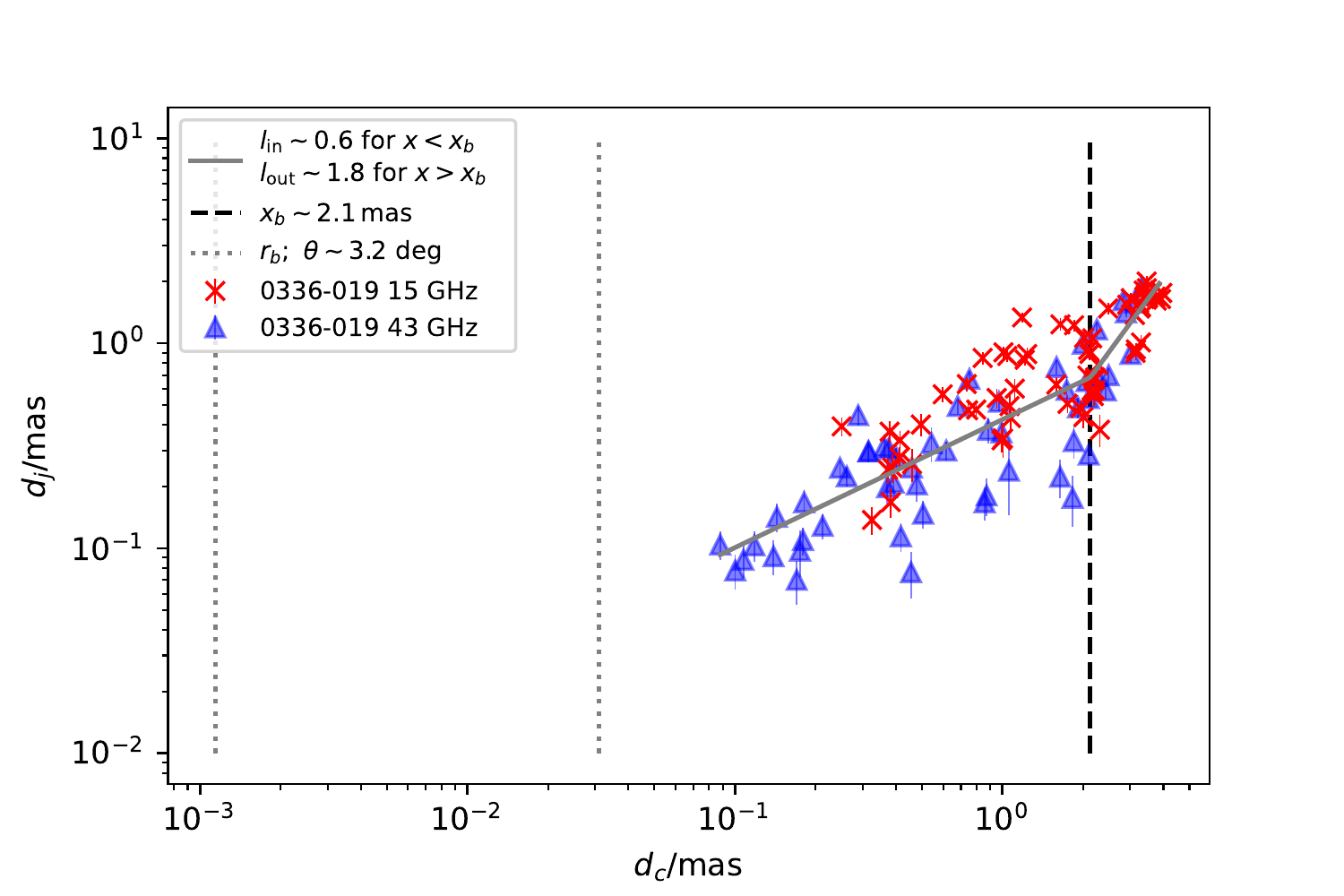}
      \caption{Jet diameter vs. core distance for CTA\,26 (0336-019). The jet shows a transition from a parabolic shape to a conical shape at $x_b \sim  2.1 \, \mathrm{mas}$ (dashed line). The dotted line indicate the estimated location of the Bondi radius.}
       \label{smooth0336}
\end{figure}

\paragraph{0336$-$019 (CTA\,26):} 
The jet diameter along the jet axis in this source is characterized by strong scattering around the best-fit model (Fig.~\ref{0336-019}). This behavior is typical for a FSRQ (see discussion above).
When combining the information from both frequencies as described in Sect.\ref{sect:method_breaks}, it is possible to constrain a geometry break point in this source (Fig.~\ref{smooth0336}). The jet is best described through a parabolic shape out to $d_c \sim 2.1 \, \mathrm{mas}$, which opens up to a cone afterwards. The  Bondi radius is smaller than the location of this break by two orders of magnitude and is therefore unlikely to be related in any way. 
The $T_{\rm B}$ profile shows an excess at $d_c \sim 1-2 \,\mathrm{mas}$ which is persistent across the epochs (indicated by the different colored components in Fig.~\ref{0336-019}), which might be associated with a standing shock at this location in the jet.

\begin{figure}
\centering
	\includegraphics[width=\hsize]{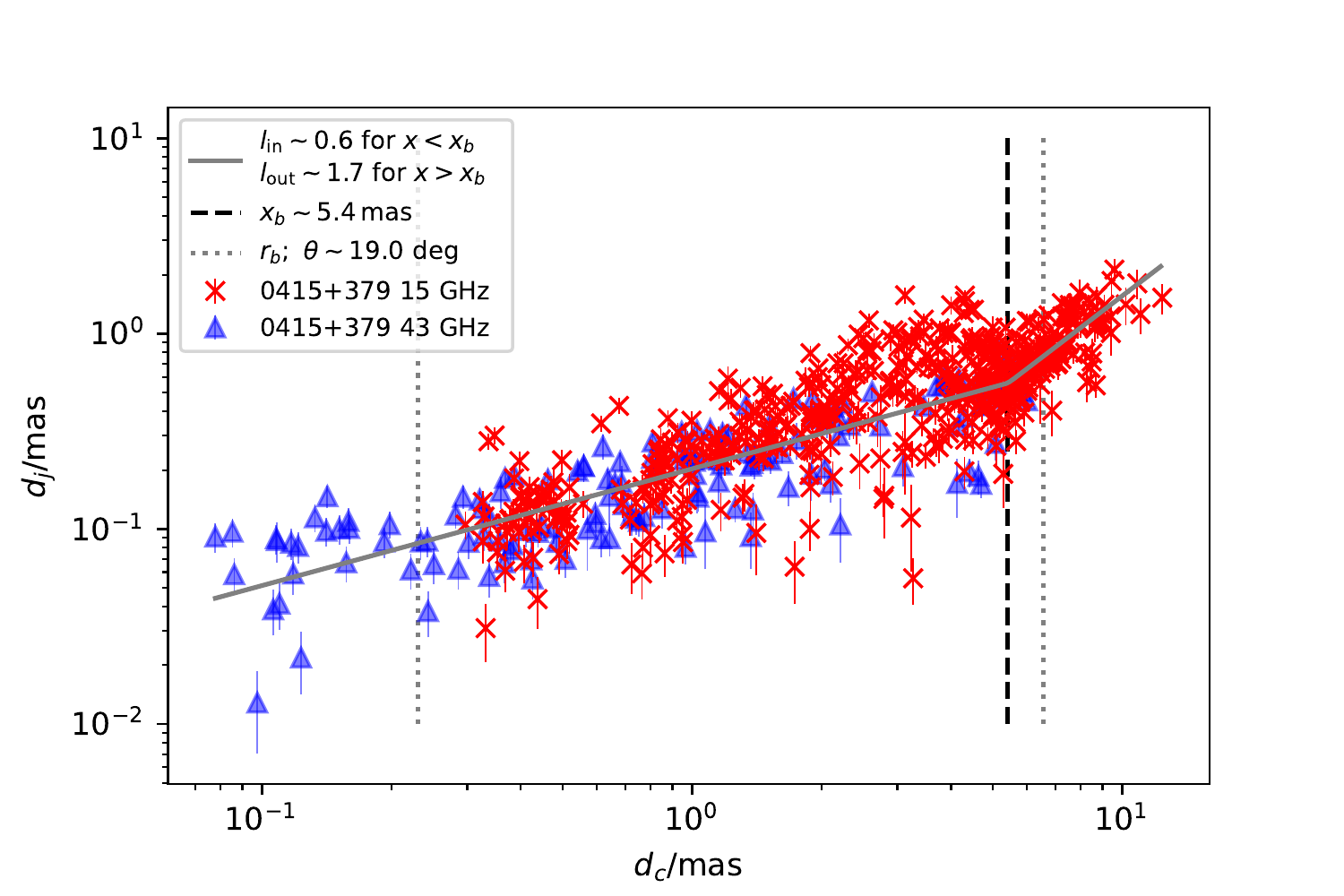}
      \caption{Jet diameter vs. core distance for 3C\,111 (0415+379). The jet shows a transition from a parabolic shape to a conical shape at $x_b \sim  5.4 \, \mathrm{mas}$ (dashed line). The dotted line indicates the estimated location of the Bondi radius.}
       \label{smooth0415}
\end{figure}

\begin{figure}
\centering
	\includegraphics[width=\hsize]{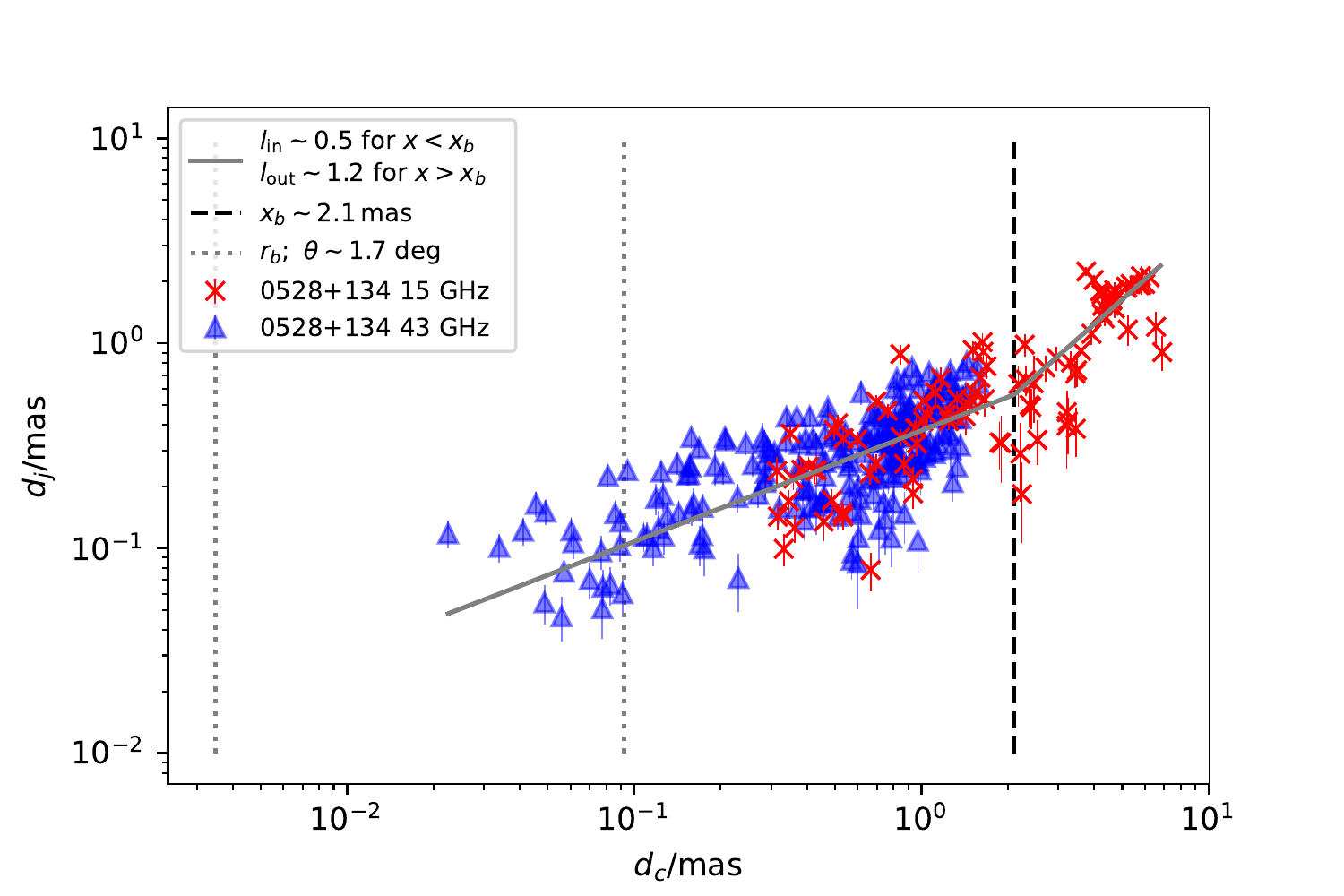}
      \caption{Jet diameter vs. core distance for PKS\,0528$+$134 (0528+134). The jet shows a transition from a parabolic shape to a conical shape at $x_b \sim  2.1 \, \mathrm{mas}$ (dashed line). The dotted line indicates the estimated location of the Bondi radius.}
       \label{smooth0528}
\end{figure}

\paragraph{0415$+$379 (3C\,111):} This is one out of only two radio galaxies in our sample. Model-fitting results for this source are shown in Fig.~\ref{0415+379}. 
The morphology and kinematics are rather complex and best described 
with a parabolic-jet model at both frequencies and relatively flat brightness temperature gradients $-1.8\lesssim s\lesssim -1.5$. \citet{Kadler2004}  fit a brightness temperature gradient of with a power law index of $s\sim -2.4$ by fitting individual tracked components. The difference can be explained by the fitting of more epochs in this work and not excluding single components for the power law fit. A break in the jet geometry  from parabolic to hyperbolic shape can be constrained at $\sim 5.4 \, \mathrm{mas}$ from the radio core, see Fig.\,\ref{smooth0415}. We adopt the inclination angle of $19\,\mathrm{deg}$ from \citet{Kadler2008_3C111} to project the position of the Bondi radius, which may coincide with the transition region for the smallest plausible gas temperatures near the SMBH. \citet{Beuchert2018} fit the brightness temperature and the diameter along the jet axis for this source with two separate power laws in the region $<3~\mathrm{mas}$ and $>3~\mathrm{mas}$ from the radio core. They find a similar switching of the jet geometry consistently with our findings. \citet{Kovalev2020} also find a transition of geometry at $x_b=(7.03\pm 0.5)\, \mathrm{mas}$. \citet{Boccardi2020} find a transition in this region.Taking the systematic errors at a 1-$\sigma$-limit into account, there is still a difference between the found break point in this study and the break point found in \citet{Kovalev2020} and \citet{Boccardi2020}. We tested the impact of different established fitting methods (least square and orthogonal distance regression) in Append~\ref{appendix} and found that different methods yield the same results with the given data within the errors. We therefore attribute this difference between our result and the \citet{Kovalev2020} and \citet{Boccardi2020} results to the use of different data sets, especially the different epochs with different time coverage of the dynamically evolving sources and with different observational setups and image noise levels (cf. Sect.~\ref{sec_sample}) and the different sets of model fits (cf. Sect.~\ref{sec_method}) parameterizing this highly complex source.

\paragraph{0430$+$052 (3C\,120):} The model fitting results for second radio galaxy in our sample
are shown in Fig.~\ref{0430+052}. We find a parabolic jet shape on all scales tested along the  entire jet length. The $s$-values are flatter than expected from a freely expanding jet which is consistent with the narrowly confined jet geometry. \citet{Kovalev2020} and \citet{Boccardi2020} find a transition of the jet shape from parabolic to conical at $\sim 1.5~\mathrm{mas}$ from the radio core. This region is also resolved by the data used in this work, however \citet{Kovalev2020} probe somewhat larger scales in their analysis of $15\, \mathrm{GHz}$ and $1.4\, \mathrm{GHz}$ data. Examining Fig.~1 in \citet{Kovalev2020} it becomes clear, that the geometry transition is seen primarily due to the use $1.4~\mathrm{GHz}$ data. Overall our findings are consistent with \citet{Kovalev2020} in the sense that closer to the physical jet base they find a parabolic shape consistent with our results. Even when rigorously searching for a geometry transition, it is not possible to clearly constrain a significant break point on the scales covered by our higher-frequency data. This is due to the large scatter of the data in the region where \citet{Kovalev2020} and \citet{Boccardi2020} find the geometry transition. 

\paragraph{0528$+$134 (PKS\,0528+134):} This FSRQ shows a prominent excess in brightness temperature between $0.3\lesssim d_{\mathrm{core}}/\mathrm{mas} \lesssim 1$ distance from the radio core at $43\, \mathrm{GHz}$ which is not seen in the $15\, \mathrm{GHz}$ data  (Fig.~\ref{0528+134}). 
Since this excess in the $43~\mathrm{GHz}$ is seen in epochs with are close in time to each other (see color coding in Fig.~\ref{0528+134}) this could be due to a radio flare in this source. 
This excess can be associated with a dip in the jet-diameter profile of the $43\, \mathrm{GHz}$ jet. This indicates a region where the jet locally contracts. The observations seem to probe the jet in the region where its geometry is still parabolic. The geometry fits at both frequencies are  indistinguishable  within their errors: $l_{43} = 0.45\pm 0.19, \, l_{15} = 0.79 \pm 0.21$. This also holds in the overlap region of both frequencies. The brightness temperature gradients however seem to be significantly flatter at $43\, \mathrm{GHz}$ than at $15\, \mathrm{GHz}$, where the latter shows the canonical value, expected from a simple \citet{BK79}-jet ($s_{43} = -1.23\pm 0.43, \, s_{15} = -2.65 \pm 0.46$), when considering the respective entire jet lengths. The difference of the $s$-values can be attributed to a variation of one of the parameters $n$ and/or $b$ on the different angular scales probed. Due to the excess in $T_b$ at $43\, \mathrm{GHz}$ and the subsequent steepening of the gradient's fit, this difference vanishes in the overlap zone of both frequencies. At high frequencies above $\sim 30\, \mathrm{GHz} $ \citep{Marscher_rev2016}, the radio core is not necessarily the region where the synchrotron radiation becomes optically thin 
as in the \citet{BK79} jet model but might rather represent a standing recollimation shock which is backed by polarization measurements in the radio core of 1803+784 at $43\, \mathrm{GHz}$ \citep{Cawthorne2013}, which they link to conical recollimation shocks \citep{Daly1988,Cawthorne_Cobb1990,Cawthorne2006}. \citet{Jorstad_2001} studied 42 $\gamma$-ray detected blazars at $22\, \mathrm{GHz}$ and $43\, \mathrm{GHz}$ of which 27 showed a sum of 45 nonmoving components besides the radio core which they also link to standing recollimation shocks if the component is close to the radio core and to jet bending accompanied by maximized Doppler boosting of the component is farther down to jet. 
We find a geometry break from parabolic to conical jet geometry in this source at $x_b\sim 2.1\,\mathrm{mas}$. The radius of the Bondi sphere is closer to the radio core than $x_b$ in this source. Combining $15\, \mathrm{GHz}$ and $1.4\, \mathrm{GHz}$, \citet{Kovalev2020}, do not find a geometry break in this source. We explain this by the additional use of $43~\mathrm{GHz}$ data in this work which allow to resolve components closer to the physical jet base due to following reasons: With the single power law fit, \citet{Kovalev2020} find a diameter gradient power law index of $\sim 0.8$ which is consistent with the value we find at $15~\mathrm{GHz}$ ($l_{15} = 0.79\pm 0.21$) along the entire jet length. In the overlap region between the $43\, \mathrm{GHz}$ and $15\, \mathrm{GHz}$ the jet shape is parabolic. This reveals that the jet features a geometry transition which can be constrained closer to the jet based, by combining the high frequency observations. 

\paragraph{0716+714 (S5\,0716+71):} This source features a  jet profile with $0.5 \leq l \leq 1$ at both frequencies
and  a quasi-canonical  $s$-value in the range of the simple \citet{BK79} jet model at both frequencies along the entire jet lengths and the overlap region (Fig.~\ref{0716+714}). 
No strong excesses in the brightness-temperature  or the jet-diameter profiles are found and no geometry break can be constrained in this source.

\paragraph{0735$+$178 (OI\,158):} The jet in this BL\,Lac object is slightly collimated  at both frequencies, along the entire jet lengths and the overlap region. Combining $15\, \mathrm{GHz}$ and $1.4\, \mathrm{GHz}$ data \citet{Kovalev2020} also find this slight collimation in this source ($l\sim 0.93$).

The $s$-values cannot be distinguished in the overlap region, however along the entire jet length the $15\, \mathrm{GHz}$ jet shows the canonical \citet{BK79}-expected value while the $43\, \mathrm{GHz}$ jet shows a flatter $T_b$-gradient ($s_{43} = -1.75\pm 0.36, \, s_{15} = -2.46 \pm 0.37$; see Fig.~\ref{0735+178}). According to Eq.~\ref{def_s} this might be attributed to a variation in the magnetic field strength gradient between the frequencies..



\paragraph{0827$+$243 (OJ\,248):} At each of the two frequencies, this quasar jet features $l$ and $s$-values which are indistinguishable within their errors, along the entire jet length and the overlap region (Fig.~\ref{0827+243}). 
In comparison, the $l_{43}$-value is consistent with a parabolic jet structure while the $l_{15}$-value  tends to be closer to a conical jet geometry.
However, due to the large uncertainties and the overall consistency of the values with each other, a geometry transition zone cannot be constrained in this source. Combining $15\, \mathrm{GHz}$ and $1.4\, \mathrm{GHz}$ the jet geometry can be fit with a cone \citep{Kovalev2020} which is consistent with the findings in this work in the $15~\mathrm{GHz}$ jets.

\paragraph{0829+046 (OJ\,049):} This BL\,Lac object shows a slightly colliamted jet at $43\,\mathrm{GHz}$ (Fig.~\ref{0829+046}). The $l$-values here are consistent with the $15\,\mathrm{GHz}$ results in the overlap region. Considering the entire jet length, the $15\,\mathrm{GHz}$ data shows a larger $l$ value indicative of a conical jet on larger scales. However, a consistent geometry break cannot be constrained due to the large scatter of the data at $43\,\mathrm{GHz}$. This scatter is largely caused by a $T_b$-excess accompanied by a recollimation of the jet which can be seen at $43\,\mathrm{GHz}$ between $0.2\lesssim d_c/\,\mathrm{mas}\lesssim 1$, which is less pronounced at $15\,\mathrm{GHz}$. This can be understood as a result of jet recollimation similar to the case of PKS\,0528+134. Investigating the $43~\mathrm{GHz}$ light curve of this source, provided on the website of thwe BU monitoring program\footnote{\url{https://www.bu.edu/blazars/VLBA_GLAST/0829.html}}, reveals, that the source showed flaring activity within the time period, analyzed in this wortk between 2008 and 2009  and in the time period between 2012 and 2013. In these time periods the MOJAVE data base provides sparse data with about one epoch per year which might be the reason that this excess is not seen in the $15~\mathrm{GHz}$ data. 
\paragraph{0836$+$710 (4C\,+71.07):} Model-fitting results for this distant ($z=2.218$) quasar are shown in Fig.~\ref{0836+710}. The jet is more collimated at $43\,\mathrm{GHz}$ than at $15\,\mathrm{GHz}$ if the full jet lengths in both data sets are considered. However, the $l$-value in the overlap region of the $15\,\mathrm{GHz}$ jet is smaller and consistent with the 43\,GHz geometry, which indicates a more collimated jet on small scales. This indicates that there is a geometry transition from parabolic to conical within this jet. The $s$-values are consistent with the canonical value of the simple \citet{BK79} model except for the overlap region at $43\,\mathrm{GHz}$. The fact that the $s$-values are consistent with the canonical value for a freely expanding jet while the jet itself features a parabolic shape at $43\,\mathrm{GHz}$, suggests that either $n$ or $b$ are causing the flattening the $s$-values. Interestingly, \citet{Orienti2020} find limb-brightened polarized flux up to $1\, \mathrm{mas}$ from the core which indicates a complex jet structure and that the magnetic field gradient changes at this point.

\begin{figure}
\centering
	\includegraphics[width=\hsize]{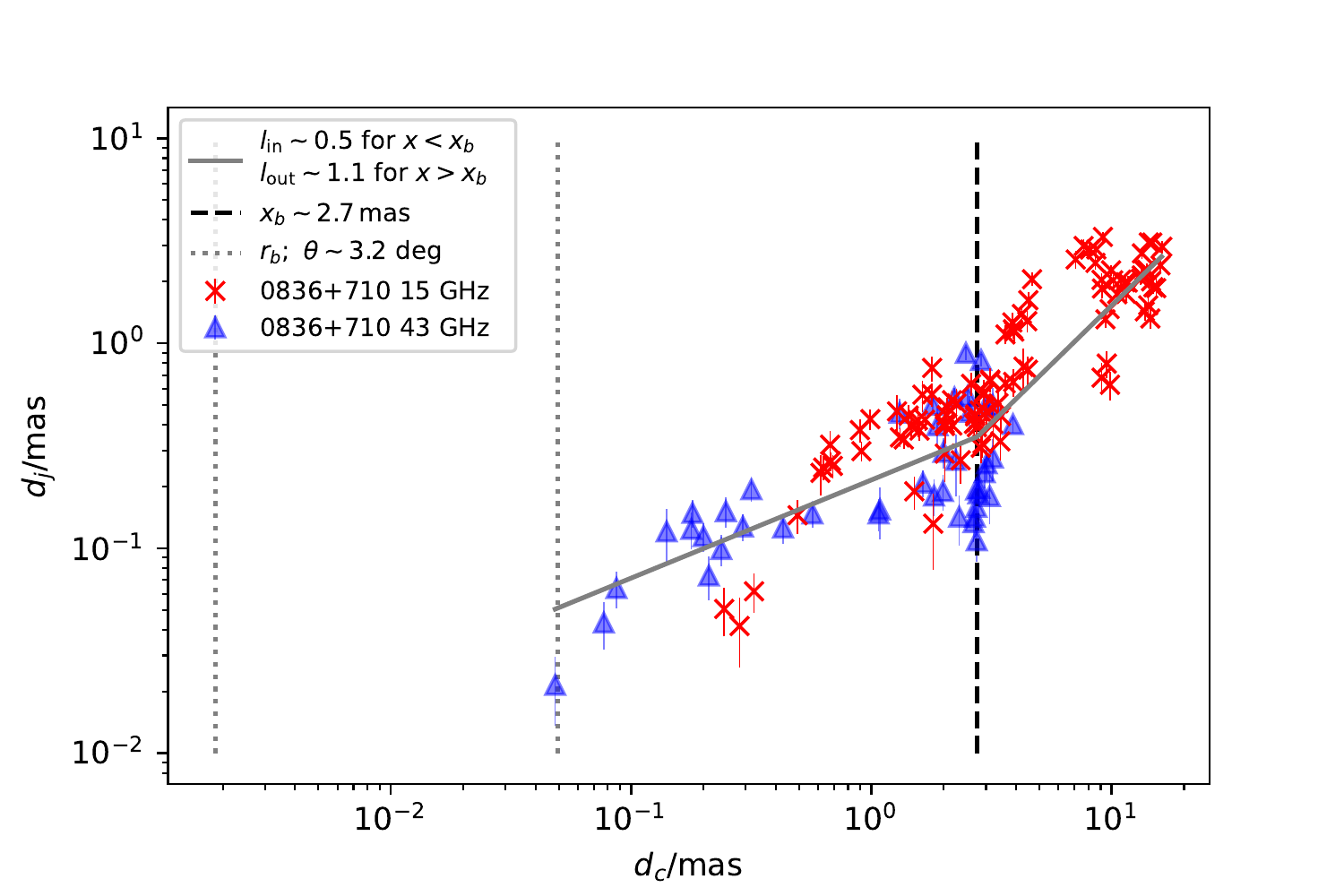}
      \caption{Jet diameter vs. core distance for 4C\,$+$71.07 (0836+710). The jet shows a transition from a parabolic shape to a conical shape at $x_b \sim  2.7 \, \mathrm{mas}$ (dashed line). The dotted line indicate the estimated location of the Bondi radius.}
       \label{smooth0836}
\end{figure}

A two-zone jet geometry model is shown in Fig.~\ref{smooth0836}. The jet features a parabolic shape up to $d_c \sim 2.7\,\mathrm{mas}$, transitioning to a conical shape further downstream. The projected Bondi radius along the jet axis is smaller than the location of the transitioning zone by almost two orders of magnitude.
The change in the magnetic-field gradient, suggested by the results of \citet{Orienti2020}  might indicate that the jet changes from magnetically dominated to particle dominated which in turn can explain the geometry transition \citep{Kovalev2020}. The complexity of this source is also well studied regarding growing Kelvin-Helmholtz instabilities within the jet at multiple scales \citep{Perucho2012,LVG2019,LVG2020}. They also show that the jet is likely particle dominated over large scales out to $\sim 100 \,\mathrm{mas}$ in jet length. 

\paragraph{0851$+$202 (OJ\,287):} This well known BL\,Lac object has been suggested as a likely candidate for a binary supermassive black hole system \cite{Sillanpaa1988,Lehto1996}. In this context, it is interesting to note that this source is among the objects with the absolute highest scatter in the jet-diameter fits of our whole sample (Fig.~\ref{0851+202}). It features $l$ and $s$-values which are indistinguishable within the uncertainties, along the entire jet length and the overlap region. On larger scales at 15\,GHz, the average jet geometry appears quasi conical, which can be interpreted as a freely expanding jet, which is consistent with the findings at lower frequencies from \citet{Kovalev2020}.

\paragraph{0954$+$658 (S4\,0954+65):} This BL\,Lac shows a jet geometry which in both frequencies is consistent with a conical jet along the entire jet length and the overlap region (Fig.~\ref{0954+658}). Also the $s$-values are indistinguishable from each other, however they tend to be steeper than the canonical value of the simple \citet{BK79} model, especially at $43\, \mathrm{GHz}$ in the overlap region, and at $15\, \mathrm{GHz}$. As the jet overall seems to feature a conical shape, this must be understood in the context of the parameters $n$ and $b$. Indeed when allowing not only toroidal magnetic field components \citep{BK79} but also poloidal magnetic field components \citep{K81}, even steeper $T_b$-gradients can be obtained.

\begin{figure}
\centering
	\includegraphics[width=\hsize]{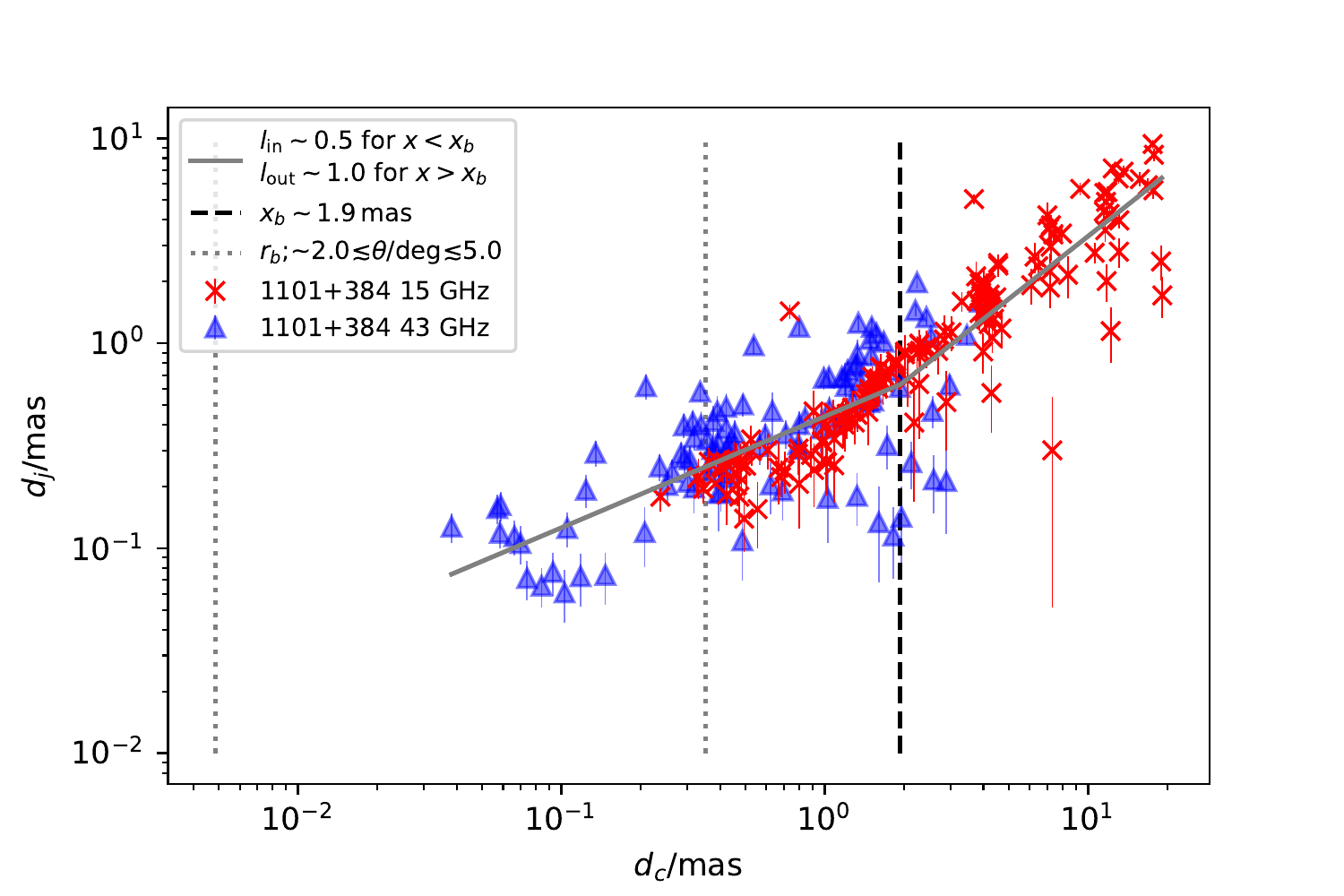}
      \caption{Jet diameter vs. core distance for Mrk\,421 (1101+384). The jet shows a transition from a parabolic shape to a conical shape at $x_b \sim  1.9 \, \mathrm{mas}$ (dashed line). The dotted line indicates the estimated location of the Bondi radius.}
       \label{smooth1101}
\end{figure}

\paragraph{1101$+$384 (Mrk\,421):} This is a nearby ($z=0.0308$) high-peaked BL\,Lac object in which we are probing the smallest linear scales within our sample (0.61\,pc/mas). The source features $s$-values which are not distinguishable within their uncertainties at a given frequency (Fig.~\ref{1101+384}). However $s_{43}$ overall shows a clear tendency of being flatter than $s_{15}$. This is also resembled in that fact, that $l_{43}$ is consistent with a parabolic jet shape while $l_{15}$ is consistent with a conical jet shape. This hints to a possible geometry transition within this source which is indeed found by our algorithm (see Sect.~\ref{sect:method_breaks}). 
A two-zone jet geometry model is shown in Fig.~\ref{smooth1101} and features a parabolic jet shape out to $d_c \sim 1.9\, \mathrm{mas}$ where it transitions to a conical shape jet further downstream.
On kpc-scales, the inclination angle of this source was estimated to be $\theta \sim 19\, \mathrm{deg}$, based on the jet to counter-jet flux density ratio \citep{Mrk421_theta}. On pc-scales, inclination angles for this source were estimated based on the Doppler factors obtained from variability analysis \citep{Hovatta2015_mrk421} to be between $0.5 \lesssim \theta/\mathrm{deg} \lesssim 4$, however they address a point made by \citet{Lico2012}, that the maximum measured component velocity $\beta _{\mathrm{app,max}}= 0.218 \pm 0.026 $ \citep{Lister_2019_betaapps} is most likely not the flow velocity of the jet because the beaming properties predicted by the flux ratio of the jet and the counter jet require viewing angles close to zero. \citet{Lico2012} estimate the inclination angle to be  $2 \lesssim \theta/\mathrm{deg} \lesssim 5$ based on the argument that the jet shows limb brightening \citep{Piner2010}, an indication for a transverse velocity structure across the jet axis. For our analysis we adopt the angle range given by \citet{Lico2012}. 
The resulting estimated location of the Bondi radius is well below the distance of the break.

\paragraph{1127$-$145 (PKS\,1127$-$14):} Model-fitting results for this quasar jet
are shown in Fig.~\ref{1127-145}. It features a parabolic jet shape at both frequencies. The $s$-values are flatter than expected for a freely expanding jet. The relatively large $\chi^2 _{\mathrm{red}}$ in the gradient fits, indicate a complex jet structure. Taking both frequencies together, a systematic geometry transition cannot be found in this source. The lower frequency data from \citet{Kovalev2020} indicate a steepening of the diameter gradient. Cmbining the $43~\mathrm{GHz}$, $15~\mathrm{GHz}$ and $1.4~\mathrm{GHz}$ data of this source might reveal a systematic geometry transition.

\begin{figure}
\centering
	\includegraphics[width=\hsize]{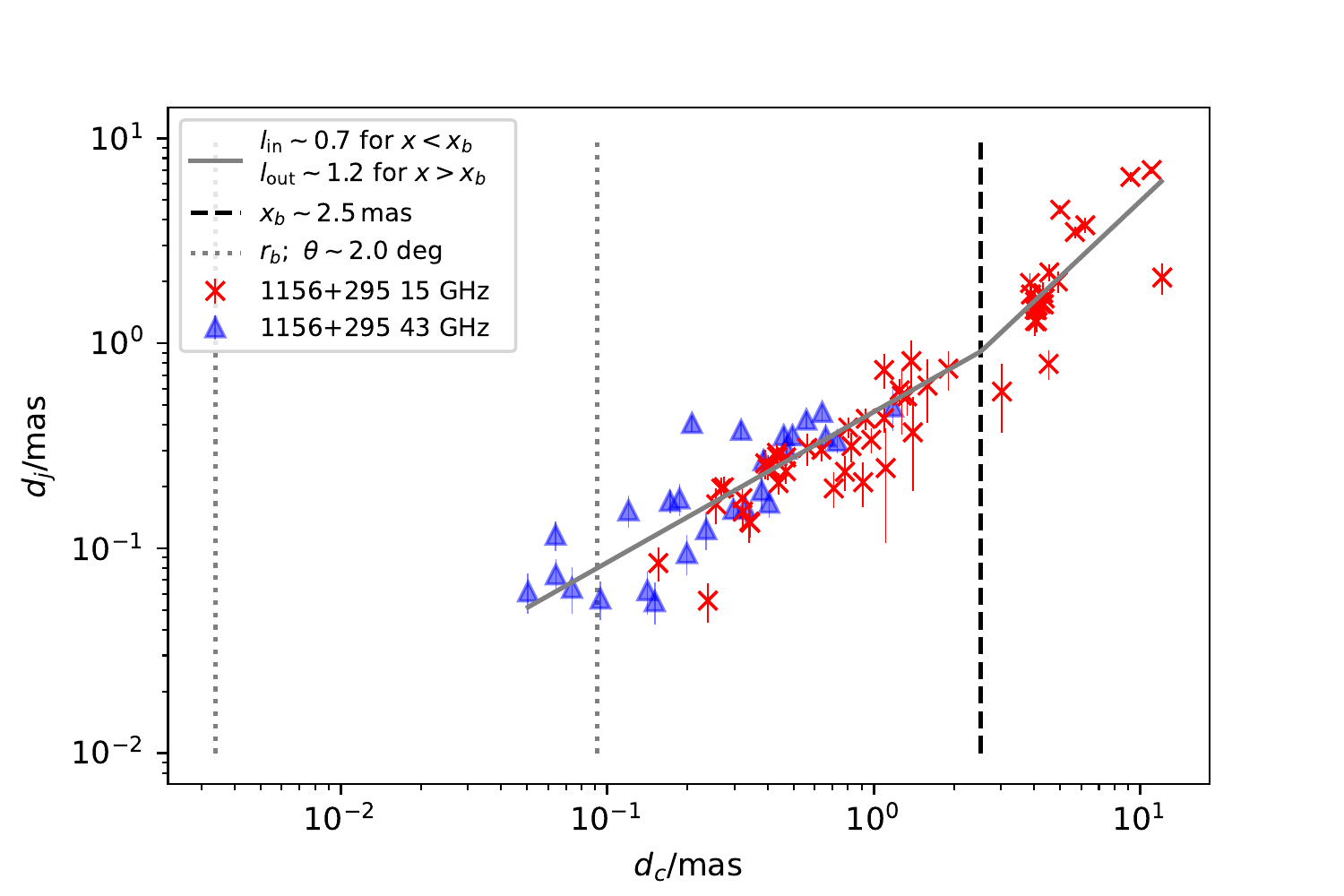}
      \caption{Jet diameter vs. core distance for 4C\,29.45 (1156+295). The jet shows a transition from a parabolic shape to a conical shape at $x_b \sim  2.5 \, \mathrm{mas}$ (dashed line). The dotted line indicate the estimated location of the Bondi radius.}
       \label{smooth1156}
\end{figure}

\begin{figure}
\centering
	\includegraphics[width=\hsize]{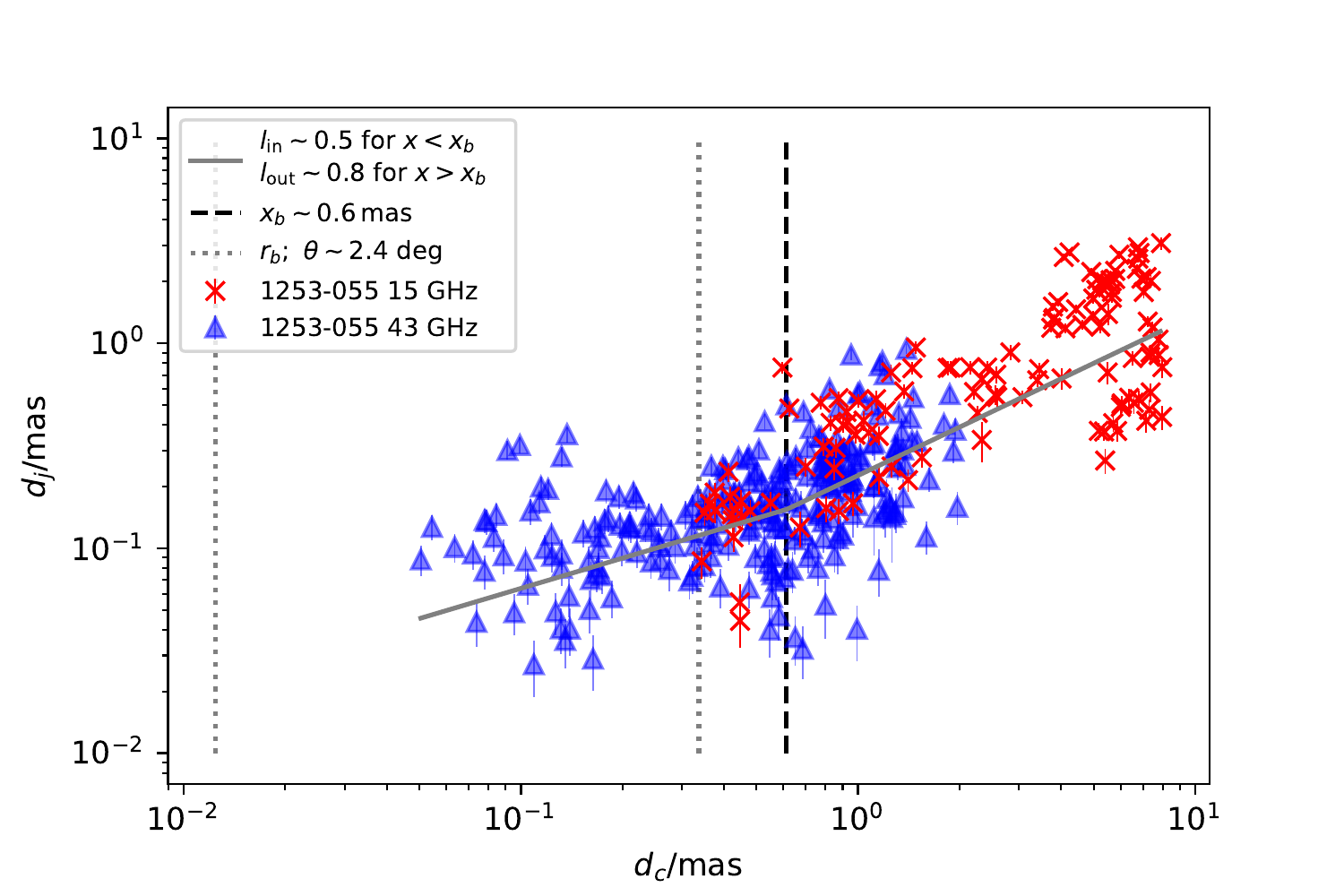}
      \caption{Jet diameter vs. core distance for 3C\,279 (1253-055). The jet shows a transition from a parabolic shape to a conical shape at $x_b \sim  0.6 \, \mathrm{mas}$ (dashed line). The dotted line indicates the estimated location of the Bondi radius.}
       \label{smooth1253}
\end{figure}

\paragraph{1156$+$295 (4C\,29.45):} This quasar features brightness temperature gradients which are consistent with a freely expanding jet (Fig.~\ref{1156+295}). The diameter gradients, however, indicate a collimation of the jet at both frequencies. A possible jet geometry transition is indicated when comparing the overlap region in the $15\, \mathrm{GHz}$ jet with the entire jet length.
The two-zone jet geometry model is shown in Fig.~\ref{smooth1156}. The jet features a parabolic shape out to $d_c \sim 2.5\,\mathrm{mas}$, transitioning to a conical shape further out. The projected Bondi radius along the jet axis is smaller than the location of the transitioning zone by almost two orders of magnitude. In this source no steady bright component can be seen in the $T_b$-gradient, which is consistent with the low scatter atypical for quasars (cf. Figs. \ref{plot:l-values_sc} and \ref{plot:s-values_sc}).
At lower frequencies \citet{Kovalev2020} do not find a geometry transition. The single power law fit they find is consistent with $l_{\mathrm{out}}\sim 1.3$ from our two zone model.

\paragraph{1219$+$285 (W\,Comae):} This source features $l$ and $s$-values which are indistinguishable within their uncertainties, along the entire jet length and the overlap region (Fig.~\ref{1219+285}). No significant break is found in the jet geometry of this source. The jet geometry overall is consistent both with a conical and parabolic jet, due to the scattering of the data. The $s$-values are consistent with the canonical parameter values of the simple \citet{BK79} model. Allowing for the extensions of \citet{K81} and in combination with the conical jet shape, this can be interpreted as a freely expanding jet. 

\paragraph{1222$+$216 (4C\,+21.35):} This is a quasar that features $T_b$ gradients which are consistent with the $s$-value expected in a canonical \citet{BK79} jet model, at both frequencies, as seen in Fig.~\ref{1222+216}. In the overlap region, the data suggest a steeper $T_\mathrm{B}$ gradient, especially at 15\,GHz.
When applying the search, discussed in Subsect.~\ref{gt_sec} it is not possible to constrain a break point between two different slopes for the geometry scaling. This might hint to a more complex, transverse jet structuring, where at $43\, \mathrm{GHz}$ in the overlap region, a more collimated substructure of the jet is measured than at $15\, \mathrm{GHz}$. 
 
\paragraph{1226$+$023 (3C\,273):} The $T_b$ gradients in 3C\,273 at both frequencies  are flatter than expected in a freely expanding jet. Indeed, the diameter gradient at both frequencies is consistent with a parabolic jet shape (Fig.~\ref{1226+023}), although the data show large scatter as it is typical for quasar jets. As it is apparent in Fig.~\ref{parameter_space}, low $l$-values tend to be associated with flat $s$ gradients.  
The complex structure of the 3C\,273 jet has been analyzed in great detail by several authors. In particular, a double-helix structure out to kpc-scales was found (see \citet{3C273_doublehelix}). 

\paragraph{1253$-$055 (3C\,279):}
Similarly to 3C\,273, the $s$-values are consistent with a canonical \citet{BK79} jet model. However also in this source, the $43\, \mathrm{GHz}$ jet diameter gradient is consistent with a parabolic jet shape and is more collimated than the $15\, \mathrm{GHz}$ jet diameter gradient which is consistent with a conical jet, in both the overlap region and the respective entire jet lengths. \citet{3C273_2018} find a transition from a parabolic to a conical jet shape however they also make use of VLBA data at $1.4\, \mathrm{GHz}$ and MERLIN data at $1.6\, \mathrm{GHz}$, which probes the jet further downstream out to kpc scales. They constrain a transition zone which in Fig.~\ref{1253-055} would appear at $\sim 1\,\mathrm{mas}$.
Our higher-frequency VLBA data are more sensitive to smaller scales and we find a transition zone at $x_b\sim 0.6 \,\mathrm{mas}$. 
In the outer fit component, we still find an $l$-value smaller than unity and a possible second break at or above 1\,mas is not probed by our data.
The radius of the Bondi sphere for this source is well below the location of both the geometry break point reported by \citet{3C273_2018}, however also may coincide with the transition region, we found for the lowest gas temperatures. With the lower frequency VLBA data, \citet{Kovalev2020} did not constrain a geometry break. The single power law fit of the jet geometry is  consistent with the $l_{15}$-value we find. 

\paragraph{1308$+$326 (OP\,313):} This quasar
features $l$ and $s$-values which are indistinguishable within their uncertainties, along the entire jet length and the overlap region (Fig.~\ref{1308+326}). The jet geometry overall appears conical and the $s$-values are consistent with the canonical parameter values from the \citet{BK79} model, which can be interpreted as a freely expanding jet. This is also consistent with \citet{Kovalev2020}.

\paragraph{1633$+$382 (4C\,+38.41):} This quasar jet
features $l$ and $s$-values which are indistinguishable within their uncertainties, along the entire jet length and the overlap region (Fig.~\ref{1633+382}). In spite of strong scatter in the data, the jet geometry appears conical and the $s$-values are consistent with the canonical parameter values the \citet{BK79} model which
can be interpreted as a freely expanding jet. Lower frequency data suggest an opening of the jet with a diameter gradient index of $l\sim 1.3$ \citep{Kovalev2020}. Also for this source it would be interesting to combine the high and low frequency data in order to constrain a possible geometry transition.

\paragraph{1652$+$398 (Mrk\,501):} This is the second closest source in our sample and compared to most other sources is covered by only a relatively small number of observing epochs. It
features diameter gradients at both frequencies which are consistent with a parabolic jet shape, as seen in Fig.~\ref{1652+398}. This is suggestive of a jet that is observed on scales where it is still accelerating rather than freely expanding. This is also consistent with the $T_b$-gradients which are flatter than those expected for a freely expanding jet. 

\paragraph{1730$-$130 (NRAO\,530):} This is a quasar jet that
features $l$ and $s$ values which are consistent with a freely expanding jet, when analyzing the overlap region at both frequencies (Fig.~\ref{1730-130}). Along the entire jet length, however, the $15\, \mathrm{GHz}$ jet seem to be more collimated and features a flatter $T_b$-gradient. This is dominated by  the influence of a rather distant component between $20\lesssim d_c/\mathrm{mas} \lesssim 30$ from the core which appears in epochs between $\sim 2006$ and  $\sim 2012$. This component is relatively compact and features comparatively high $T_b$-values with respect to the distance from the core. This might be suggestive of a shocked region where the jet recollimates. \citet{Lu2011} find this component to be stationary. At lower frequenceie \citet{Kovalev2020} also do not constrain a geometry transition. The scaling of the jet diameter is consisntent with what is found in this study.

\paragraph{1749$+$096 (OT\,081):} Model-fit results for this BL\,Lac object are shown in Fig.~\ref{1749+096}. Its diameter gradients are consistent with a conical jet shape while the brightness temperature gradients are steeper than expected within the simple \citet{BK79} jet model. 
This suggests that the parameters $n$ and $b$ are responsible for the steep $T_\mathrm{B}$ gradients. A poloidal magnetic field component could explain a steepening of the brightness temperature gradients while maintaining a conical jet shape.

\begin{figure}
\centering
	\includegraphics[width=\hsize]{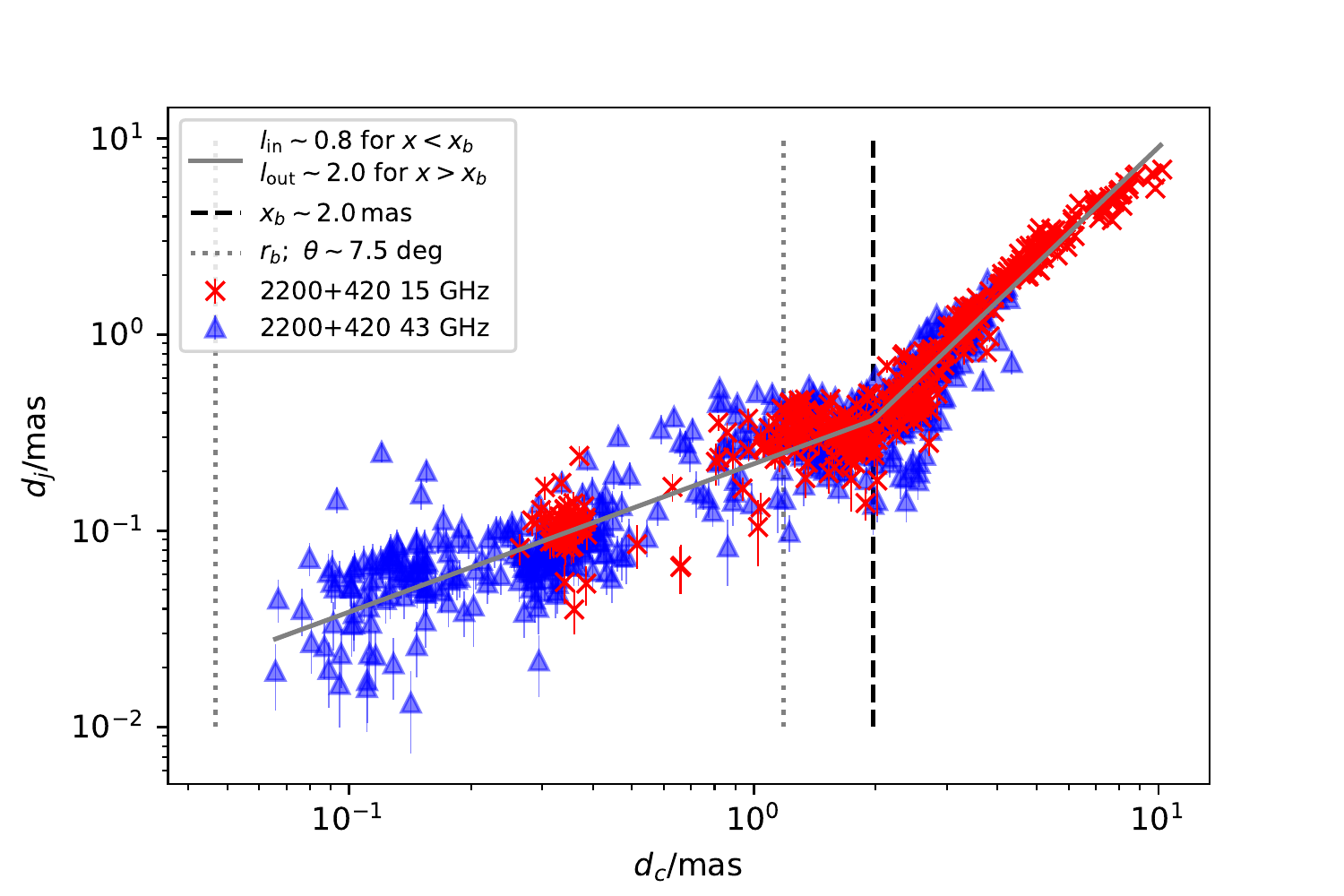}
	\includegraphics[width=\hsize]{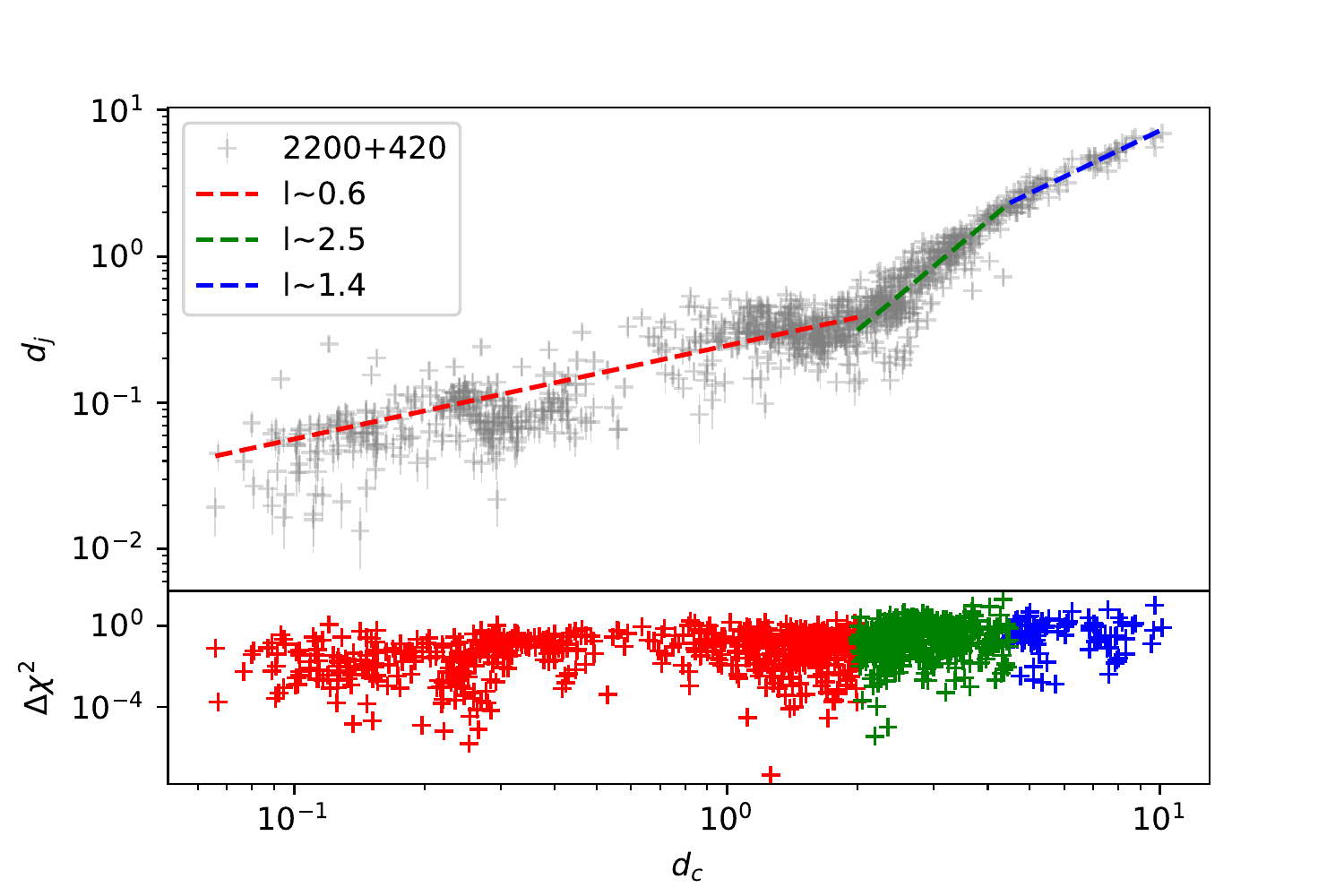}
      \caption{Jet diameter vs. core distance for BL Lac (2200+420). \textit{Top:} Single transition zone model (rigorous fit). The shape transitions at $x_b \sim 1.8 \, \mathrm{mas}$ from parabolic to conical. \textit{Bottom:} Two-transition model where the break points are 'fit by eye' parametrizes the complex structure.}
       \label{smooth2200}
\end{figure}

\paragraph{2200$+$420 (BL\,Lacertae):}\label{par.:2200}
The inner region of this prototypical BL\,Lac object features an overall jet shape which is consistent with a cone which is also confirmed by \citet{Casadio2021} by additionally taking GMVA data at $86\, \mathrm{GHz}$ into account. If fit with a simple power-law model (Fig.~\ref{2200+420}), the $s$-values are consistent with the canonical value expected for a simple \citet{BK79} jet model. At $\sim 1.5\, \mathrm{mas}$ from the respective cores, we find a region where the jet locally shows a dip in its  diameter and excess $T_b$ values which is also consistent with the findings of \citet{Casadio2021}. They associate this region with a recollimation shock which is connected to the changing external pressure gradient close to the Bondi radius \citep{Kovalev2020}.
The large $\chi ^2 _{\mathrm{red}}$ values suggest that the overall jet structure is complex and is not sufficiently represented by a single power law.
The two-zone diameter-gradient model is shown in Fig.~\ref{smooth2200} (top) and features a parabolic jet shape out to $d_c \sim 2.0\, \mathrm{mas}$ transitioning to a hyperbolic shape jet further downstream which is consistent with the findings of \citet{Kovalev2020}. 
Analyzing the jet in more detail it becomes clear that the jet of BL\,Lacertae is even more complex and still not fully described by a two-zone diameter-gradient model. To emphasize this we also show Fig.~\ref{smooth2200} (bottom). Here the break points are not found rigorously but are a 'fit by eye'. In this representation, the jet has a parabolic region up to roughly the point where the transition was fitted in the single-transition zone model, opening up to an extreme jet geometry ($l\sim 2.5$), before transitioning again at $\sim 5 \, \mathrm{mas}$ into a quasi-conical jet. The latter transition is roughly coinciding with the estimated location of the Bondi radius. This seems to contradict the findings of \citet{Casadio2021} where they do not find a geometry transition, additionally analyzing $86 \, \mathrm{GHz}$ GMVA data. \citet{Casadio2021} argue that the region between $0.5\lesssim r/\mathrm{mas}\lesssim 1.5$ should be excluded from a fit because in this region the jet shows a different expansion rate accompanied by a decrease in brightness which we also find in the brightness-temperature data (Fig.~\ref{2200+420}). Not excluding this region may yield a parabolic geometry of this jet. This tentative contradiction can be resolved by the jet's complex structure where clearly different regions with different behavior in the jet can be found. This would not affect the overall jet shape, measured over several decades in the distance from the respective radio cores, but would allow local geometry transitions of the jet.\newline
We also tested the effect of the measured core shift of $\sim 0.1\,\mathrm{mas}$, reported by \citet{Osullivan2009}. The resulting jet geometry and brightness temperature gradients along the entire jet length are indistinguishable within the errors from the case where the core shift is not considered. Also the found geometry break position and the inner and outer diameter gradient do not differ. 

\paragraph{2223$-$052 (3C\,446):} This quasar jet
features $s$ values which are indistinguishable regarding the two frequencies and are well described by the \citet{BK79} jet model. The diameter gradients are also indistinguishable, however there seems to be peculiarity within the $15\, \mathrm{GHz}$ data. While the $43\, \mathrm{GHz}$ jet shape is consistent with a cone, the $15\, \mathrm{GHz}$ jet shape is suggestive of  a parabolic shape. 
This might be explained by the large scatter (as it is found to be characteristic for quasars) and the relatively low number of observing epochs for this source.
A systematic jet shape transition however cannot be found in this source. The overall parabolic jet shape is consistent with the findings of \citet{Kovalev2020}. 

\paragraph{2230$+$114 (CTA\,102):} This quasar \citep[see e.g., ][]{fromm2011,fromm2013a,fromm2013b}
features $s$-values which are consistent with a parabolic jet in equipartition at both frequencies. The jet geometry can consistently be fit with a parabola at both frequencies, however along the entire jet axis at $15\,\mathrm{GHz}$, the $l$-value becomes also consistent with a cone due to the larger uncertainties introduced by the larger scatter of the data beyond $d_c \sim 2\,  \mathrm{mas}$. A transition in this source cannot be constrained.

\paragraph{2251$+$158 (3C\,454.3):}
Similarly to 1222$+$216 (4C\,+21.35) and 1253$-$055 (3C\,279), the $s$-vlaues are consistent with a \citet{K81} jet model. However also in this source, the $43\, \mathrm{GHz}$ jet diameter gradient is consistent with a parabolic jet shape and is more collimated than the $15\, \mathrm{GHz}$ jet diameter gradient which is consistent with a conical jet, in both the overlap region and the respective entire jet lengths. Also here when rigorously searching for possible break points and an inner and outer slope, it cannot be constrained properly, due to the large scatter of the data points between $0.2 \lesssim d_c/\mathrm{mas} \lesssim 3 $ in the $43\, \mathrm{GHz}$ jet. This jet also might be a candidate where at higher frequencies, more collimated, inner structures of a multilayered jet are observed. The overall parabolic jet shape is consistent with the findings of \citet{Kovalev2020}.

%% file: conclusion_2.tex
\section{Summary and conclusions}
{
Using dual-frequency VLBI observational data for a sample of 15 FSRQs, 11 BL Lacs, and 2 radio galaxies at $15\, \mathrm{GHz}$ and $43\, \mathrm{GHz}$, the geometry and  brightness temperature gradients along the jet axes has been determined and interpreted.  Our main observational findings are the following:

\begin{itemize}
    \item The geometry of the VLBI jets in our sample can generally be described as power laws with indices $l$ in the range between 0.5 and 1, i.e., collimated jet geometries. 
	\item At $43\,\mathrm{GHz}$, the jets tentatively show a stronger collimation than at $15\,\mathrm{GHz}$. This is a significant difference if the full jet extent seen at both frequencies is considered, but not significant when the analysis is restricted to the overlap region where the jet is imaged at both frequencies (Table~\ref{tab3})
	\item FSRQs show stronger collimation than BL Lacs at both frequencies, but again this is not significant if restricted to the overlap region (Figs.~\ref{plot:l-values_sc} and \ref{plot:s-values_sc}). 
	\item FSRQs show a more complex structure than BL Lac objects (with more local variations of the brightness temperature gradients leading to higher $\chi^2$ values of the power law fits in Figs.~\ref{plot:l-values_sc} and \ref{plot:s-values_sc}). 
	This is true for the jet geometry, indicating regions where the jets collimate and widen locally, as well as for the brightness temperature gradients indicating regions in the jet where the flux density is locally higher than in other regions. 
	\item We found parabolic-to-conical geometry transitions in 5 FSRQs: CTA\,26 (0336$-$019), PKS\,0528$+$134, 4C\,+71.07 (0836$+$710), 4C\,+29.45 (1156$+$295), 3C\,279 (1253$-$055). In addition, we found parabolic-to-conical geometry transitions in 3 BL Lacs: 3C\,66A (0219$+$428), Mrk\,421 (1101$+$384), and BL\,Lacertae (2200$+$420); and in a radio galaxy: 3C\,111 (0415$+$379).
	\item The jet radius in the transition zone is smaller in BL Lac objects than in FSRQs and the geometry transitions occurs further down the jet for FSRQs which is likely due to systematically smaller SBMH masses in BL Lacs. 	
	\item The FSRQs CTA\,26 (0336$-$019) and 4C\,+71.07 (0836$+$710) show a stationary component downstream the jet past the Bondi radius. 
	\item BL\,Lacertae (2200$+$420) shows a jet structure of substantially greater complexity than a simple transition from a parabolic to a conical shape due to local effects like pressure gradient change and locally different expansion rates between $0.5\lesssim r/\mathrm{mas}\lesssim 1.5$, \citep{Casadio2021} 
\end{itemize}

The high-frequency VLBI observations apparently resolve jets on physical scales where in many cases their shapes change from parabolic (collimated) to conical (free) morphologies.  
This observational finding is in agreement with relativistic magnetohydrodynamical jet models, {\it e.g.,} \citet{2006ApJ...651..272F,2011ApJ...737...42P}.  
These models show that
the accretion disk in blazars with high accretion rates (FSRQs) flattens the magnetization profile at the jet base compared to jets from disks with low accretion rates (BL Lacs), either by advection of magnetic flux with the disk or by generating it with a disk dynamo. 
The radially flattened magnetization profile and the strong disk wind envelope are predicted to lead to (i) a collimation of the 
toroidal-field dominated jet extending to large distances from the core, (ii) morphological complexity as these jets are prone to produce axially propagating knots, and (iii) high terminal speeds. By contrast, jets in BL Lacs with steeper magnetization profiles at their base and weaker wind envelopes are expected more uniform, slower, and widen closer to the core \citep{Potter_Cotter_2015}, which is in good agreement with our observational results.  The acceleration of the jets is driven by the conversion of Poynting flux to kinetic energy flux of particles entrained from the surrounding disk wind or injected by reconnection events or electrostatic gaps at the jet base \citep{2021A&A...646A.115W}.
The breakdown of the collimation (due to the kink instability) first affects the low-frequency sheath of the jet, in accord with the observational results reported here.  On the kinetic level of description, the energy conversion is likely to be associated with turbulence, anomalous resistivity, and ultrarelativistic particle acceleration. These dissipative processes involve complex 3D-substructures in the jet plasma leading to fast flux variability at high energies. In this scenario, no direct correlation with the extent of the Bondi radius around the central supermassive black hole is predicted, which is in agreement with our results. Further improvements of this type of studies are possible: adding polarimetric analyses and increasing the sample size, the dynamical range, and the accessible bandwidth of VLBI observations, holds the promise to further advance our understanding of the physical processes involved in the blazar jet phenomenon.}


%% file: Appendix.tex
\begin{appendix}
\onecolumn
\section{Single power law fits}
\label{appendix_gradfit}
\input{./gradient_results.tex}

\twocolumn
\section{Comparing different fit methods}
\label{appendix}
\begin{figure}[htpb]
\centering
    \includegraphics[width=0.47\textwidth]{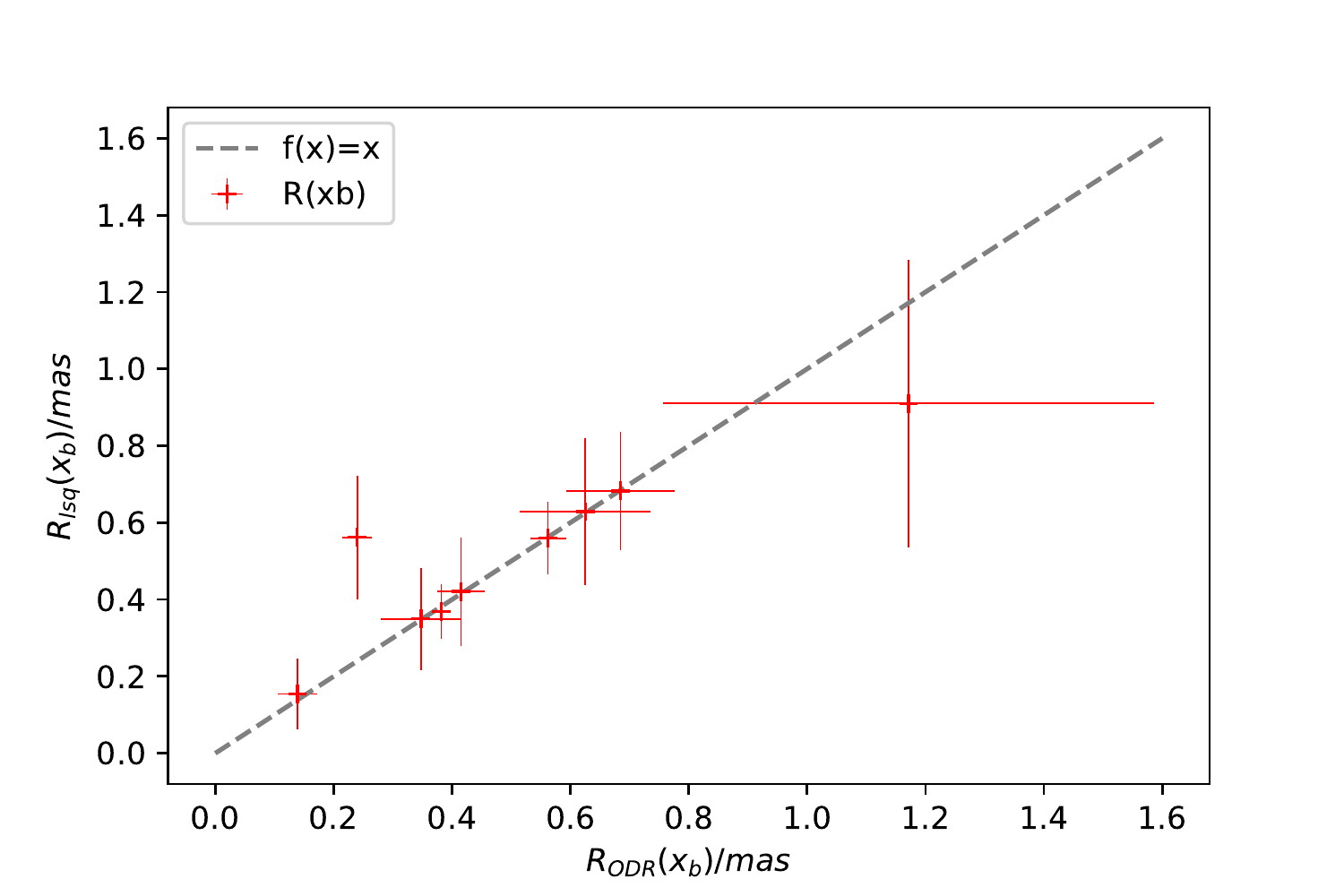}\hfill
	\includegraphics[width=0.47\textwidth]{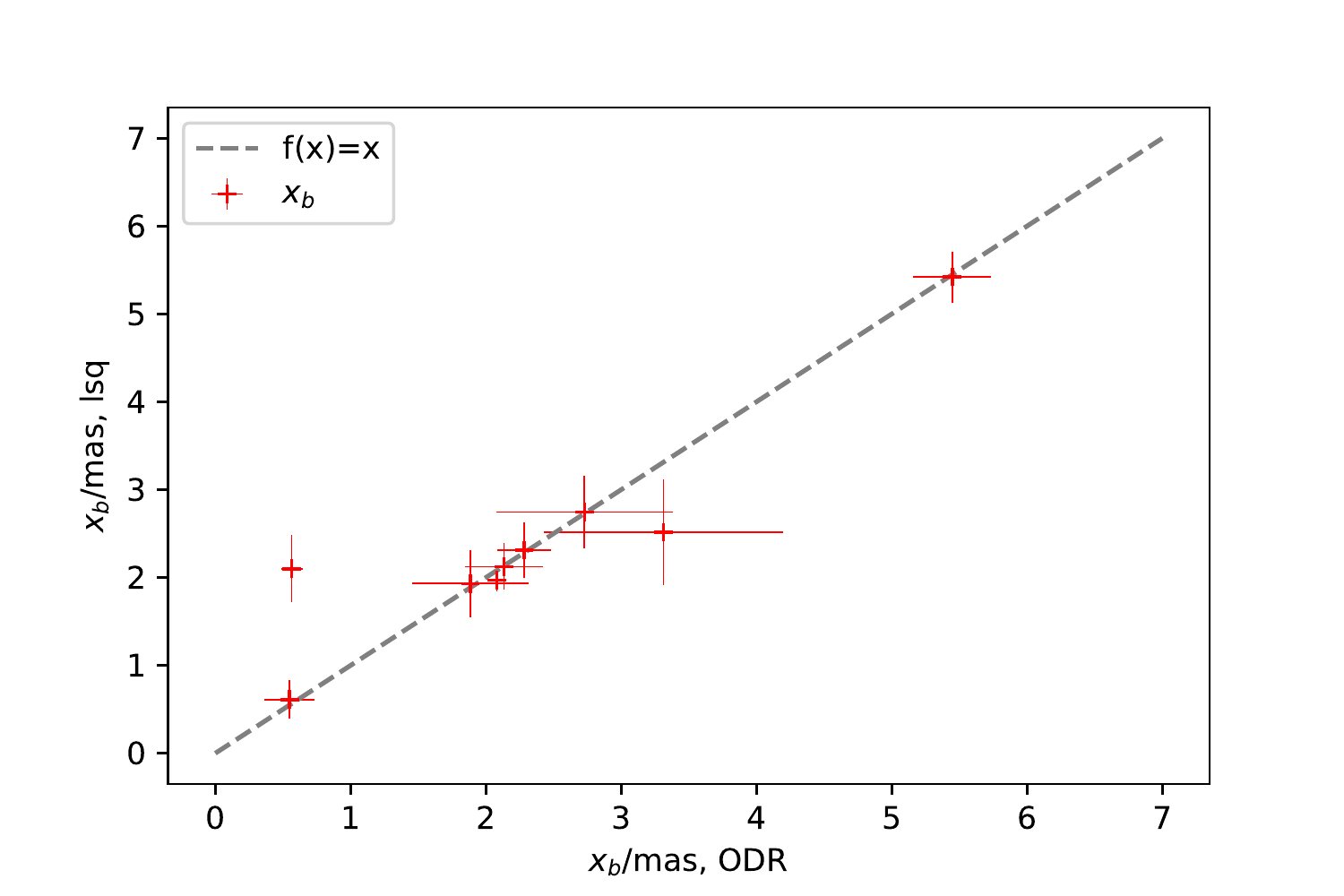}\hfill
	 \includegraphics[width=0.47\textwidth]{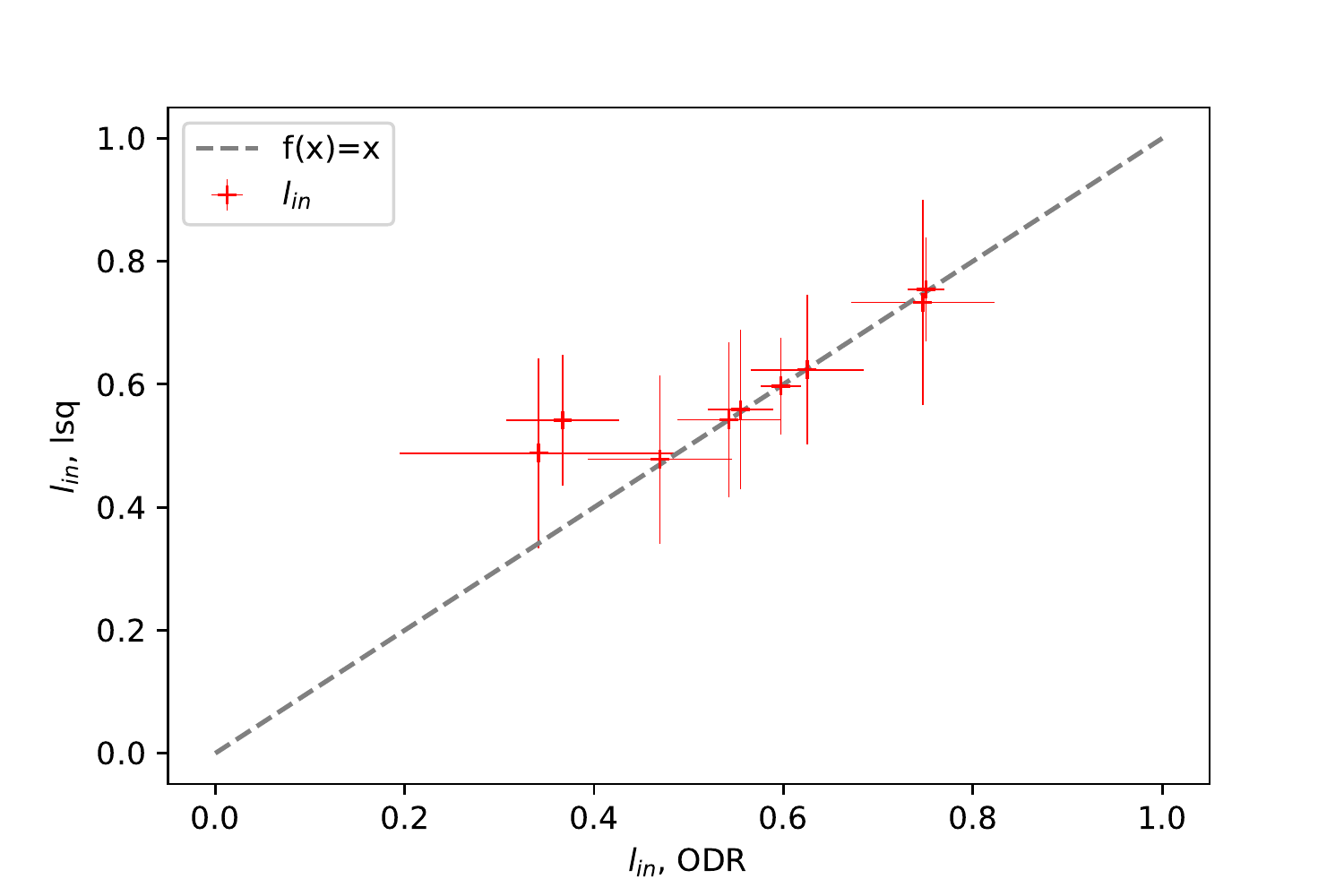}\hfill
	 \includegraphics[width=0.47\textwidth]{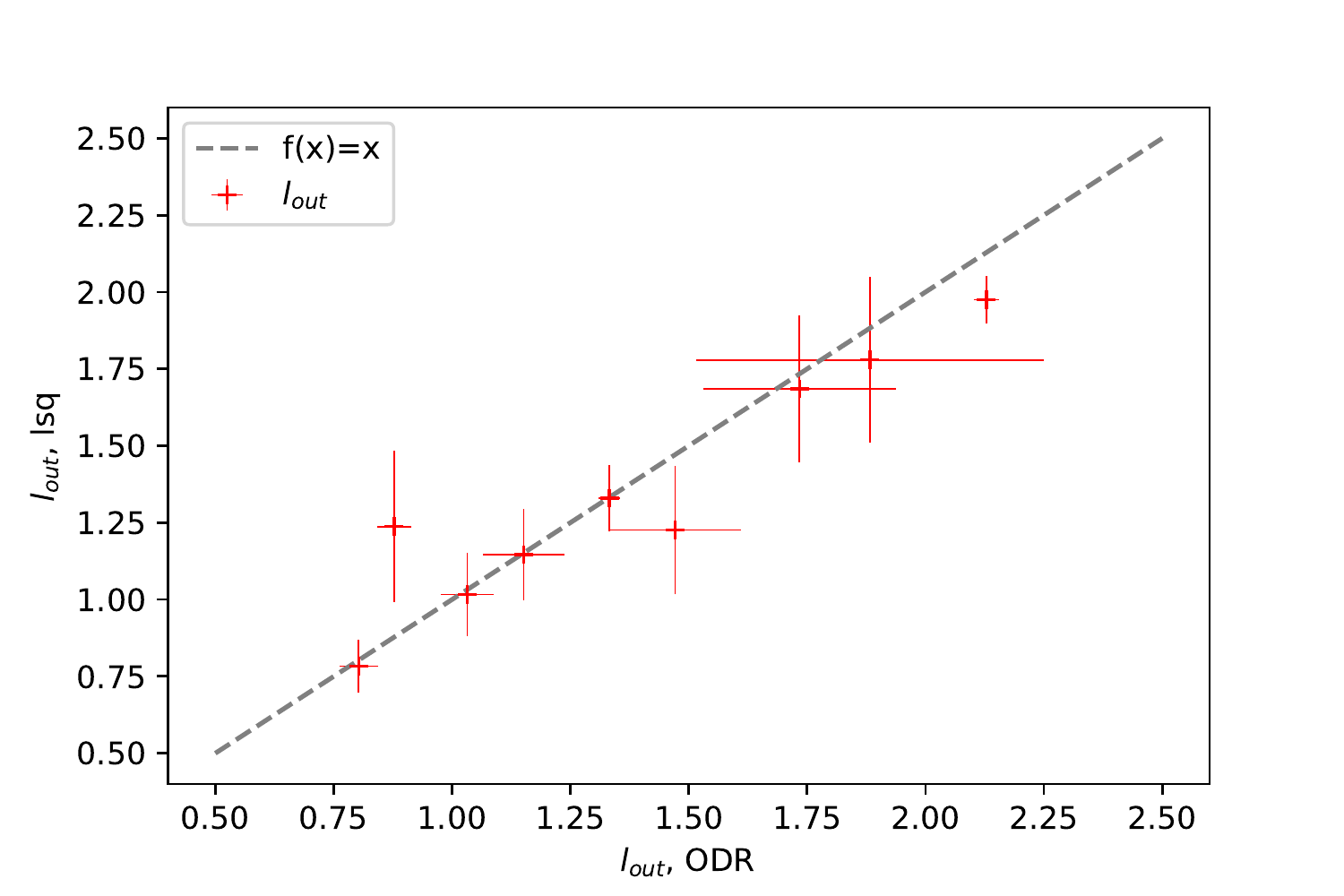}\hfill
    \caption{Comparison of the orthogonal distance regression and the least square fitting method. Overall the fitting results for the different parameters are in agreement within the 1-$\sigma$ error bars. \textit{Top left: }Jet radius at the break point; \textit{Top right: } break point; \textit{Bottom left: } inner diameter gradient; \textit{Bottom right: } outer diameter gradient.}
    \label{plot:ODR_lsq}
\end{figure}

In this study we made use of the orthogonal distance regression (ODR) as used in \citet{baczko2016}, \citet{Baczko2019_tb} and \citet{Baczko2021_arx} in order to fit the smoothly broken power law in the geometry transition sources. For the discussion in this paper, we use the least square implementation in scipy with the trust-region reflective algorithm in order to fit the smoothly broken power law which allows us to set bounds on each parameter which we get from the numerical method when searching for a possible break in geometry in each jet. Figure~\ref{plot:ODR_lsq} shows the fit parameters acquired from the ODR method and the least square method. Both well established methods yield results that are undistinguishable within the error bars.
\newpage
\end{appendix}

%% file: gradient_results.tex
\begin{figure*}[!b]
\centering
	 \includegraphics[width=0.55\hsize,clip]{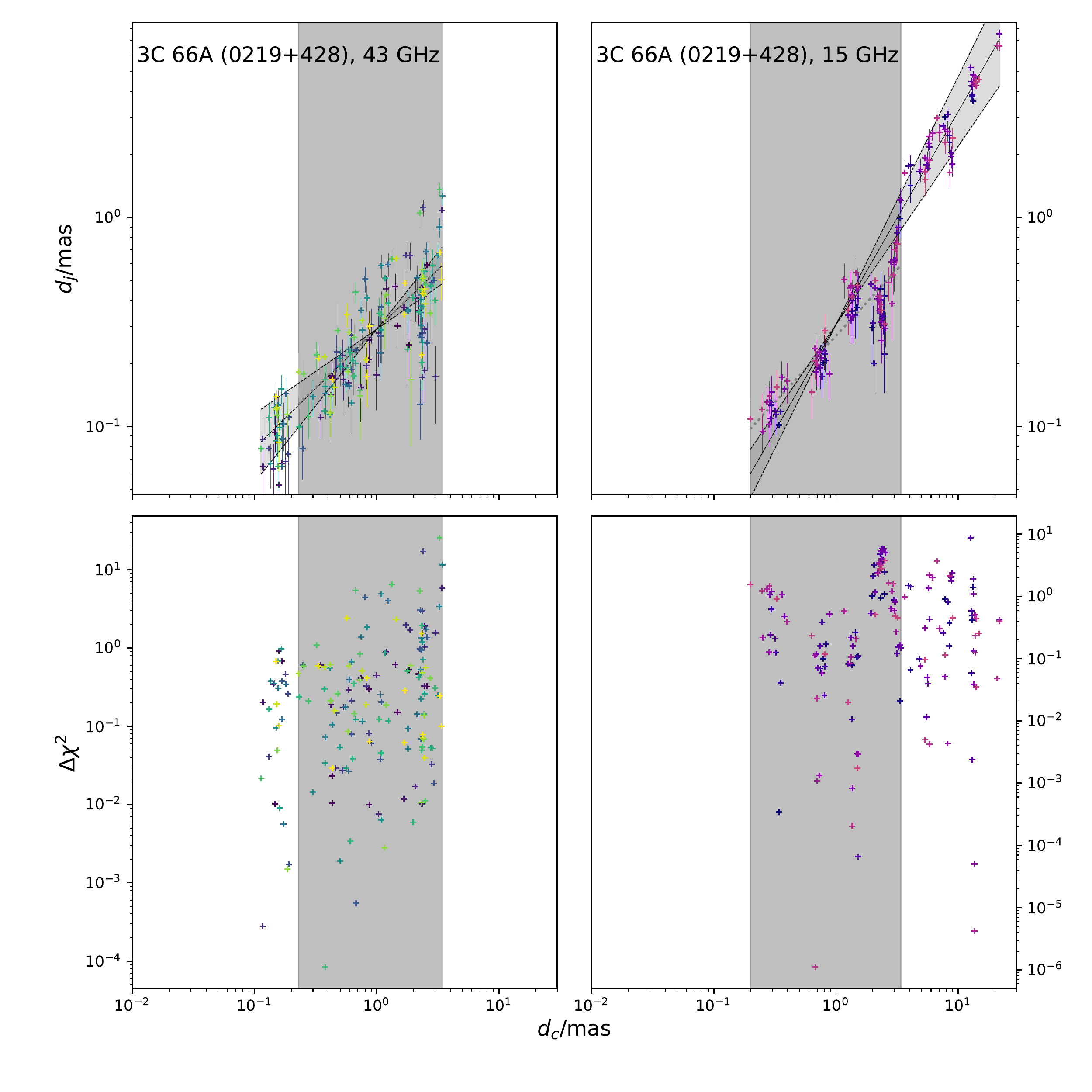}
		\includegraphics[width=0.55\hsize,clip]{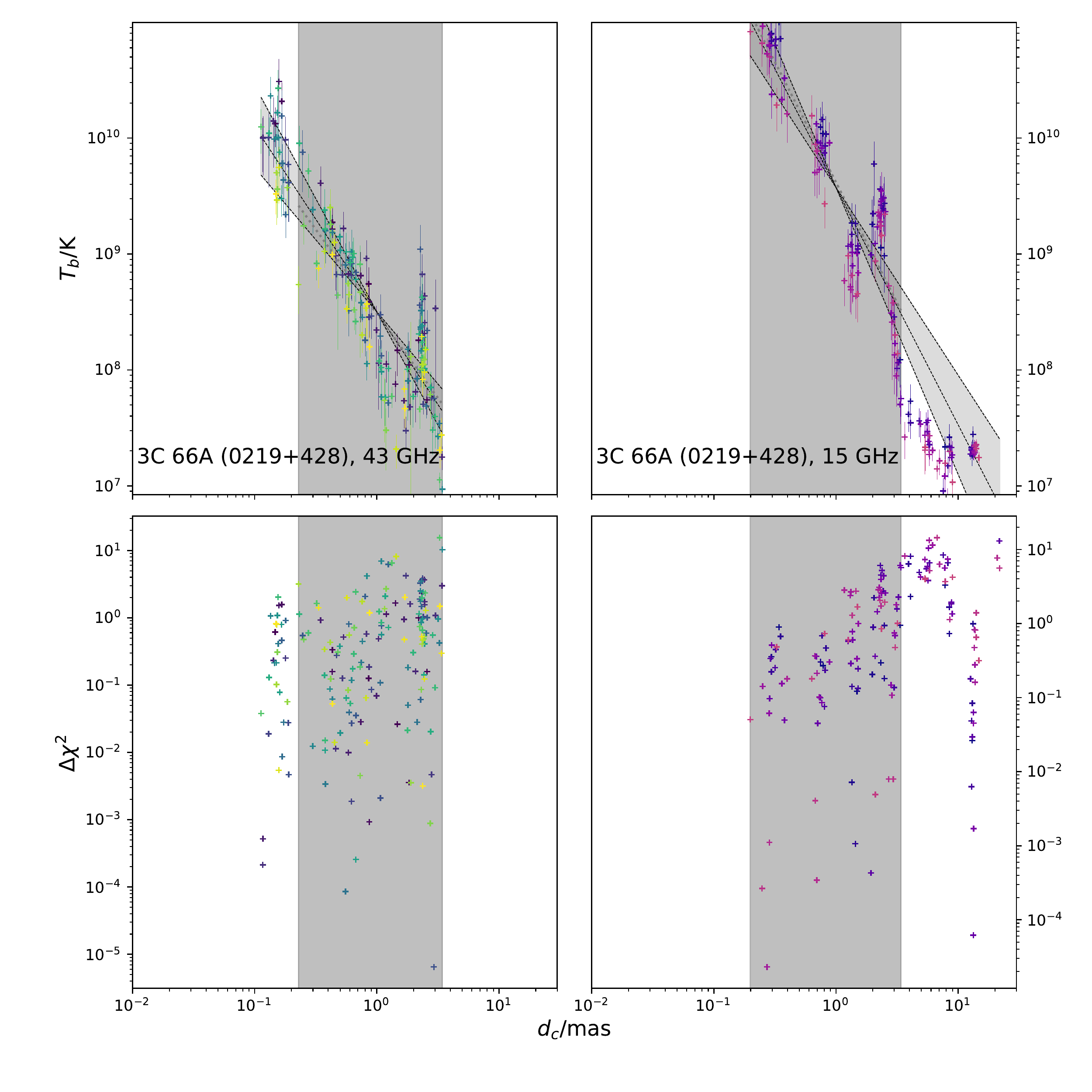}
      \caption{0219+428 jet diameter $d_{\rm j}$ as a function of the distance from the core $d_{\rm c}$ and the brightness temperature $T_{\rm B}$ as function of $d_{\rm c}$. The color coding for both frequencies indicates the epoch, where lighter colors indicate more recent epochs. }
       \label{0219+428}
\end{figure*}

\begin{figure*}
\centering
	 \includegraphics[width=0.65\hsize,clip]{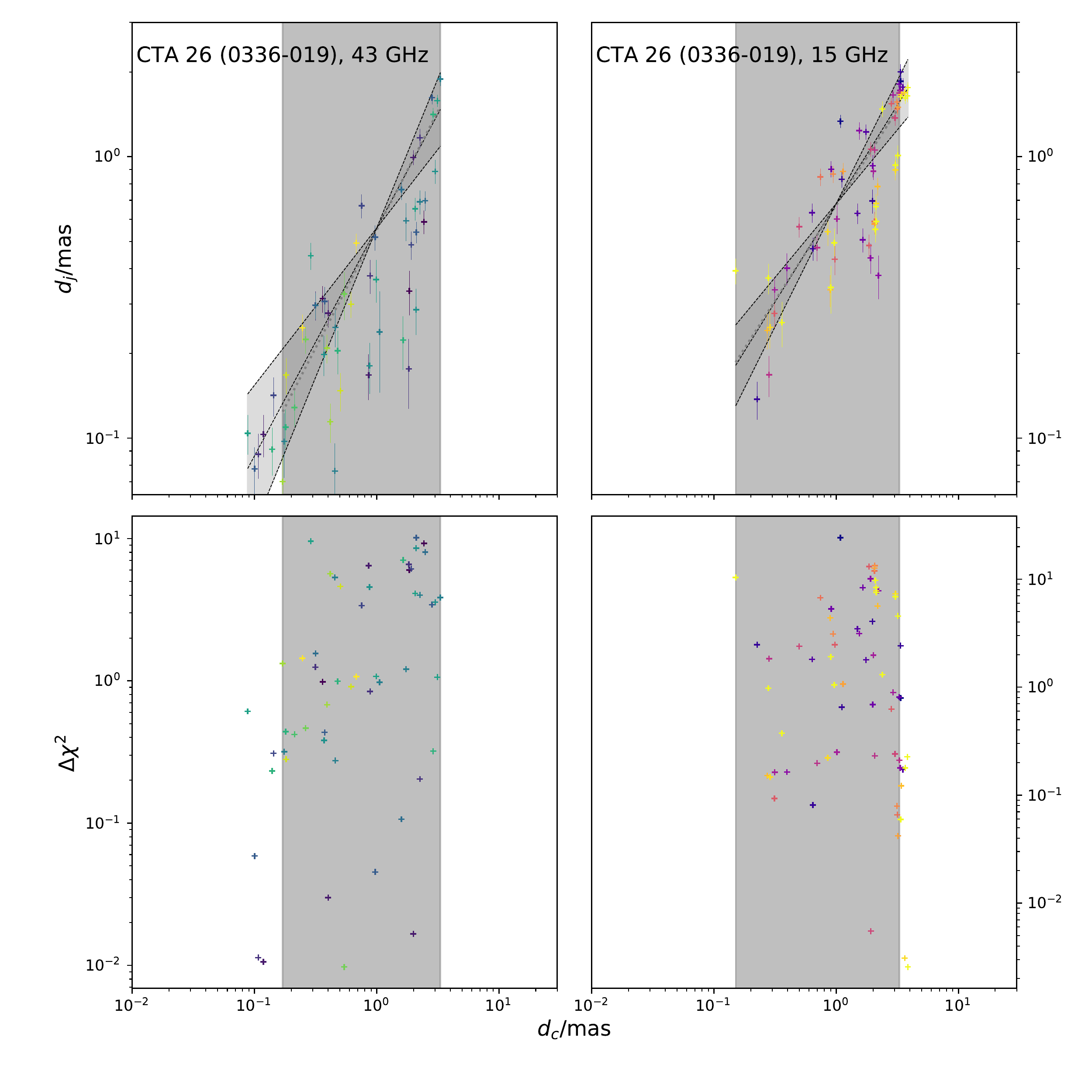}
		\includegraphics[width=0.65\hsize,clip]{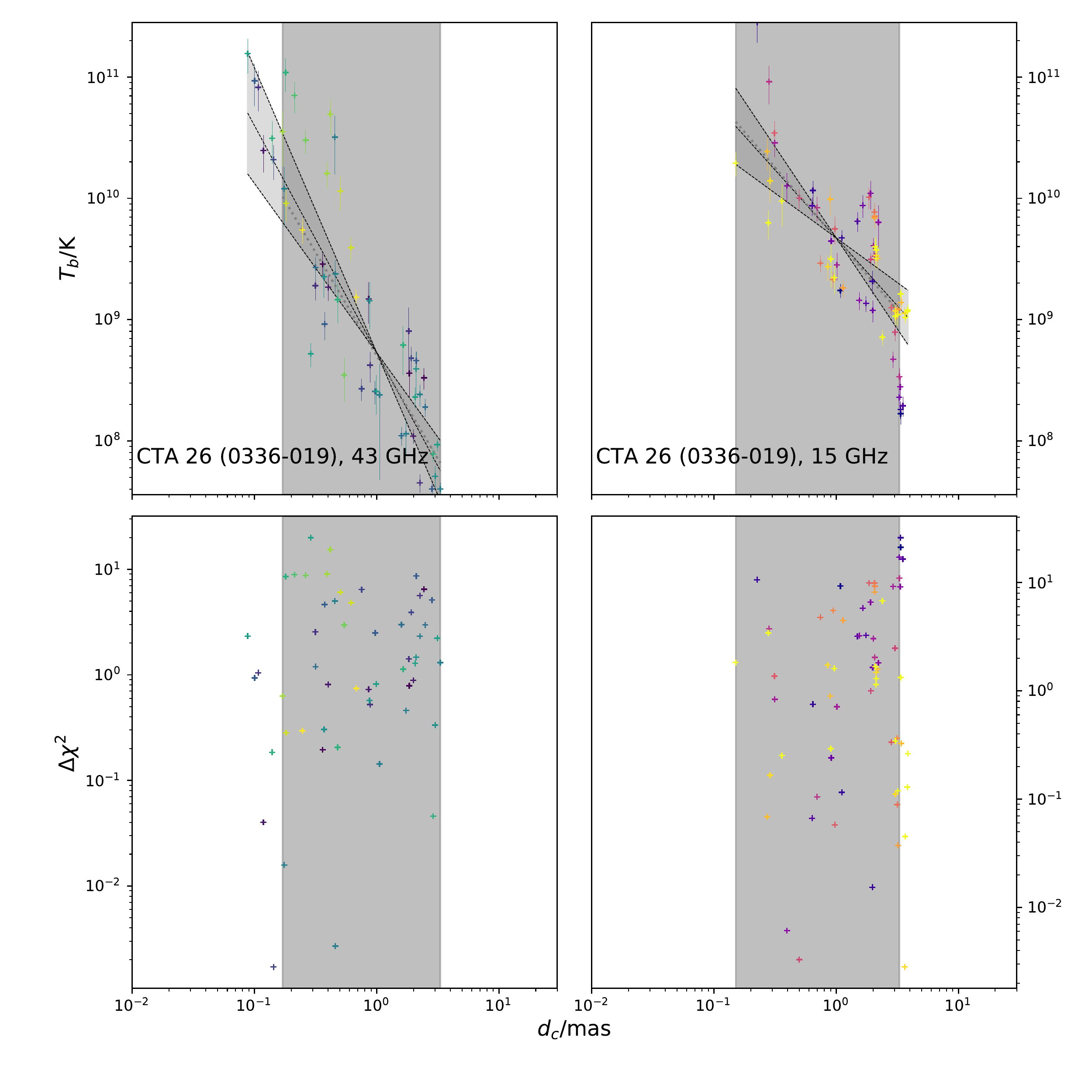}
      \caption{0336-019 jet diameter $d_{\rm j}$ as a function of the distance from the core $d_{\rm c}$ and the brightness temperature $T_{\rm B}$ as function of $d_{\rm c}$. The color coding for both frequencies indicates the epoch, where lighter colors indicate more recent epochs.}
       \label{0336-019}
\end{figure*}

\begin{figure*}
\centering
	 \includegraphics[width=0.65\hsize,clip]{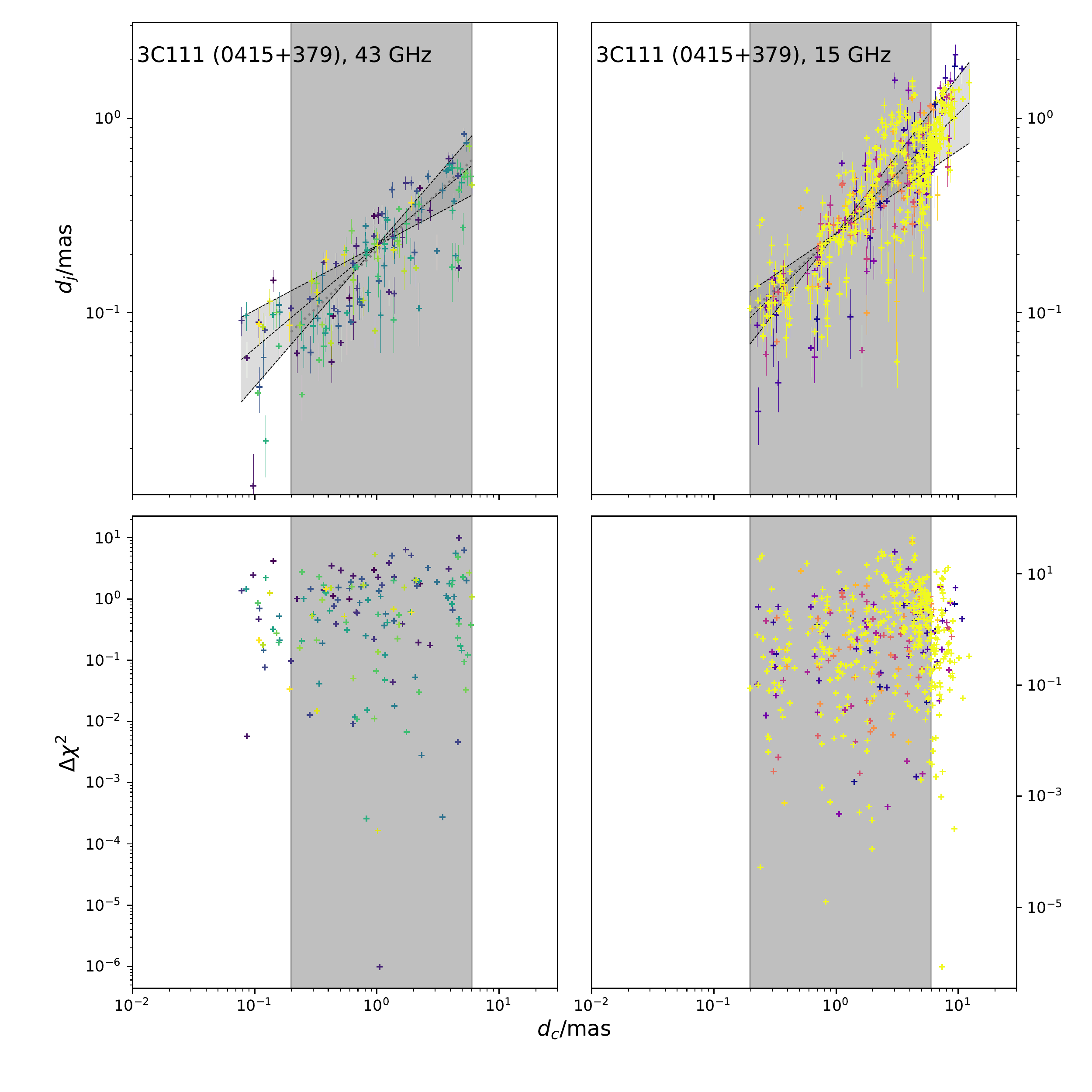}
		\includegraphics[width=0.65\hsize,clip]{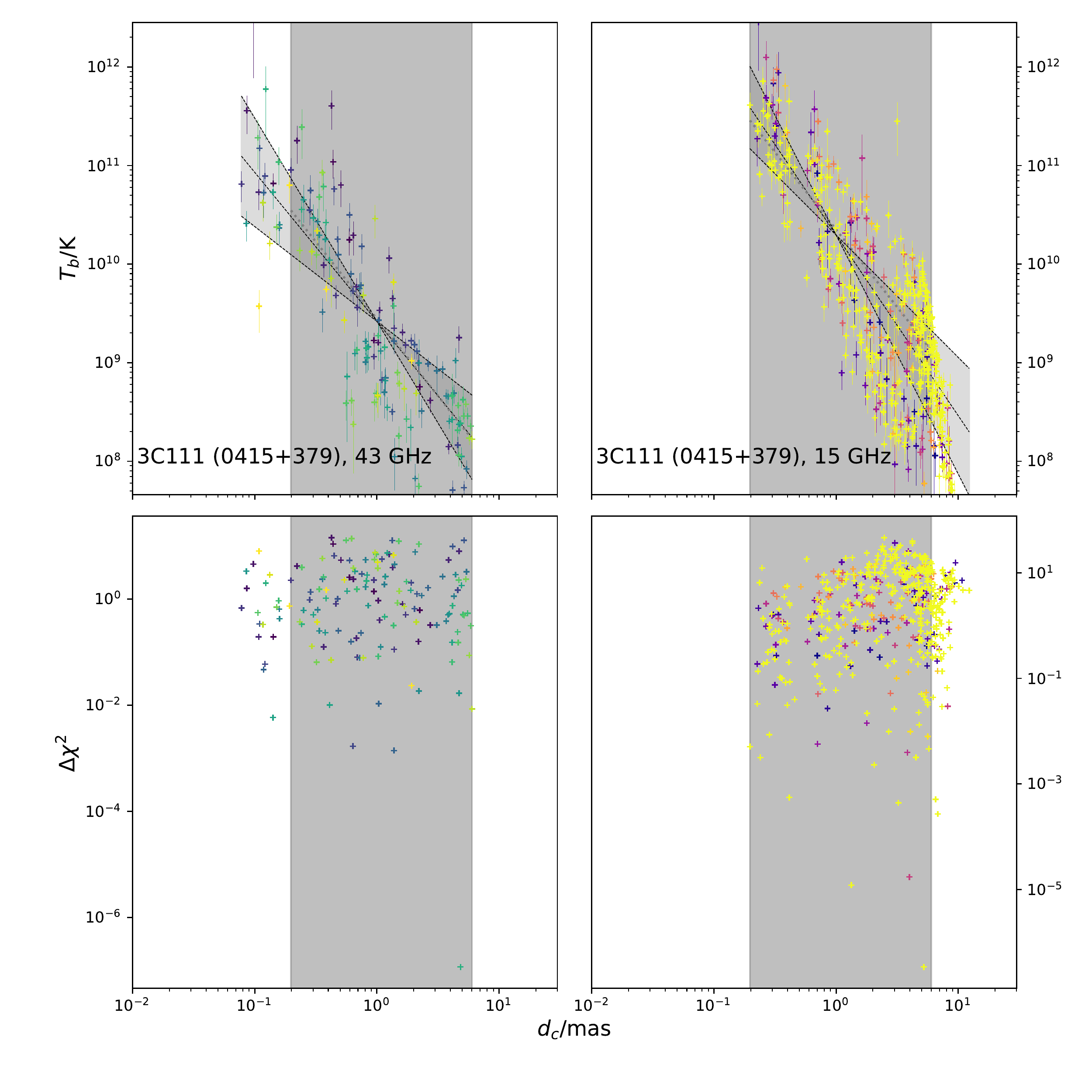}
      \caption{0415+379 jet diameter $d_{\rm j}$ as a function of the distance from the core $d_{\rm c}$ and the brightness temperature $T_{\rm B}$ as function of $d_{\rm c}$. The color coding for both frequencies indicates the epoch, where lighter colors indicate more recent epochs.}
       \label{0415+379}
\end{figure*}

\begin{figure*}
\centering
	 \includegraphics[width=0.65\hsize,clip]{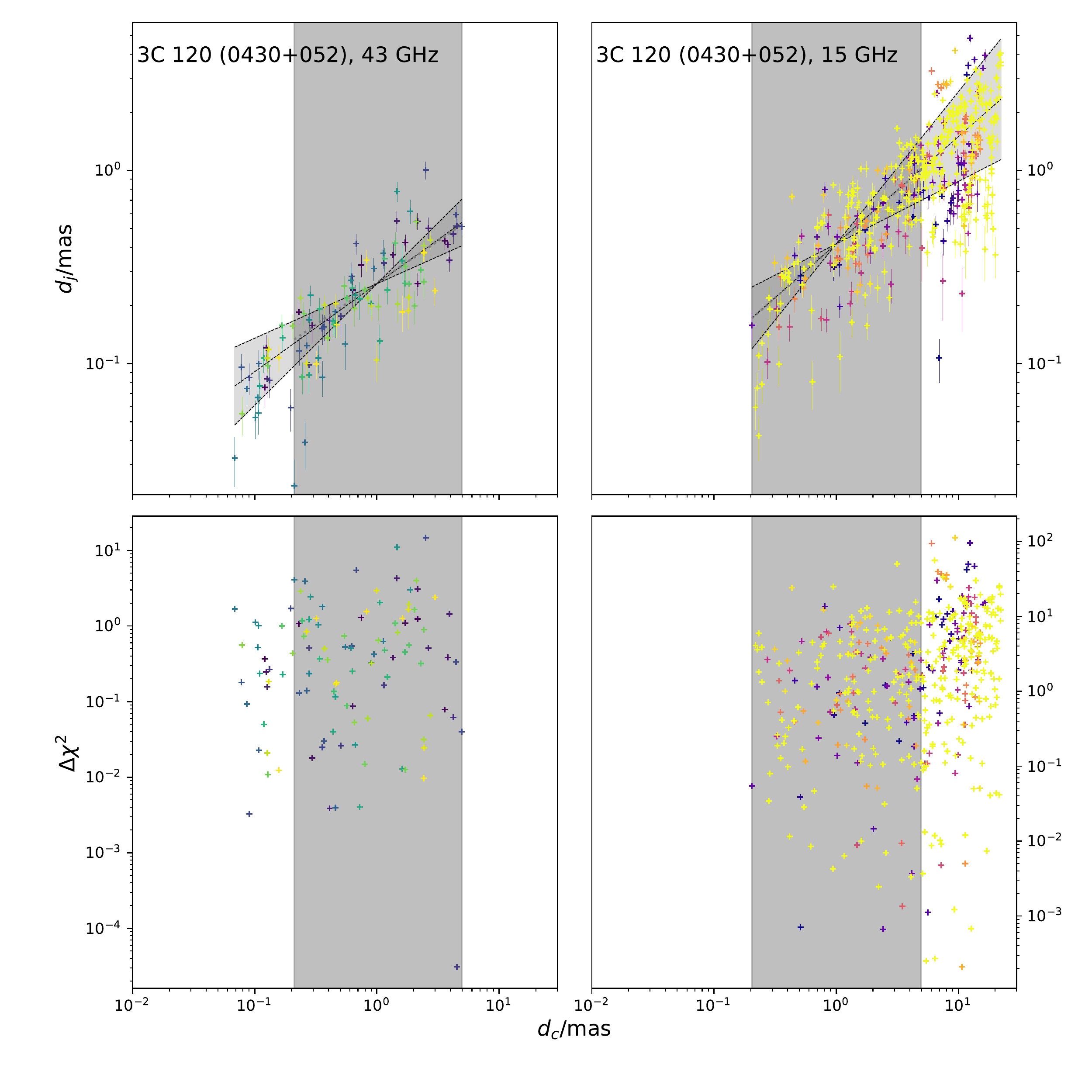}
		\includegraphics[width=0.65\hsize,clip]{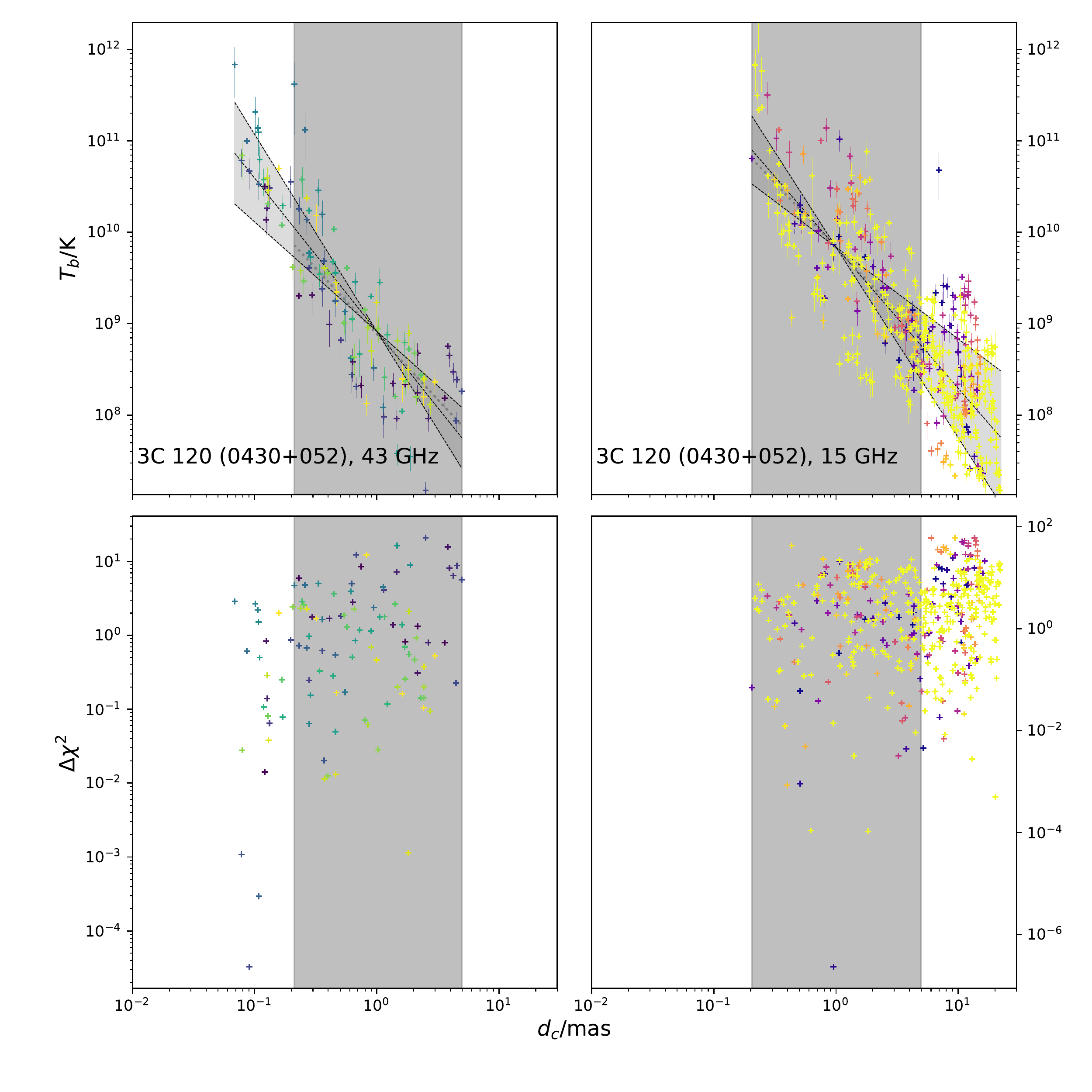}
      \caption{0430+052 jet diameter $d_{\rm j}$ as a function of the distance from the core $d_{\rm c}$ and the brightness temperature $T_{\rm B}$ as function of $d_{\rm c}$. The color coding for both frequencies indicates the epoch, where lighter colors indicate more recent epochs.}
       \label{0430+052}
\end{figure*}

\begin{figure*}
\centering
	 \includegraphics[width=0.65\hsize,clip]{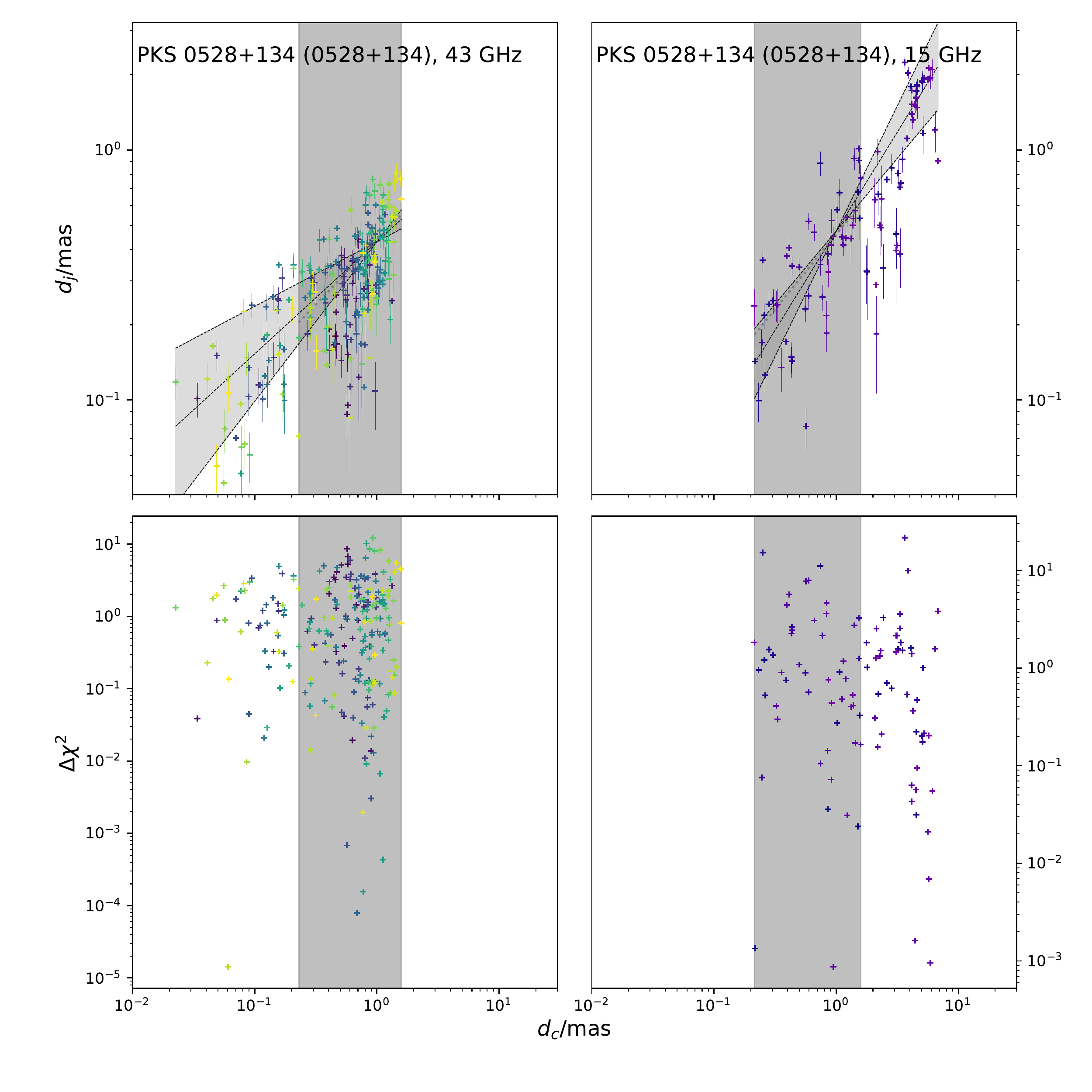}
		\includegraphics[width=0.65\hsize,clip]{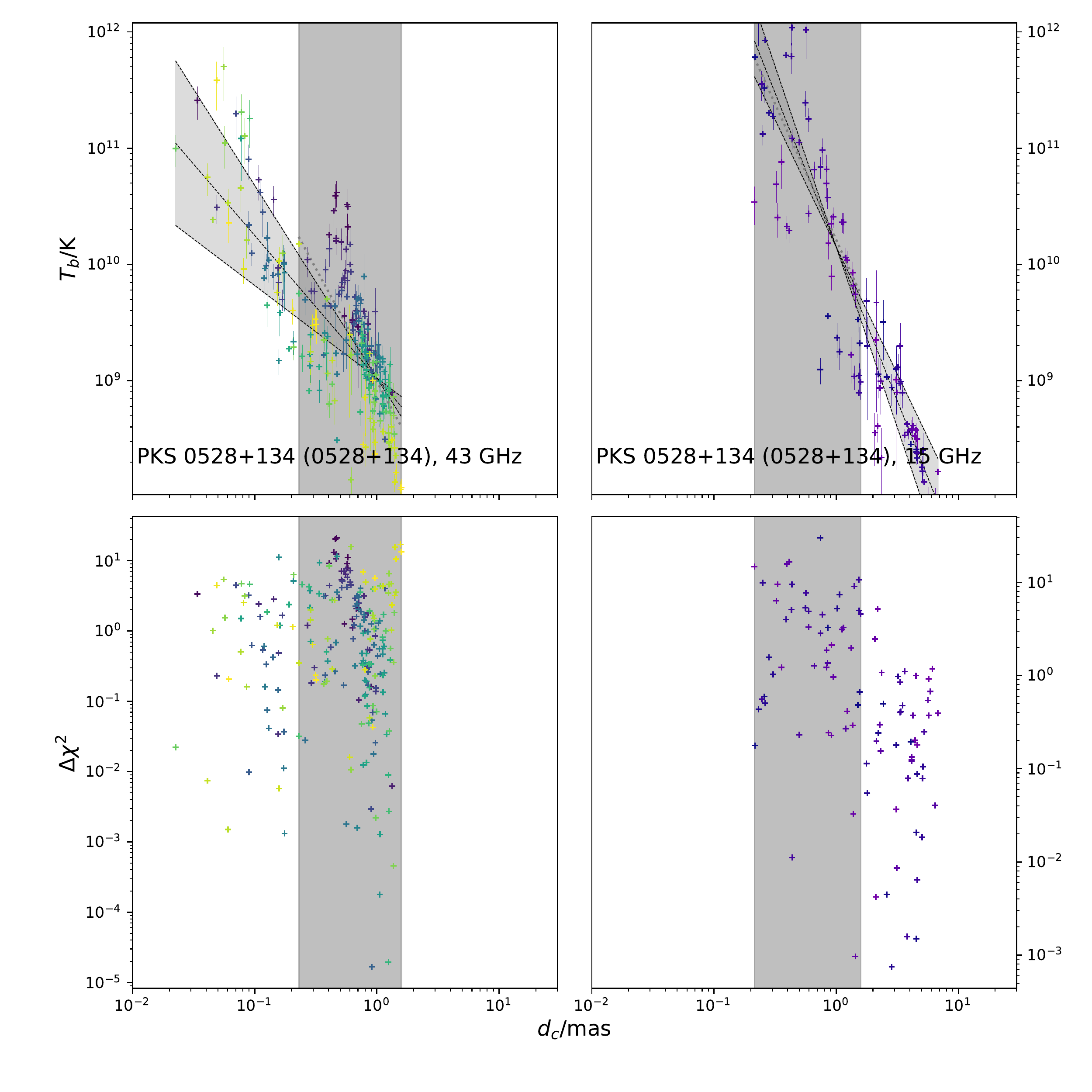}
      \caption{0528+134 jet diameter $d_{\rm j}$ as a function of the distance from the core $d_{\rm c}$ and the brightness temperature $T_{\rm B}$ as function of $d_{\rm c}$. The color coding for both frequencies indicates the epoch, where lighter colors indicate more recent epochs.}
       \label{0528+134}
\end{figure*}
\begin{figure*}
\centering
	 \includegraphics[width=0.65\hsize,clip]{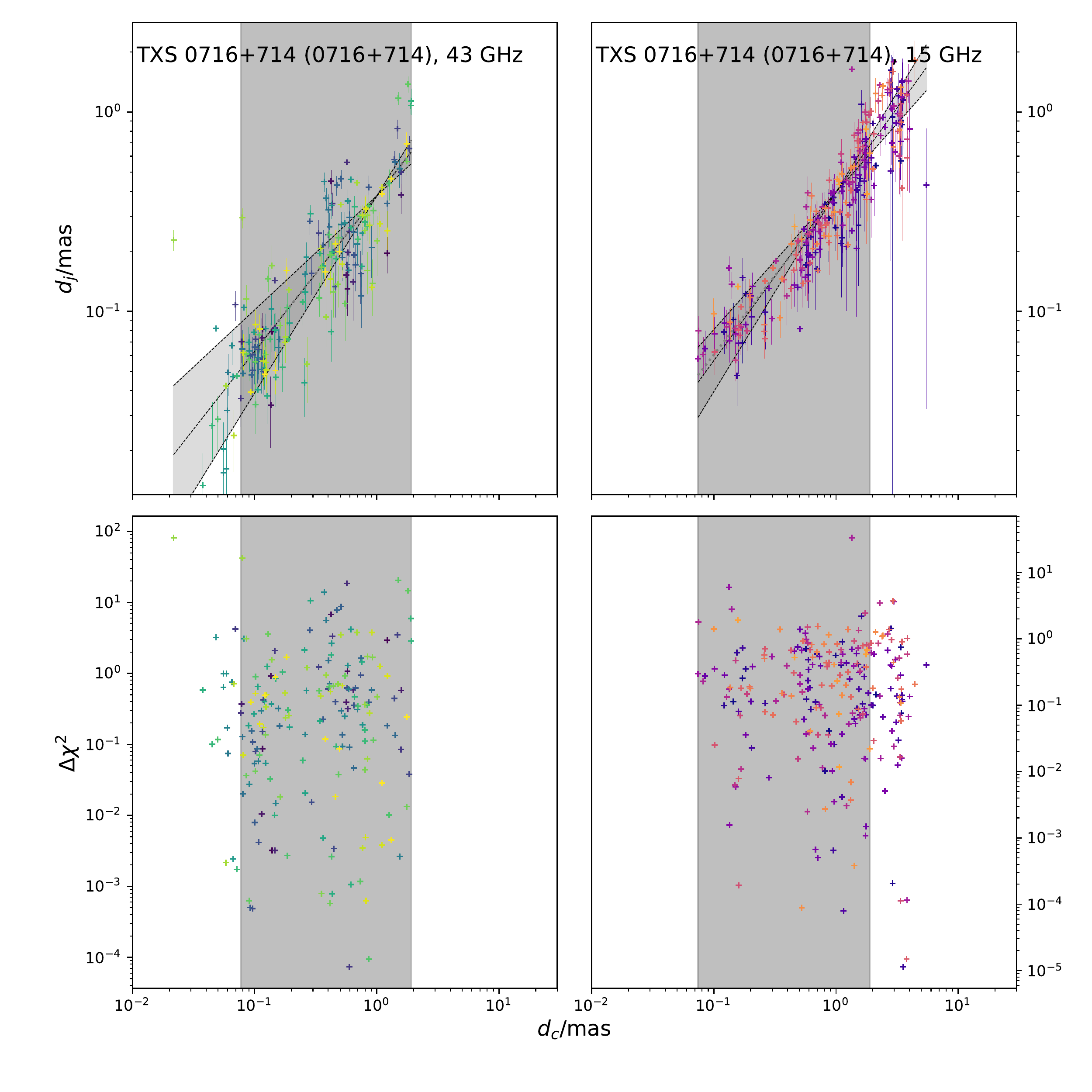}
		\includegraphics[width=0.65\hsize,clip]{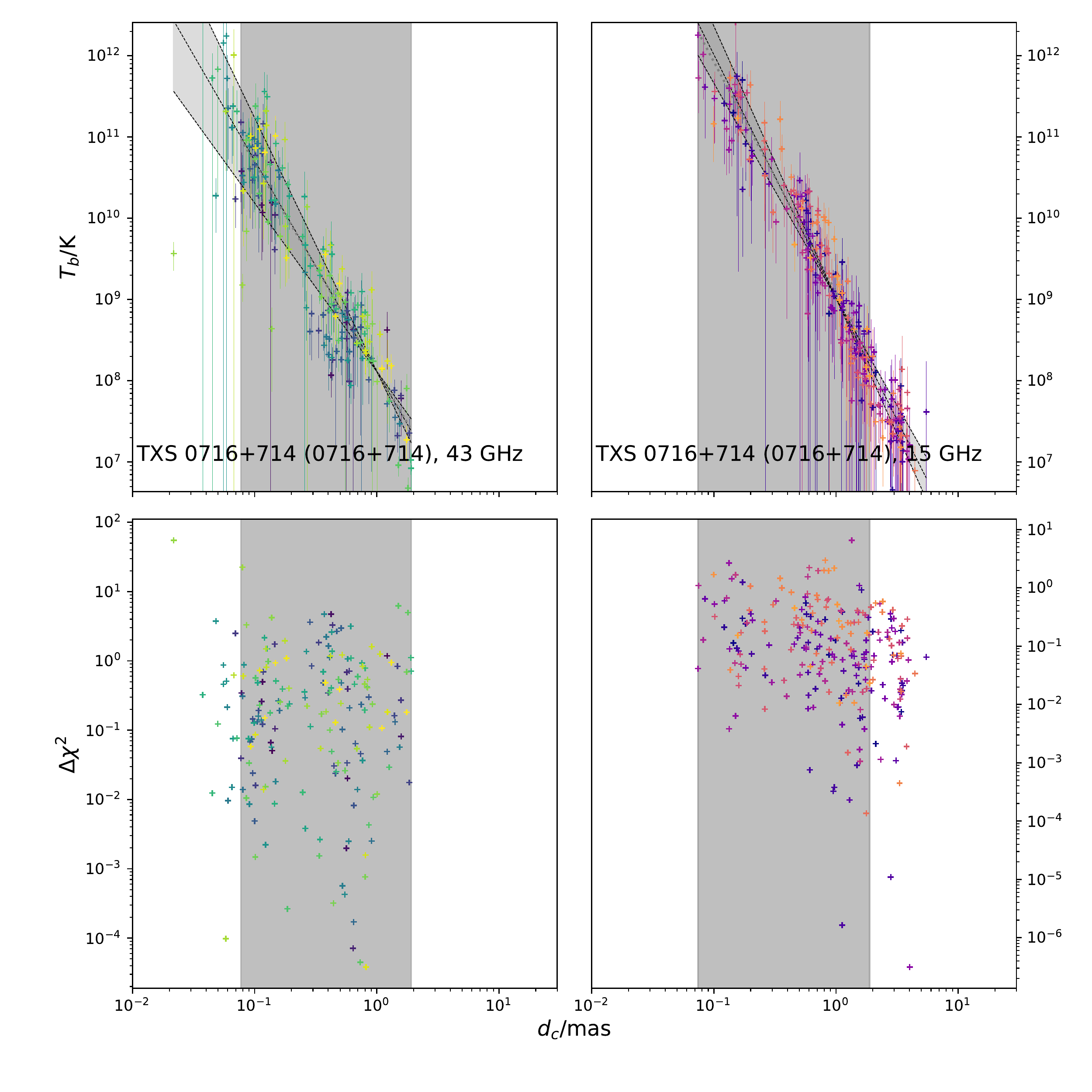}
      \caption{0716+714 jet diameter $d_{\rm j}$ as a function of the distance from the core $d_{\rm c}$ and the brightness temperature $T_{\rm B}$ as function of $d_{\rm c}$. The color coding for both frequencies indicates the epoch, where lighter colors indicate more recent epochs.}
       \label{0716+714}
\end{figure*}

\begin{figure*}
\centering
	 \includegraphics[width=0.65\hsize,clip]{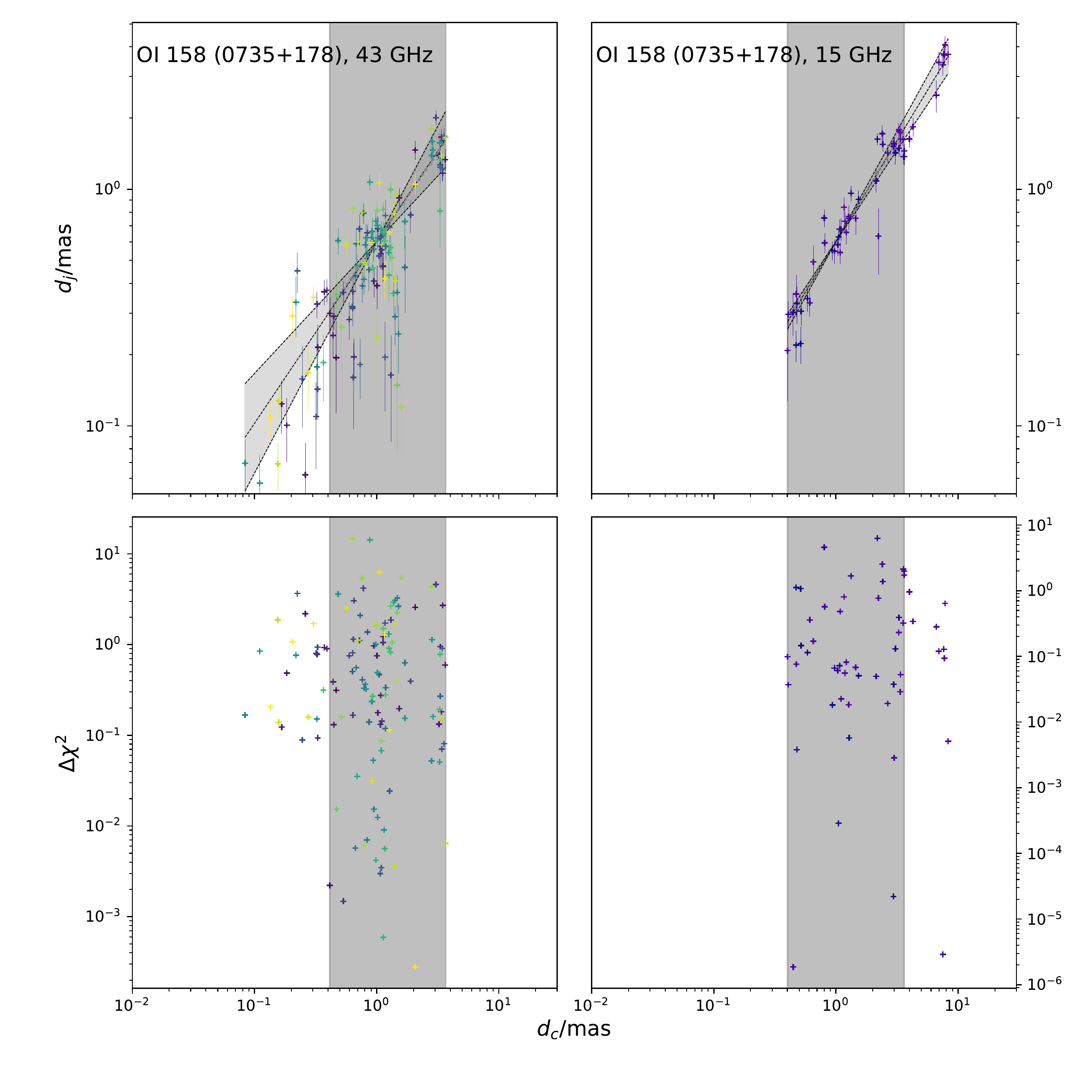}
		\includegraphics[width=0.65\hsize,clip]{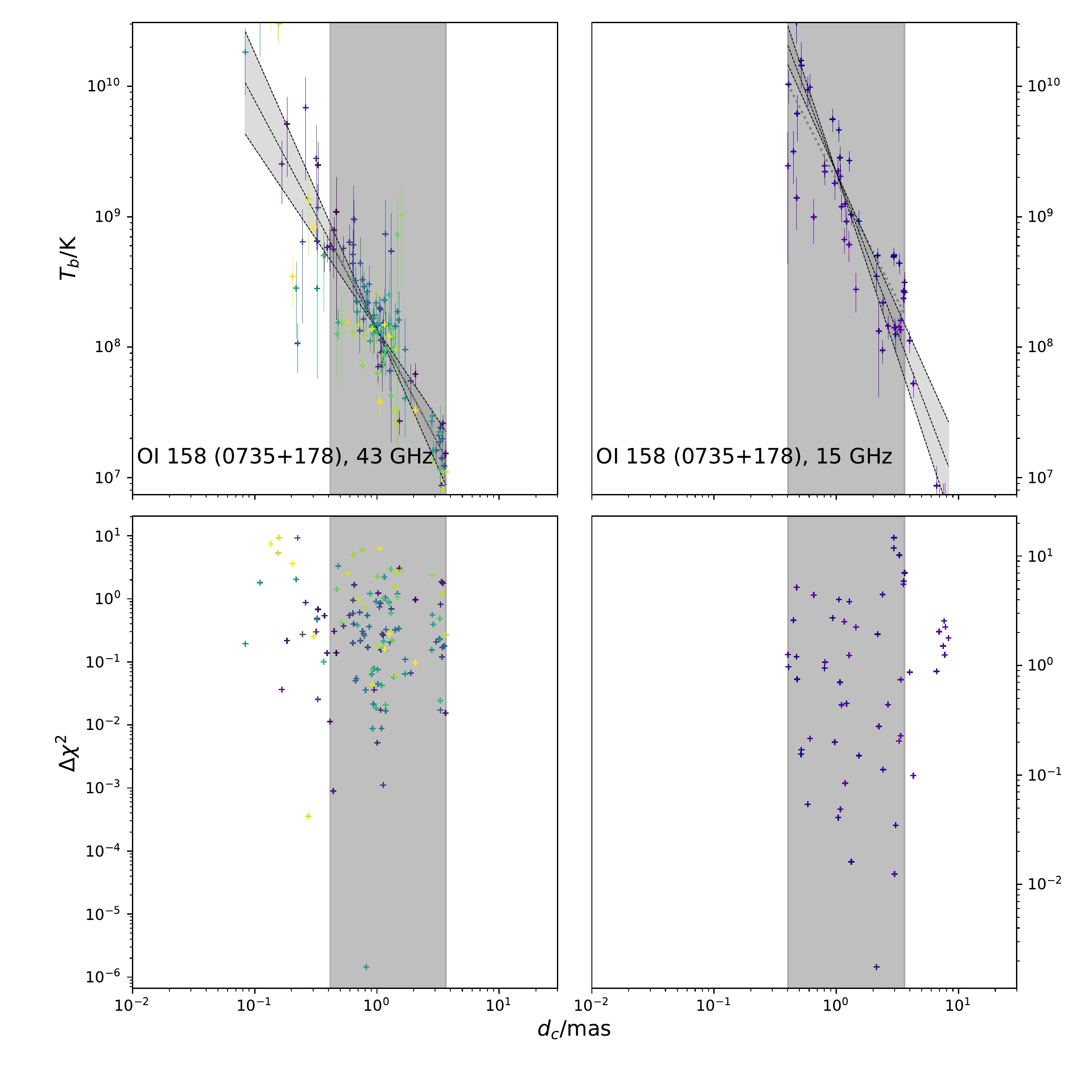}
      \caption{0735+178 jet diameter $d_{\rm j}$ as a function of the distance from the core $d_{\rm c}$ and the brightness temperature $T_{\rm B}$ as function of $d_{\rm c}$. The color coding for both frequencies indicates the epoch, where lighter colors indicate more recent epochs.}
       \label{0735+178}
\end{figure*}

\begin{figure*}
\centering
	 \includegraphics[width=0.65\hsize,clip]{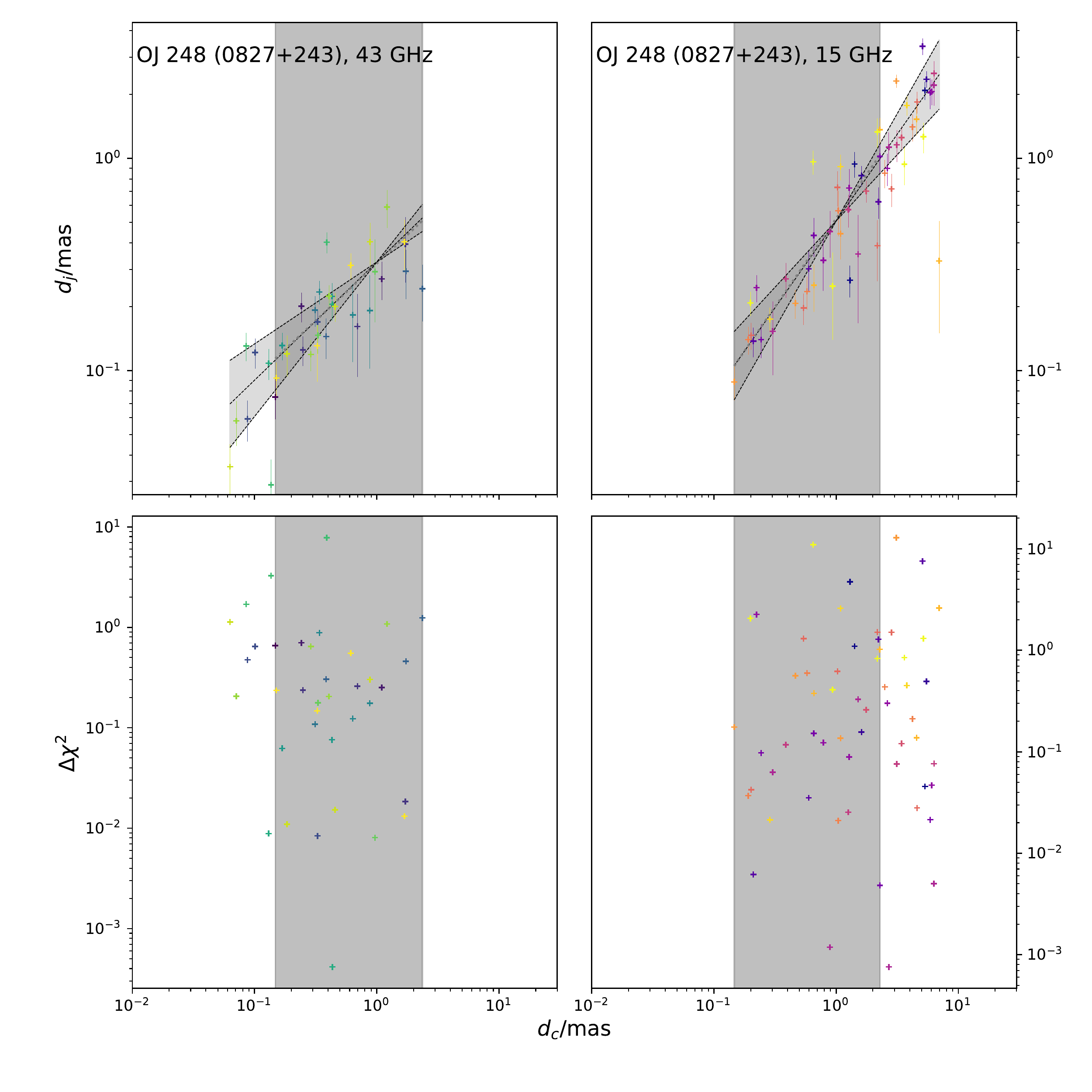}
		\includegraphics[width=0.65\hsize,clip]{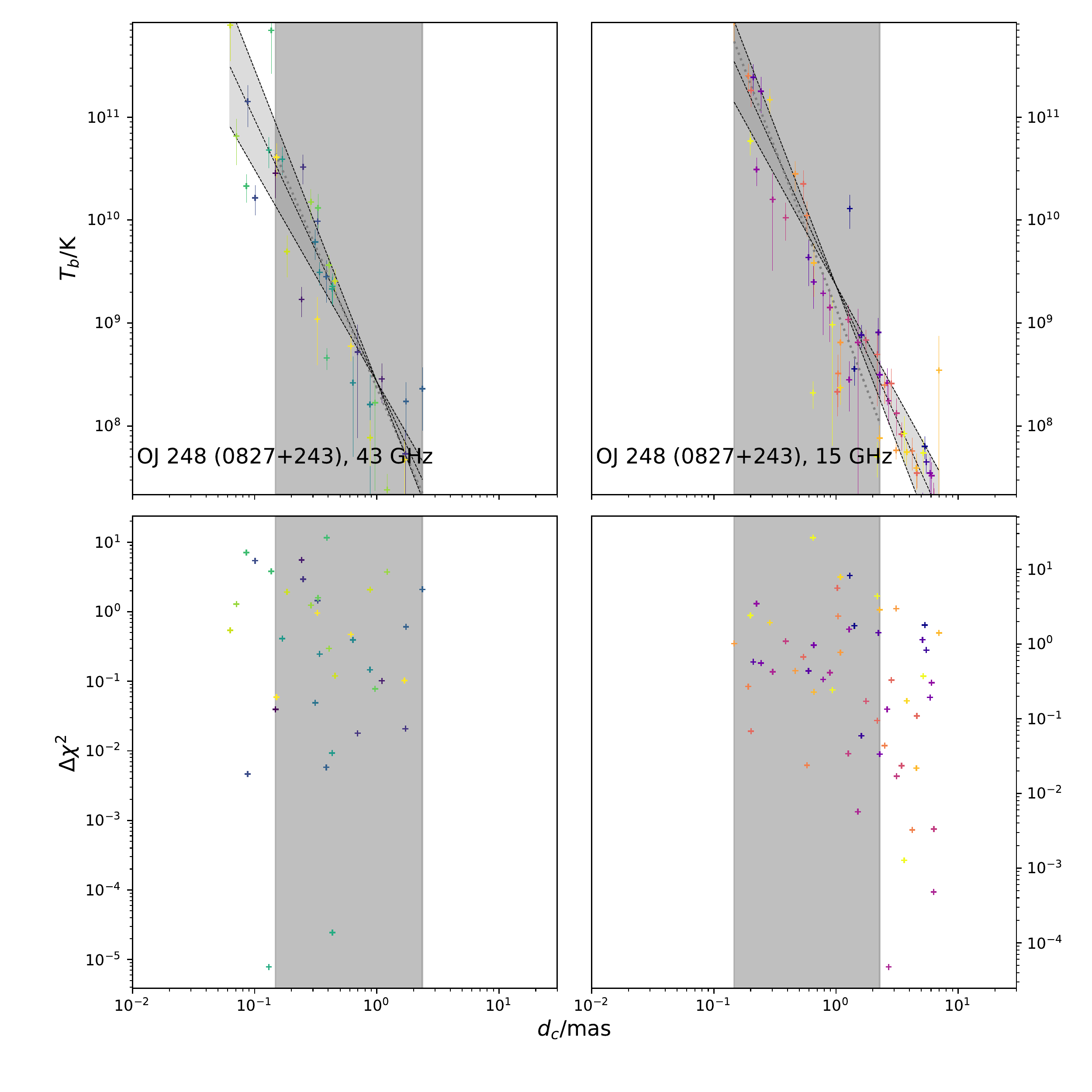}
      \caption{0827+243 jet diameter $d_{\rm j}$ as a function of the distance from the core $d_{\rm c}$ and the brightness temperature $T_{\rm B}$ as function of $d_{\rm c}$. The color coding for both frequencies indicates the epoch, where lighter colors indicate more recent epochs.}
       \label{0827+243}
\end{figure*}

\begin{figure*}
\centering
	 \includegraphics[width=0.65\hsize,clip]{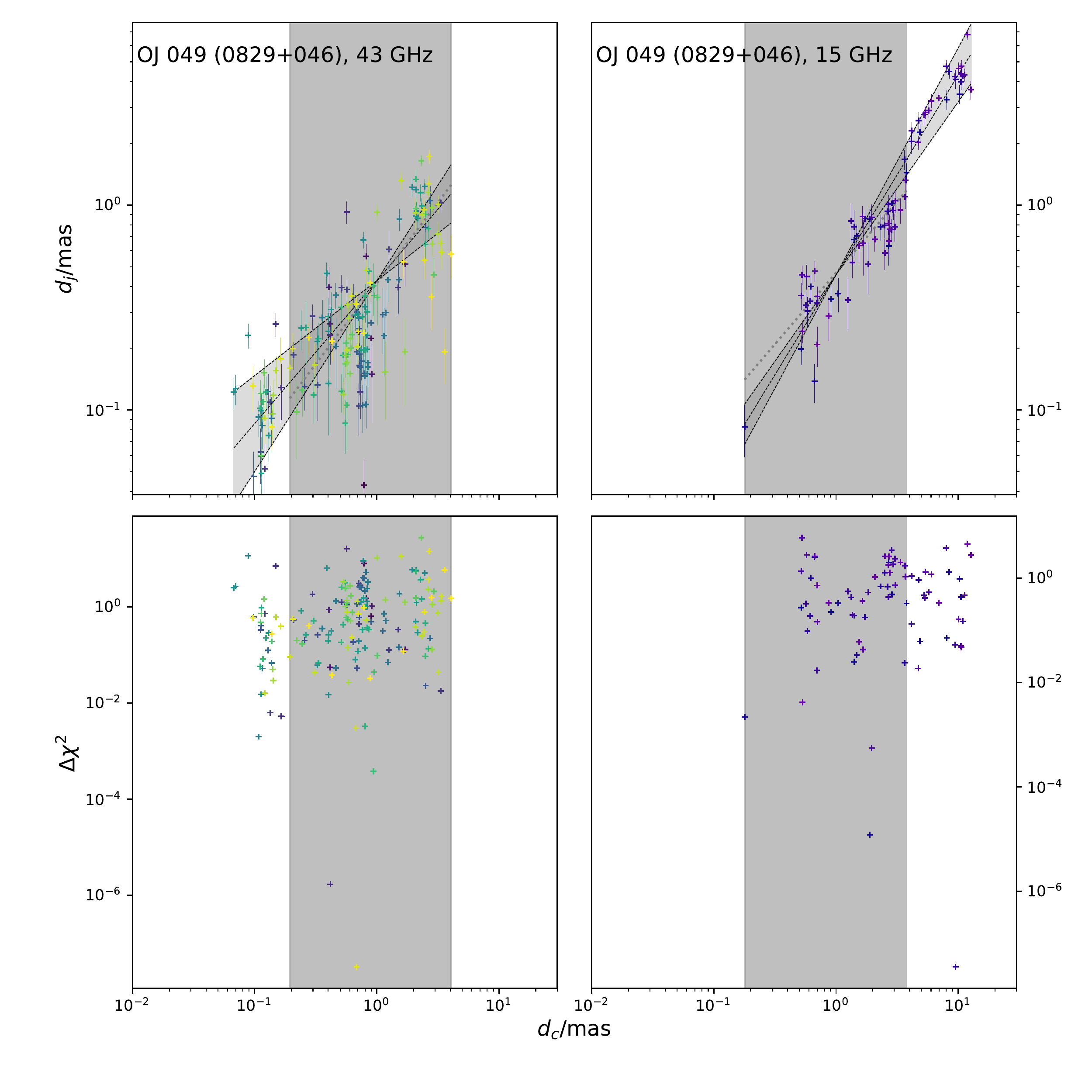}
		\includegraphics[width=0.65\hsize,clip]{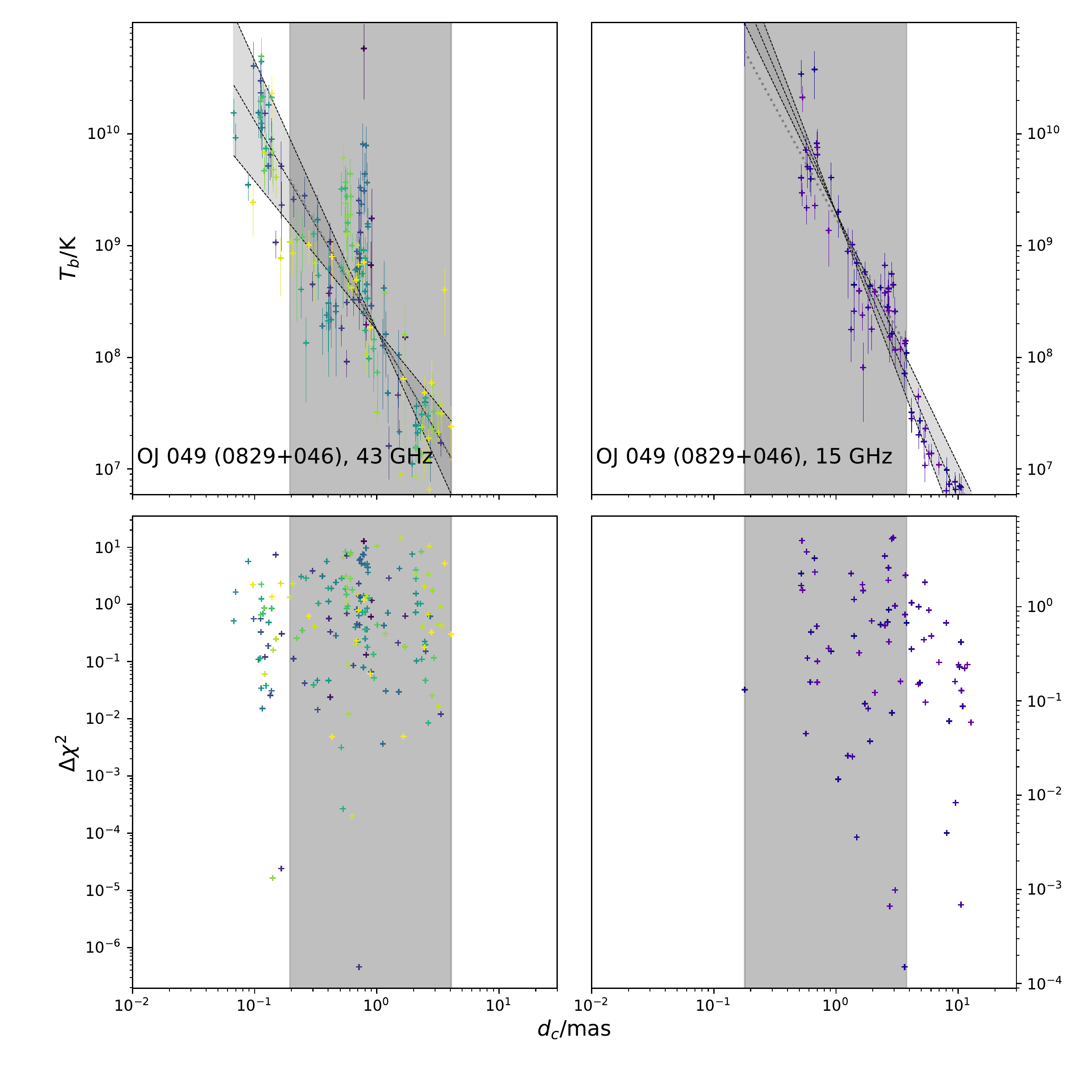}
      \caption{0827+243 jet diameter $d_{\rm j}$ as a function of the distance from the core $d_{\rm c}$ and the brightness temperature $T_{\rm B}$ as function of $d_{\rm c}$. The color coding for both frequencies indicates the epoch, where lighter colors indicate more recent epochs.}
       \label{0829+046}
\end{figure*}

\begin{figure*}
\centering
	 \includegraphics[width=0.65\hsize,clip]{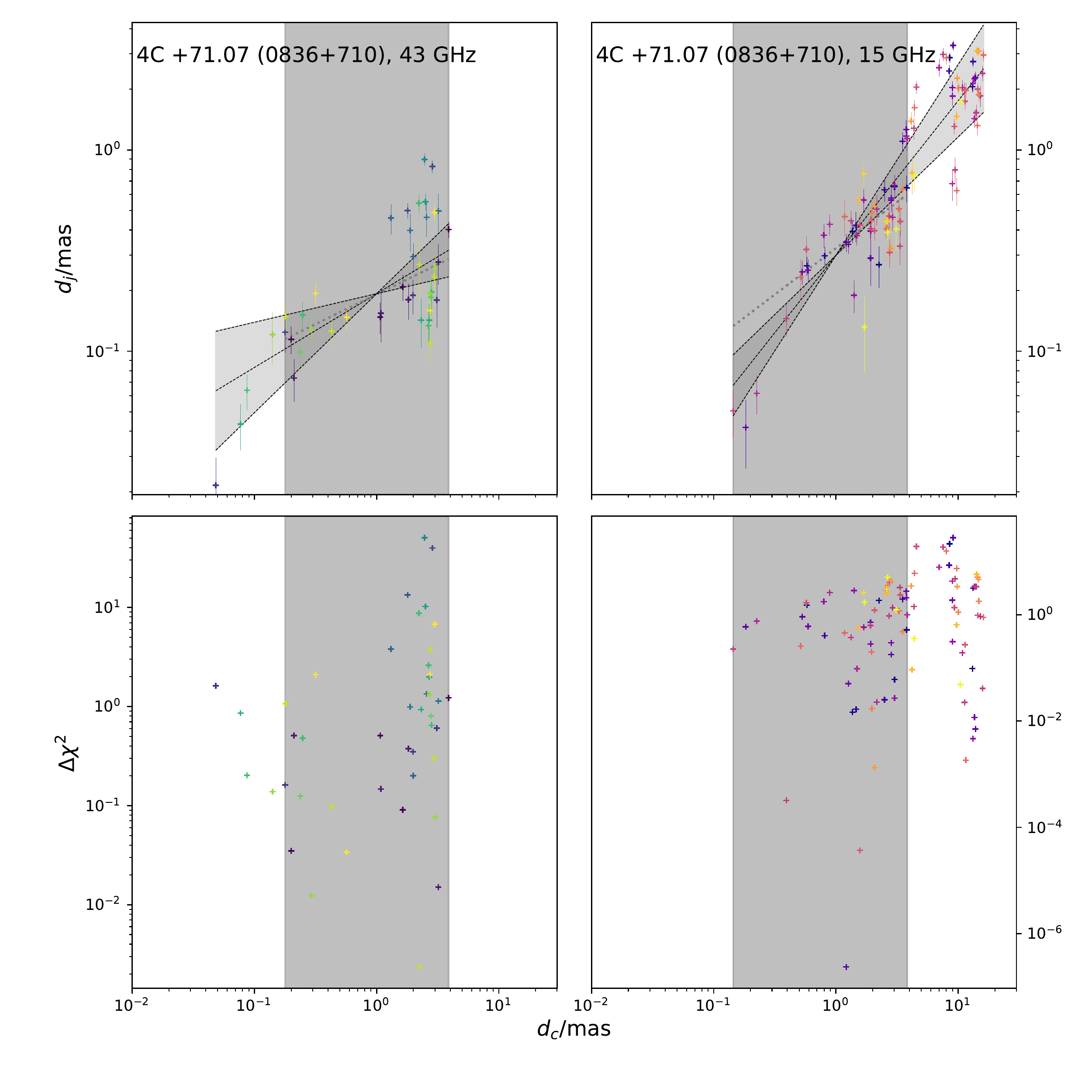}
		\includegraphics[width=0.65\hsize,clip]{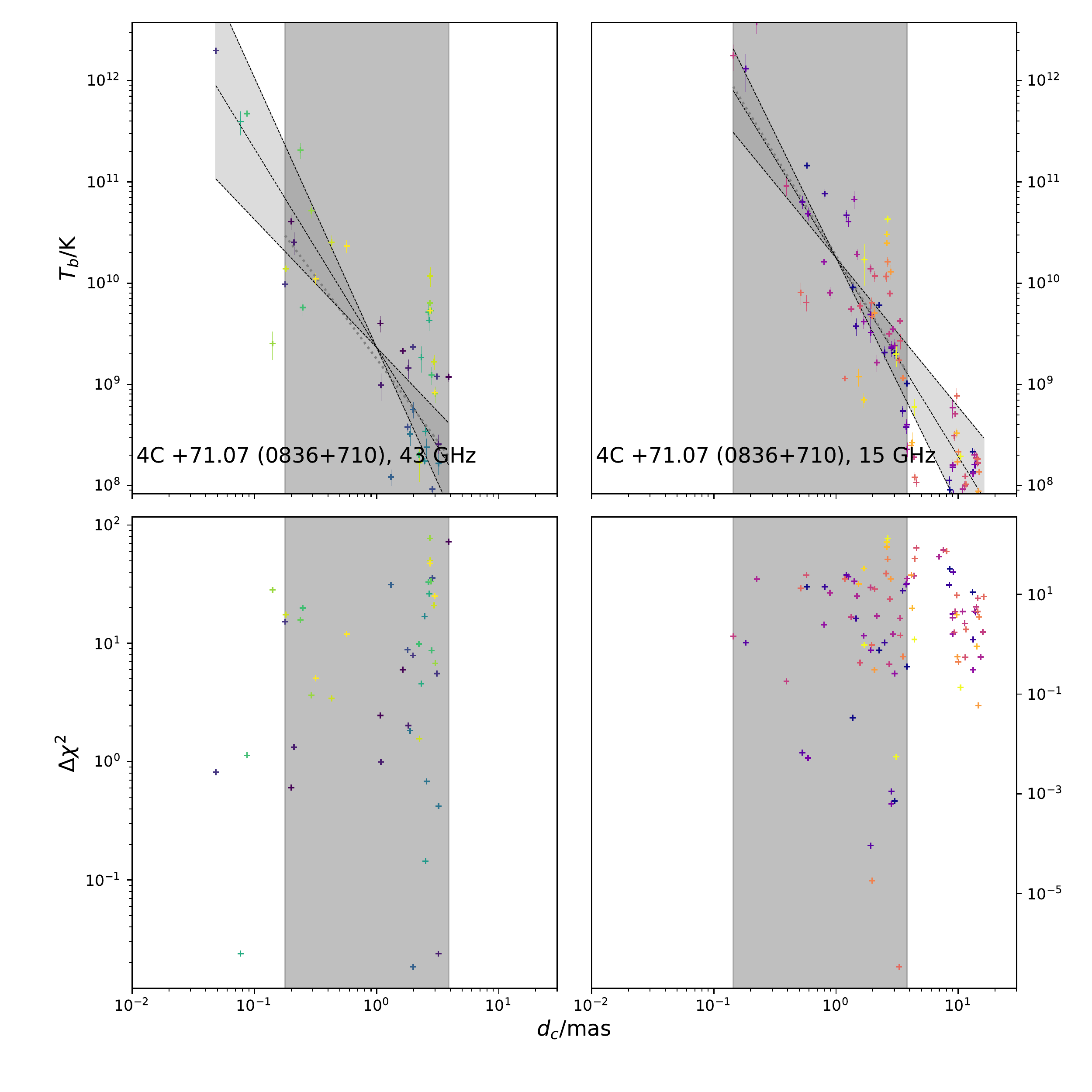}
      \caption{0836+710 jet diameter $d_{\rm j}$ as a function of the distance from the core $d_{\rm c}$ and the brightness temperature $T_{\rm B}$ as function of $d_{\rm c}$. The color coding for both frequencies indicates the epoch, where lighter colors indicate more recent epochs.}
       \label{0836+710}
\end{figure*}

\begin{figure*}
\centering
	 \includegraphics[width=0.65\hsize,clip]{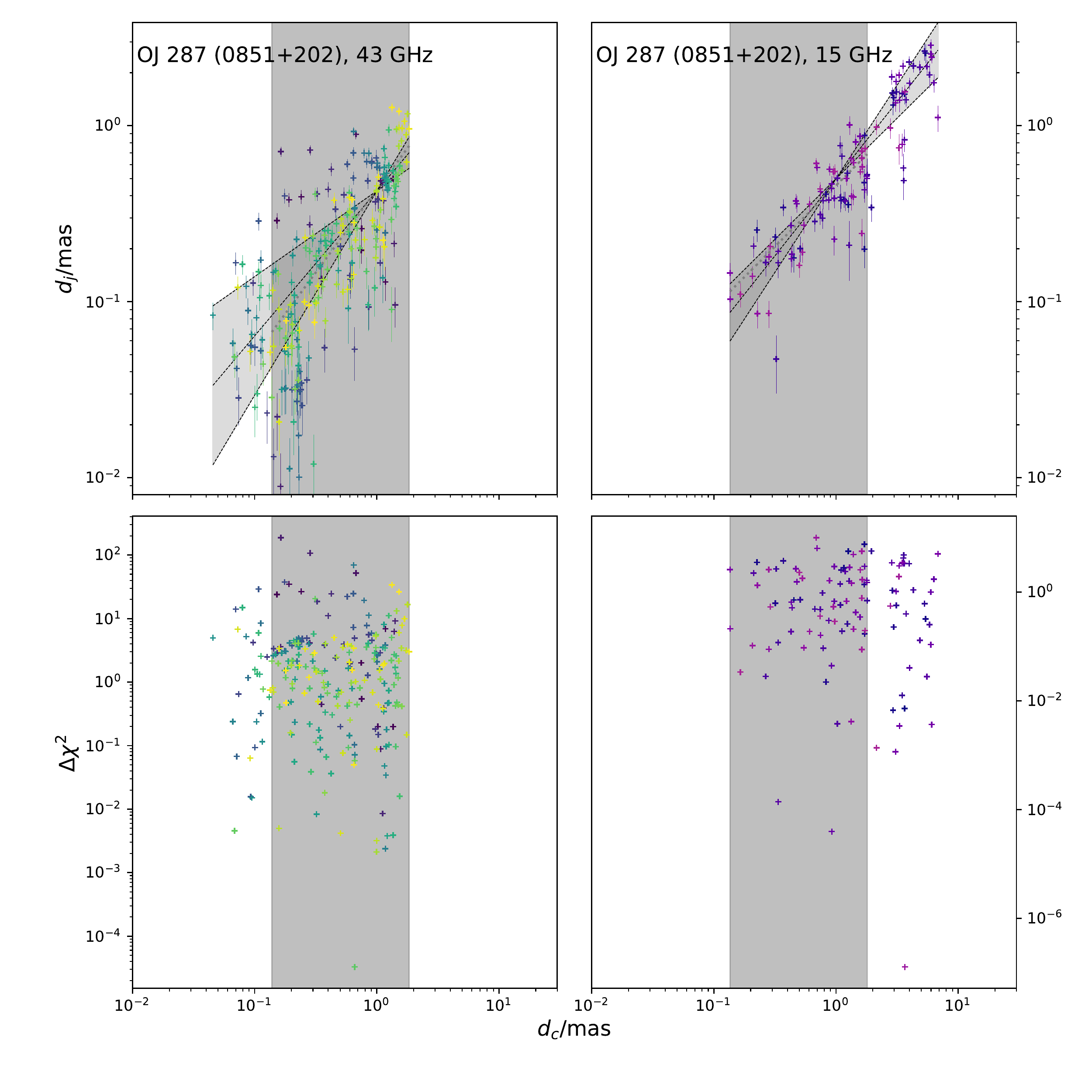}
		\includegraphics[width=0.65\hsize,clip]{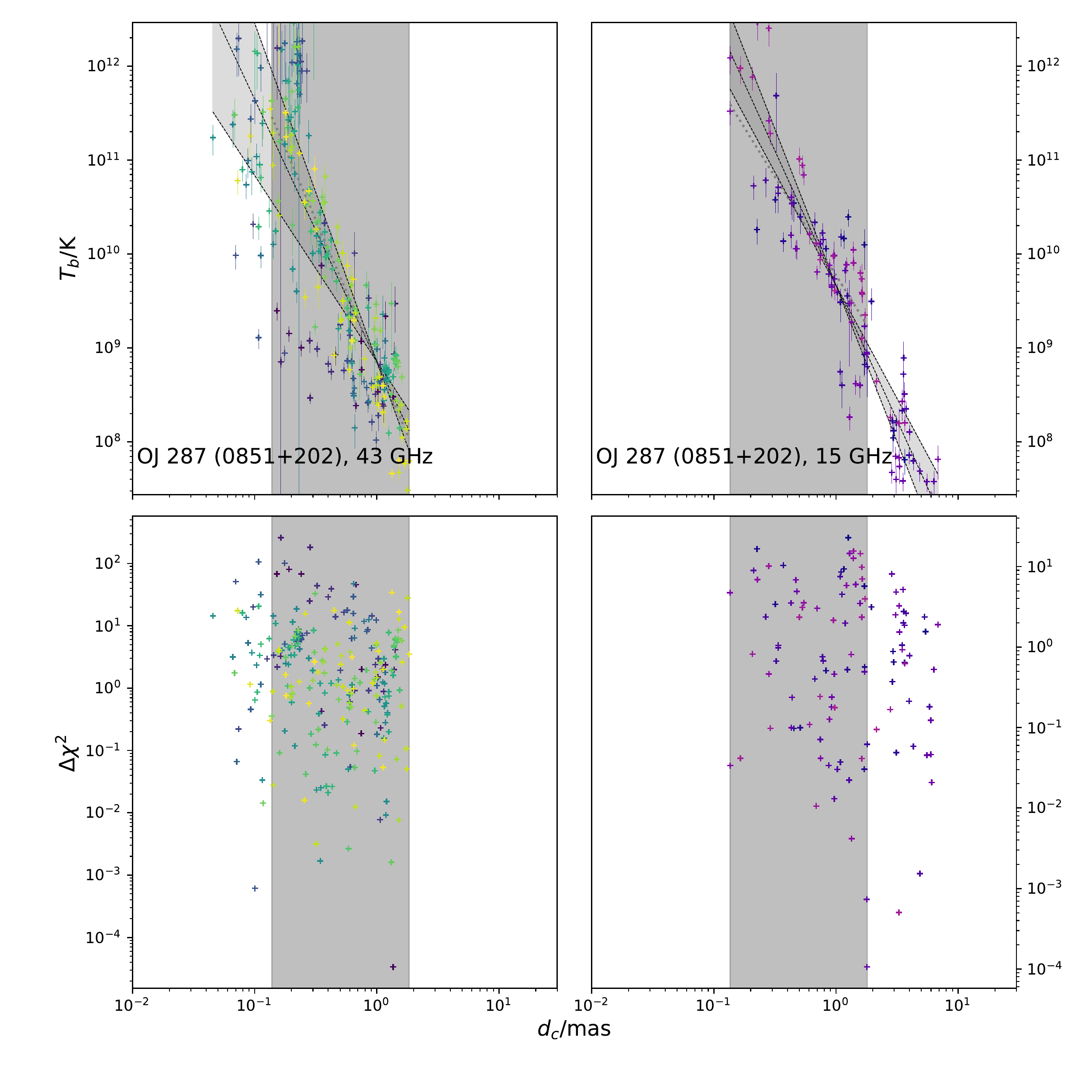}
      \caption{0851+202 jet diameter $d_{\rm j}$ as a function of the distance from the core $d_{\rm c}$ and the brightness temperature $T_{\rm B}$ as function of $d_{\rm c}$. The color coding for both frequencies indicates the epoch, where lighter colors indicate more recent epochs.}
       \label{0851+202}
\end{figure*}

\begin{figure*}
\centering
	 \includegraphics[width=0.65\hsize,clip]{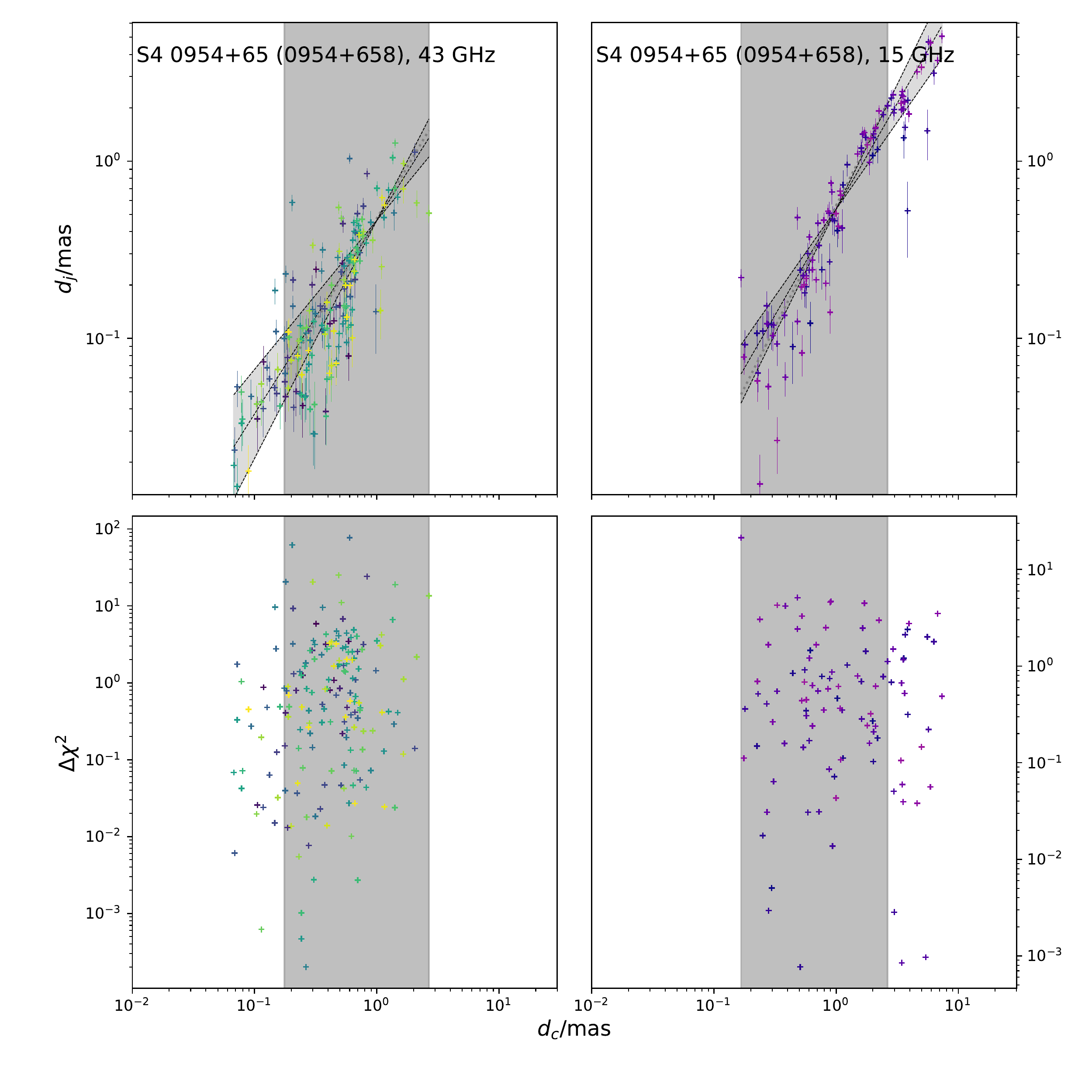}
		\includegraphics[width=0.65\hsize,clip]{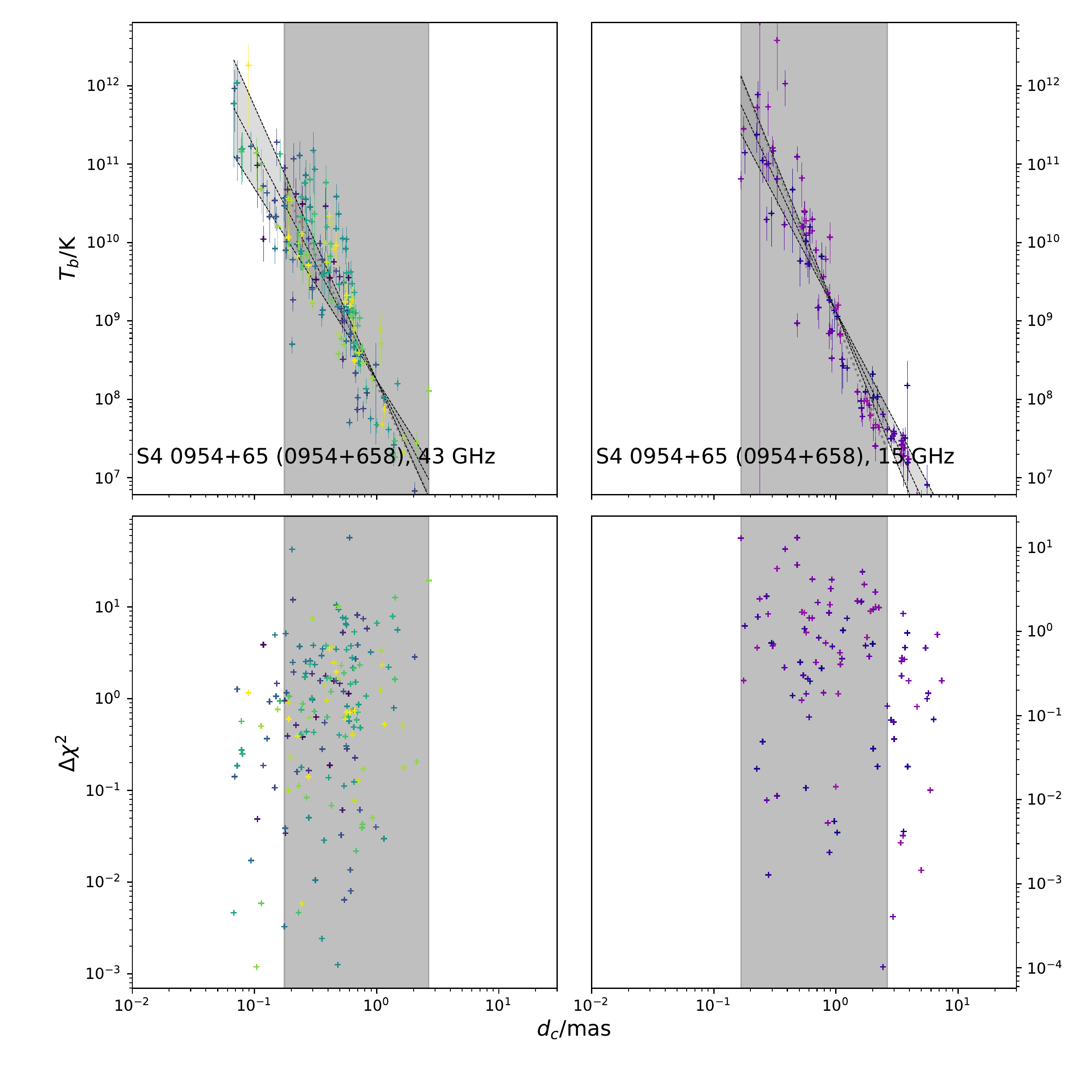}
      \caption{0954+658 jet diameter $d_{\rm j}$ as a function of the distance from the core $d_{\rm c}$ and the brightness temperature $T_{\rm B}$ as function of $d_{\rm c}$. The color coding for both frequencies indicates the epoch, where lighter colors indicate more recent epochs.}
       \label{0954+658}
\end{figure*}

\begin{figure*}
\centering
	 \includegraphics[width=0.65\hsize,clip]{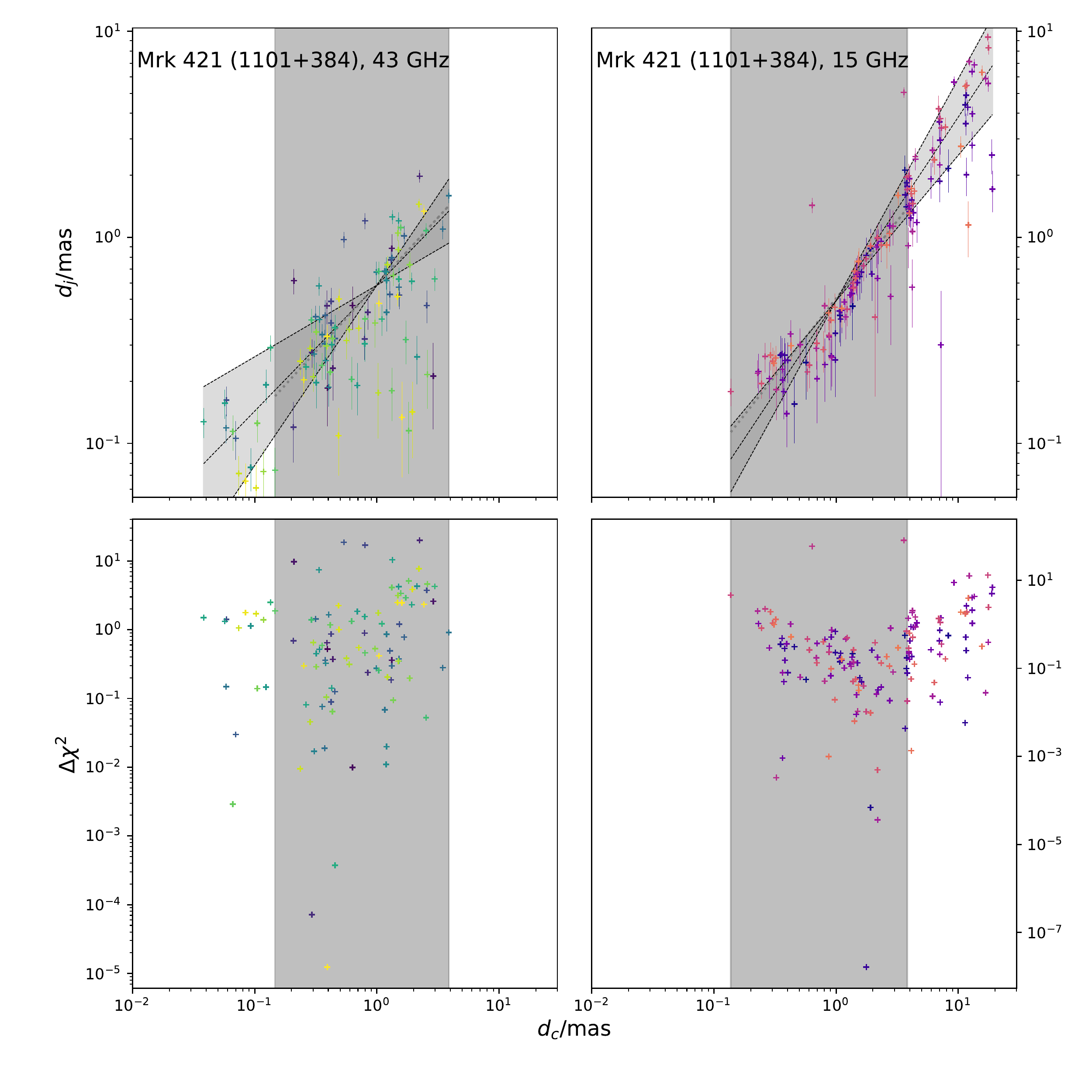}
		\includegraphics[width=0.65\hsize,clip]{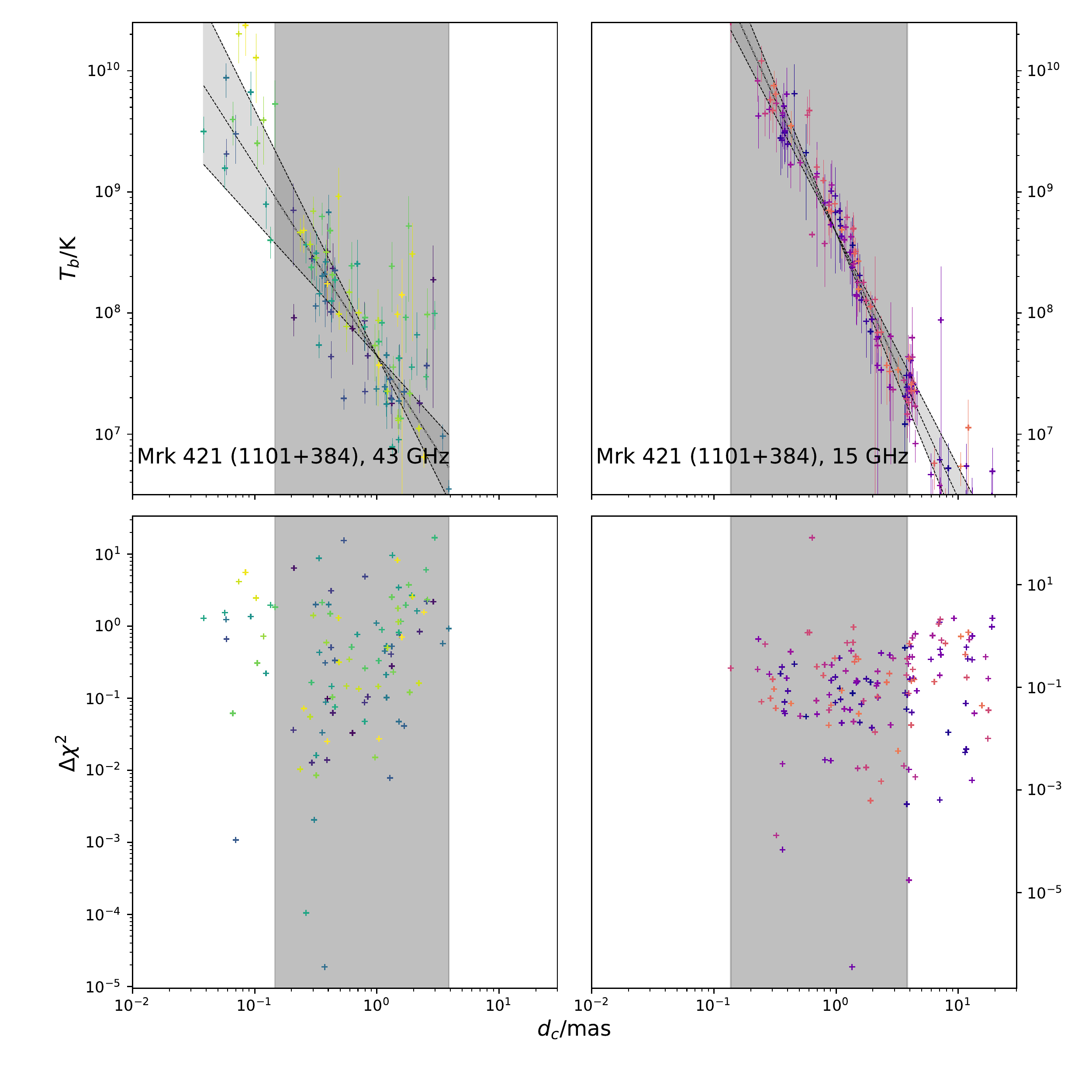}
      \caption{1101+384 jet diameter $d_{\rm j}$ as a function of the distance from the core $d_{\rm c}$ and the brightness temperature $T_{\rm B}$ as function of $d_{\rm c}$. The color coding for both frequencies indicates the epoch, where lighter colors indicate more recent epochs.}
       \label{1101+384}
\end{figure*}

\begin{figure*}
\centering
	 \includegraphics[width=0.65\hsize,clip]{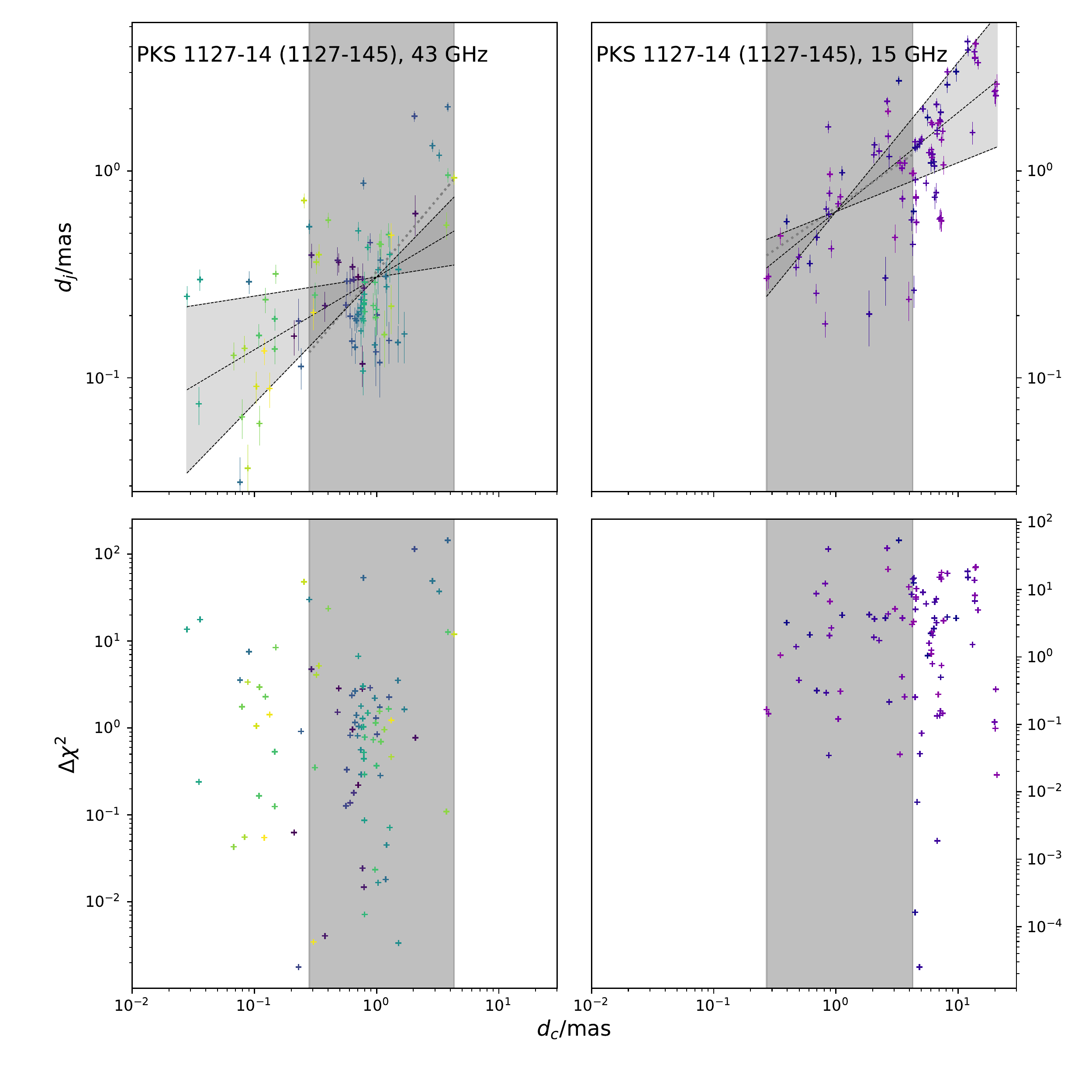}
		\includegraphics[width=0.65\hsize,clip]{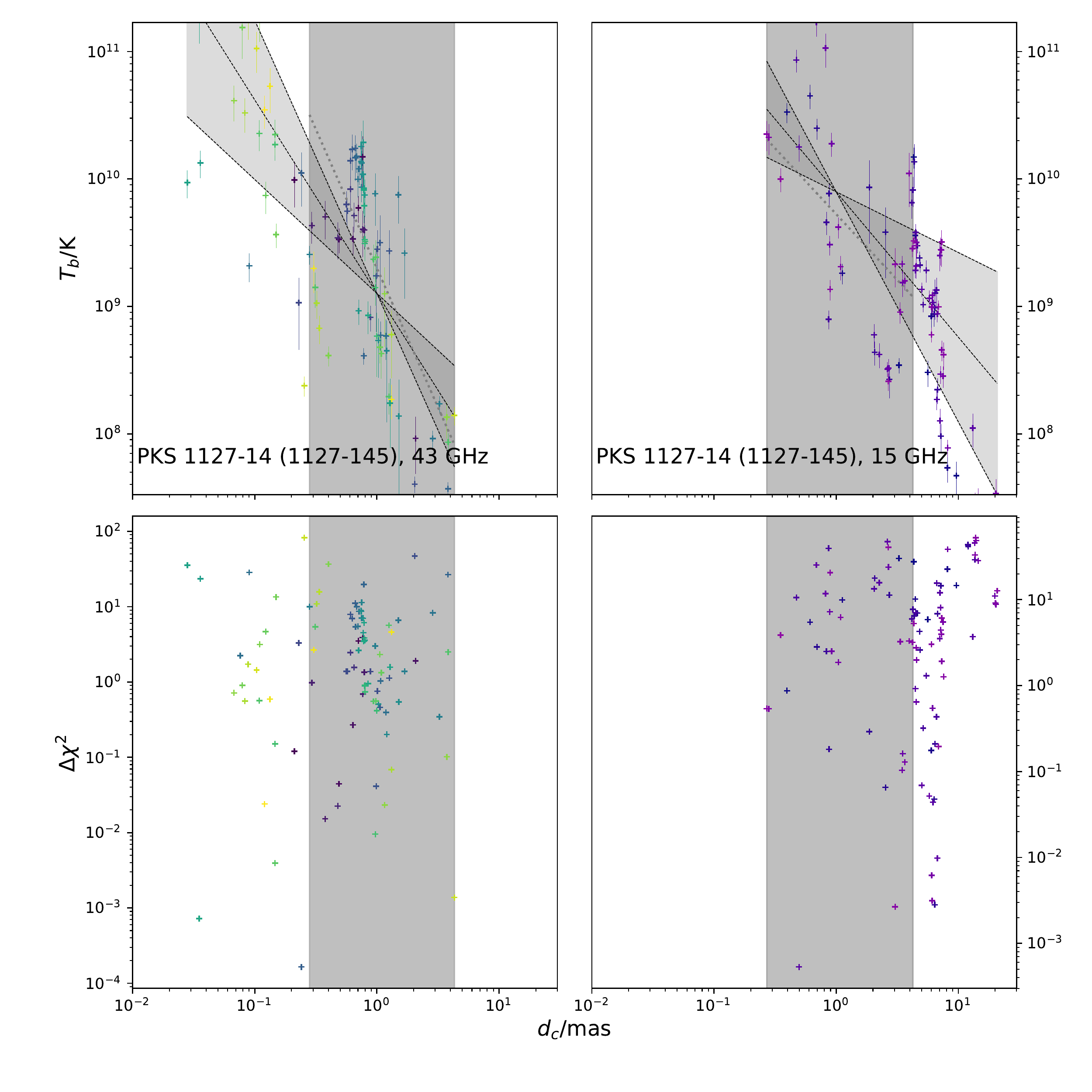}
      \caption{1127-145 jet diameter $d_{\rm j}$ as a function of the distance from the core $d_{\rm c}$ and the brightness temperature $T_{\rm B}$ as function of $d_{\rm c}$. The color coding for both frequencies indicates the epoch, where lighter colors indicate more recent epochs.}
       \label{1127-145}
\end{figure*}

\begin{figure*}
\centering
	 \includegraphics[width=0.65\hsize,clip]{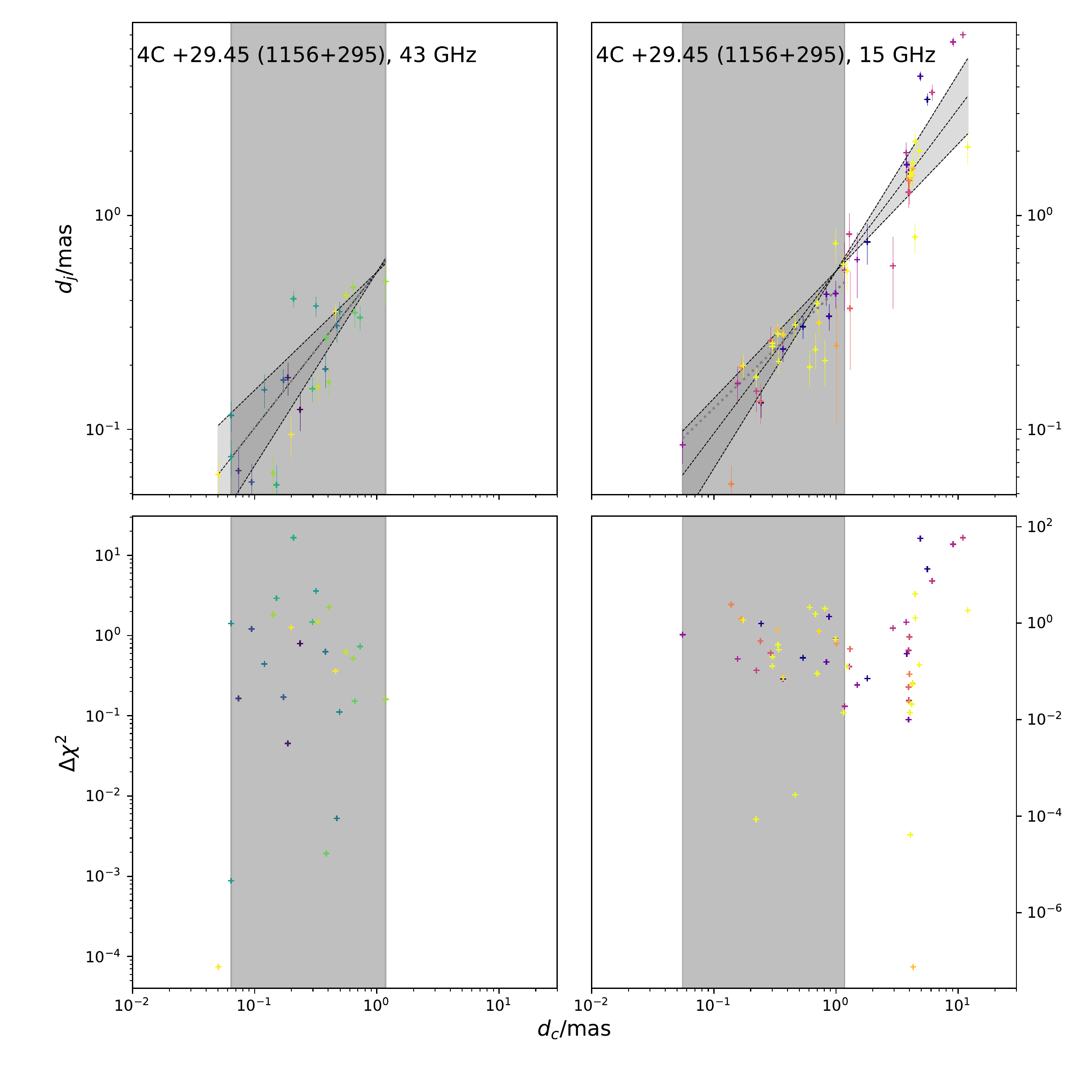}
		\includegraphics[width=0.65\hsize,clip]{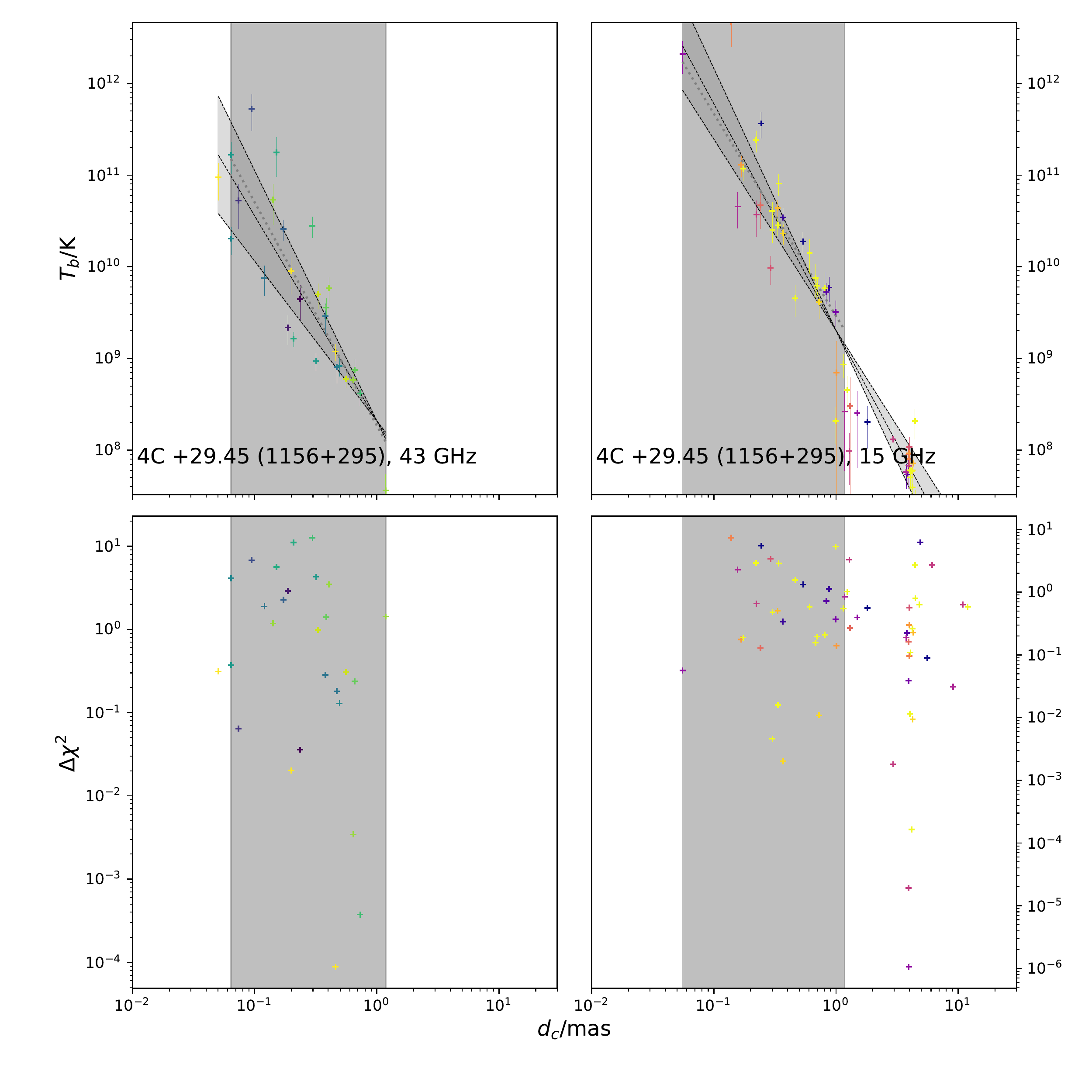}
      \caption{1156+295 jet diameter $d_{\rm j}$ as a function of the distance from the core $d_{\rm c}$ and the brightness temperature $T_{\rm B}$ as function of $d_{\rm c}$. The color coding for both frequencies indicates the epoch, where lighter colors indicate more recent epochs.}
       \label{1156+295}
\end{figure*}

\begin{figure*}
\centering
	 \includegraphics[width=0.65\hsize,clip]{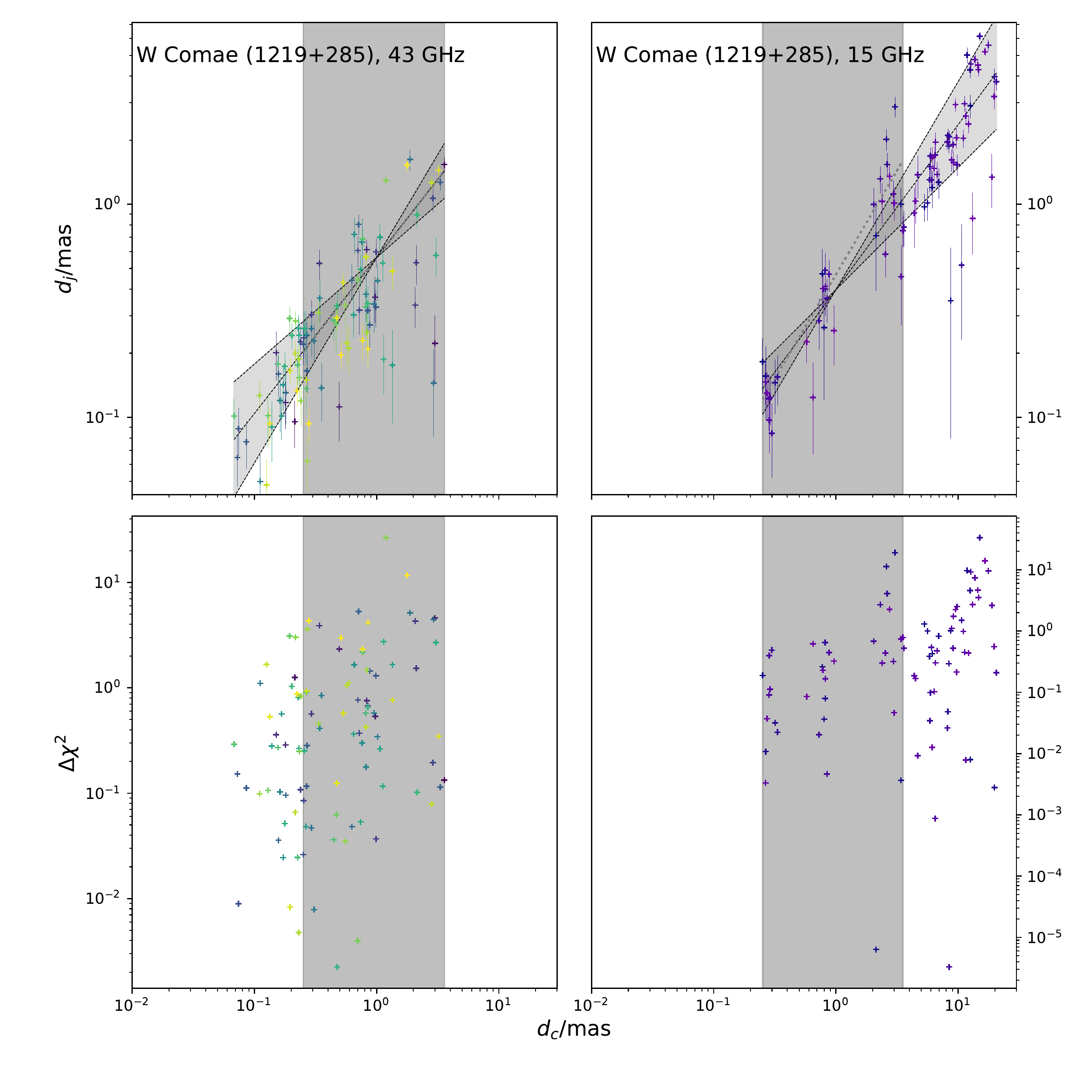}
		\includegraphics[width=0.65\hsize,clip]{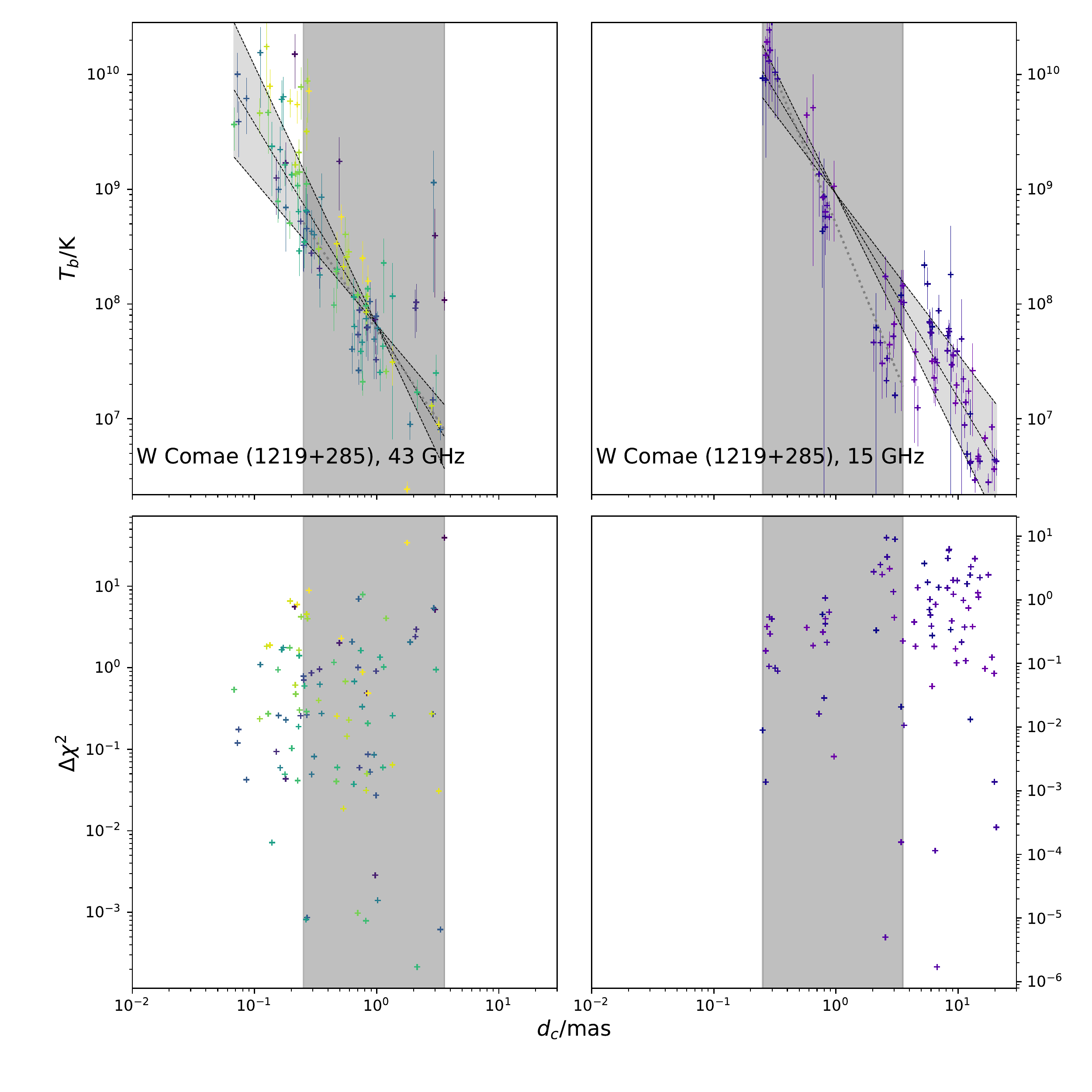}
      \caption{1219+285 jet diameter $d_{\rm j}$ as a function of the distance from the core $d_{\rm c}$ and the brightness temperature $T_{\rm B}$ as function of $d_{\rm c}$. The color coding for both frequencies indicates the epoch, where lighter colors indicate more recent epochs.}
       \label{1219+285}
\end{figure*}

\begin{figure*}
\centering
	 \includegraphics[width=0.65\hsize,clip]{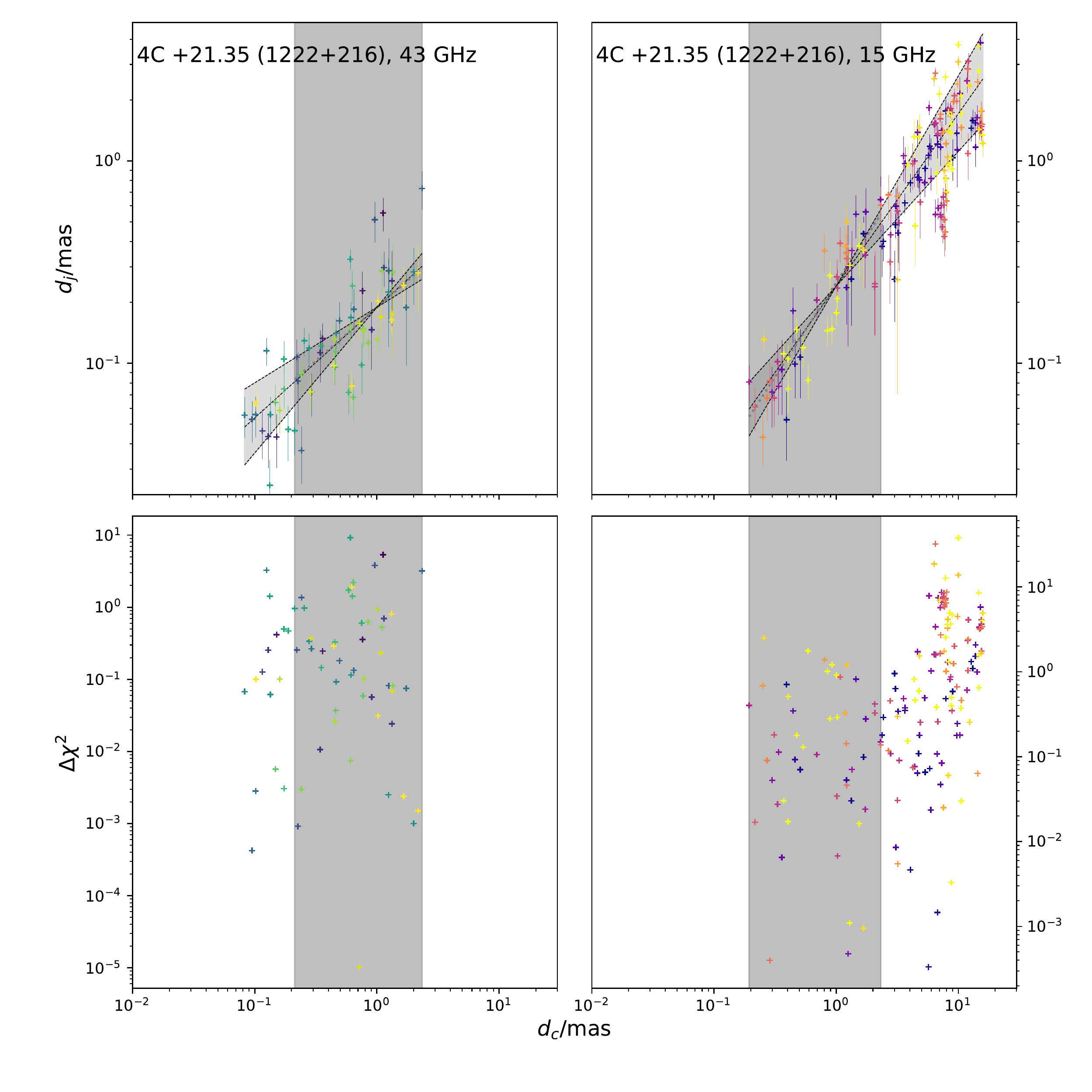}
		\includegraphics[width=0.65\hsize,clip]{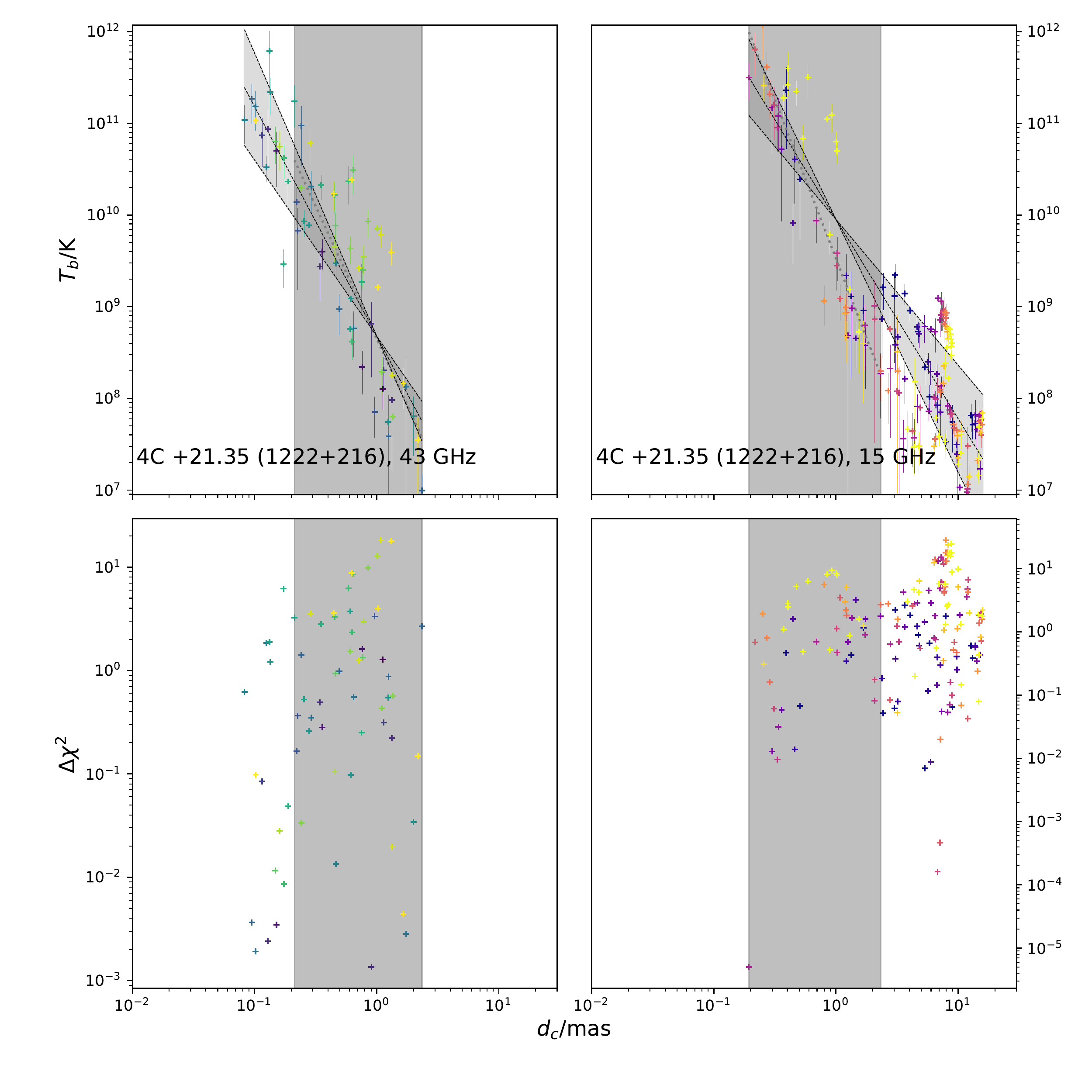}
      \caption{1222+216 jet diameter $d_{\rm j}$ as a function of the distance from the core $d_{\rm c}$ and the brightness temperature $T_{\rm B}$ as function of $d_{\rm c}$}. The color coding for both frequencies indicates the epoch, where lighter colors indicate more recent epochs.
       \label{1222+216}
\end{figure*}

\begin{figure*}
\centering
	 \includegraphics[width=0.65\hsize,clip]{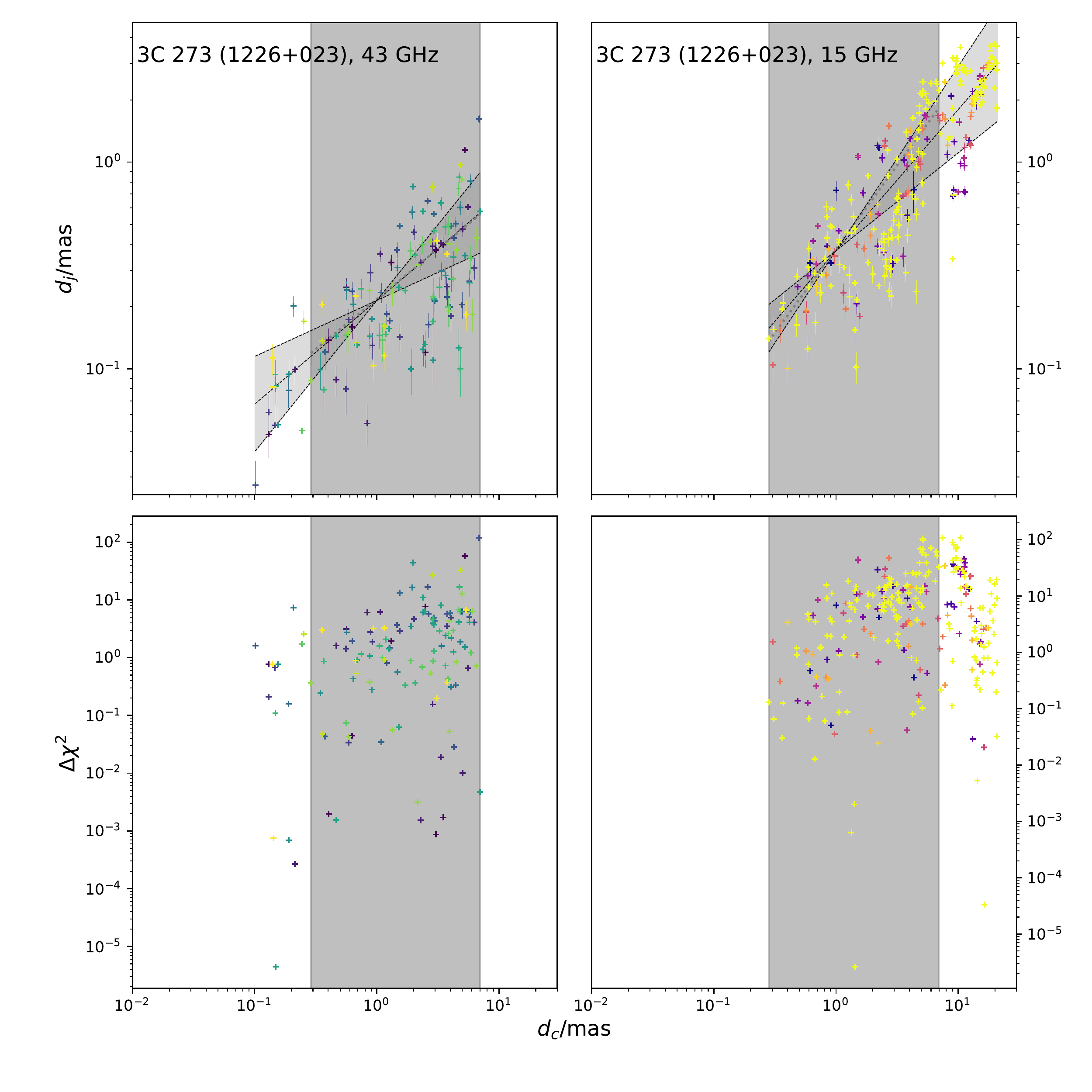}
		\includegraphics[width=0.65\hsize,clip]{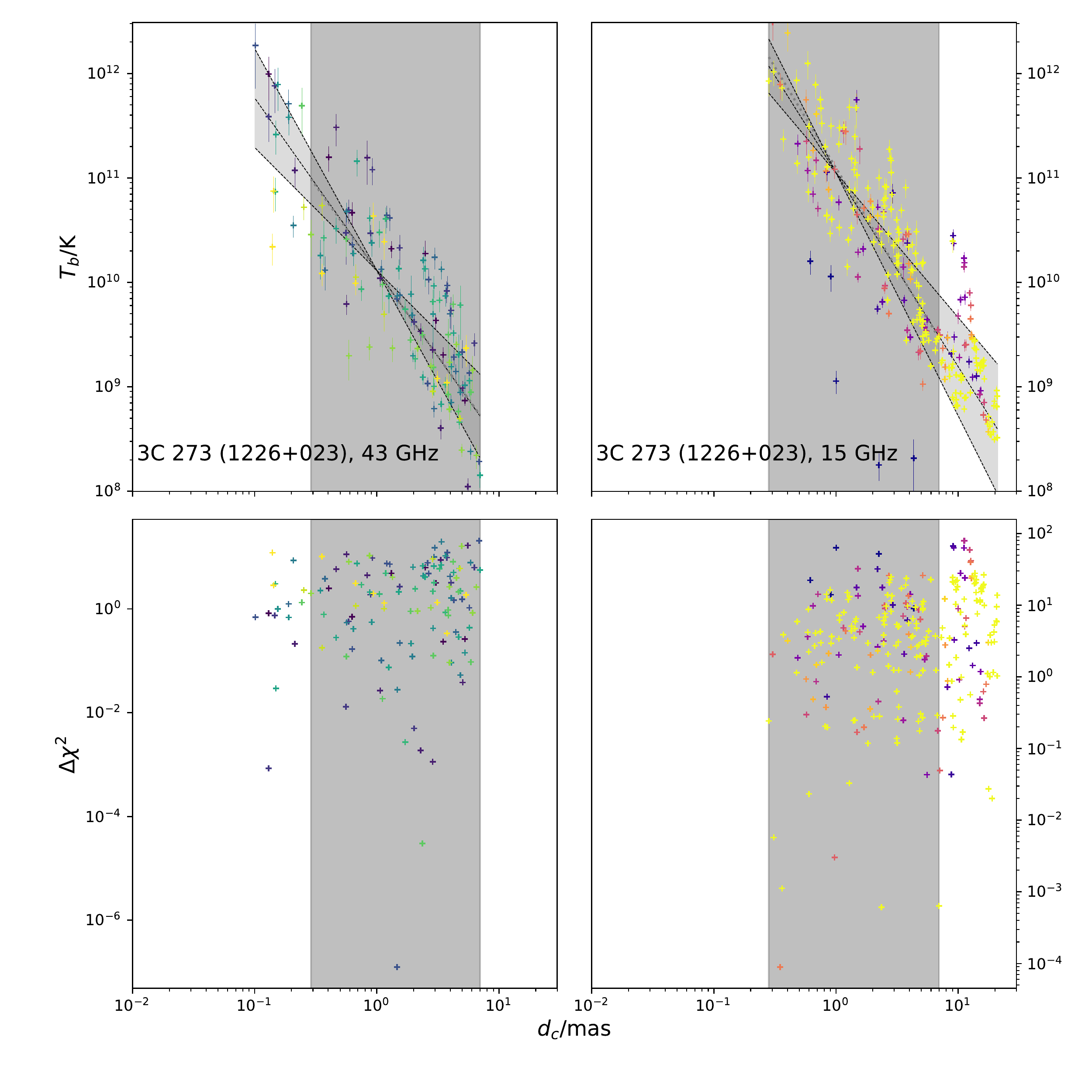}
      \caption{1226+023 jet diameter $d_{\rm j}$ as a function of the distance from the core $d_{\rm c}$ and the brightness temperature $T_{\rm B}$ as function of $d_{\rm c}$. The color coding for both frequencies indicates the epoch, where lighter colors indicate more recent epochs.}
       \label{1226+023}
\end{figure*}

\begin{figure*}
\centering
	 \includegraphics[width=0.65\hsize,clip]{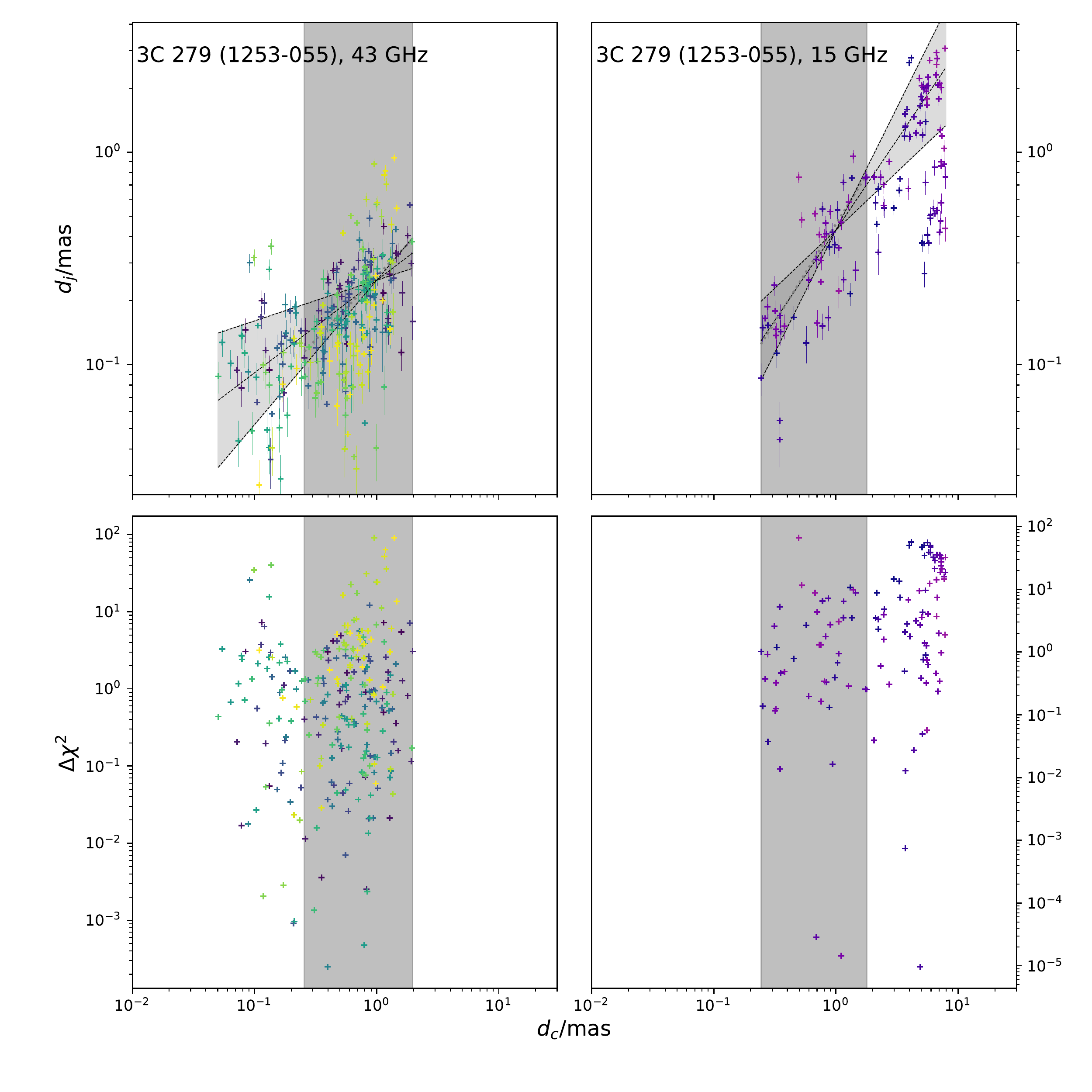}
		\includegraphics[width=0.65\hsize,clip]{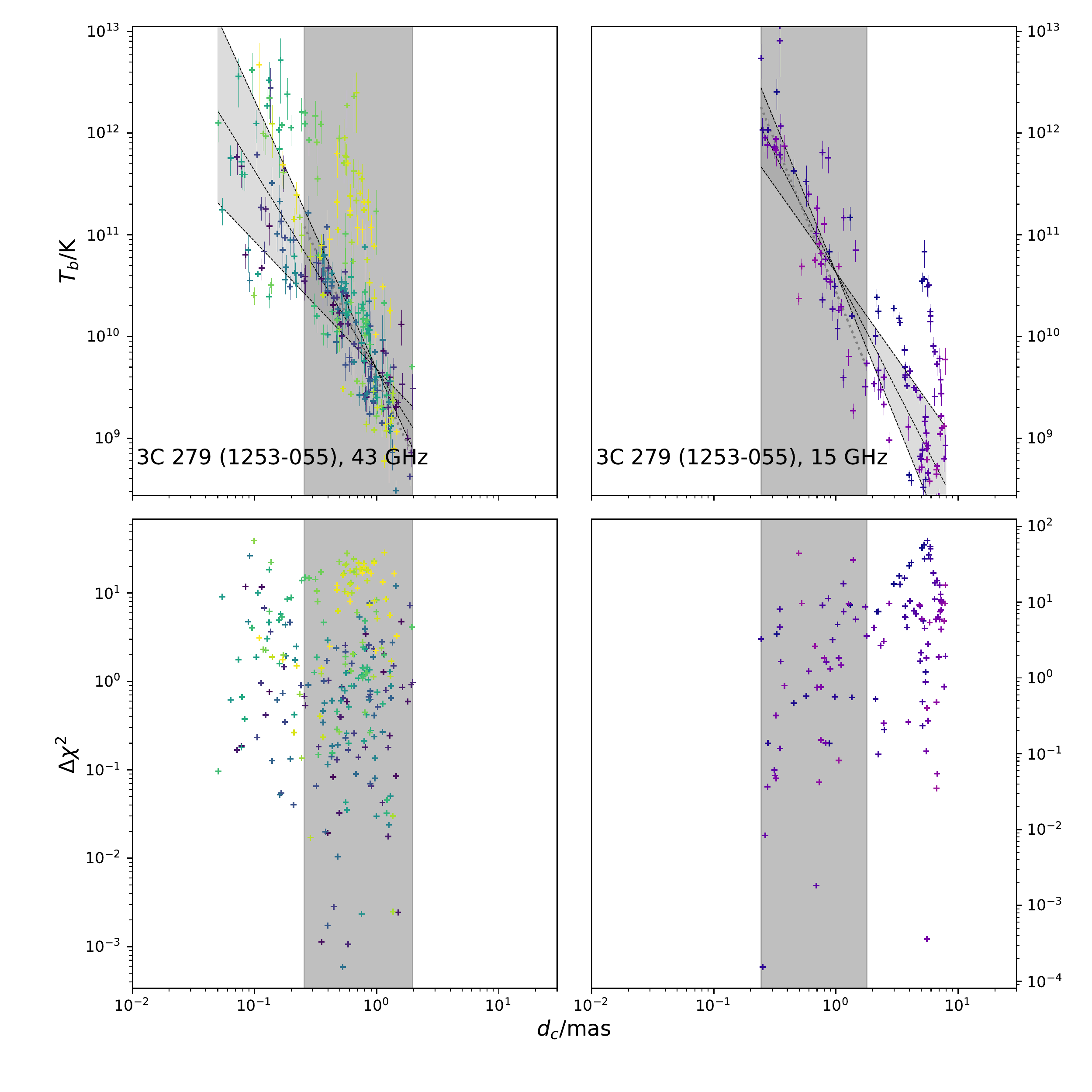}
      \caption{1253-055 jet diameter $d_{\rm j}$ as a function of the distance from the core $d_{\rm c}$ and the brightness temperature $T_{\rm B}$ as function of $d_{\rm c}$. The color coding for both frequencies indicates the epoch, where lighter colors indicate more recent epochs.}
       \label{1253-055}
\end{figure*}

\begin{figure*}
\centering
	 \includegraphics[width=0.65\hsize,clip]{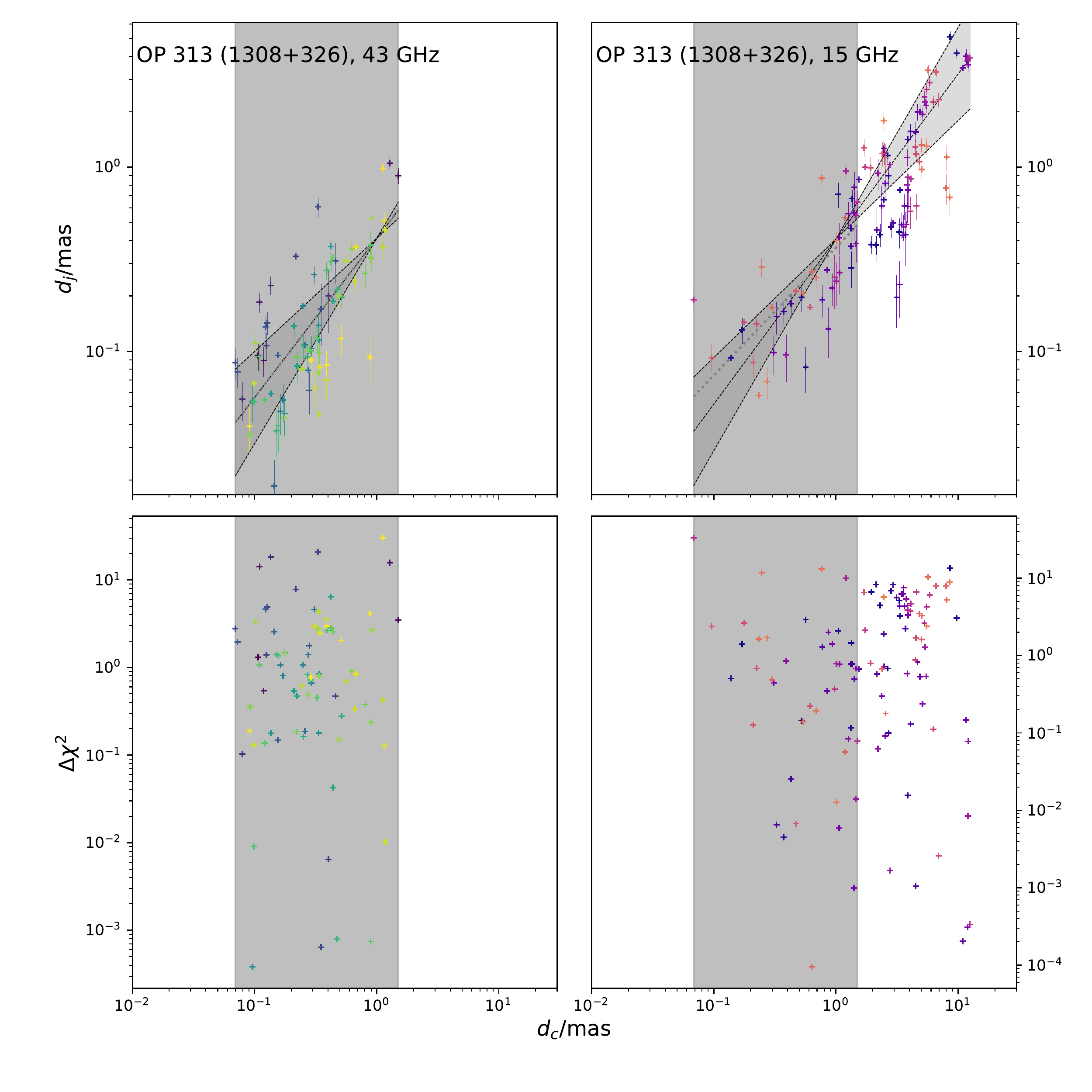}
		\includegraphics[width=0.65\hsize,clip]{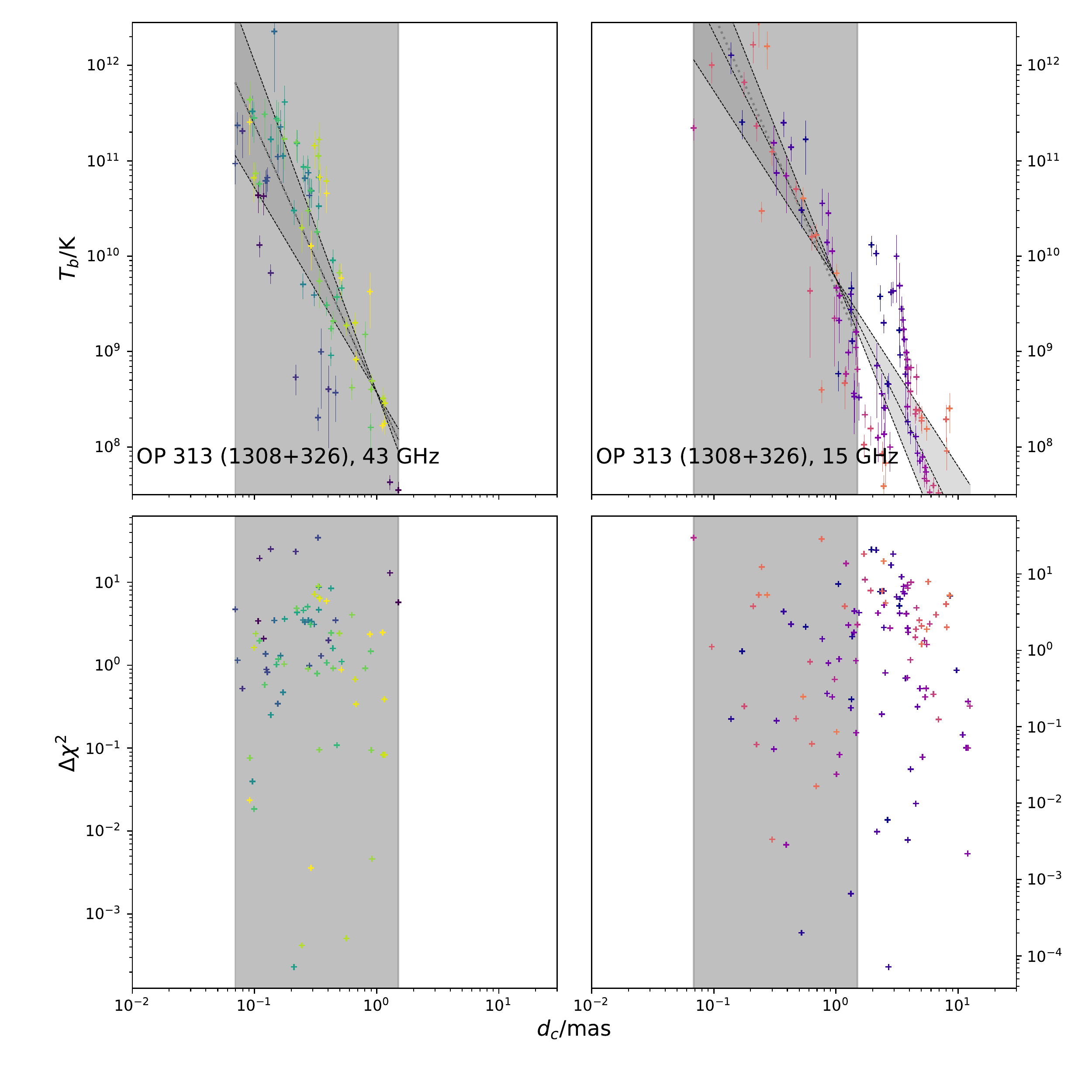}
      \caption{1308+326 jet diameter $d_{\rm j}$ as a function of the distance from the core $d_{\rm c}$ and the brightness temperature $T_{\rm B}$ as function of $d_{\rm c}$. The color coding for both frequencies indicates the epoch, where lighter colors indicate more recent epochs.}
       \label{1308+326}
\end{figure*}

\begin{figure*}
\centering
	 \includegraphics[width=0.65\hsize,clip]{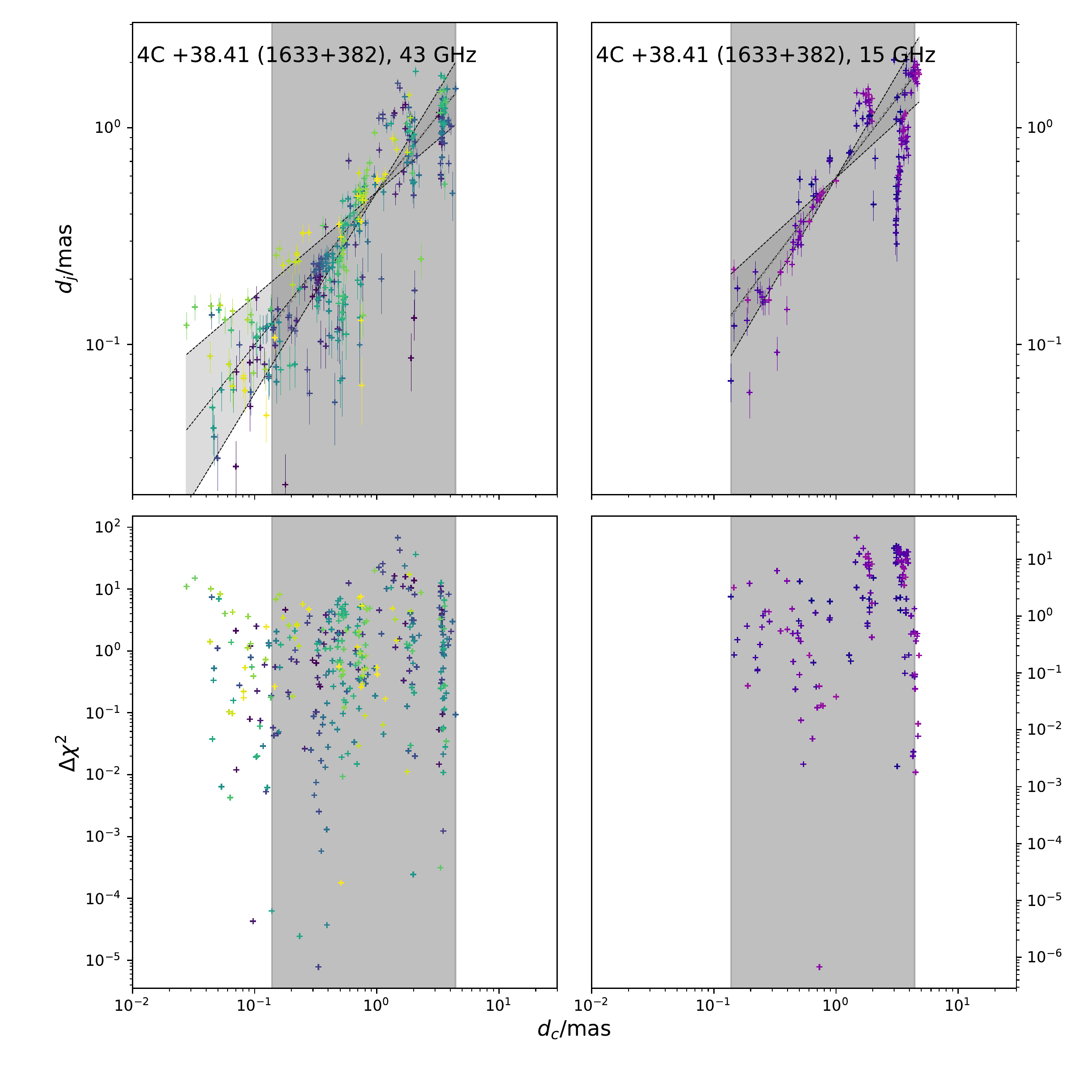}
		\includegraphics[width=0.65\hsize,clip]{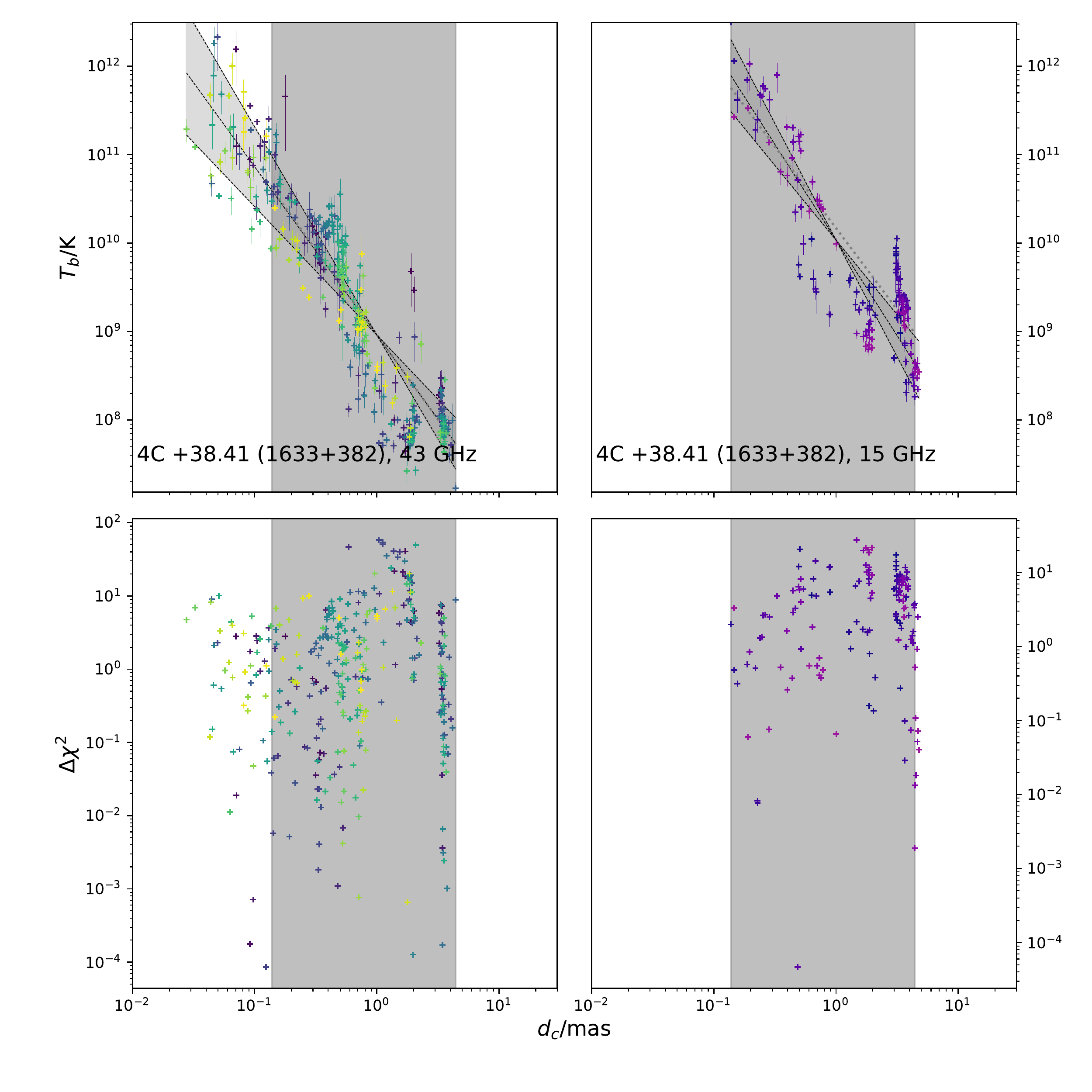}
      \caption{1633+382 jet diameter $d_{\rm j}$ as a function of the distance from the core $d_{\rm c}$ and the brightness temperature $T_{\rm B}$ as function of $d_{\rm c}$. The color coding for both frequencies indicates the epoch, where lighter colors indicate more recent epochs.}
       \label{1633+382}
\end{figure*}

\begin{figure*}
\centering
	 \includegraphics[width=0.65\hsize,clip]{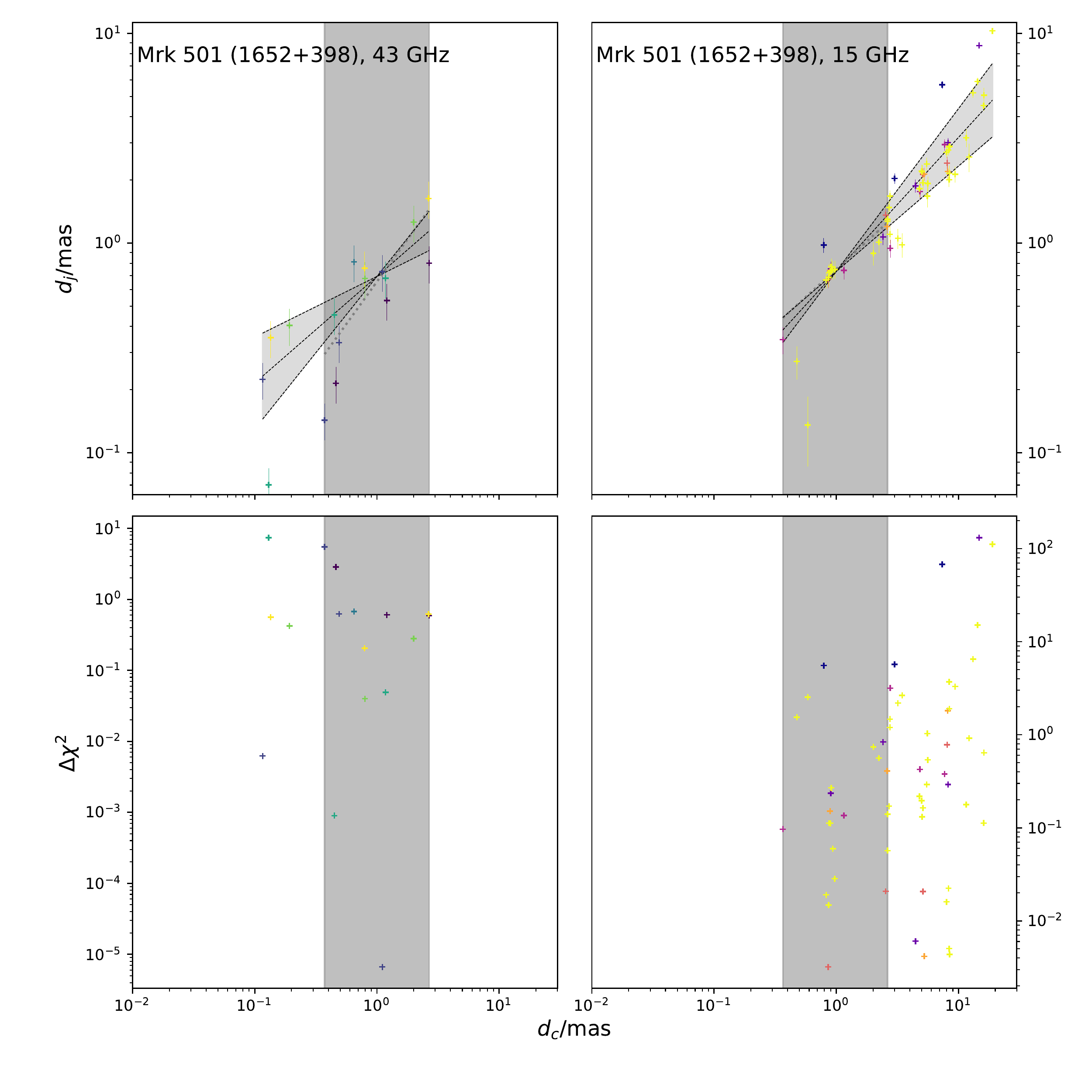}
		\includegraphics[width=0.65\hsize,clip]{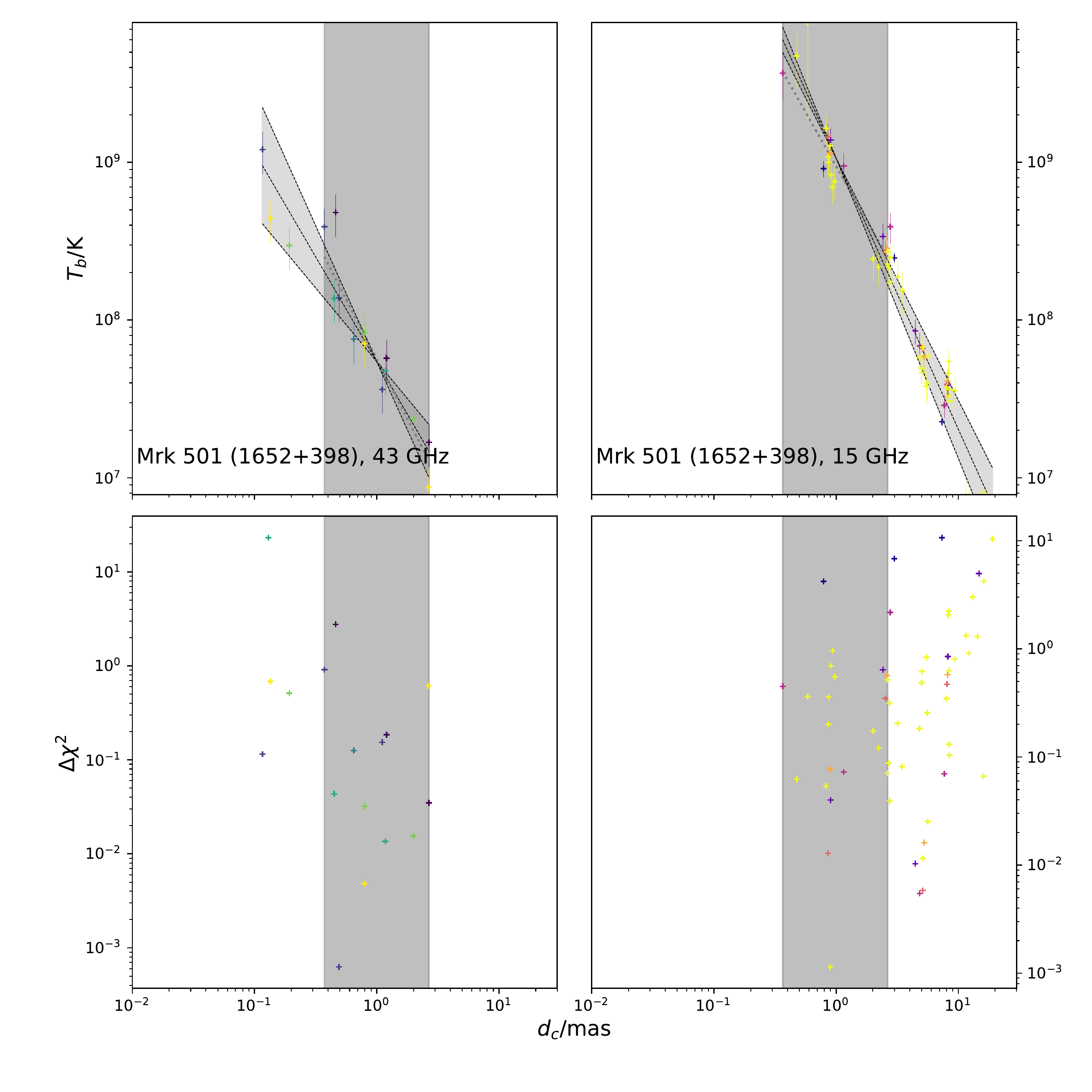}
      \caption{1652+398 jet diameter $d_{\rm j}$ as a function of the distance from the core $d_{\rm c}$ and the brightness temperature $T_{\rm B}$ as function of $d_{\rm c}$. The color coding for both frequencies indicates the epoch, where lighter colors indicate more recent epochs.}
       \label{1652+398}
\end{figure*}

\begin{figure*}
\centering
	 \includegraphics[width=0.65\hsize,clip]{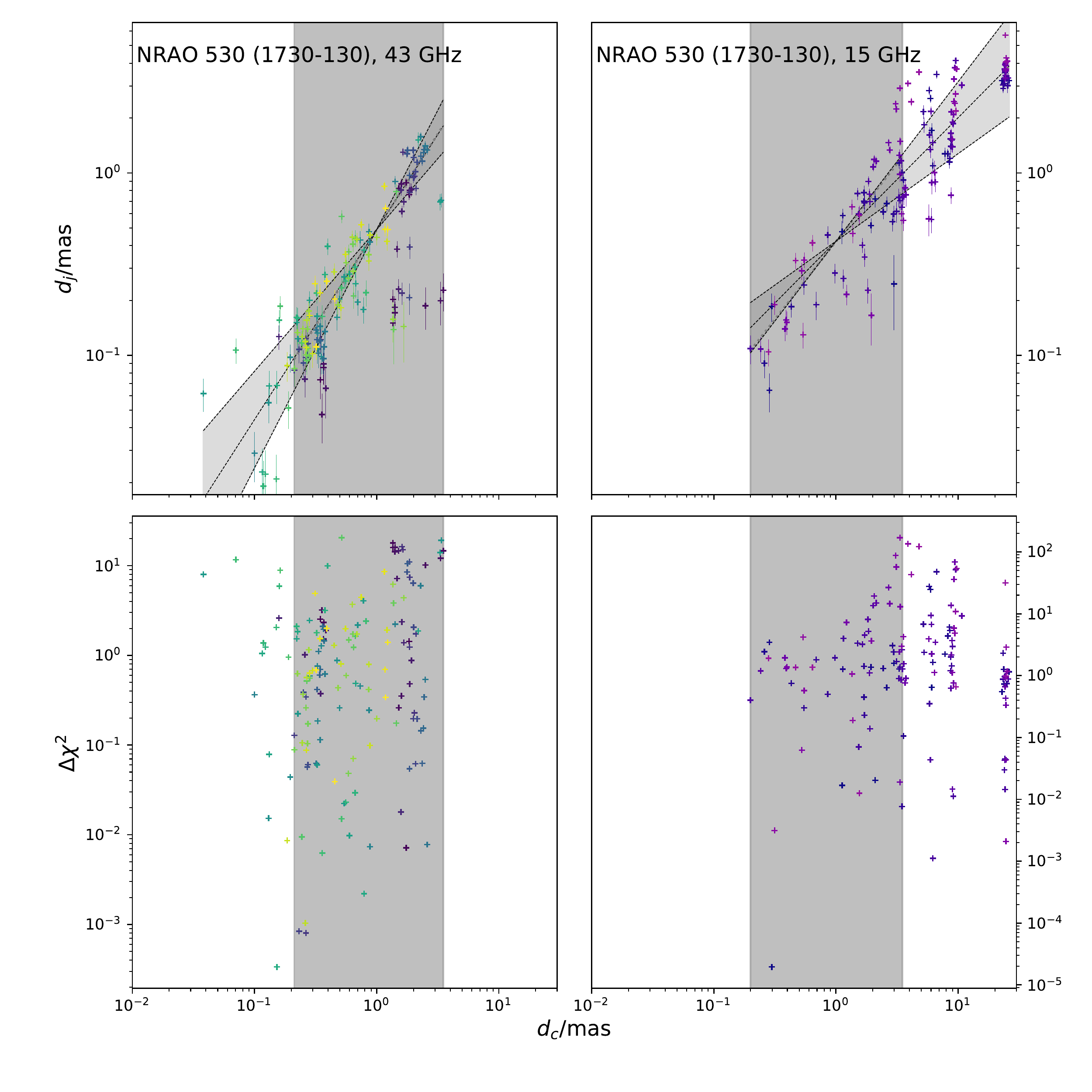}
		\includegraphics[width=0.65\hsize,clip]{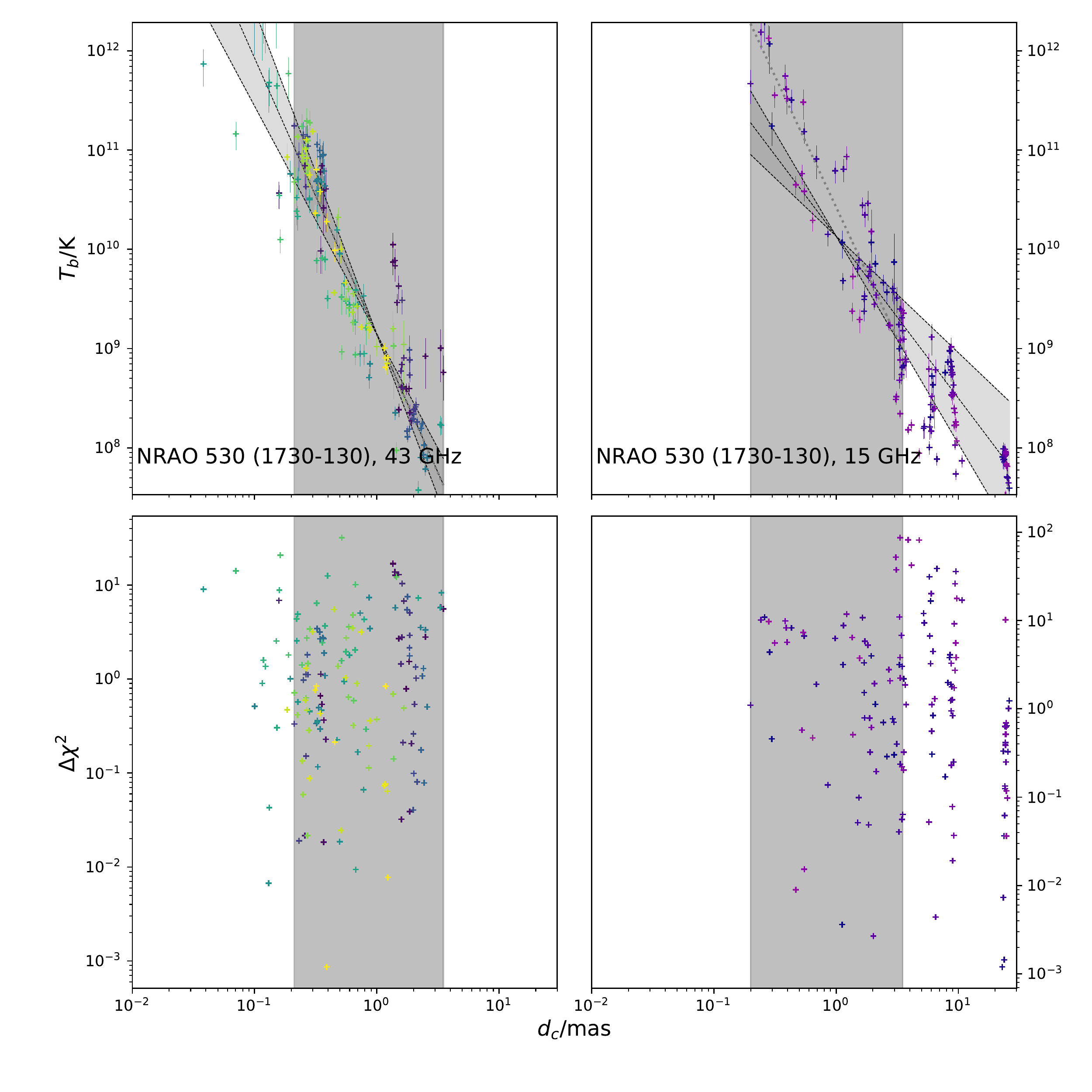}
      \caption{1730-130 jet diameter $d_{\rm j}$ as a function of the distance from the core $d_{\rm c}$ and the brightness temperature $T_{\rm B}$ as function of $d_{\rm c}$. The color coding for both frequencies indicates the epoch, where lighter colors indicate more recent epochs.}
       \label{1730-130}
\end{figure*}

\begin{figure*}
\centering
	 \includegraphics[width=0.65\hsize,clip]{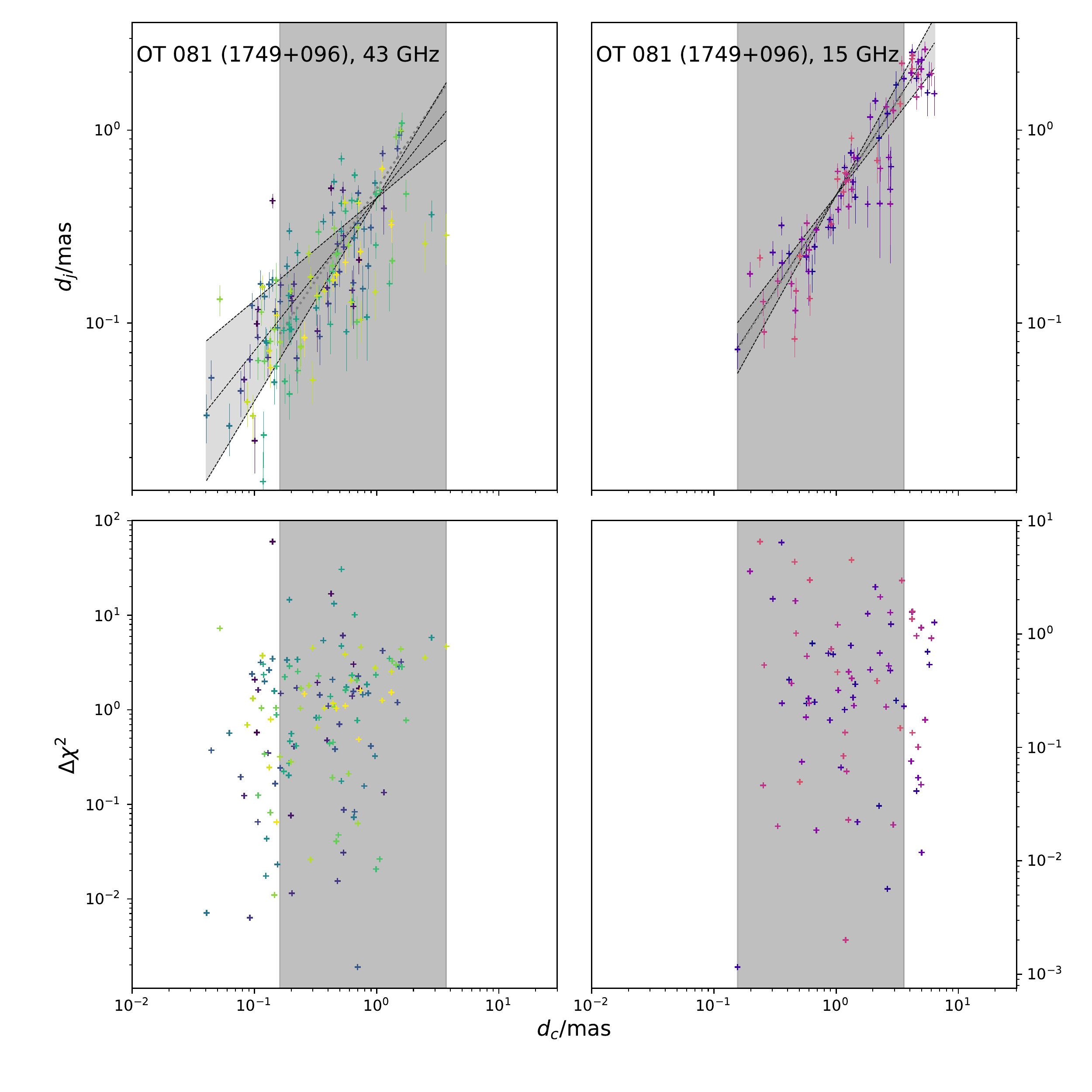}
		\includegraphics[width=0.65\hsize,clip]{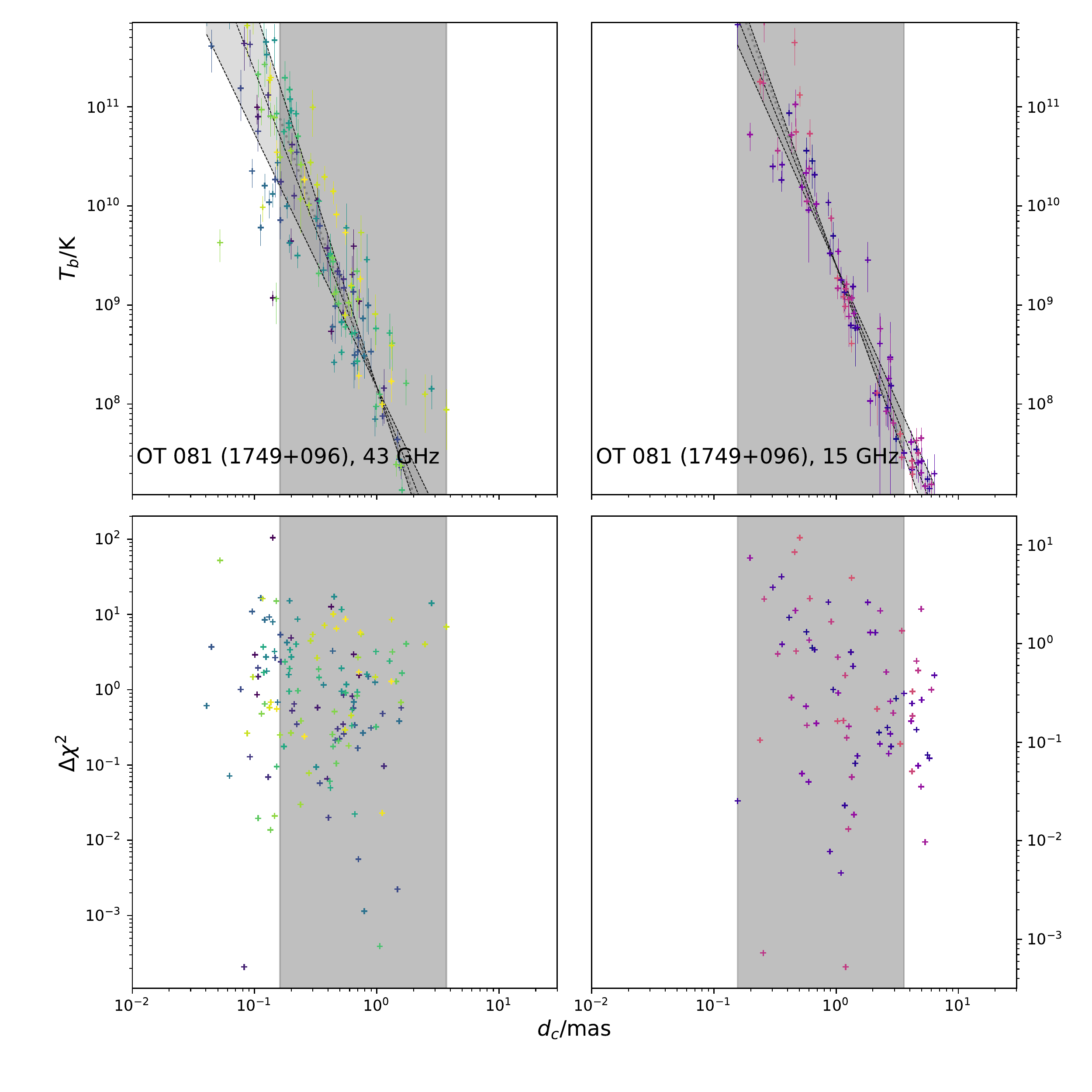}
      \caption{1749+096 jet diameter $d_{\rm j}$ as a function of the distance from the core $d_{\rm c}$ and the brightness temperature $T_{\rm B}$ as function of $d_{\rm c}$. The color coding for both frequencies indicates the epoch, where lighter colors indicate more recent epochs.}
       \label{1749+096}
\end{figure*}

\begin{figure*}
\centering
	 \includegraphics[width=0.65\hsize,clip]{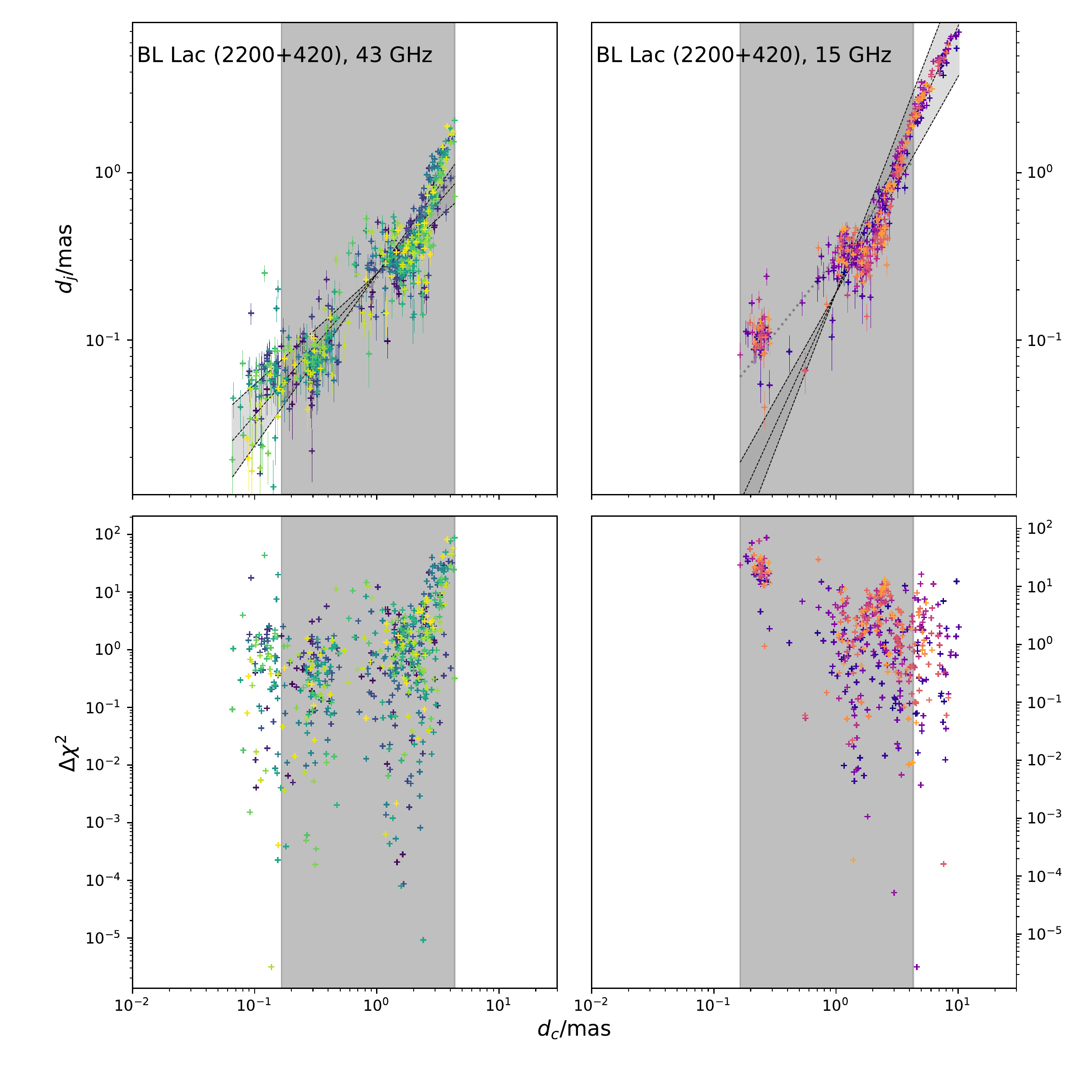}
		\includegraphics[width=0.65\hsize,clip]{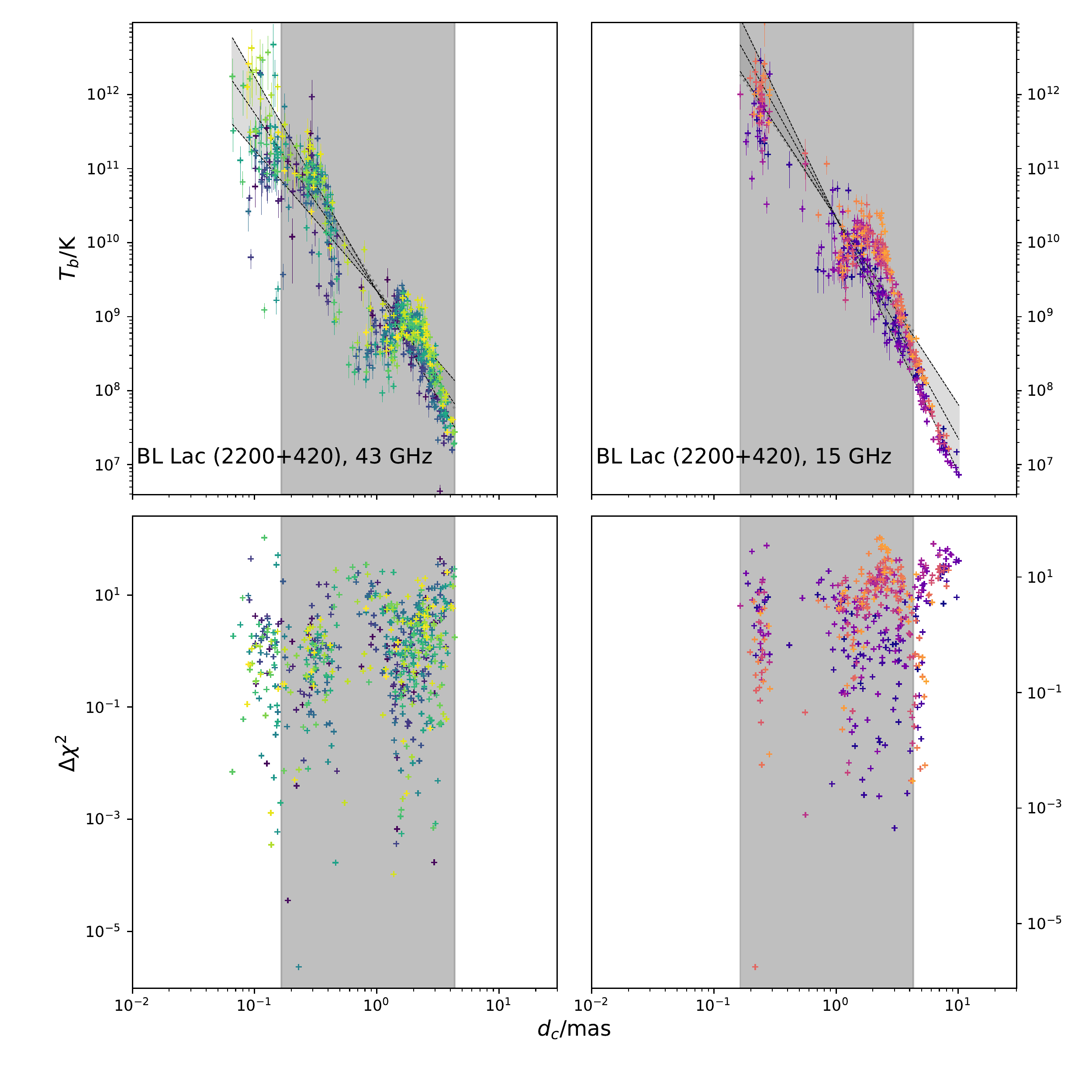}
      \caption{2200+420 jet diameter $d_{\rm j}$ as a function of the distance from the core $d_{\rm c}$ and the brightness temperature $T_{\rm B}$ as function of $d_{\rm c}$. The color coding for both frequencies indicates the epoch, where lighter colors indicate more recent epochs.}
       \label{2200+420}
\end{figure*}

\begin{figure*}
\centering
	 \includegraphics[width=0.65\hsize,clip]{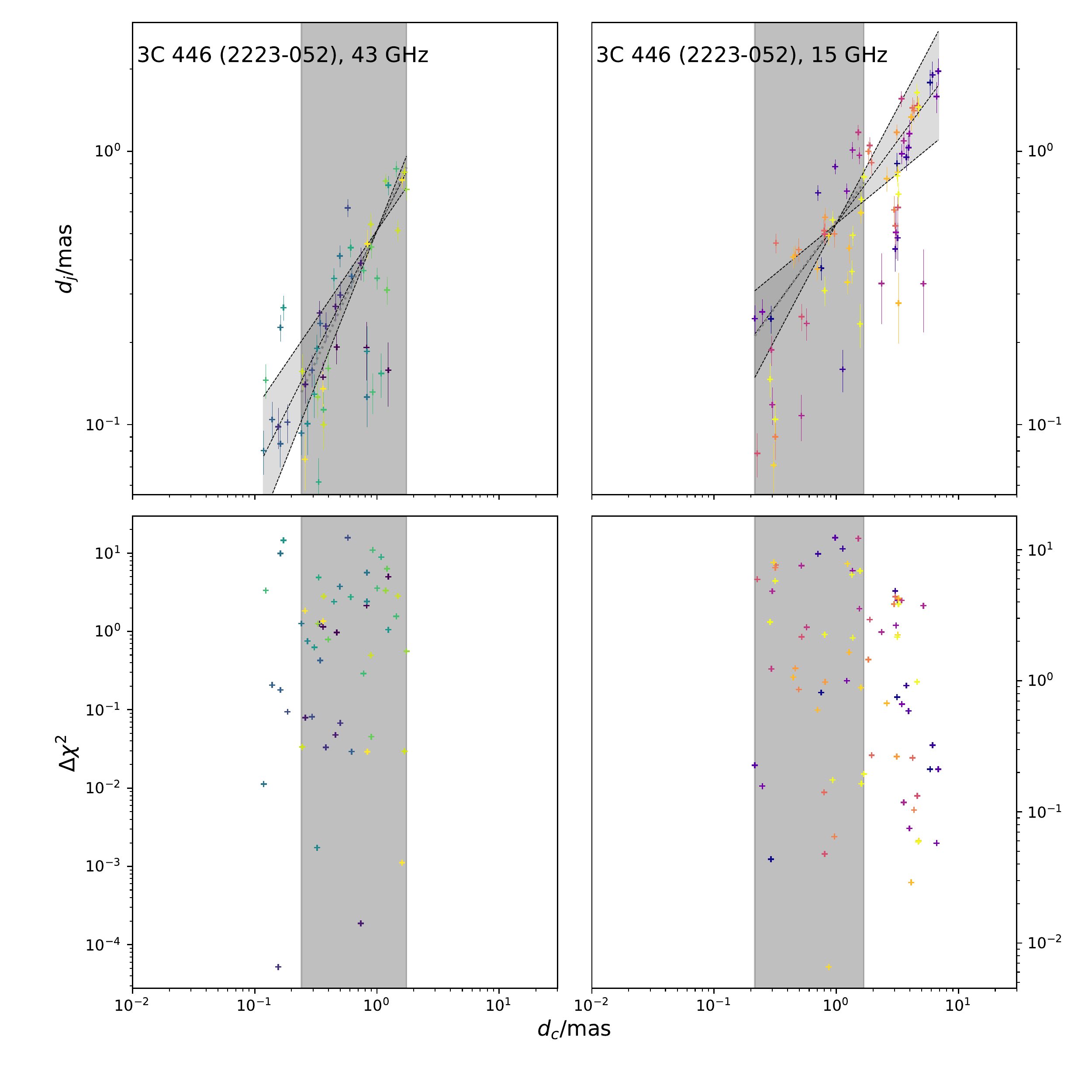}
		\includegraphics[width=0.65\hsize,clip]{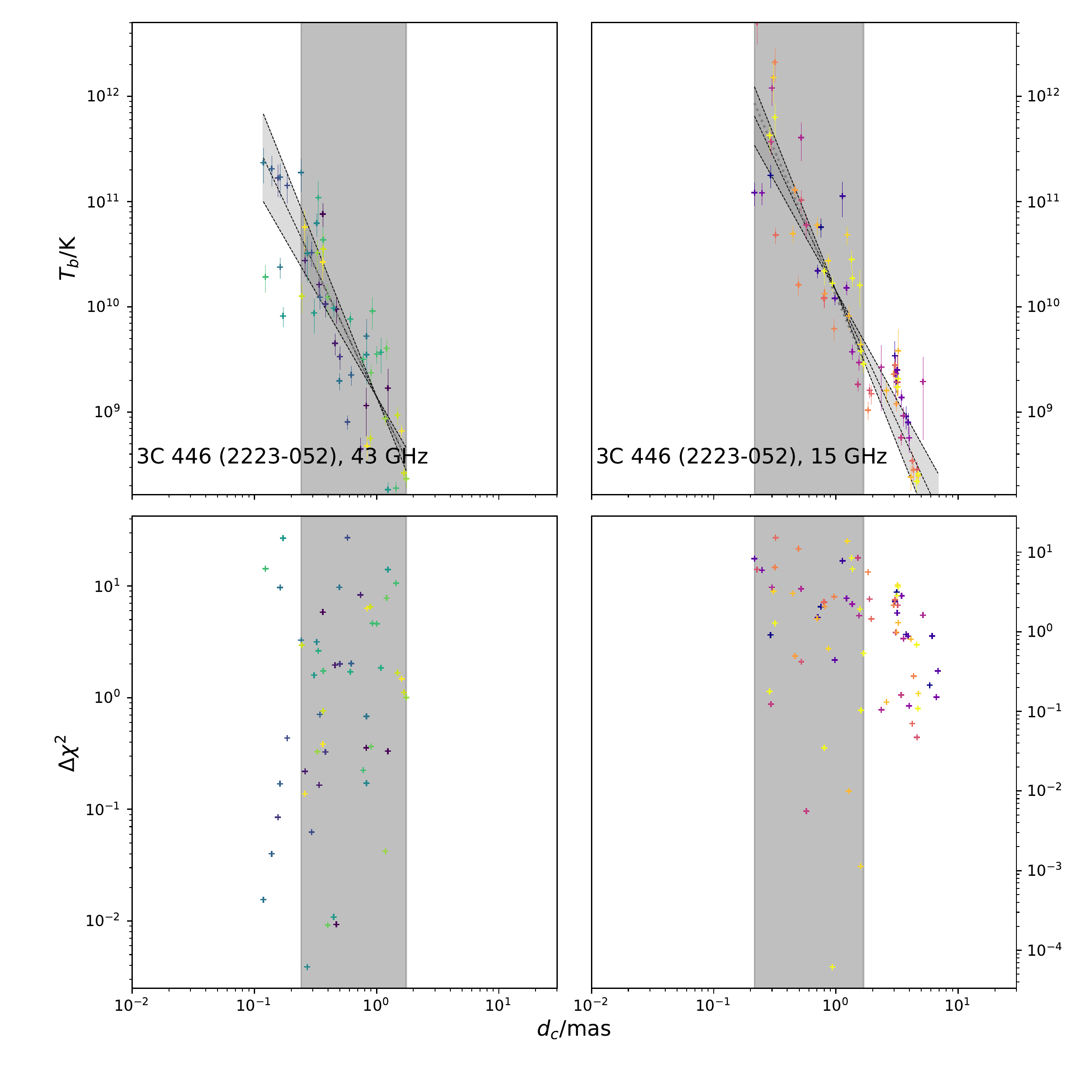}
      \caption{2223-052 jet diameter $d_{\rm j}$ as a function of the distance from the core $d_{\rm c}$ and the brightness temperature $T_{\rm B}$ as function of $d_{\rm c}$. The color coding for both frequencies indicates the epoch, where lighter colors indicate more recent epochs.}
       \label{2223-052}
\end{figure*}

\begin{figure*}
\centering
	 \includegraphics[width=0.65\hsize,clip]{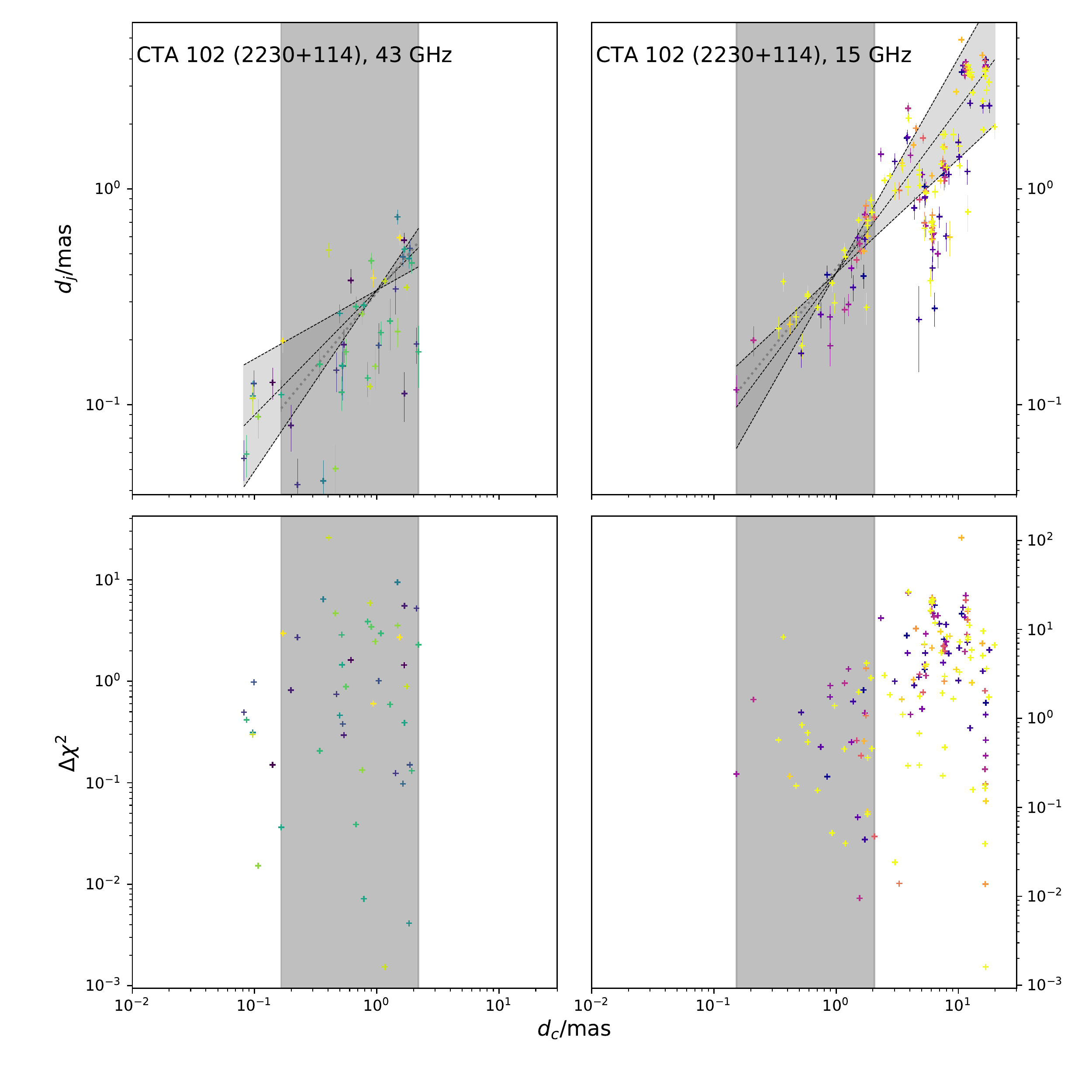}
		\includegraphics[width=0.65\hsize,clip]{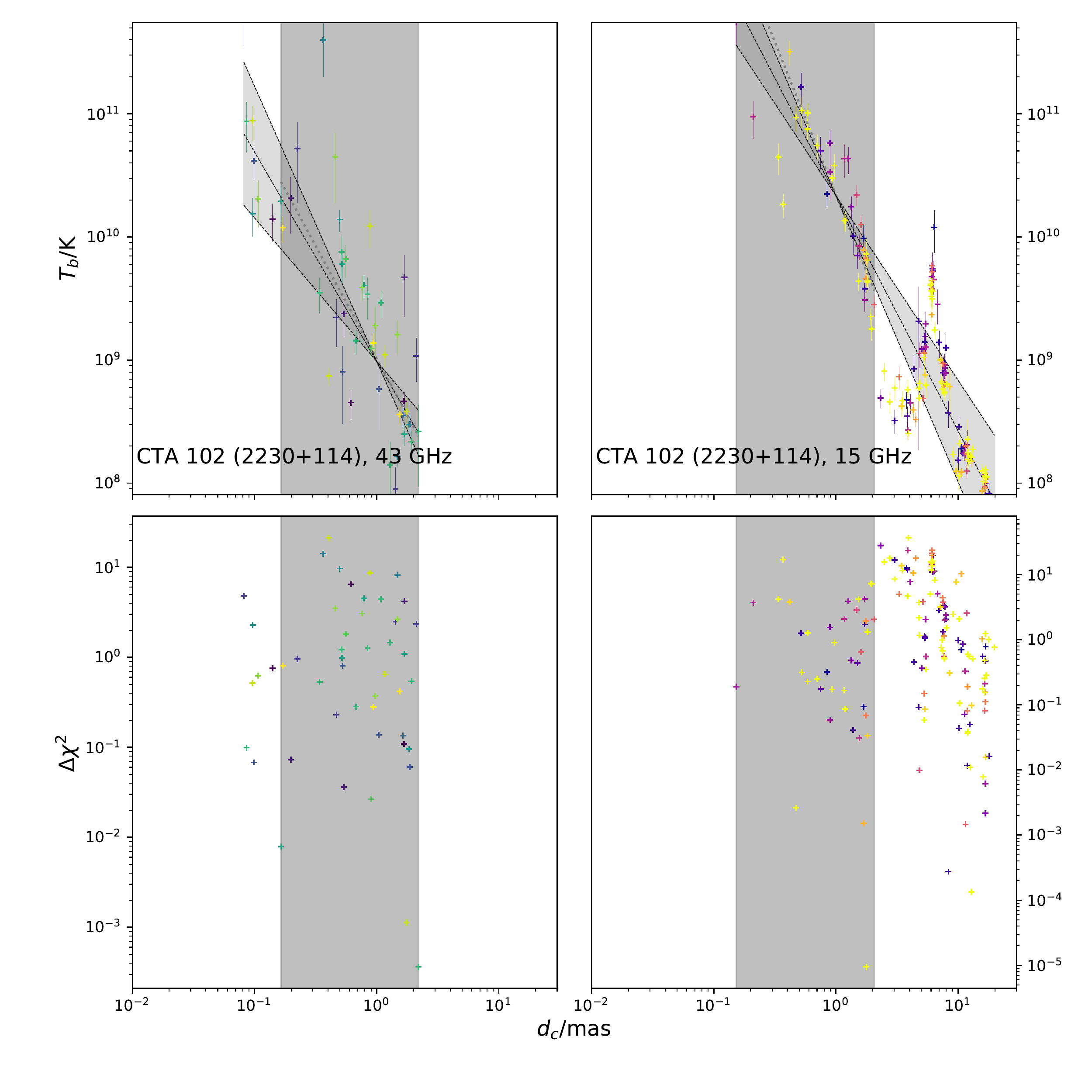}
      \caption{2230+114 jet diameter $d_{\rm j}$ as a function of the distance from the core $d_{\rm c}$ and the brightness temperature $T_{\rm B}$ as function of $d_{\rm c}$. The color coding for both frequencies indicates the epoch, where lighter colors indicate more recent epochs.}
       \label{2230+114}
\end{figure*}

\begin{figure*}
\centering
	 \includegraphics[width=0.65\hsize,clip]{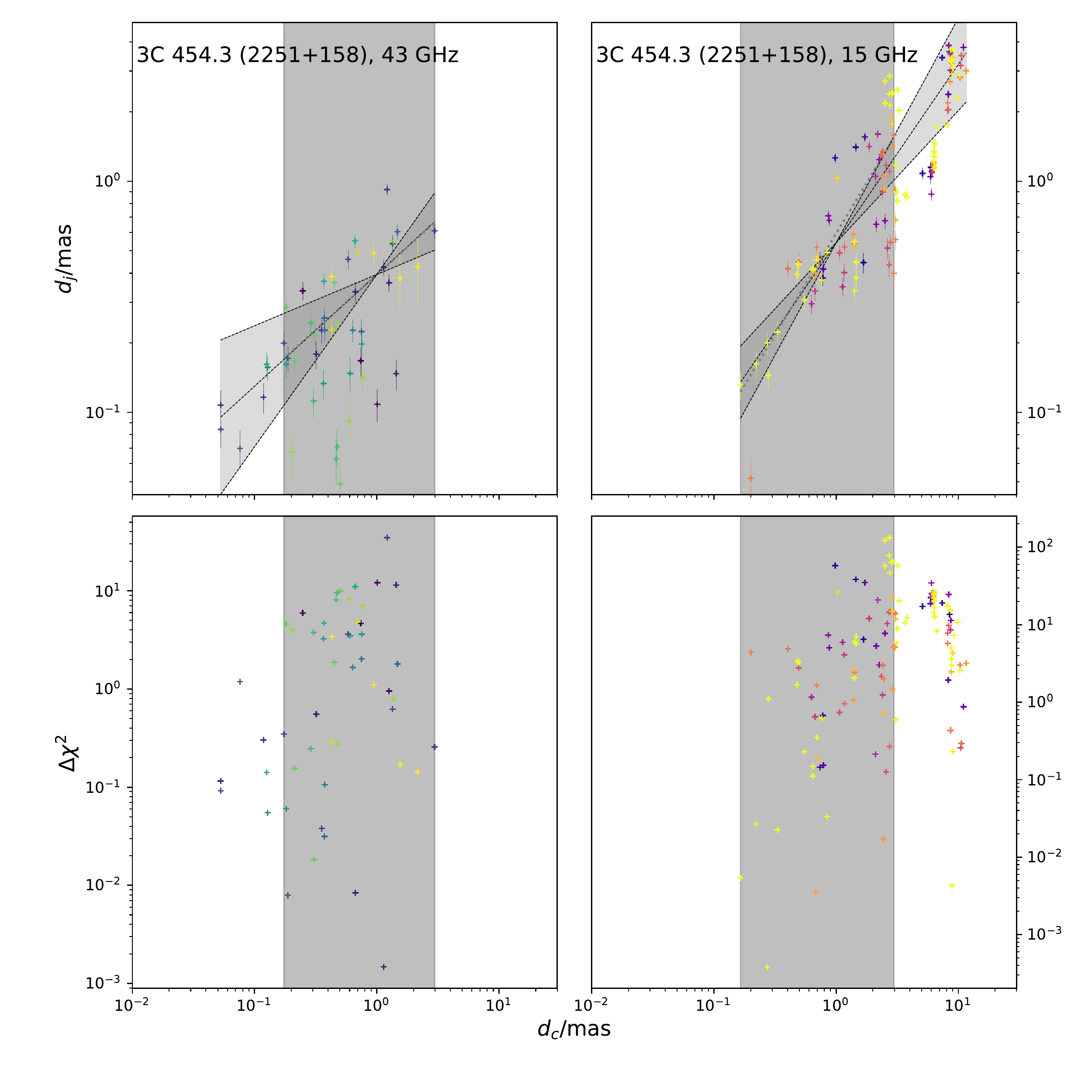}
		\includegraphics[width=0.65\hsize,clip]{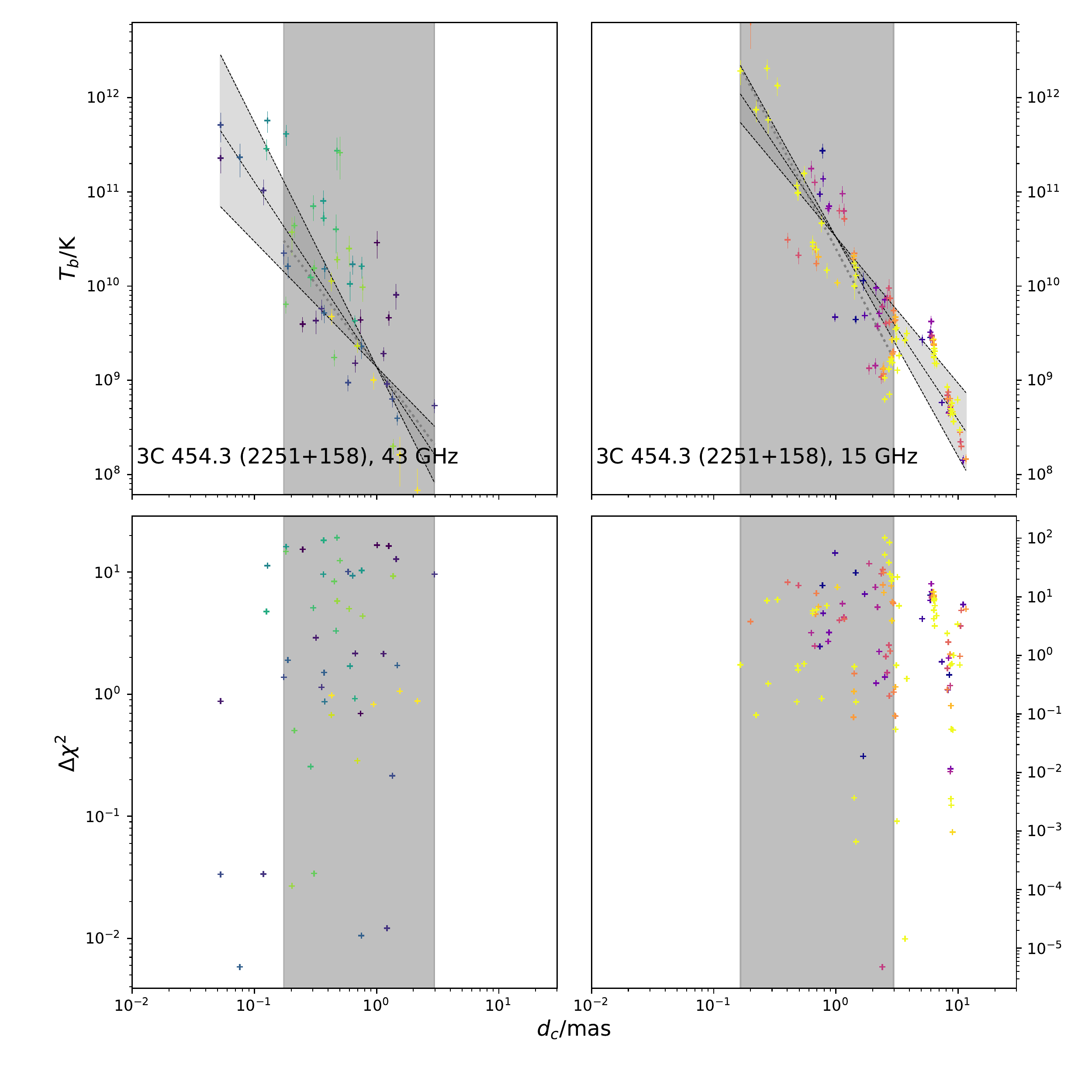}
      \caption{2251+158 jet diameter $d_{\rm j}$ as a function of the distance from the core $d_{\rm c}$ and the brightness temperature $T_{\rm B}$ as function of $d_{\rm c}$. The color coding for both frequencies indicates the epoch, where lighter colors indicate more recent epochs.}
       \label{2251+158}
\end{figure*}